\documentclass[aps,rmp,twocolumn,floatfix,longbibliography]{revtex4-2}
\pdfoutput=1
\usepackage{standalone, gensymb, tensor, amsmath,amsfonts,amssymb,amsthm,bbm,bm, braket}
\usepackage[T1]{fontenc}
\usepackage{hyperref}
\usepackage[inline]{enumitem}
\usepackage{scalerel, physics}
\usepackage{wasysym}
\usepackage{comment}
\usepackage{enumerate}
\usepackage{graphicx}
\usepackage{mathtools}
\usepackage{soul}
\usepackage{xparse}
\usepackage{dsfont}
\usepackage{ifthen}
\usepackage{tensor}
\usepackage{tikz}
\usepackage{pgfplots}
\usepackage{tikz-network}
\usepackage{simplewick} 
\usepackage{ mathrsfs }

\usetikzlibrary{patterns,decorations.pathreplacing}
\theoremstyle{break}        

\theoremstyle{break}

\usepackage{color}
\definecolor{OliveGreen}{RGB}{85,107,47}
\definecolor{NavyBlue}{RGB}{0,0,128}
\usepackage{tensor}
\usepackage{xifthen}
\usepackage{xargs}

\definecolor{blue1}{RGB}{0, 0, 255}
\definecolor{blue2}{RGB}{0, 150, 255}
\definecolor{blue3}{RGB}{0, 71, 171}
\definecolor{blue4}{RGB}{100, 149, 237}
\definecolor{blue5}{RGB}{93, 63, 211}
\definecolor{blue6}{RGB}{101,147,245}
\definecolor{blue7}{RGB}{176,223,229}
\definecolor{blue8}{RGB}{0,0,128}
\definecolor{blue9}{RGB}{0,108,255}
\definecolor{blue10}{RGB}{101,147,245}
\definecolor{blue11}{RGB}{115,194,251}
\definecolor{blue12}{RGB}{87,160,211}
\definecolor{blue13}{RGB}{137,207,240}
\definecolor{blue14}{RGB}{29,41,81}
\definecolor{blue15}{RGB}{14,77,146}
\definecolor{blue16}{RGB}{15,82,186}

\definecolor{red1}{RGB}{238, 75, 43}
\definecolor{red2}{RGB}{233, 116, 81}
\definecolor{red3}{RGB}{222, 49, 99}
\definecolor{red4}{RGB}{250, 160, 160}
\definecolor{red5}{RGB}{236, 88, 0}
\definecolor{red6}{RGB}{232,102,102}
\definecolor{red7}{RGB}{202,52,51}
\definecolor{red8}{RGB}{205,92,92}
\definecolor{red9}{RGB}{178,34,34}
\definecolor{red10}{RGB}{164,90,82}
\definecolor{red11}{RGB}{255,8,0}
\definecolor{red12}{RGB}{202,52,51}
\definecolor{red13}{RGB}{66,13,9}
\definecolor{red14}{RGB}{141,2,31}
\definecolor{red15}{RGB}{250,128,114}
\definecolor{red16}{RGB}{237,41,57}

\definecolor{yellow1}{RGB}{254,220,86}
\definecolor{yellow2}{RGB}{255,229,180}
\definecolor{yellow3}{RGB}{238,220,130}
\definecolor{yellow4}{RGB}{253,165,15}
\definecolor{yellow5}{RGB}{255,195,11}
\definecolor{yellow6}{RGB}{218,165,32}
\definecolor{yellow7}{RGB}{255,211,0}
\definecolor{yellow8}{RGB}{248,222,126}
\definecolor{yellow9}{RGB}{245,245,220}
\definecolor{yellow10}{RGB}{248,228,115}

\definecolor{grey1}{RGB}{98,98,98}
\definecolor{grey2}{RGB}{211,211,211}
\definecolor{grey3}{RGB}{192,192,192}
\definecolor{grey4}{RGB}{169,169,169}
\definecolor{grey5}{RGB}{246,246,246}
\definecolor{grey6}{RGB}{32,32,32}
\definecolor{grey7}{RGB}{64,64,64}
\definecolor{grey8}{RGB}{96,96,96}
\definecolor{grey9}{RGB}{128,128,128}
\definecolor{grey10}{RGB}{160,160,160}
\definecolor{grey11}{RGB}{224,224,224}
\definecolor{grey12}{RGB}{180,180,180}

\definecolor{green1}{RGB}{80, 180, 152}

\definecolor{orange}{RGB}{255, 116, 23}
\definecolor{orange2}{RGB}{244, 174, 114}

\newcommandx{\fineq}[5][1=-.8ex,2=1,3=1,5=0]{
	\begin{tikzpicture}[baseline={([yshift=#1]current  bounding  box.center)}, scale = #2, every node/.style={scale = #3},rotate around={#5:(0,0)},every node/.style={transform shape}]
		#4
	\end{tikzpicture}
}

\newcommandx{\tikzdiagup}{
	\tikz {\draw[thick] (0,0)--(0.15,0.15); \draw (0,0) rectangle (0.15,0.15);}
}

\newcommandx{\gatecross}[1][1=0.5]{
	\pgfmathparse{#1/2.0}
	\let\x\pgfmathresult
	\draw[thick] (-\x,-\x) -- (\x,\x);
	\draw[thick] (\x,-\x) -- (-\x,\x);
}
\newcommandx{\gatesqu}[2][1=0.25,2=]{
	\pgfmathparse{#1/2.0}
	\let\x\pgfmathresult
	\ifthenelse{\equal{#2}{}}{
		\draw[thick, fill=white, rounded corners=2pt] (-\x,\x) rectangle (\x,-\x);
	}{
		\draw[thick, fill=#2, rounded corners=2pt] (-\x,\x) rectangle (\x,-\x);
	}
}

\newcommandx{\gatemark}[2][1=0.075,2=tr]{
	\pgfmathparse{#1}
	\let\l\pgfmathresult
	\ifthenelse{\equal{#2}{topleft}}{
		\draw[thick] (0,\l) -- ++(-\l,0) --++ (0,-\l);
	}{}
	\ifthenelse{\equal{#2}{topright}}{
		\draw[thick] (0,\l) -- ++(\l,0) --++ (0,-\l);
	}{}
	
	\ifthenelse{\equal{#2}{bottomleft}}{
		\draw[thick] (0,-\l) -- ++(-\l,0) --++ (0,\l);
	}{}
	\ifthenelse{\equal{#2}{bottomright}}{
		\draw[thick] (0,-\l) -- ++(\l,0) --++ (0,\l);
	}{}
	\ifthenelse{\equal{#2}{right}}{
		\draw[thick] (-\l/2,\l/2) -- ++(\l,0) --++ (0,-\l);
	}{}
	\ifthenelse{\equal{#2}{left}}{
		\draw[thick] (\l/2,\l/2) -- ++(-\l,0) --++ (0,-\l);
	}{}
	\ifthenelse{\equal{#2}{rightf}}{
		\draw[thick] (-\l/2,-\l/2) -- ++(\l,0) --++ (0,\l);
	}{}
	\ifthenelse{\equal{#2}{leftf}}{
		\draw[thick] (\l/2,-\l/2) -- ++(-\l,0) --++ (0,\l);
	}{}
	
}

\newcommandx{\squaregate}[3][1=0,2=0,3=white]
{
	\begin{scope}[shift={(#1,#2)},rounded corners= 2pt]
		\draw[thick,fill=#3] (-.13,-.13) rectangle (.15,.15);
	\end{scope}
}

\newcommandx{\roundgate}[6][1=0,2=0,3=1,4=topright,5=white,6=-1]{
	\pgfmathparse{#3}
	\let\l\pgfmathresult
	\begin{scope}[shift={(#1,#2)}]
		\gatecross[\l]
			\pgfmathparse{\l/2.0}
		\let\s\pgfmathresult
		\gatesqu[\s][#5]
		\pgfmathparse{\l*0.15}
		\let\m\pgfmathresult
	\ifthenelse{\equal{#6}{-1}}{		\gatemark[\m][#4]
	}{	\node at ({0},{0}) {\scalebox{1}{$#6$}};}
\end{scope}
}

\newcommandx{\wcirc}[2]{\begin{scope}
		\draw[fill=white] (#1,#2) circle (0.15);	\end{scope}} 
\newcommandx{\wcircc}[2]{\begin{scope}
		\draw[fill=white] (#1,#2) circle (0.13);	\end{scope}} 
\newcommandx{\wsqr}[2]{\begin{scope}
		\draw[fill=white,shift={(#1,#2)}] (-.13,.13) rectangle (.13,-.13);	\end{scope}} 
\newcommandx{\wsqrr}[2]{\begin{scope}
		\draw[fill=white,shift={(#1,#2)}] (-.11,.11) rectangle (.11,-.11);	\end{scope}}
\newcommandx{\bcirc}[2]{\begin{scope}
		\draw[fill=black] (#1,#2) circle (0.15);	\end{scope}} 
\newcommandx{\thetastate}[4][1=0,2=0,3=1,4=]{
	\pgfmathparse{#3/2}
	\let\l\pgfmathresult
	\pgfmathparse{\l*0.15}
	\let\m\pgfmathresult
	\begin{scope}[shift={(#1,#2)}]
		\draw[thick] (0,0)--(\l,\l);
		\draw[thick] (0,0)--(-\l,\l);
		\ifthenelse{\equal{#4}{}}{
			\draw[fill=white] (0,0) circle (0.15);
		}{
			\draw[thick, fill=#4] (0,0) circle (0.15);
		}
	\end{scope}
}

\newcommandx{\thetastateflipped}[4][1=0,2=0,3=1,4=]{
	\pgfmathparse{#3/2}
	\let\l\pgfmathresult
	\pgfmathparse{\l*0.15}
	\let\m\pgfmathresult
	\begin{scope}[shift={(#1,#2)}]
		\draw[thick] (0,0)--(\l,-\l);
		\draw[thick] (0,0)--(-\l,-\l);
		\ifthenelse{\equal{#4}{}}{
			\draw[fill=white] (0,0) circle (0.15);
		}{
			\draw[thick, fill=#4] (0,0) circle (0.15);
		}
	\end{scope}
}

\newcommandx{\vertgate}[5][1=0,2=0,3=4,4=orange,5=topright]
{
	\begin{scope}[shift={(#1,#2)}]
		\ifthenelse{\equal{#3}{1}}{
			\roundgate[0][0][1][#5][#4]
		}{
			\foreach \n[evaluate=\n as \y using {2*\n-2}] in {1,...,#3}{
				\roundgate[0][\y][1][#5][#4]
			}
		}
	\end{scope}
}

\newcommandx{\circgate}[4][1=0,2=0,3=white,4]
{

         \let\m\pgfmathresult
	\begin{scope}[shift={(#1,#2)}]
	\draw[thick,fill=#3] (0,0) circle (.125);
	\gatemark[.125*\m][#4]
	\end{scope}
	
}

\newcommandx{\tsfmatV}[8][1=0,2=0,3=l,4=4,5=tr,6=init,7=orange,8=topright]{
	\begin{scope}[shift={(#1,#2)}]
		\ifthenelse{\equal{#3}{l}}{
			\pgfmathsetmacro{\flag}{0}
		}{
			\pgfmathsetmacro{\flag}{1}
		}
		
		\foreach \y[evaluate=\y as \x using {mod(\y+\flag,2)}] in {1,...,#4}{
			\roundgate[\x][\y][1][#8][#7]
		}
		\ifthenelse{\equal{#5}{tr}}{
			\foreach \y[evaluate=\y as \x using {mod(\y+\flag,2)}] in {#4}{
				\draw [fill=white] (\x-0.5,\y+0.5) circle (0.15);
				\draw [fill=white] (\x+0.5,\y+0.5) circle (0.15);
			}
		}{}
		\ifthenelse{\equal{#6}{init}}{
			\thetastate[\flag][0][1][#7]
		}{}
	\end{scope}
}

\newcommandx{\leftriangle}[5][1=0,2=0,3=4,4=orange,5=topright]{
	\begin{scope}[shift={(#1,#2)}]
		\pgfmathsetmacro{\t}{#3}
		\pgfmathsetmacro{\steps}{ceil(\t/2)}
		\foreach \i[evaluate=\i as \x using -\t+2*\i-1, evaluate=\i as \ylim using \t-2*\i+2] in {1,...,\steps}{
			\foreach \y[evaluate=\y as \thisx using {\x+\y-1}] in {1,...,\ylim}{
				\roundgate[\thisx][\y][1][#5][#4]
			}
		}
	\end{scope}
}

\newcommandx{\rightriangle}[5][1=0,2=0,3=4,4=orange,5=topright]{
	\begin{scope}[shift={(#1,#2)}]
		\pgfmathsetmacro{\t}{#3}
		\pgfmathsetmacro{\steps}{ceil(\t/2)}
		\foreach \i[evaluate=\i as \x using -\t+2*\i-1, evaluate=\i as \ylim using \t-2*\i+2] in {1,...,\steps}{
			\foreach \y[evaluate=\y as \thisx using {-\x-\y+1}] in {1,...,\ylim}{
				\roundgate[\thisx][\y][1][#5][#4]
			}
		}
	\end{scope}
}

\newcommandx{\eigenVL}[8][1=0,2=0,3=l,4=5,5=tr,6=init,7=orange,8=topright]{
	\begin{scope}[shift={(#1,#2)}]
		\pgfmathsetmacro{\t}{#4}
		\leftriangle[0][0][\t][#7][#8]
		
		\ifthenelse{\equal{#6}{init}}{
			\drawinitstate[0][0][l][\t][#7]
		}{}
		
		\ifthenelse{\equal{#5}{tr}}{
			\draw[fill=white] \foreach \x in {0,...,\t} {(\x-0.5-\t,0.5+\x) circle (0.15)};
			\ifthenelse{\equal{#3}{r}}{
				\draw[fill=white] (0.5,\t+0.5) circle (0.15);
			}{}
		}{}
		\ifthenelse{\equal{#5}{parttr}}{
			\draw[fill=white] \foreach \x in {0,...,\t} {(\x-0.5-\t,0.5+\x) circle (0.15)};
		}{}
	\end{scope}
}

\newcommandx{\eigenVR}[8][1=0,2=0,3=l,4=5,5=tr,6=init,7=orange,8=topright]{
	\begin{scope}[shift={(#1,#2)}]
		\pgfmathsetmacro{\t}{#4}
		\rightriangle[0][0][\t][#7][#8]
		
		\ifthenelse{\equal{#6}{init}}{
			\drawinitstate[0][0][r][\t][#7]
		}{}
		
		\ifthenelse{\equal{#5}{tr}}{
			\draw[fill=white] \foreach \x in {0,...,\t}{(-\x+0.5+\t,0.5+\x) circle (0.15)};
			\ifthenelse{\equal{#3}{l}}{
				\draw[fill=white] (-0.5,\t+0.5) circle (0.15);
			}{}
		}{}
	\end{scope}
}

\newcommandx{\tra}[2][1]{\underset{#1}{\text{tr}}\left[#2\right]}

\newcommandx{\tsfmatDgate}[7][1=0,2=0,3=l,4=4,5=tr,6=orange,7=topright]
{
	\begin{scope}[shift={(#1,#2)}]
		\ifthenelse{\equal{#3}{l}}{
			\pgfmathsetmacro{\flag}{-1}
		}{
			\pgfmathsetmacro{\flag}{1}
		}
		\pgfmathsetmacro{\t}{#4}
		\foreach \i[evaluate=\i as \x using {\flag*\i}, evaluate=\i as \y using \i] in {1,...,\t}{
			\roundgate[\x][\y][1][#7][#6]
		}
		
		\ifthenelse{\equal{#5}{tr}}{
			\foreach \i[evaluate=\i as \x using {\flag*\i}, evaluate=\i as \y using \i] in {\t}{
				\draw [fill=white] (\x-0.5,\y+0.5) circle (0.15);
				\draw [fill=white] (\x+0.5,\y+0.5) circle (0.15);
			}  
		}{}
	\end{scope}
	
}

\newcommandx{\tsfmatD}[8][1=0,2=0,3=l,4=4,5=tr,6=init,7=orange,8=topright]{
	\begin{scope}[shift={(#1,#2)}]
		\ifthenelse{\equal{#6}{init}}{
			\thetastate[0][0][1][#7]
		}{}
		
		\ifthenelse{\equal{#3}{l}}{
			\pgfmathsetmacro{\flag}{-1}
		}{
			\pgfmathsetmacro{\flag}{1}
		}
		
		\pgfmathsetmacro{\t}{#4}
		\foreach \i[evaluate=\i as \x using {\flag*\i}, evaluate=\i as \y using \i] in {1,...,\t}{
			\roundgate[\x][\y][1][#8][#7]
		}
		
		\ifthenelse{\equal{#5}{tr}}{
			\foreach \i[evaluate=\i as \x using {\flag*\i}, evaluate=\i as \y using \i] in {\t}{
				\draw [fill=white] (\x-0.5,\y+0.5) circle (0.15);
				\draw [fill=white] (\x+0.5,\y+0.5) circle (0.15);
			}  
		}
		\ifthenelse{\equal{#5}{parttr}}{
			\foreach \i[evaluate=\i as \x using {\flag*\i}, evaluate=\i as \y using \i] in {\t}{
				\draw [fill=white] (\x+0.5*\flag,\y+0.5) circle (0.15);
			}  
		}
		{}
	\end{scope}
}

\newcommandx{\drawinitstate}[5][1=0,2=0,3=l,4=4,5=orange]{
	\pgfmathsetmacro{\t}{#4}
	\begin{scope}[shift={(#1,#2)}]
		\pgfmathsetmacro{\steps}{ceil((\t-1)/2)}
		\ifthenelse{\equal{#3}{l}}{
			\foreach \i[evaluate=\i as \x using -\t+2*\i] in {0,...,\steps}{
				\thetastate[\x][0][1][#5]
			}
		}{
			\foreach \i[evaluate=\i as \x using -\t+2*\i] in {0,...,\steps}{      
				\thetastate[-\x][0][1][#5]
			}
		}
	\end{scope}
}

\newcommandx{\drawinitstateflipped}[5][1=0,2=0,3=l,4=4,5=orange]{
	\pgfmathsetmacro{\t}{#4}
	\begin{scope}[shift={(#1,#2)}]
		\pgfmathsetmacro{\steps}{ceil((\t-1)/2)}
		\ifthenelse{\equal{#3}{l}}{
			\foreach \i[evaluate=\i as \x using -\t+2*\i] in {0,...,\steps}{
				\thetastateflipped[\x][0][1][#5]
			}
		}{
			\foreach \i[evaluate=\i as \x using -\t+2*\i] in {0,...,\steps}{      
				\thetastateflipped[-\x][0][1][#5]
			}
		}
	\end{scope}
}

\newcommandx{\eigenDL}[6][1=0,2=0,3=l,4=4,5=orange,6=topright]{
	\begin{scope}[shift={(#1,#2)}]
		\pgfmathsetmacro{\t}{#4}
		\ifthenelse{\equal{#3}{l}}{
			\eigenVL[0][0][l][\t][tr][init][#5][#6]
			\pgfmathsetmacro{\t}{#4-1}
			\rightriangle[1][0][\t][#5][#6]
			\drawinitstate[1][0][r][\t][#5]
		}{
			\begin{scope}[shift={(-0.5,0.5)}]
				\foreach \i[evaluate=\i as \x using \i, evaluate=\i as \y using \i] in {0,...,\t}{      
					\draw (\x,\y)--++(0.5,0);
					\draw[fill=white] (\x,\y) circle (0.15);
				}
			\end{scope}
		}
	\end{scope}
}

\newcommandx{\eigenDR}[6][1=0,2=0,3=l,4=4,5=orange,6=topright]{
	\begin{scope}[shift={(#1,#2)}]
		\pgfmathsetmacro{\t}{#4}
		\ifthenelse{\equal{#3}{r}}{
			\eigenVR[0][0][r][\t][tr][init][#5][#6]
			\pgfmathsetmacro{\t}{#4-1}
			\leftriangle[-1][0][\t][#5][#6]
			\drawinitstate[-1][0][l][\t][#5]
		}{
			\begin{scope}[shift={(0.5,0.5)}]
				\foreach \i[evaluate=\i as \x using \i, evaluate=\i as \y using \t-\i] in {0,...,\t}{      
					\draw (\x,\y)--++(0.5,0);
					\draw[fill=white] (\x+0.5,\y) circle (0.15);
				}
			\end{scope}
		}
	\end{scope}
}

\newcommandx{\idonpurity}[2][1=0,2=0]
{
	\begin{scope}[shift={(#1,#2)}]
		\draw[thick] (-0.5,0)--++(-0.1,0.1)--++(0,0.2)--++(0.1,-0.1);
		\draw[thick] (-0.5,0.4)--++(-0.1,0.1)--++(0,0.2)--++(0.1,-0.1);
		\draw[thick] (0.5,0)--++(0.1,0.1)--++(0,0.2)--++(-0.1,-0.1);
		\draw[thick] (0.5,0.4)--++(0.1,0.1)--++(0,0.2)--++(-0.1,-0.1);
	\end{scope}
}

\newcommandx{\swaponpurity}[2][1=0,2=0]
{
	\begin{scope}[shift={(#1,#2)}]
		\draw[thick] (-0.5,0)--++(-0.2,0.2)--++(0,0.6)--++(0.2,-0.2);
		\draw[thick] (-0.5,0.2)--++(-0.075,0.075)--++(0,0.2)--++(0.075,-0.075);
		\draw[thick] (+0.5,0)--++(+0.2,0.2)--++(0,0.6)--++(-0.2,-0.2);
		\draw[thick] (+0.5,0.2)--++(+0.075,0.075)--++(0,0.2)--++(-0.075,-0.075);
	\end{scope}
}

\newcommandx{\hook}[4][1=0,2=0,3=t,4=l]{
	\begin{scope}[shift={(#1,#2)}]
		\ifthenelse{\equal{#3}{t}}{
			\ifthenelse{\equal{#4}{l}}{\draw[thick] (0.5,-0.5) arc (45:-90:0.15);}{\draw[thick] (0.5,-0.5) arc (45:270:0.15);}
		}{\ifthenelse{\equal{#4}{l}}{\draw[ thick] (0.5,-0.5) arc (-45:90:0.15);}{\draw[ thick] (0.5,-0.5) arc (315:90:0.15);}
		}
	\end{scope}
}

\newcommandx{\hhook}[4][1=0,2=0,3=t,4=l]{
	\begin{scope}[shift={(#1,#2)}]
		\ifthenelse{\equal{#3}{t}}{
			\ifthenelse{\equal{#4}{l}}{\draw[thick] (0.5,-0.5) arc (-45:175:0.15);}{\draw[thick] (0.5,-0.5) arc (225:0:0.15);}
		}{\ifthenelse{\equal{#4}{l}}{\draw[ thick] (0.5,-0.5) arc (-45:180:-0.15);}{\draw[ thick] (0.5,-0.5) arc (45:-180:0.15);}
		}
	\end{scope}
}

\newcommandx{\Pproj}[3][3=$P_\Lambda$]{
\begin{scope}[shift={(#1-.5,#2-1)}]
\draw[thick,fill=white] (0,0)rectangle (1,2);
\draw[thick] (0,1.5)--(-.5,1.5);
\draw[thick] (1,1.5)--(1.5,1.5);
\draw[thick] (0,.5)--(-.5,.5);
\draw[thick] (1,.5)--(1.5,.5);
\node[scale=2] at (.5,1) {#3};
\end{scope}}

\definecolor{FcolU}{rgb}{0.71,0.78,0.91}
\definecolor{colLines}{rgb}{0.31,0.31,0.31}
\definecolor{colVMPSLines}{rgb}{0.11,0.11,0.11}
\definecolor{IcolUc}{rgb}{0.71,0.41,0.42}
\definecolor{IcolU}{rgb}{0.71,0.8,0.76}
\definecolor{IcolVMPSc}{rgb}{0.73,0.69,0.7}
\definecolor{IcolVMPS}{rgb}{0.81,0.77,0.78}
\definecolor{colObs}{rgb}{1.,1.,1.}

\def\dx{0.3}
\def\r{0.08}

\newcommand\TECsheet[2]{
	\draw [thick,colLines,fill=red6,rounded corners=0.5] ({(#2)*\dx},{-(#1)*\dx}) rectangle ({((#2)+8)*\dx},{(-(#1)+6.75)*\dx});
	\draw [thick,colVMPSLines,fill=IcolVMPSc,rounded corners=0.5] ({((#2)-0.01)*\dx},{(-(#1)-0.01)*\dx}) rectangle ({((#2)+8+0.01)*\dx},{(-(#1)+0.31)*\dx});
}

\newcommand\TEsheetrotated[2]{
	\draw [thick,colLines,fill=blue6,rounded corners=0.5] ({(#2)*\dx},{-(#1)*\dx}) rectangle ({((#2)+8)*\dx},{(-(#1)+6.25)*\dx});
	\draw [thick,colVMPSLines,fill=IcolVMPSc,rounded corners=0.5] ({((#2)-0.01)*\dx},{(-(#1)+6.25-0.31)*\dx}) rectangle ({((#2)+8+0.01)*\dx},{(-(#1)+6.25+0.01)*\dx});
}

\newcommand\sSquare[3]{
	\draw [thick,rounded corners=0.5,colLines,fill=#3] ({\dx*#2-0.5*\r},{-\dx*#1-0.5*\r}) rectangle ({\dx*#2+0.5*\r},{-\dx*#1+0.5*\r});
}

\newcommandx{\eightlegs}[2][1=0,2=0]{
	\begin{scope}[shift={(#1,#2)}]
		\foreach \x in {1,...,8}{
			\draw (\x, 0)--++(0,0.25);
			\draw[fill] (\x,0) circle (0.05);
		}
		\foreach \x in {1,3}{
			\pgfmathsetmacro\result{2*\x-1} 
			\node () at (\result,-0.5) {$i_{\x}$};
			\pgfmathsetmacro\result{2*\x}
			\node () at (\result,-0.5) {$j_{\x}$};	
		}
		\foreach \x in {2,4}{
			\pgfmathsetmacro\result{2*\x} 
			\node () at (\result,-0.5) {$i_{\x}$};
			\pgfmathsetmacro\result{2*\x-1}
			\node () at (\result,-0.5) {$j_{\x}$};	
		}
	\end{scope}
}

\newcommandx{\MPSinitialstate}[6][1=0,2=0,3=orange,4=topright,5=-1,6=1]{
\begin{scope}[shift={(#1,#2)},rounded corners=1.5pt]
	\draw[black,thick,fill=#3] 
	(-0.25*#6,.25*#6)--++(.5*#6,0)--++(0,-.3*#6)--++(-.5*#6,0)--cycle;
	\draw[thick] (-.25*#6,.25*#6)--++(-.25*#6,.25*#6);
	\draw[thick] (.25*#6,.25*#6)--++(.25*#6,.25*#6);
	\draw[very thick] (-.5*#6,.-.05*#6)--++(1*#6,0);
\ifthenelse{\equal{#5}{-1}}{
	\ifthenelse{\equal{#4}{topright}}{\draw[thick,rounded corners=0.3]
	(-.1*#6,.15*#6)--++(.2*#6,0)--++(0,-0.1*#6); }{}
	\ifthenelse{\equal{#4}{topleft}}{\draw[thick,rounded corners=0.3]
	(.1*#6,.15*#6)--++(-.2*#6,0)--++(0,-0.1*#6); }{}
	\ifthenelse{\equal{#4}{bottomleft}}{\draw[thick,rounded corners=0.3]
	(.1*#6,-.15*#6)--++(-.2*#6,0)--++(0,0.1*#6); }{}	\ifthenelse{\equal{#4}{bottomright}}{\draw[thick,rounded corners=0.3]
	(-.1*#6,-.15*#6)--++(.2*#6,0)--++(0,0.1*#6); }{}}{			\node at ({0},{0.085*#6}) {\scalebox{1.}{{$#5$}}};}
\end{scope}
}

\newcommandx{\Cmatrix}[6][1=0,2=0,3=2,4=orange,5=,6=topright]{
	\pgfmathsetmacro\result{#3-1} 
	\begin{scope}[shift={(#1,#2)}]
		\foreach \i in {0,...,\result}
		{\foreach \j in {0,...,\i}
			{\roundgate[\i+\j][\i-\j][1][#6][#4]}
		}
		\ifthenelse{\equal{#5}{init}}{
			\foreach \i in {0,...,#3}
			{
				\MPSinitialstate[-1+2*\i][-1][#4]
			}
		}{}
	\end{scope}
}

\newcommand{\msqr}{\fineq[-0.6ex][0.7][1]{\sqrstate[0][0][]}}

\newcommand{\mcirc}{\fineq[-0.6ex][0.7][1]{\cstate[0][0][]}}
\renewcommand{\bcirc}{\fineq[-0.6ex][0.7][1]{\cstate[0][0][][black]}}

\newcommandx{\cstate}[4][1=0,2=0,3= ,4=white]{
	\begin{scope}[shift={(0,0)}]
				\draw[fill=#4,thick] (#1,#2) circle (0.13);
				\node[scale=1.1] at (#1,#2) {$#3$};
\end{scope}
}
\newcommandx{\sqrstate}[4][1=0,2=0,3= ,4=white]{
	\begin{scope}[shift={(#1,#2)}]
		\draw[thick,fill=#4] (-0.12,-0.12) rectangle (0.12,0.12) ;
		\node[scale=1.1] at (0,0) {$#3$};
	\end{scope}
}
\newcommandx{\cstatep}[4][1=0,2=0,3= ,4=white,]{
	\begin{scope}[shift={(0,0)}]
				\draw[fill=#4,thick] (#1,#2) circle (0.15);
				\node[scale=1.1] at (#1,#2) {\scalebox{.85}{$#3$}};
\end{scope}
}

\newcommandx{\pairproduct}[2][1=0,2=0]
{\begin{scope}[shift={(#1 ,#2)}]
\draw[thick] (-.5,.5) arc(-135:-45:1/1.414);
\sqrstate[0][.5-.1414][][black]	
\end{scope}
}
\newcommandx{\bellpair}[2][1=0,2=0]
{\begin{scope}[shift={(#1 ,#2)}]
		\draw[thick] (-.5,.5) arc(-135:-45:1/1.414);
	\end{scope}
}
\newcommandx{\charge}[3][1=0,2=0,3=black]
{
	\ifthenelse{\equal{#3}{blue}}{\def \chargecolor{grey6}
	}{
	\ifthenelse{\equal{#3}{red}}{\def \chargecolor{grey6}}{\def \chargecolor {#3}}}
\begin{scope}[shift={(#1 ,#2)}]
	\draw[ fill=\chargecolor] circle (0.13);        
\end{scope}
}
\newcommandx{\trianglediag}[7][1=0,2=0,3=1,4=orange,5= ,6=-1,7=topright]
{\begin{scope}[shift={(#1 ,#2)}]
	\foreach \i in {0,...,#3}
	{	\foreach \j in {0,...,\i}
		{	\roundgate[-\j+2*\i][\j][1][topright][#4][#6]
		}
	}
	\foreach \i in {-1,...,#3}
	{\ifthenelse{\equal{#5}{bellpair}}{\bellpair[\i*2+1][-1]}{\ifthenelse{\equal{#5}{pairproduct}}{\pairproduct[\i*2+1][-1]}{\MPSinitialstate[\i*2+1][-1][#4][#7][#6]}}}
\end{scope}
}
\newcommandx{\projectorleg}[4][1=0,2=0,3=R,4=left]
{
	\begin{scope}[shift={(#1 ,#2)}]
		{\ifthenelse{\equal{#3}{R}}{		\draw[thick] (-.25,-.25)--++(.5,.5);}{\ifthenelse{\equal{#3}{L}}{		\draw[thick] (-.25,.25)--++(.5,-.5);}{\draw[thick] (-.25,0)--++(.5,0);}}
		}
	\draw[thick, fill=white] circle (0.13);
	\ifthenelse{\equal{#4}{right}}{
		\draw[thick] (.0,.07)--++(.07,0)--++(0,-.07);
		\node[scale=0.5] at (0,-.02) {$\alpha$};
	}{}
	\ifthenelse{\equal{#4}{left}}{	
	\draw[thick] (0,.07)--++(-.07,0)--++(0,-.07);
	\node[scale=0.5] at (0,-.03) {$\beta$};
	}{}
	\end{scope}
}
\usetikzlibrary{patterns,decorations.pathreplacing}

\usepackage{graphicx} 
\usepackage{color}
\usepackage[normalem]{ulem}
\usepackage{graphicx}
\usepackage{amsmath}
\usepackage{amsfonts}
\usepackage{amssymb}
\usepackage{verbatim}
\usepackage{float}
\usepackage{bbold}
\usepackage{bbm}
\usepackage{braket}

\newtheorem*{lemma}{Lemma}
\newtheorem*{theorem}{Theorem}

\usepackage{hyperref}
\hypersetup{
  colorlinks   = true, 
  urlcolor     = blue, 
  linkcolor    = blue, 
  citecolor   = blue 
}

\newcommand{\be}{\begin{equation}}
\newcommand{\ee}{\end{equation}}
\newcommand{\bruno}[1]{{\color{red}[Bruno: #1]}}

\newcommand{\tomaz}[1]{{\color{brown}[Tomaz: #1]}}

\newcommand{\lt}[0]{{({\ell})}}
\newcommand{\rt}[0]{{({r})}}
\newcommand{\ltb}[0]{\ell}
\newcommand{\rtb}[0]{r}

\begin{document}
\title{Exactly solvable many-body dynamics from space-time duality}
\author{Bruno Bertini}
\affiliation{School of Physics and Astronomy, University of Birmingham, Edgbaston, Birmingham, B15 2TT, UK}
\author{Pieter W. Claeys}
\affiliation{Max Planck Institute for the Physics of Complex Systems, 01187 Dresden, Germany}
\author{Toma\v{z} Prosen}
\affiliation{Faculty~of~Mathematics~and~Physics,~University~of~Ljubljana, 
Jadranska~19,~SI-1000~Ljubljana,~Slovenia}
\affiliation{Institute~of~Mathematics, Physics~and~Mechanics, 
Jadranska~19,~SI-1000~Ljubljana,~Slovenia}

\begin{abstract}
Recent years have seen significant advances, both theoretical and experimental, in our understanding of quantum many-body dynamics. Given this problem's high complexity, it is surprising that a substantial amount of this progress can be ascribed to exact analytical results. Here we review dual-unitary circuits as a particular setting leading to exact results in quantum many-body dynamics. Dual-unitary circuits constitute minimal models in which space and time are treated on an equal footings, yielding exactly solvable yet possibly chaotic evolution. They were the first in which current notions of quantum chaos could be analytically quantified, allow for a full characterisation of the dynamics of thermalisation, scrambling, and entanglement (among others), and can be experimentally realised in current quantum simulators. Dual-unitarity is a specific fruitful implementation of the more general idea of space-time duality in which the roles of space and time are exchanged to access relevant dynamical properties of quantum many-body systems. 
\end{abstract}

\maketitle

\tableofcontents

\section{Introduction}
\label{sec:introduction}

Exact solutions are exceptionally rare in quantum many-body physics and, when available, are often limited to ground states or equilibrium states of simple models. In a nutshell, this limitation arises because performing real-time evolution of extended interacting quantum many-body systems is fundamentally more complicated than performing imaginary time evolution for annealing towards the ground state: contrary to its imaginary-time counterpart, real-time dynamics typically produces an exponentially fast growth of complexity as time elapses. This growth of complexity is attested, for instance, by the linear temporal growth of entanglement in generic quantum many-body systems out-of-equilibrium, indicating an exponential growth of the number of states required to capture the dynamics. 
The same diagnosis can be obtained from any similar measure of complexity and quantum correlations in time-dependent states or operators. As a result, the task of accessing time-evolving observables or time-dependent correlation functions appears daunting, both analytically and via classical computations. This issue is essentially unavoidable in the presence of nontrivial interactions, e.g., it also arises in interacting integrable systems where the many-body eigenstates are in principle accessible by Bethe ansatz methods. On the other hand, precisely because of it being unfeasible by any other method, probing many-body quantum dynamics represents a promising setting for the emerging quantum computers and simulators to show advantage~\cite{arute2019quantum}. 

A breakthrough in this highly intricate problem has recently come from a statistical mechanical perspective. From this vantage point evaluating the dynamics of observables in one-dimensional systems with local interactions corresponds to computing the partition function of a two-dimensional statistical model, albeit with complex weights (a similar mapping holds in higher-dimensional lattice systems). This kind of calculation is typically performed using the transfer matrix formalism, where one has two choices for how to perform the required contractions. One can either contract the partition function in the temporal direction, which amounts to iterating the unitary many-body propagator, or in the spatial direction by defining an appropriate spatial transfer matrix. As first realised by~\cite{banuls2009matrix}, this innocent observation presents a very useful tool for evaluating dynamical observables. While it was originally proposed as a computational tool, the idea of finding information about the system's dynamics by exchanging the roles of space and time --- space-time duality --- has recently emerged as a new way to think about quantum many-body dynamics and even to find exact results in certain special cases. 

This change of paradigm originated from the analysis of a simple spin model known as the kicked Ising chain~\cite{Prosen2000}. As shown by~\cite{Gutkin}, this model features a line in parameter space where both space and time propagators are unitary, such that the space evolution can be thought of as time evolution. This property allowed~\cite{bertini_exact_2018, bertini2019entanglement} to use space-time duality to exactly compute highly non-trivial non-equilibrium properties of the system, despite this model being strongly interacting. This surprising development coincided with the simultaneous rise of quantum circuits --- many-body systems evolved in a discrete time via local unitary gates --- as minimal models for many-body quantum dynamics~\cite{nahum2017quantum, von_keyserlingk_operator_2018, khemani2018operator, chan2018solution}. Putting these two ingredients together, \cite{bertini_exact_2019, gopalakrishnan_unitary_2019} showed that in quantum circuits the property of retaining unitarity under space-time duality could be identified on the level of the constituting gates. This led to the definition of a class of circuits, \emph{dual-unitary circuits}, which can be treated exactly via space-time duality.

The advent of dual-unitary circuits generated a burst of exciting theoretical activity aimed at characterising their dynamical properties and spectral statistics. This quest has produced, for instance, exact (or quasi-exact) results for dynamical correlations, out-of-time-ordered correlators (OTOCs), entanglement dynamics, operator spreading, and the spectral form factor, to name just a few. Dual-unitary circuits were also quickly experimentally realised in digital quantum computing setups, probing the predicted dynamical behaviours and using the latter to benchmark the simulators themselves. 

Dual-unitary circuits are in many ways complementary to Bethe ansatz integrable systems since they represent a platform for exactly solvable many-body quantum chaos. It is tempting to compare these quantum systems to the textbook examples of solvable classical chaos, such as the Baker map or Arnold cat maps. In the latter context the availability of exact solutions offered a unique ground for the demonstration and potential classification of chaotic phenomena. In the same vein, dual-unitary circuits allowed for, e.g., direct connections between the universality and complexity of gate-based quantum computation and ergodicity in many-body quantum dynamics. Nevertheless, it is important to stress that dual-unitary dynamics can cover both the chaotic/ergodic and integrable domain and supports various forms of ergodicity breaking.

The aim of this review is to give a pedagogical introduction to the concept of dual-unitarity and present an overview of the numerous exact results obtained for these systems to date. We give an overview of experimental realisations as well as the various extensions and generalisations of dual-unitarity that have appeared since its introduction. Finally, we discuss how the broader idea of space-time duality can be used to characterise quantum many-body dynamics beyond the example of dual-unitary circuits.

\section{Setting}
\label{sec:setting}

Although systems defined in discrete space (e.g.~on a lattice) are a standard concept in physics since at least the nineteenth century, when thinking about dynamics we typically have in mind continuous time evolution generated by a suitable Hamiltonian. Over the last decade, however, there has been a strong paradigmatic shift and increasingly more works started to consider the dynamics of lattice systems where the evolution proceeds in discrete steps (i.e.,~also restricting time to a lattice). In essence, this happened for two main reasons: (i) Lattice systems with discrete time evolution are natively realised in the much hyped digital quantum computers, see, e.g.,~\cite{arute2019quantum}. (ii) Systems where both space and time are discrete, even when strongly interacting, are amenable to more effective analytical treatments compared to their continuous-time counterparts. This is ultimately because they treat space and time on a symmetric footing while avoiding the mathematical complications associated with continuous space-time theories aka field theories. The discussion of how this duality between space and time can be exploited to find exact results on the dynamics of interacting many-body quantum systems is the central theme of this review. 

Specifically, we here consider a special kind of lattice systems in discrete time known as \emph{brickwork quantum circuits}. In these systems each lattice point hosts a quantum variable with $q$ internal states --- a \emph{qudit} --- (qubit for $q=2$) and time evolution couples nearest neighbours through the staggered application of a unitary matrix $U$ typically referred to as the \emph{local gate}. A full time step is composed by different layers, eventually coupling all nearest neighbours together. For example, in the one-dimensional setting the time evolution operator for a full time step can be expressed as 
\be
\mathbb{U} =\mathbb{U}_e\mathbb{U}_o,\,\, \mathbb{U}_e=\bigotimes_{x=0}^{L-1} U_{x,x+1/2},\,\,\mathbb{U}_o=\bigotimes_{x=1}^{L} U_{x-1/2,x}\,.
\label{eq:floquetoperator}
\ee
Here the sites are labelled by half integers, $U_{a,b}$ acts as the $q^2\times q^2$ matrix $U$ on a pair of qudits located at lattice sites $a$ and $b$, and we denoted by $2 L$ the size of the system (number of qudits). Moreover, we take periodic boundary conditions, so that sites $0$ and $L$ coincide. 
Within this dynamics each site is periodically coupled to its left and right neighbour through $U$.
In fact, one can generalise this setting to describe systems with spatial disorder and aperiodic driving in time by replacing 
\be
U \mapsto U(x,t)\,.
\label{eq:drivendisordered}
\ee
Namely, one can use different local gates at each position and time. Unless we explicitly state otherwise, the results discussed in this review will apply to the more general setting in Eq.~\eqref{eq:drivendisordered}, however, for the sake of clarity, we will use the translationally invariant case to illustrate the derivations. In this case the evolution operator in Eq.~\eqref{eq:floquetoperator} is also referred to as the Floquet operator.
By proceeding as above one can define brickwork quantum circuits in any dimension, but here we will focus on the one-dimensional case in Eq.~\eqref{eq:floquetoperator} for the sake of simplicity. Many of the results we discuss are directly generalised to higher dimensions and we will explicitly emphasise when this is the case.  

As shown in Sec.~\ref{sec:corr_functions}, the staggered form of the evolution operator in Eq.~\eqref{eq:floquetoperator} implies that correlations lie within a finite (geometric) light cone. Namely, any pair of time-dependent local operators $a_x(t)$ and $b_y(t)$, 
where $O(t)\equiv \mathbb U^t O\mathbb U^{-t}$, satisfy 
\be
[a_x(t),b_y(t)] = 0,\qquad |\lceil x \rceil - \lceil y\rceil | > 2v_{\rm max}t, 
\ee 
where $\lceil \cdot \rceil $ denotes the ceiling function (smallest integer larger than or equal to the argument) and our choice of units gives $v_{\rm max}=1$. This bound can be considered as an extreme version of the Lieb--Robinson bound~\cite{lieb1972finite}, where the exponentially small leaking of information from the light cone is here forced to be exactly zero, and suggests that the dynamics of brickwork quantum circuits should be qualitatively similar to that generated by Hamiltonians with local interactions. This statement can be made more quantitative by obtaining a brickwork quantum circuit from a given local Hamiltonian through Suzuki--Trotter decomposition~\cite{trotter1959product,suzuki1991general} or more refined decompositions involving a finite number of Trotter steps~\cite{osborne2006efficient, haah2023quantum}. Here, however, we consider brickwork quantum circuits as the fundamental dynamical systems of interest and make no reference to an underlying Hamiltonian dynamics.

Despite representing an idealisation, the dynamics generated by the quantum circuit in Eq.~\eqref{eq:floquetoperator} for generic choices of the local gate $U$ is still far too complicated to be amenable to a quantitative characterisation and one has to introduce further simplifications. To this end, two main strategies have emerged as viable. The first one is inspired by random matrix theory, and is based on the idea of removing all additional structure besides the locality enforced by the circuit setting and perhaps a few more basic symmetries. This is achieved by taking the local gate $U$ to be a random matrix drawn from a specific ensemble~\cite{nahum2017quantum, khemani2018operator, chan2018solution, bertini2024quantum}. The second is to impose additional constraints, or symmetries, on the dynamics. While the first approach is discussed in the recent reviews~\cite{potter2022entanglement, fisher_random_2023}, here we are concerned with the second approach. In particular, we follow the proposal of~\cite{bertini_exact_2019,gopalakrishnan_unitary_2019} and consider choices of $U$ that generate unitary dynamics also when the roles of space and time are exchanged. More formally, we introduce the space-time-swapped gate $\widetilde{U}$ defined through the following matrix element reshuffling 
\be
\bra{ij}\widetilde{U}\ket{kl}\equiv \mel{jl}{U}{ik},
\label{eq:Utilde}
\ee	
and impose it to be unitary alongside ${U}$. Namely, we impose the following two conditions on the matrix $U$	
\be
\begin{aligned}
& UU^\dagger = U^\dagger U=\mathds{1}_{q^2},\\ 
& \widetilde{U}\widetilde{U}^\dagger = \widetilde{U}^\dagger\widetilde{U}=\mathds{1}_{q^2},
\end{aligned}
\label{eq:DU}
\ee
where $\mathds{1}_{x}$ represents the identity operator in $\mathbb C^{x}$. Matrices fulfilling this condition are referred to as \emph{dual-unitary} matrices~\cite{bertini_exact_2019} and quantum circuits composed of gates with this property are referred to as \emph{dual-unitary circuits}~\cite{bertini_exact_2019}. Before analysing the solutions to the condition in Eq.~\eqref{eq:DU} and its general physical implications in Secs.~\ref{sec:localproperties} and \ref{sec:corr_functions}, we proceed to introduce a convenient diagrammatic representation of quantum circuits that greatly facilitates their interpretation.

\subsection{Diagrammatic representation}	
\label{sec:diagrams}

Dynamical properties of quantum circuits can be illustrated through a diagrammatic representation (Penrose notation) borrowed from tensor network theory, see, e.g.~\cite{cirac2021matrix}. This diagrammatic approach is fully rigorous --- each diagram is in one-to-one correspondence with a mathematical equation --- but grants more physical intuition and often aids the analytical calculations. The idea is to represent matrix elements of tensors in the computational basis of the qudits
\be
\label{eq:computationalbasis}
\{\ket{i}, \quad i=1,\ldots, q\}, 
\ee
by geometrical shapes with a number of legs corresponding to the number of indices. For example, one represents the matrix elements of the local gate and its complex conjugate in the computational basis as follows  	
\begin{align}
	\mel{kl}{U}{ij}=\fineq[-0.8ex][1][1]{
		\tsfmatV[0][-0.5][r][1][][][red6]
		\node at (-0.5,-0.2) {$i$};
		\node at (0.5,-0.15) {$j$};
		\node at (-0.5,1.2) {$k$};
		\node at (0.5,1.25) {$l$};},\quad \mel{kl}{U^*}{ij}=\fineq[-0.8ex][1][1]{
		\tsfmatV[0][-0.5][r][1][][][blue6]		\node at (-0.5,-0.2) {$i$};
		\node at (0.5,-0.15) {$j$};
		\node at (-0.5,1.2) {$k$};
		\node at (0.5,1.25) {$l$};}.
\end{align}
We choose the convention that the evolution in time goes from bottom to top. Matrix multiplication is represented by connecting the legs of indices summed over. For example, the dual-unitarity conditions (cf.\ Eq.~\eqref{eq:DU}) are, respectively, expressed as 
\be
\begin{aligned}
\fineq[-0.8ex][1][1]{
\node at (-0.5,.3) {$i$};
\node at (0.5,.25) {$j$};
\node at (-0.5,3) {$k$};
\node at (0.5,3) {$l$};
\tsfmatV[0][1.25][r][1][][][red6][topright]
\tsfmatV[0][0][r][1][][][blue6][bottomright]
\draw[ thick] (-0.5,1.5) -- (-0.5,1.75);
\draw[ thick] (0.5,1.5) -- (0.5,1.75);}
& = 
\fineq[-0.8ex][1][1]{
\node at (-0.5,.3) {$i$};
\node at (0.5,.25) {$j$};
\node at (-0.5,3) {$k$};
\node at (0.5,3) {$l$};
\tsfmatV[0][1.25][r][1][][][blue6][bottomright]
\tsfmatV[0][0][r][1][][][red6][topright]
\draw[ thick] (-0.5,1.5) -- (-0.5,1.75);
\draw[ thick] (0.5,1.5) -- (0.5,1.75);}=
\fineq[-0.8ex][1][1]{	
\node at (-0.5,.3) {$i$};
\node at (0.5,.25) {$j$};
\node at (-0.5,3) {$k$};
\node at (0.5,3) {$l$};
\draw[ thick] (-0.5,.5) -- (-0.5,2.75);
\draw[ thick] (0.5,.5) -- (0.5,2.75);}\,, \\	
\!\!\!\fineq[-0.8ex][1][1]{
\node at (-0.5,.3) {$i$};
\node at (1.75,.25) {$j$};
\node at (-0.5,1.75) {$k$};
\node at (1.75,1.75) {$l$};
\tsfmatV[0][0][r][1][][][red6][topright]
\tsfmatV[1.25][0][r][1][][][blue6][topleft]
\draw[ thick] (.5,1.5) -- (.75,1.5);
\draw[ thick] (.5,0.5) -- (.75,0.5);}
& \!\!=\!\! \fineq[-0.8ex][1][1]{
\node at (-0.5,.3) {$i$};
\node at (1.75,.25) {$j$};
\node at (-0.5,1.75) {$k$};
\node at (1.75,1.75) {$l$};
\tsfmatV[0][0][r][1][][][blue6][topleft]
\tsfmatV[1.25][0][r][1][][][red6][topright]
\draw[ thick] (.5,1.5) -- (.75,1.5);
\draw[ thick] (.5,0.5) -- (.75,0.5);}
\!\!=\!\!
\fineq[-0.8ex][1][1]{
\node at (-0.25,.5) {$i$};
\node at (1.25,.45) {$j$};
\node at (-0.25,1.55) {$k$};
\node at (1.25,1.55) {$l$};
\draw[ thick] (0,1.5) -- (1,1.5);
\draw[ thick] (0,0.5) -- (1,0.5);}\,,
\end{aligned}
\label{eq:dualunitaritynonfolded}
\ee
where we represented the identity operator as a straight line. When the indices are not reported the diagram represents the full matrix rather than its elements.

Many body operators are obtained by composing local tensors via the index summation rule. For example, the time evolution operator in Eq.~\eqref{eq:floquetoperator} for a system of $2L=8$ qudits is represented as
\begin{align}\label{eq:floquetoperator_graphical}
	\mathbb{U}=\fineq[-0.8ex][.75][1]{	
		\foreach \i in {0,...,3}
		{
			\roundgate[2*\i][0][1][topright][red6][-1]
			\roundgate[2*\i+1][1][1][topright][red6][-1]
		}	
	}.
\end{align}
It is also useful to introduce the so called ``folded representation'', which helps representing quantities involving both $\mathbb{U}$, which implements forward evolution, and its inverse implementing backward evolution. The idea is to simplify the graphical representation by folding the diagram for the backward evolution (blue) underneath the one for the forward evolution (red). One then introduces a symbol for stacks of local gates and its complex conjugate, i.e.,  
\begin{align}
U\otimes U^* =\fineq{
\foreach \i in {0,...,0}
{\roundgate[\i*0.3][\i*0.15][1][topright][blue6]
\roundgate[\i*0.3-.15][\i*0.15-.075][1][topright][red6]
}
}=\fineq{\roundgate[0][0][1][topright][orange][1]}.
\label{eq:foldedgatepicturen1}
\end{align}
It is useful to introduce a special symbol to denote a connection between forward and backward evolution --- which in the unfolded language would just be a line representing the identity operator. Specifically we define
\begin{align}
\ket{\mcirc}= \sum_{i =1}^q \ket{i,i} \equiv\fineq[-0.4ex]{\draw[thick] (0,0)--++(-.5,0);
\cstate[0][0]}\,.
\label{eq:idstaten1}
\end{align}
Using Eqs.~\eqref{eq:foldedgatepicturen1} and \eqref{eq:idstaten1} we can represent the dual-unitarity conditions in Eq.~\eqref{eq:dualunitaritynonfolded} more compactly as 
\begin{align}
	\!\!\!\!\!\!\!\!\!\fineq[-0.8ex][1][1]{
	\roundgate[1][1][1][topright][orange][1]
	\cstate[1.5][1.5][]
	\cstate[.5][1.5][]
}&=
\fineq[-0.8ex][1][1]{
	\draw[thick] (.5,1.5)--++(0,-1);
	\draw[thick] (1.5,1.5)--++(0,-1);
	\cstate[1.5][1.5][] 
	\cstate[.5][1.5][] },\quad\!\!
\fineq[-0.8ex][1][1]{
	\roundgate[1][1][1][topright][orange][1]
	\cstate[1.5][.5][] 
	\cstate[.5][.5][]
	 }=\fineq[-0.8ex][1][1]{
	\draw[thick] (.5,1.5)--++(0,-1);
	\draw[thick] (1.5,1.5)--++(0,-1);
	\cstate[.5][.5][] 
	\cstate[1.5][.5][]
	 },\label{eq:unitarityfoldeddiagram}\\
\!\!\!\!\!\!\!\!\!\fineq[-0.8ex][1][1]{
	\roundgate[1][1][1][topright][orange][1]
	\cstate[1.5][1.5][] 
	\cstate[1.5][.5][]
}&=
\fineq[-0.8ex][1][1]{
	\draw[thick] (1.5,1.5)--++(-1,0);
	\draw[thick] (1.5,.5)--++(-1,0);
	\cstate[1.5][1.5][] 
	\cstate[1.5][.5][]
	 },\qquad\!\!\!\!\!\!
\fineq[-0.8ex][1][1]{
	\roundgate[1][1][1][topright][orange][1]		
	\cstate[.5][1.5][]
	\cstate[.5][.5][]
	 }=\fineq[-0.8ex][1][1]{
	\draw[thick] (1.5,.5)--++(-1,0);
	\draw[thick] (1.5,1.5)--++(-1,0);
	\cstate[.5][1.5][]
	\cstate[.5][.5][] 
	}.
	\label{eq:spaceunitarityfoldeddiagram}
\end{align}
More generally, it is useful to introduce a compact representation for multi-replica quantities needed, e.g., for operator dynamics and entanglement~(cf.\ Secs.~\ref{sec:corr_functions} and \ref{sec:quantumdynamics}). In this case it is convenient to fold each copy of backward evolution underneath a forward one. We then introduce a symbol for stacks of $n$ local gates and its complex conjugate generalising Eq.~\eqref{eq:foldedgatepicturen1}, i.e.,  
\begin{align}
\left(U\otimes U^*\right)^{\otimes n}=\fineq{
\foreach \i in {0,...,-2}
{\roundgate[\i*0.3][\i*0.15][1][topright][blue6]
\roundgate[\i*0.3-.15][\i*0.15-.075][1][topright][red6]
}
\draw [decorate, decoration = {brace}]   (-1.35,0.1)--++(1*0.9,.5*0.9);
\node[scale=1.25] at (-1.135,.5) {${}_{2n}$};
}=\fineq{\roundgate[0][0][1][topright][orange][n]}.
\label{eq:foldedgatepicture}
\end{align}
Finally, we introduce special symbols to represent the states in the replicated space that correspond to contractions between the replicas. Specifically, labelling the $2n$ replicas as $(1,1^*, 2, 2^*,\ldots, n, n^*)$, we can define a state indexed by a permutation $\sigma$ of $n$ elements as
\begin{align}\label{eq:permutation_states}
\ket{\sigma}= \sum_{i_1,\ldots,i_n =1}^q \ket{i_1,i_{\sigma(1)},\ldots, i_n, i_{\sigma(n)}} \equiv\fineq[-0.4ex]{\draw[thick] (0,0)--++(-.5,0);
\cstatep[0][0][\sigma]}\,.
\end{align}
A contraction with this state consists of contracting replica $i$ with replica $\sigma(i)^*$, $\forall i=1\dots n$.
Using these states we represent the $n!$ dual-unitarity conditions fulfilled by the $n$-folded gates as 
\begin{align}
	\!\!\!\!\!\!\!\!\!\fineq[-0.8ex][1][1]{
	\roundgate[1][1][1][topright][orange][n]
	\cstatep[1.5][1.5][\sigma]
	\cstatep[.5][1.5][\sigma]
}&=
\fineq[-0.8ex][1][1]{
	\draw[thick] (.5,1.5)--++(0,-1);
	\draw[thick] (1.5,1.5)--++(0,-1);
	\cstatep[1.5][1.5][\sigma] 
	\cstatep[.5][1.5][\sigma] },\quad\!\!
\fineq[-0.8ex][1][1]{
	\roundgate[1][1][1][topright][orange][n]
	\cstatep[1.5][.5][\sigma] 
	\cstatep[.5][.5][\sigma]
	 }=\fineq[-0.8ex][1][1]{
	\draw[thick] (.5,1.5)--++(0,-1);
	\draw[thick] (1.5,1.5)--++(0,-1);
	\cstatep[.5][.5][\sigma] 
	\cstatep[1.5][.5][\sigma]
	 },\label{eq:unitaritynfoldeddiagram}\\
\!\!\!\!\!\!\!\!\!\fineq[-0.8ex][1][1]{
	\roundgate[1][1][1][topright][orange][n]
	\cstatep[1.5][1.5][\sigma] 
	\cstatep[1.5][.5][\sigma]
}&=
\fineq[-0.8ex][1][1]{
	\draw[thick] (1.5,1.5)--++(-1,0);
	\draw[thick] (1.5,.5)--++(-1,0);
	\cstatep[1.5][1.5][\sigma] 
	\cstatep[1.5][.5][\sigma]
	 },\qquad\!\!
\fineq[-0.8ex][1][1]{
	\roundgate[1][1][1][topright][orange][n]		
	\cstatep[.5][1.5][\sigma]
	\cstatep[.5][.5][\sigma]
	 }=\fineq[-0.8ex][1][1]{
	\draw[thick] (1.5,.5)--++(-1,0);
	\draw[thick] (1.5,1.5)--++(-1,0);
	\cstatep[.5][1.5][\sigma]
	\cstatep[.5][.5][\sigma] 
	}.
	\label{eq:spaceunitaritynfoldeddiagram}
\end{align}
In particular, we introduce two special symbols for two particularly useful permutations 
\begin{align}
\ket{\rm id}\equiv \ket{\mcirc}&\equiv\fineq{\draw[thick] (0,0)--++(-.5,0);\cstate[0][0]}\,,
\label{eq:circlestate}\\
\ket{\pi}\equiv\ket{\msqr}&\equiv \fineq{\draw[thick] (0,0)--++(-.5,0);\sqrstate[0][0]}\,,
\label{eq:squarestate}
\end{align}
where ${\rm id}$ corresponds to the identity permutation and $\pi$ to a cyclic shift by one replica to the right, i.e., 
\be
\pi = \begin{pmatrix}
1 & 2 & 3 & 4 & \cdots & n-1 & n \\
n & 1 & 2 & 3  & \cdots & n-2 &  n-1\\
\end{pmatrix}\,.
\ee
Note that Eq.~\eqref{eq:circlestate} is a multi-replica generalisation of Eq.~\eqref{eq:idstaten1}.

\section{Dual-Unitary Local Gates}
\label{sec:localproperties}

In this section we discuss some of the basic properties of dual-unitary local gates, i.e., two-qudit gates solving Eq.~\eqref{eq:DU}. In particular, in Sec.~\ref{sec:parameterisation} we provide some explicit parameterisations, in Sec.~\ref{sec:relations} we briefly discuss the relations between the dual-unitarity condition and other related concepts appeared in previous mathematical and quantum information theoretical literature, while in Sec.~\ref{sec:entpow} we discuss their entangling power. 

\subsection{Parameterisation}
\label{sec:parameterisation}

A complete parameterisation of the solutions to Eq.~\eqref{eq:DU} was presented in \cite{bertini_exact_2019} for $q=2$, while no exhaustive parameterisation is currently known for $q>2$. Many different families of dual-unitary gates with $q>2$ have appeared in the recent literature~\cite{gutkin_exact_2020, rather_creating_2020, bertini2021random, aravinda_dual-unitary_2021, claeys_ergodic_2021, prosen2021many,singh_diagonal_2022, rather_construction_2022,borsi_construction_2022}. To discuss these constructions, we begin by noting that the dual-unitarity property is not affected by local unitaries, i.e., for any given dual-unitary gate $V$ one can construct an orbit of dual-unitary gates as follows 
\be
V \mapsto U = (v_1\otimes v_2) V (v_3\otimes v_4), 
\label{eq:orbit}
\ee
with arbitrary $v_1, v_2, v_3, v_4 \in \mathrm{U}(q)$ (note that different elements of a given orbit will generally have very different physical properties). For this reason we only report the explicit form of the minimal two-site gate. For $q=2$ the most general dual unitary gate can be obtained by plugging in Eq.~\eqref{eq:orbit} a $V$ of the form~\cite{bertini_exact_2019}
\be
V = S e^{i J \sigma^{z}\otimes \sigma^{z}}, \qquad J \in \mathbb [0,\pi/4],
\label{eq:general_para_q2}
\ee
where $S$ is the SWAP gate and $\{\sigma^{x},\sigma^{y},\sigma^{z}\}$ denote the Pauli matrices. Direct parameter counting gives 14 free parameters for the most general dual-unitary gate, which is two fewer than the most general $4\times4$ unitary matrix. Using the Pauli algebra, however, one can further reduce the number of independent parameters to 12~\cite{prosen2021many}. We stress that without dressing the gate in Eq.~\eqref{eq:general_para_q2} with single-site unitary matrices, $U = V$ implements integrable dynamics. It corresponds to a special curve in the parameter space of the integrable Trotterisation of the XXZ spin-chain~\cite{ljubotina2019ballistic}. Dressing with generic one-site unitaries however results in non-integrable dynamics.

The family in Eq.~\eqref{eq:general_para_q2} can be directly extended to any $q> 2$ replacing $\sigma^{z}/2$ with the $q$-dimensional representation of the $z$ component of the angular momentum operator~\cite{bertini2021random} (the SWAP gate is defined for any $q$). One, however, is allowed much more freedom. \cite{claeys_ergodic_2021} showed that any $V$ of the form 
\be
V= S D,
\label{eq:paraSD}
\ee 
where $D$ is diagonal and unitary, fulfils the dual-unitary condition. This family can be further extended by replacing the SWAP with a dual-unitary permutation matrix~\cite{borsi_construction_2022}. 

\begin{table}
\begin{ruledtabular}
\begin{tabular}{ccccc}
$q$ & $q'$ & ${\rm dim}\,{\rm U}(q q')=(q q')^2$ &
${\rm dim}\,{\rm dualU}(q,q')$ \\
\hline
2 & 2 & 16 & 12 (12)  \\
3 & 3 & 81 & 45 (45,43,41)  \\
4 & 4 & 256 & 112 (94)  \\
5 & 5 & 625 & 225 (97)  \\
6 & 6 & 1296 & 396 (141) \\
7 & 7 & 2401 & 637 (193) \\
3 & 2 & 36 & 24 (24) \\
4 & 2 & 64 & 40 (40) \\
5 & 2 & 100 & 60 (60) \\
6 & 2 & 144 & 84 (84) \\
7 & 2 & 196 & 112 (112) \\
4 & 3 & 144 & 72 (66,64) \\
5 & 3 & 225 & 105 (77) \\
\end{tabular}
\end{ruledtabular}
\caption{Local dimensions of the manifold of dual unitary gates ${\rm dualU}(q,q')$. The dimensions of local Hilbert spaces at even and odd lattice sites ($q$ and $q'$) can differ. The results in the table are estimated as the dimensions of the tangent spaces at randomly sampled solutions of the dual-unitarity constraints in Eq.~\eqref{eq:DU} (numbers within brackets). For comparison we also show the dimensions of tangent spaces at random instances of the explicit parametrization in Eq.~\eqref{eq:parq3} (unbracketed numbers). Adapted from \cite{prosen2021many}.
\label{table1}}
\end{table}

Other families of dual-unitary gates for $q> 2$ can be obtained by considering direct sums~\cite{rather_creating_2020, aravinda_dual-unitary_2021, prosen2021many, borsi_construction_2022}. For example, \cite{prosen2021many, borsi_construction_2022} showed that the gates 
\be
V = S \left(\sum_{j=1}^q \ketbra{j}{j}\otimes u^{(j)}\right), \qquad u^{(j)}\in \mathrm{U}(q), 
\label{eq:parq3}
\ee
(and their left-right swapped counterparts) are dual-unitary. This family contains a number of parameters scaling as $q^3$, which matches the scaling found from numerical estimations of the maximal dimensionality of the dual-unitary manifold~\cite{prosen2021many}, see Table~\ref{table1}.

Other families have been constructed using finite rings~\cite{borsi_construction_2022}, diagonal compositions of dual-unitary gates~\cite{borsi_construction_2022}, solutions to the set-theoretical Yang-Baxter equation~\cite{borsi_construction_2022}, and complex Hadamard matrices~\cite{aravinda_dual-unitary_2021, borsi_construction_2022,singh_diagonal_2022,claeys_emergent_2022}. A numerical algorithm to iteratively generate dual-unitary gates for arbitrary dimensionality was proposed in~\cite{rather_creating_2020}. The convergence properties of this algorithm were further studied in \cite{rather_construction_2022}, which additionally proposed parameterisations based on orthogonal (quantum) Latin squares.

A possible approach to determine the dimensionality of the dual-unitary manifold is to construct tangent spaces around a known set of solutions (e.g.\ Eq.~(\ref{eq:parq3})). In doing that, in order to increase generality, one may relax the condition that local Hilbert spaces at even and odd lattice sites are isomorphic. Instead, one can assume that wires which go along southwest-northeast diagonals carry spaces of dimension $q$, while wires along northwest-southeast diagonals carry spaces of dimension $q'$. See Tab.~\ref{table1} for a collection of empirical results from~\cite{prosen2021many}. These empirical dimensions are described by the following conjectural formula 
\begin{align}
 {\rm dim}\,{\rm dualU}(q\ge q',q') =\, & (q^2+{q'}^2)q' +(4q-5q')q' \notag\\
 &+ 6(q'-q)\,,
\end{align}
with $q$ and $q'$ swapped if ${q < q'}$. For ${q=q'}$ the formula gives ${\rm dim}\,{\rm dual U}(q,q) = (2q-1)q^2$.

\subsection{Related Concepts}
\label{sec:relations}

As pointed out by~\cite{borsi_construction_2022} the dual-unitarity property appeared long ago in the context of pure mathematics --- more specifically in the study of planar algebras~\cite{krishnan1996OnBP, jones1999planaralgebrasi, jones1997introduction} --- under the name \emph{biunitarity}. In fact, biunitarity is broader than dual-unitarity as we defined it in Eq.~\eqref{eq:DU} and has recently been used by \cite{claeys2023dual} to unify different constructions of dual-unitary circuits beyond the brickwork setting (see Sec.~\ref{sec:generalisations}). 

On the quantum information theoretical side dual-unitarity is instead related to the concept of \emph{absolutely maximally entangled} (AME) states~\cite{helwig2012absolute, helwig2013absolutelymaximallyentangledqudit}, also called \emph{perfect maximally multipartite entangled states} by~\cite{facchi2008maximally}, \emph{perfect tensors}~\cite{pastawski2015holographic}, and $k$-unitary matrices by~\cite{goyeneche2015absolutely}. The latter are states on a $2n$-qubit Hilbert space that have maximal entanglement across all possible bipartitions of the qubits. From this point of view, dual-unitary gates are states on a $4$-qubit Hilbert space where all \emph{contiguous} bipartitions of the 4 qubits are maximally entangled. This means that they are similar to but less restricted than the absolutely maximally entangled states. States with this property have been recently introduced in~\cite{berger2018perfecttangles}, where they are called \emph{perfect tangles}, and \cite{harris20018calderbank}, where they are referred to as \emph{block perfect tensors}. They have been also called \emph{planar maximally entangled states} by \cite{dorudiani2020planar}. We note here that numerical algorithms developed in the study of dual-unitarity have found direct applications in constructing AME states~\cite{rather2022thirty,rather_absolutely_2023,zyczkowski_9_2023,pozsgay_tensor_2024,rather_construction_2024}.
The construction of AME states can be reformulated in terms of properties of subalgebras, see e.g.~\cite{gross_thirty-six_2025}, and the dual-unitary condition appeared in this context in~\cite{petz2007complementarity}, with the resulting unitaries termed \emph{useful unitaries}\footnote{We acknowledge Suhail A. Rather for this observation.}.

Dual-unitary matrices also previously appeared in the construction of hyperinvariant tensor networks as models for holography~\cite{evenbly_hyperinvariant_2017}, relaxing the previous use of perfect tensors from~\cite{pastawski2015holographic}, and were termed \emph{doubly unitary} in this context.

\subsection{Entangling power}
\label{sec:entpow}

Different dual-unitary gates are characterised by very different entangling properties (which, as we review in Sec.~\ref{sec:quantumdynamics}, are also visible at the many-body level). To discuss them it is useful to introduce the \emph{entangling power}~\cite{zanardi2000entangling, aravinda_dual-unitary_2021}. The entangling power of a gate is a measure of the average amount of entanglement produced when the gate is applied on pure product state. Specifically, the entangling power of a general two-site gate $U$ is defined as
\begin{align}
e_P(U) = C_q\, \mathbb E_{\rm Haar}\left[{\mathcal{E}(U \ket{\phi_1} \otimes \ket{\phi_2})}\right]\,,
\end{align} 
where $\mathbb E_{\rm Haar}\left[\cdot\right]$ represents the average over $\ket{\phi_1}$ and $\ket{\phi_2}$ distributed according to the Haar measure, $C_q$ is an unimportant scale factor, and we introduced the linear entropy
\be
\label{eq:linearentropy}
\mathcal{E}(\ket{\phi_{12}}) = 1 - \tr_1((\tr_2\ketbra{\phi_{1,2}})^2).
\ee
The average can be evaluated using standard methods of random matrix theory to return
\begin{align}
\label{eq:def_entanglingpower}
e_P(U) = \frac{1}{E(S)} \left[E(U)+E(US)-E(S)\right]\,,
\end{align}
where we introduced the linear entropy of the gate seen as a 4-qudit state (also called \emph{operator entanglement} by~\cite{rather_creating_2020,aravinda_dual-unitary_2021})
\be
\label{eq:linearentropy_U}
E(U) = 1 - \frac{1}{q^4}\tr[(\tilde U^\dag\tilde U)^2] = 1 - \frac{1}{q^4} \fineq[-0.8ex][1][1]{
	\roundgate[1][1][1][topright][orange][2]		
	\cstate[1.5][1.5]
	\cstate[1.5][.5]
	\sqrstate[.5][1.5]
	\sqrstate[.5][.5]}. 
\ee
A simple bound for the latter quantities can be obtained by noting that the eigenvalues of $\tilde U^\dag\tilde U$ are non-negative and sum to $q^2$: this gives $E(U)\in[0,1-{1}/{q^2}]$ and $e_P(U)\in[0,1]$. The upper bound $e_P(U)=1$ is attained by perfect tensors (cf.\ Sec.~\ref{sec:relations}), which are known to exist for all $q\ge 3$~\cite{huber2018bounds, rather2022thirty}. Instead, for $q=2$ a direct calculation using the most general form of a two-qubit gate~\cite{kraus2001optimal, vatan2004optimal} gives a maximum entangling power $e_P(U)|_{{\rm max}, q=2}=2/3$.

The operator entanglement was argued to be an appropriate measure for the `interaction strength' of local gates \cite{hahn_absence_2024}. For dual-unitary gates the operator entanglement is maximized, $E(U)=E(S) = 1-1/q^2$, and the entangling power simplifies to~\cite{rather_creating_2020}
\be
\label{eq:def_entanglingpowerDU}
e_P(U)\bigl |_{\rm DU} = \frac{E(US)}{E(S)}.
\ee
For example, the SWAP gate itself does not create entanglement when acting on product states and, accordingly, Eq.~\eqref{eq:def_entanglingpowerDU} gives $e_P(S)=0$. For the specific case of $q=2$, the explicit form in Eq.~\eqref{eq:general_para_q2} gives 
\be
e_P(U)\bigl |_{{\rm DU}, q=2}\, = \frac{2}{3} \cos^2\left(\frac{\pi}{2}-2J\right), \quad J \in \mathbb [0,\pi/4].
\ee 
The entangling power was extended to the \emph{operator space entangling power} in~\cite{jonnadula2020entanglement, andreadakis_operator_2024} as the average operator entanglement generated when performing a unitary transformation of product operators, which can be directly related to probes of operator growth such as out-of-time-order correlations and local operator entanglement (Sec.~\ref{subsec:operator_growth}).

\section{Dynamical correlations}
\label{sec:corr_functions}

One of the most striking effects of dual-unitarity can be observed in the behaviour of spatiotemporal correlation functions of observables acting non-trivially only on one site, which we will refer to as 1-local from now on. Indeed, as shown in \cite{bertini_exact_2019}, these dynamical correlations are identically zero everywhere except on the edge of the causal light cone of the initial operator. 

To see that we consider a basis for the space of local operators $\{a^{\alpha}\}_{\alpha=0}^{q^2-1}$. We take these operators to be Hilbert-Schmidt orthonormalised as $\tr[(a^{\alpha})^{\dagger}a^{\beta}] = q \delta_{\alpha \beta}$ and choose ${a^0=\mathbbm{1}}$ such that all other $a^{\alpha}$ are traceless, i.e., ${\tr[a^{\alpha}]=0}$ if $\alpha \neq 0$. For a local two-dimensional Hilbert space, representing e.g. a qubit or a spin-$1/2$ degree of freedom, such an operator basis is provided by the Pauli matrices. The operators $a_x^{\alpha}$ provide a basis of local operators at site $x$, acting as $a^{\alpha}$ on site $x$ and as the identity everywhere else.

We now consider dynamical correlation functions of these local operators in the infinite temperature state. The latter are defined as
\begin{align}\label{eq:corr_functions_def}
c_{\alpha \beta}(x,y,t) =  \langle a_x^{\alpha} \mathbb{U}^{-t} a_y^{\beta} \mathbb{U}^{t}\rangle\,,
\end{align}
where the expectation value is taken w.r.t. the infinite-temperature state, $\expval{\bullet} \equiv \tr[\bullet ]/\tr[\mathbbm{1}]$.  The infinite temperature state is the natural choice of equilibrium state for unitary circuits since they do not support a description in terms of a local Hamiltonian, such that there is no conservation of energy. 

Let us consider a translationally invariant system where all gates are chosen to be identical, although the present result directly extends to the case of a circuit that is inhomogeneous in both space and time~\cite{bertini_exact_2019}. In this case the correlation functions simplify as 
\be
c_{\alpha \beta}(x,y,t)=\begin{cases}
c^{(++)}_{\alpha \beta}(x-y,t) & 2x\,\, {\rm even}; 2y\,\, {\rm even}\\ 
c^{(+-)}_{\alpha \beta}(x-y,t) & 2x\,\, {\rm odd}; 2y\,\, {\rm even}\\
c^{(-+)}_{\alpha \beta}(x-y,t) & 2x\,\, {\rm even}; 2y\,\, {\rm odd}\\ 
c^{(--)}_{\alpha \beta}(x-y,t) & 2x\,\, {\rm odd}; 2y\,\, {\rm odd}
\end{cases}\!\!,
\ee
where we accounted for the fact that, since the problem is invariant only under two-site shifts, correlations among integers and half-odd-integer sites can be different. For definiteness, from now on we will mostly focus on the case $x,y\in \mathbb Z$. 

\subsection{Correlations outside the causal light cone} 

Let us first illustrate how unitary circuits support a strict causal light cone. This property has also been referred to as \emph{quantum causal influence} in \cite{cotler_quantum_2019, vidal_class_2008,hastings_light-cone_2009,evenbly_algorithms_2009}.
The time-evolved operator $a_x^{\alpha}$ can be graphically represented as
\begin{align}
\label{eq:timeevolvingoperator}
\mathbb{U}^{t}a_x^{\alpha} \mathbb{U}^{-t} = \fineq[-0.8ex][.6][1]{
\foreach \j in {0,...,2}
{	
		\foreach \i in {0,...,3}
		{
			\roundgate[2*\i][2*\j][1][bottomright][blue6][-1]
			\roundgate[2*\i+1][2*\j+1][1][bottomright][blue6][-1]
		}
}
\foreach \i in {0,...,7}{
\draw[thick] (\i+.5,6)--(\i+.5,5.5);
\draw[thick] (\i+.5,6)--(\i+.5,5.5);
}		
\foreach \j in {3.75,...,5.75}
{	
		\foreach \i in {0,...,3}
		{
			\roundgate[2*\i][2*\j][1][topright][red6][-1]
			\roundgate[2*\i+1][2*\j-1][1][topright][red6][-1]
		}
}	
\charge[2.5][5.75][black]
\node[scale=1.7] at (3,5.75){$a^{\alpha}$};
}
\end{align}
where the loose lines on the left and right boundaries are connected to each other because of the periodic boundary conditions. This expression can be simplified by introducing the folded representation discussed in Sec.~\ref{sec:diagrams}, i.e., folding the backward sheet (blue) underneath the forward one (red) 
\begin{align}
\label{eq:folded_sigmat}
\mathbb{U}^{t}a_x^{\alpha} \mathbb{U}^{-t}=   
\fineq[-0.8ex][.6][1]{
\foreach \j in {3.75,...,5.75}
{	
		\foreach \i in {0,...,3}
		{
			\roundgate[2*\i][2*\j][1][bottomright][orange][1]
			\roundgate[2*\i+1][2*\j-1][1][bottomright][orange][1]
		}
}
\foreach \i in {0,...,7}{
\draw[thick] (\i+.5,6)--(\i+.5,5.75);
\draw[thick] (\i+.5,6)--(\i+.5,5.75);
\cstate[\i+.5][5.75]
}			
\charge[2.5][5.75][black]
\node[scale=1.5] at (3,5.75){$a^{\alpha}$};
},
\end{align}
where we introduced the local state associated to a local operator $a$ 
\be
\label{eq:statea}
\fineq[-0.8ex][1][1]{
\draw[thick] (-3.5,0.35) -- (-3.5,-0.35);
\charge[-3.5][0][black]; 
\node[scale=1] at (-3.2,0){$a$};
}
\longmapsto
\fineq[-0.8ex][1][1]{
\draw[thick] (-2,0.25) -- (-2,-0.251);
\draw[thick] (-1.5,0.4) -- (-1.5,-0.101);
\draw[thick] (-2,-0.25) to[out=-85,in=-80] ( (-1.505,-0.1);
\draw[thick, fill=black] (-1.75,-0.3) circle (0.1cm); 
\node[scale=1] at (-1.75,-0.55){$a$};
}
=
\fineq[-.4ex][1][1]{
\draw[very thick] (-0.15,0.25) -- (-0.15,-0.251);
\draw[thick, fill=black] (-.15,-0.25) circle (0.1cm); 
\node[scale=1] at (-.15,-0.55){$a$};
}\equiv \ket{a}\,.
\ee
We see that this folding corresponds to an operator-to-state mapping or vectorization. 
Repeatedly applying the unitarity relations in Eq.~\eqref{eq:unitarityfoldeddiagram}, we find that Eq.~\eqref{eq:folded_sigmat} simplifies to
\begin{align}
\mathbb{U}^{t}a_x^{\alpha} \mathbb{U}^{-t}= 
\hspace{-.75cm}
\fineq[-0.8ex][.6][1]{
\foreach \j [evaluate=\j as \jplus using {\j+1}] in {0,...,2}
{	
\foreach \i in {-\j,...,\j}
{
\roundgate[2*\i][2*\j][1][bottomright][orange][1]
}
\foreach \i  [evaluate=\i as \ieval using {\i+.5}] in {-\jplus,...,\j}
{
\roundgate[2*\ieval][2*\j+1][1][bottomright][orange][1]
}
}
\foreach \i in {0,...,5}{
\cstate[\i+.5][\i-.5]
\cstate[-\i-.5][\i-.5]
}			
\charge[-.5][-.5][black]
\node[scale=1.8] at (0,-.75){$a^{\alpha}$};
}
\hspace{-.75cm}.
\end{align}
The operator only acts nontrivially inside of a support that grows linearly with the number of time steps and acts as the identity (not reported in the diagram) everywhere outside the support of the causal light cone $|x-y| \leq t$. As a direct consequence, correlations outside this light cone for which $|x-y| > t$ factorise
\begin{align}
c^{(\pm\pm)}_{\alpha \beta}(x-y,t) = \frac{1}{q^{2L}}\tr\smash{\big[\mathbb{U}^{t}a_{0}^{\alpha} \mathbb{U}^{-t} \big]}\tr\smash{\big[a_{x-y}^{\beta} \big]} = 0,
\end{align}
vanishing identically whenever $\alpha \neq 0$ or $\beta \neq 0$. Note that for $\alpha=0$ or $\beta=0$ the dynamics is trivial. In the thermodynamic limit and at finite times, the dynamics is effectively restricted to a finite system size $L=2t$.
This property holds irrespective of dual-unitarity, and for generic unitary circuits the resulting geometric light cone $|x-y|=t$ presents an upper bound to the more physical Lieb-Robinson light cone.

\subsection{Correlations inside the causal light cone}  
\label{sec:correlationswithinLC}

As a consequence of the additional graphical identities in Eq.~\eqref{eq:spaceunitarityfoldeddiagram}, for dual-unitary circuits all correlations vanish \emph{also} whenever $|x-y| \leq t-1$. 
To see, this consider for concreteness the following correlation function 
\begin{align}
\hspace{-.25cm} {q^{4t}} c^{(++)}_{\alpha \beta}(t-1,t) = 
\hspace{-1.25cm}
\fineq[-0.8ex][.6][1]{
\foreach \j [evaluate=\j as \jplus using {\j+1}] in {0,...,2}
{	
\foreach \i in {-\j,...,\j}
{
\roundgate[2*\i][2*\j][1][bottomright][orange][1]
}
\foreach \i  [evaluate=\i as \ieval using {\i+.5}] in {-\jplus,...,\j}
{
\roundgate[2*\ieval][2*\j+1][1][bottomright][orange][1]
}
}
\foreach \i in {0,...,5}{
\cstate[\i+.5][\i-.5]
\cstate[-\i-.5][\i-.5]
}
\foreach \i in {-2,...,3}{
\cstate[2*\i-1.5][5.5]
\cstate[2*\i-0.5][5.5]
}			
\charge[-.5][-.5][black]
\node[scale=1.8] at (0,-.75){$a^{\alpha}$};
\charge[3.5][5.5][black]
\node[scale=1.8] at (4.05,5.75){$a^{\beta}$};
}
\hspace{-.75cm}.
\label{eq:corrdiag}
\end{align}
where the rightmost unitary gate can be removed using the spatial unitarity condition~Eq.~\eqref{eq:spaceunitarityfoldeddiagram} as
\begin{align}
\hspace{-.25cm} {\frac{q^{4t}}{q}} c^{(++)}_{\alpha \beta}(t-1,t) = 
\hspace{-1.25cm} 
\fineq[-0.8ex][.6][1]{
\foreach \j [evaluate=\j as \jplus using {\j+1}] in {0,...,1}
{	
\foreach \i in {-\j,...,\j}
{
\roundgate[2*\i][2*\j][1][bottomright][orange][1]
}
\foreach \i  [evaluate=\i as \ieval using {\i+.5}] in {-\jplus,...,\j}
{
\roundgate[2*\ieval][2*\j+1][1][bottomright][orange][1]
}
}	
\foreach \i in {-2,...,2}
{
\roundgate[2*\i][2*2][1][bottomright][orange][1]
}
\foreach \i  [evaluate=\i as \ieval using {\i+.5}] in {-3,...,1}
{
\roundgate[2*\ieval][4+1][1][bottomright][orange][1]
}
\foreach \i in {0,...,4}{
\cstate[\i+.5][\i-.5]
\cstate[-\i-.5][\i-.5]
}
\foreach \i in {-2,...,2}{
\cstate[2*\i-1.5][5.5]
\cstate[2*\i-0.5][5.5]
}			
\charge[-.5][-.5][black]
\node[scale=1.8] at (0,-.75){$a^{\alpha}$};
\charge[3.5][5.5][black]
\node[scale=1.8] at (4.05,5.75){$a^{\beta}$};
\cstate[4.5][4.5]
\cstate[-5.5][4.5]
}
\hspace{-.25cm}.
\end{align}
Repeatedly applying this identity along the rightmost edge, the full diagram simplifies to
\begin{align}
\hspace{-.25cm} \frac{q^{4t}}{q} c^{(++)}_{\alpha \beta}(t-1,t) = 
\hspace{-1.25cm} 
\fineq[-0.8ex][.6][1]{
\foreach \j [evaluate=\j as \jplus using {\j+1}] in {0,...,1}
{	
\foreach \i in {-\j,...,\j}
{
\roundgate[2*\i][2*\j][1][bottomright][orange][1]
}
\foreach \i  [evaluate=\i as \ieval using {\i+.5}] in {-\jplus,...,\j}
{
\roundgate[2*\ieval][2*\j+1][1][bottomright][orange][1]
}
}	
\foreach \i in {-2,...,2}
{
\roundgate[2*\i][2*2][1][bottomright][orange][1]
}
\foreach \i in {0,...,4}{
\cstate[\i+.5][\i-.5]
\cstate[-\i-.5][\i-.5]
}
\foreach \i in {-2,...,2}{
\cstate[2*\i-.5][4.5]
\cstate[2*\i+0.5][4.5]
}			
\charge[4.5][4.5][black]
\node[scale=1.8] at (5.05,4.75){$a^{\beta}$};
\charge[-.5][-1][black]
\node[scale=1.8] at (-1,-1){$a^{\alpha}$};
\draw[thick] (-.5,-1)--(.5,-1);
\cstate[.5][-1]
}
\hspace{-.75cm}.
\end{align}
Here the contraction between $\circ$ and $\bullet$ indicates $\tr[a^{\alpha}]=0$, causing the whole expression to vanish. 

\begin{figure}[t]
  \includegraphics[width=0.6\columnwidth]{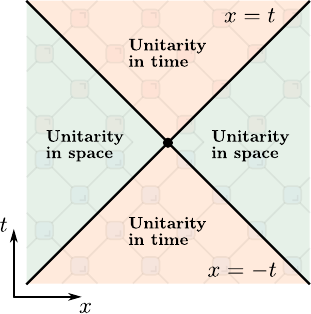}
  \caption{Dynamical correlations vanish outside a causal light cone along the vertical direction (orange) due to unitarity in time, and outside a causal light cone along the horizontal direction (green) due to unitarity in space, such that ultralocal correlations in dual-unitary circuits can only be nonzero on the overlapping edges of both causal light cones.
  \label{fig:lightcone}}
\end{figure}

Intuitively, unitarity along the time direction enforces a causal light cone along the vertical direction and unitarity along the space direction enforces a causal light cone along the horizontal direction, and all ultralocal correlations can only be nonzero on the (overlapping) edges of both causal light cones, see Fig.~\ref{fig:lightcone}.

\subsection{Correlations on the causal light cone} 

Let us now consider the case where $|x-y|=t$. Using unitarity and starting the contractions from the top left, we find that
\begin{align}\label{eq:temp_corr_onlc_0}
c^{(++)}_{\alpha \beta}(t,t) = \frac{1}{{q^{2t+1}} }\hspace{-.25cm}
\fineq[-0.8ex][.6][1]{
\foreach \i in {0,...,5}{
\roundgate[\i][\i][1][bottomright][orange][1]
\cstate[\i+.5][\i-.5]
\cstate[\i-.5][\i+.5]
}			
\charge[-.5][-.5][black]
\node[scale=1.8] at (0,-.75){$a^{\alpha}$};
\charge[5.5][5.5][black]
\node[scale=1.8] at (6.05,5.75){$a^{\beta}$};
}
\hspace{-.75cm}\,.
\end{align}
This expression does not vanish, but is rather the diagrammatic representation of 
\begin{align}
c^{(++)}_{\alpha \beta}(t,t) =\frac{1}{q} \tr\left[\mathcal{M}_+^{2t}(a^{\alpha})\, a^{\beta}\right],
\end{align}
in which we have defined the quantum channel
\begin{align}
\mathcal{M}_+(a)= \frac{1}{q}\tr_1\left[U^{\dagger}(a \otimes \mathbbm{1})\,U\right].
\label{eq:mapMplus}
\end{align}
The definition of this channel follows from unfolding Eq.~\eqref{eq:temp_corr_onlc_0}
\begin{align}
 \frac{1}{q} 
\fineq[-0.8ex][1][1]{
\foreach \i in {0,...,0}{
\roundgate[\i][\i][1][bottomright][orange][1]
\cstate[\i+.5][\i-.5]
\cstate[\i-.5][\i+.5]
}			
\charge[-.5][-.5][black]
\node[scale=1.1] at (-.75,-.65){$a$};
} = \frac{1}{q}\,\, \fineq[-0.8ex][1][1]{
\roundgate[0][.65][1][topright][red6][-1]
\roundgate[0][-.65][1][bottomright][blue6][-1]
\charge[-.5][0][black]
\draw[thick] (-.5,-.15)--(-.5,.15);
\draw[thick] (.5,-.15)--(.5,.15);
\draw[thick, rounded corners] (-.5,-1.15) -- (-.75,-1.15)--(-.75,1.15) -- (-.5,1.15);
\node[scale=1.1] at (-0.2,0){$a$};
}\,.
\end{align}
The map $\mathcal{M}_+(\cdot)$ is completely positive and trace-preserving, therefore it represents a proper quantum channel~\cite{nielsen2010quantum}. Furthermore, unitarity implies that this quantum channel maps the identity operator to itself, $\mathcal{M}_+(\mathbbm{1}) = \mathbbm{1}$, meaning that it is unital or doubly-stochastic~\cite{nielsen2010quantum}. Physically this channel accounts for the fact that all information that `leaks' inside the causal light cone becomes inaccessible to future light-cone correlation functions and is hence traced out. A similar expression written in terms of the channel
\be
\mathcal{M}_-(a)= \frac{1}{q}\tr_2\left[U^{\dagger}(\mathbbm{1}\otimes a )\,U\right],
\label{eq:mapMminus}
\ee
is obtained for correlations among operators at half-odd-integer sites, which move on the left edge of the causal light cone~\cite{bertini_exact_2019}, i.e., 
\be
c^{(--)}_{\alpha \beta}(-t,t)=\frac{1}{q} \tr\left[\mathcal{M}_-^{2t}(a^{\alpha})\, a^{\beta}\right].
\ee
The channels $\mathcal{M}_\pm(a)$ can be equivalently thought of as $q^2 \times q^2$ matrices via the folding mapping. When thinking about them as matrices we denote them as $\mathcal{M}_\pm$, with no explicit dependence on their argument.

Since any operator $a$ can be expanded in the (generalised) eigenvectors of  $\mathcal{M}_\pm$, the long-time dynamics of correlation functions is determined by their eigenvalues $\{\lambda_{\gamma}\}_{\gamma=0}^{q^2-1}$. For instance, assuming that $\mathcal{M}_+$ is diagonalisable we can write it, in vectorized notation, as 
\begin{align}
\mathcal{M}_+ = \sum_{\gamma=0}^{q^2-1} \lambda_{\gamma} \ket{R_{\gamma}}\bra{L_{\gamma}},
\end{align}
where $\ket{R_{\gamma}}$ and $\bra{L_{\gamma}}$ are the right and left eigenoperators corresponding to eigenvalue $\lambda_{\gamma}$. The light-cone correlations \eqref{eq:temp_corr_onlc_0} are then given by
\begin{align}
c^{(++)}_{\alpha \beta}(t,t) = \frac{1}{q}\sum_{\gamma=0}^{q^2-1} \lambda_{\gamma}^{2t} \braket{a^{\beta}}{R_{\gamma}}\braket{L_{\gamma}}{a^{\alpha}}\,,
\end{align}
where $\braket{A}{B}=\tr[A^\dagger B]$.
Due to unitality, $\mathcal{M}_+$ is guaranteed to have a trivial eigenvalue $\lambda_0=1$, with corresponding left and right eigenoperators given by the vectorisation of the identity, $\ket{R_0} = \ket{\mcirc} / \sqrt{q}$ and $\bra{L_0} = \bra{\mcirc}/\sqrt{q}$. The factors $\sqrt{q}$ here fix the normalization as $\braket{L_0}{R_0} = \tr[\mathbb{1}]/q = 1$. Graphically,
\be
\frac{1}{q}\,
\fineq[-0.8ex][1][1]{
\roundgate[0][0][1][bottomright][orange][1]
\cstate[.5][-.5]
\cstate[-.5][.5]			
\cstate[-.5][-.5]
}=
\fineq[-0.8ex][1][1]{		
\draw[thick] (-.25,-.25)--(.25,.25);
\cstate[-.25][-.25]
}, 
\qquad\frac{1}{q}\,
\fineq[-0.8ex][1][1]{
\roundgate[0][0][1][bottomright][orange][1]
\cstate[.5][-.5]
\cstate[-.5][.5]			
\cstate[.5][.5]
}=
\fineq[-0.8ex][1][1]{		
\draw[thick] (-.25,-.25)--(.25,.25);
\cstate[.25][.25]
}\,.
\ee
For this eigenvalue the left and right eigenoperators are identical, but this is generally not the case. Note that $\mathcal{M}_+$ can also be non-diagonalisable, giving rise to Jordan blocks and exceptional points as explicitly discussed in~\cite{hu_exact_2024}, but the long-time dynamics remains governed by the eigenvalues~\cite{bertini_exact_2019}.

Since the maps $\mathcal{M}_\pm(\cdot)$ are non-expanding all their eigenvalues are constrained to lie on the unit disc, i.e. $|\lambda_{\gamma}| \leq 1$. Note that $\mathcal{M}_\pm$ is generally neither Hermitian nor unitary and its eigenvalues are not constrained to be real or lie on the unit circle. If all nontrivial eigenvalues satisfy $|\lambda_{\gamma}| < 1$, we find that
\begin{align}
\lim_{t \to \infty} \mathcal{M}^{2t}(a) = \frac{1}{q} \ket{\mcirc} \braket{\mcirc}{a} = \frac{\mathbbm{1}}{q}  \tr[a],
\end{align}
such that for an initially traceless operator all correlations decay to zero, consistent with ergodic and mixing dynamics. 

\subsection{Ergodicity-based classification of dual-unitary gates}
\label{sec:ergodicitycorrelations}

Dual-unitary circuits can be classified based on the different degrees of ergodicity of 1-local observables, as set by the ergodicity of the corresponding quantum channels. We will only consider the behaviour along $x=t$, governed by the eigenvalues of $\mathcal{M}_+$, but this classification can be immediately extended to $x=-t$ and $\mathcal{M}_-$. In fact, since the underlying gates are generally not parity-invariant, one can find instances of `chiral' behaviour where $\mathcal{M}_{\pm}$ have different numbers of nontrivial eigenvalues and corresponding non-decaying eigenoperators. 

\begin{itemize}
\item[1.] \emph{Non-interacting}: All $q^2$ eigenvalues equal 1. All dynamical correlations remain constant.
\item[2.] \emph{Non-ergodic and interacting}: More than one but less than $q^2$ eigenvalues are equal to 1, some dynamical correlations decay to a non-zero constant.
\item[3.] \emph{Ergodic and non-mixing}: All nontrivial eigenvalues are different from 1, but there exists at least one nontrivial eigenvalue with unit magnitude. All correlations oscillate around a time-averaged value corresponding to the ergodic value.
\item[4.] \emph{Ergodic and mixing}: All nontrivial eigenvalues lie within the unit disc and all dynamical correlations decay (exponentially) to their ergodic value.
\item[5.] \emph{Maximally ergodic}: All $q^2-1$ nontrivial eigenvalues equal zero. Dynamical correlations decay to the ergodic value in finitely many steps.
\end{itemize}

\begin{figure}[tb!]
\includegraphics[width=\columnwidth]{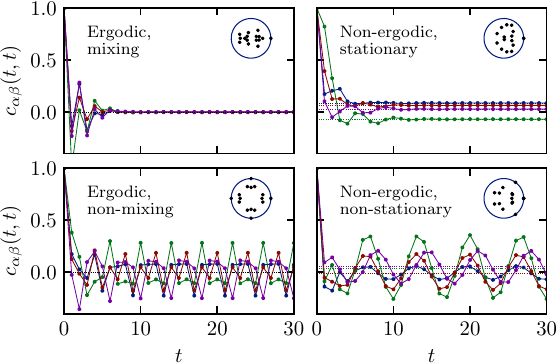}
\caption{Evolution of correlation functions $c_{\alpha \beta}(t,t)$, where $\sigma_{\alpha,\beta} \in \mathbbm{C}^{q \times q}$ are randomly generated matrices with $\tr(\sigma_{\alpha})=1$ and $\tr(\sigma_{\beta})=0$ leading to a thermal value $c_{\alpha \beta}(t,t) \to 0$. Local Hilbert space dimension $q=4$ and $4$ different operators are considered. Insets show the eigenspectrum of $\mathcal{M}_+$ on the unit disc. Based on a similar figure in \cite{claeys_ergodic_2021}.
\label{fig:dynamicalcorrelations}}
\end{figure}

Dual-unitary gates belonging to each of these different classes can be systematically constructed for general $q$~\cite{claeys_ergodic_2021}: some representative examples for the behaviour of the correlations are illustrated in Fig.~\ref{fig:dynamicalcorrelations}. Non-ergodic dynamics is realised by the Trotterised XXZ gate in Eq.~\eqref{eq:general_para_q2}, which reduces to non-interacting dynamics when this gate reduces to the SWAP gate. For `generic' dual-unitary circuits, i.e. without any fine-tuning in the parameterisation, the dynamics is typically ergodic and mixing. 

The maximally ergodic case is also known as \emph{Bernoulli dynamics} and was first identified in~\cite{aravinda_dual-unitary_2021}. A sufficient condition for attaining maximally ergodic dynamics is to take the local gate to be a perfect tensor (see Sec.~\ref{sec:relations}). Namely, impose the gate to satisfy 
\be
\fineq[-0.8ex][1][1]{
\roundgate[0][0][1][bottomright][orange][n]
\cstatep[.5][-.5][\sigma]
\cstatep[-.5][.5][\sigma]		
}=
\fineq[-0.8ex][1][1]{		
\draw[thick] (-.5,-.5)--(-.25,-.25);
\draw[thick] (.5,.5)--(.25,.25);
\cstatep[.175][.175][\sigma] 
\cstatep[-.175][-.175][\sigma] 
}, 
\qquad
\fineq[-0.8ex][1][1]{
\roundgate[0][0][1][bottomright][orange][n]
\cstatep[-.5][-.5][\sigma] 
\cstatep[.5][.5][\sigma] 
}=
\fineq[-0.8ex][1][1]{		
\draw[thick] (-.5,.5)--(-.25,.25);
\draw[thick] (.5,-.5)--(.25,-.25);
\cstatep[.175][-.175][\sigma] 
\cstatep[-.175][.175][\sigma] 
}\,,
\label{eq:diagonalunitarityfoldeddiagram}
\ee
in addition to Eqs.~\eqref{eq:unitarityfoldeddiagram} and~\eqref{eq:spaceunitarityfoldeddiagram}. 

The first identity directly implies that $\mathcal{M}_+$ is a projector on $\ket{\mcirc}$, such that the dynamical correlation functions reach their steady-state value after a single application of this quantum channel, and the second (equivalent) identity implies the same for $\mathcal{M}_-$. Note that these equations are too restrictive for $q=2$ (no two-qubit perfect tensors exists), but perfect tensors exist for any $q\geq 3$~\cite{huber2018bounds, rather2022thirty}. 
Still, even though there are no perfect tensors, maximally ergodic gates also exist for $q=2$: one can construct examples of maximally ergodic gates where all the non-trivial eigenvalues of $\mathcal{M}_+$ vanish but they are associated to a non-trivial Jordan block~\cite{aravinda_dual-unitary_2021}. This means that Eq.~\eqref{eq:diagonalunitarityfoldeddiagram} does not hold despite the maximally mixing property of the gate.  

Interestingly,~\cite{singh_ergodic_2024} have shown that all the ergodic classes discussed above can be realised considering only dual-unitary gates characterised by `local diagonal orthogonal invariance'~\cite{singh_diagonal_2022}, i.e., dual-unitary gates $U$ fulfilling
\be
U = (o\otimes o) U  (o^T\otimes o^T), 
\ee
where $o\in \mathrm{O}(d)$ is diagonal. Moreover, as pointed out by~\cite{aravinda_dual-unitary_2021}, for general dual-unitary circuits there is an interesting connection between the ergodicity of dynamical correlations and the entangling power of the local gate discussed in Sec.~\ref{sec:entpow}. If a gate has a sufficiently large entangling power, it is necessarily mixing. Defining the critical entangling power as
\begin{align}
e_P^* = \frac{q^2-2}{q^2-1},
\end{align}
any dual-unitary circuit which satisfies $e_P(U) > e_P^*$ is guaranteed to be mixing. For $e_P(U) \leq e_P^*$ the circuit can be mixing, ergodic, or nonergodic. Furthermore, if $e_P(U) = 0$ the circuit cannot be mixing, whereas $e_P(U) = 1$ if and only if $U$ is a perfect tensor and results in Bernoulli dynamics.
For qubits $q=2$ and $e_P^* = 2/3$. In this case the bound is trivial, however, since no two-qubit gates exist with entangling power larger than $2/3$. For $q>2$ gates with entangling power $0 \leq e_P(U) \leq 1$ can however be realised and this bound is nontrivial.

This bound can be proven by relating the entangling power to the Frobenius norm of the matrix representation of the quantum channel governing the correlations. Defining $\widetilde{\mathcal{M}}_+ = \mathcal{M}_+-\ket{\mcirc}\bra{\mcirc}/q$, the operator norm of this channel follows as
\begin{align}
\|\widetilde{\mathcal{M}}_+\|_{2}^2 = \|\mathcal{M}_+\|_2^2-1 = (q^2-1)(1-e_P(U)),
\end{align} 
where we denoted the Frobenius norm by $\|\cdot\|_{2}$, recalled Eq.~\eqref{eq:linearentropy} to set
\be
\|\mathcal{M}_+\|_2^2 = q^2(1-E[SU]),
\ee
and used Eq.~\eqref{eq:def_entanglingpowerDU}. The Frobenius norm bounds the largest eigenvalue $|\lambda_1|$, and we find that
\begin{align}
|\lambda_1| \leq \sqrt{1-e_P(U)} \sqrt{q^2-1}.
\end{align}
For an entangling power $e_P(U) > e_P^*$ the largest nontrivial eigenvalue satisfies $|\lambda_1|<1$ such that the circuit is necessarily mixing according to the above classification. In fact,~\cite{aravinda_dual-unitary_2021} noted that averaging over local unitaries one finds $|\lambda_1| \approx \sqrt{1-e_P(U)}$, removing the $\sqrt{q^2-1}$ factor. This also corresponds to the upper bound given (for a similar but not identical channel) by the `single-ring' theorem~\cite{vijaywargia2025quantum}.

To conclude, we mention that this way of calculating correlation functions on the light cone is not particular to dual-unitary circuits, but holds for general unitary circuits~\cite{claeys_maximum_2020} and in a wide class of kicked spin chains~\cite{prosen2007chaos,gutkin_exact_2020}. In non-dual-unitary circuits, however, light cone correlations are not the only non-vanishing ones.

\subsection{Correlations of extended operators}

The discussion of the above subsections can be generalised to consider extended operators. In particular, using the superscript $[s]$ to indicate that a given operator has support on $s$ qudits, we note that any extended operator of the form
\be
\ell_x^{[2s+1]} = \mathbbm{1}^{\otimes (2 x+1)} \otimes a^{[1]} \otimes O^{[2s]} \otimes \mathbbm{1}^{\otimes (2(L - x- s - 1))}, 
\ee
with $O^{[2s]}$ arbitrary and $a^{[1]}$ traceless, is \emph{left moving} in dual-unitary circuits. Namely we have 
\be
\!\!\!\langle \ell_x^{[2s+1]} \mathbb{U}^{-t} \ell_0^{[2s+1]}\mathbb{U}^{t}\rangle \!=\! \delta_{x+t,0} \langle \ell_{-t}^{[2s+1]} \mathbb{U}^{-t} \ell_0^{[2s+1]}\mathbb{U}^{t}\rangle\,.
\label{eq:leftmovinggen}
\ee 
Analogously  
\be
r_x^{[2s+1]} = \mathbbm{1}^{\otimes (2 x)} \otimes O^{[2s]} \otimes a^{[1]} \otimes \mathbbm{1}^{\otimes (2(L - x- s)-1)}, 
\ee
is \emph{right moving}, i.e., 
\be
\!\!\!\langle r_x^{[2s+1]} \mathbb{U}^{-t} r_0^{[2s+1]}\mathbb{U}^{t}\rangle \!=\! \delta_{x-t,0} \langle r_{t}^{[2s+1]} \mathbb{U}^{-t} r_0^{[2s+1]}\mathbb{U}^{t}\rangle\,.
\label{eq:rightmovinggen}
\ee
These facts can be readily established using the graphical simplifications described above. In fact, proceeding in this way one can show that the correlation functions in Eqs.~\eqref{eq:leftmovinggen} and~\eqref{eq:rightmovinggen} are respectively determined by powers of the many-body transfer matrices (quantum channels) $\mathcal{M}_{-,2s+1}$ and $\mathcal{M}_{+,2s+1}$, where we set  
\be
\begin{aligned}    
\mathcal{M}_{-,x} = &\fineq[1.5ex][.7][1]{
\begin{scope}[rotate around={-45:(0,0)}]
\foreach \i in {0,...,5}{
\foreach \j in {0,...,0}{
\roundgate[\i+\j][\i-\j][1][bottomright][orange][1]
}
}
\foreach \i in {0}{
\cstate[\i-.5][-\i-.5]	
\cstate[\i+5.5][-\i+5.5]
}
\end{scope}
\draw [decorate, thick, decoration = {brace}]   (7.25,-.75)--++(-7.5,0);
\node[scale=1.5] at (3.5,-1.25) {$x$};
},\\
\mathcal{M}_{+,x} = &\fineq[1.5ex][.7][1]{
\begin{scope}[rotate around={45:(0,0)}]
\foreach \i in {0,...,0}{
\foreach \j in {0,...,5}{
\roundgate[\i+\j][\i-\j][1][bottomright][orange][1]
}
}
\cstate[-.5][.5]	
\cstate[5.5][-5.5]
\end{scope}
\draw [decorate, thick, decoration = {brace}]    (7.25,-.75)--++(-7.5,0);
\node[scale=1.5] at (3.5,-1.25) {$x$};
},
\end{aligned}
\label{eq:AC}
\ee
such that $\mathcal{M}_{\pm,1}=\mathcal{M}_{\pm}$ (cf.\ Eqs.~\eqref{eq:mapMplus} and \eqref{eq:mapMminus}). 

Characterising these matrices becomes impractical for large $x$, as their dimension grows exponentially with $x$, however, one can again show that these matrices are non-expanding, i.e., their spectrum lies on the unit disc. Moreover,~\cite{foligno2024entanglement} showed that whenever a circuit with $q=2$ is generated by a DU gate drawn at random, these matrices have a unique eigenvalue one with probability one, i.e., they have a finite gap. Recalling the ergodicity based classification of Sec.~\ref{sec:ergodicitycorrelations}, this means that for $q=2$ dual-unitary circuits are almost always \emph{ergodic and mixing for all operators with finite support}.

\begin{figure}
\centering
\begin{tikzpicture}[baseline={([yshift=-0.6ex]current bounding box.center)},scale=.65]
\def\shift{8}
\def\shiftx{7}	
\draw[thick] (-.5,4.5) -- (-0.5,5.5);
\draw[thick] (-.5,3.5) -- (-0.5,2.5);
\draw[thick] (-.5,1.5) -- (-0.5,.5);
\draw[thick] (-.5,-.5) -- (5,5) -- (4.5,5.5);
\draw[thick] (1.5,-.5) -- (5,3) -- (2.5,5.5);
\draw[thick] (3.5,-.5) -- (5,1) -- (0.5,5.5);
\draw[thick] (0,0+2) -- (3.5,5.5);
\draw[thick] (0,0+4) -- (1.5,5.5);
\draw[thick] (0,0+2) -- (2.5,-0.5);
\draw[thick] (0,0+4) -- (4.5,-0.5);
\foreach \i in {0,...,2}{
\roundgate[0][2*\i][1][bottomright][green1][1]}
\draw[thick] (-.5,-1.5+\shift) -- (-0.5,-.5+\shift);
\draw[thick] (-.5,3.5+\shift) -- (-0.5,2.5+\shift);
\draw[thick] (-.5,1.5+\shift) -- (-0.5,.5+\shift);
\draw[thick] (-.5,-.5+\shift) -- (4.5,4.5+\shift);
\draw[thick] (.5,-1.5+\shift) -- (5,3+\shift) -- (3.5,4.5+\shift);
\draw[thick] (2.5,-1.5+\shift) -- (5,1+\shift) -- (1.5,4.5+\shift);
\draw[thick] (0,0+2+\shift) -- (2.5,4.5+\shift);
\draw[thick] (0,0+4+\shift) -- (0.5,4.5+\shift);
\draw[thick] (0,\shift) -- (1.5,-1.5+\shift);
\draw[thick] (0,0+2+\shift) -- (3.5,-1.5+\shift);
\draw[thick] (0,0+4+\shift) -- (5,-1+\shift) -- (4.5,-1.5+\shift);
\foreach \i in {0,...,2}{
\roundgate[0][2*\i+\shift][1][bottomright][green1][1]}
\foreach \i in {0,...,4}{
\cstate[\i+.5][-.5]}
\foreach \i in {0,...,4}{
\cstate[\i+.5][4.5+\shift]}
\charge[-0.5][4.5+\shift][black]
\charge[-0.5][-.5][black]
\node[scale=1.25] at (4.5,6.25){$\vdots$};
\node[scale=1.25] at (-.5,6.25){$\vdots$};
\node[scale=1.25] at (-.5,-1){$a$};
\node[scale=1.25] at (-0.5,5+\shift){$b$}; 
\node[scale=1.25] at (5.5,6){$=$}; 
\begin{scope}[rotate around={-45:(\shiftx,9)}]
\foreach \i in {0,...,3}{
\foreach \j in {0,...,4}{
\roundgate[\i+\j+\shiftx][\i-\j+9][1][bottomright][blue9][1]
}
}
\foreach \i in {0,...,4}{
\charge[5-.5+\shiftx][-5+.5+9][black]
\cstate[\i-.5+\shiftx][-\i-.5+9]	
\cstate[\i+3.5+\shiftx][-\i+3.5+9]	
}
\end{scope}
\draw[thick] (1.415+\shiftx,0+9-6.35) -- (1.415+\shiftx-1.415,0+8-6.35);
\draw[thick] (2*1.415+\shiftx,0+9-6.35) -- (1.415+\shiftx,0+8-6.35);
\draw[thick] (3*1.415+\shiftx,0+9-6.35) -- (2*1.415+\shiftx,0+8-6.35);
\charge[3*1.415+\shiftx][0+9-6.35-1][black]
\foreach \i in {0,...,3}{
\draw[thick] (\i*1.415+\shiftx,0+9-6.35-1) -- (\i*1.415+\shiftx,0+9-6.35-1.3);
}
\foreach \i in {0,...,3}{
\draw[thick] (\i*1.415+\shiftx,0+9+1.71-1) arc (180:0:0.15);
\draw[thick] (\i*1.415+\shiftx,0+9-6.65-1) arc (180:360:0.15);
}
\node[scale=1.25] at (9,11.5){$=$}; 
\roundgate[8][11.5][1][bottomright][blue9][1]
\roundgate[10][11.5+0.5][1][bottomright][green1][1]
\draw[thick] (9.5,11-1+0.5) -- (10.5,12-1+0.5);
\draw[thick] (9.5,12-1+0.5) -- (10.5,11-1+0.5);
\node[scale=1.25] at (7,2.25){$a$};
\node[scale=1.25] at (11.35,2.25){$b$};
\end{tikzpicture}
\caption{`Boundary Chaos': Equivalence between a boundary-perturbed SWAP circuit of size $L$ and duration $t$ and a $2$-dimensional network on a helix of size $t/L \times L$. We show an example of a 2-point time correlator between local boundary observables $a$ and $b$ in the operator-space/folded representation. Adapted from~\cite{fritzsch_boundary_2021}.}
\label{fig:BC}
\end{figure}

We remark that transfer matrices like those in Eq.~\eqref{eq:AC} also control the correlation functions of 1-local operators inside the light cone in general (non-dual unitary) unitary circuits. In this case $x$ is the distance from the light cone edge. Away from the dual-unitary point there are no general results characterising the gap of these matrices, however, a characterisation can be achieved for certain special choices of the gates. For instance~ \cite{claeys_correlations_2022} have shown that whenever the gates composing $\mathcal{M}_{\pm,x}$ are integrable one can use integrability techniques to characterise their spectrum. 

We also note that matrices like those in Eq.~\eqref{eq:AC} have been shown to control \emph{finite-volume} correlations in a setting known as {\em boundary chaos}~\cite{fritzsch_boundary_2021}. In the latter, one considers a brickwork circuit of SWAP gates $S$, with a single nontrivial 2-qudit gate $U$ placed at one of the boundaries --- this single gate generically suffices to make the whole circuit behave chaotically, as shown through a calculation of the spectral form factor in~\cite{boundary_chaos_2024}. In this setting, \cite{fritzsch_boundary_2021} have shown that a dynamical correlator of the form   
\be
\tr(b\,\mathbb{U}^{-t} a\,\mathbb{U}^t),
\ee 
can be mapped to a partition function on a ${t}/{L}\times L$ helix where $\mathcal{M}_{\pm,x}$ play the role of transfer matrices --- see Fig.~\ref{fig:BC} for a graphical depiction of the mapping. The fact that one can `integrate out' the free evolution between subsequent `collisions' with the boundary, and reduce the `quantum volume' of the circuit by a factor $L$ from $t\times L$ to $t$, is reminiscent of the idea of introducing the Poincar\'e map, a standard tool in studies of single-particle chaos (specifically in the so-called billiard systems). A similar approach can be used to access entanglement dynamics in these circuits~\cite{fritzsch_boundary_2023}.

\subsection{Spectral functions}
 
Autocorrelation functions of relevant observables, and their Fourier transforms known as \emph{spectral functions}, give information about the distribution of matrix elements in the basis of eigenstates of the time-evolution operator playing an important role in, e.g., the Eigenstate Thermalisation Hypothesis (ETH)~\cite{dalessio_quantum_2016}. For a given operator $A$ one defines the spectral function as 
\begin{align}
|f_A(\omega)|^2 = \int dt\, e^{i \omega  t} \expval{A(t)A},
\end{align}
where the expectation value is again taken w.r.t.\ the infinite-temperature state. The spectral function can alternatively be written in Lehmann representation as
\begin{align}
|f_A(\omega)|^2 = \sum_{m,n} |\langle m|A|n \rangle|^2\, \delta(\theta_m-\theta_n-\omega),
\end{align}
where the summation runs over the eigenstates of the unitary evolution operator, satisfying $\mathbb{U}\ket{n} = e^{i \theta_n}\ket{n}$.

This spectral function was studied for dual-unitary circuits in \cite{fritzsch_eigenstate_2021}.
For dual-unitary dynamics the autocorrelation function of 1-local observables vanishes identically after a single time step, i.e. $\expval{a_x^{\alpha}(t) a_x^{\alpha}} = \delta_{t,0}\expval{(a_x^{\alpha})^2}$. The resulting spectral function is constant. A more interesting behaviour is found when considering extensive sums of 1-local observables of the form
\begin{align}\label{eq:extensive_sum}
A_\nu= \frac{1}{\sqrt{L}} \sum_{x=1}^L a_{x+\nu/2}.
\end{align}
This operator acts nontrivially on either the integer ($\nu=0$) or half-odd-integer ($\nu = 1$) sites. For 1-local observables $a$ normalised as $\tr[a^{\dagger} a ]=1$, the initial operator is normalised as $\tr[A_\nu^{\dagger}A_\nu]=1$. The autocorrelation function of this operator in dual-unitary circuits follows directly as
\begin{align}
\expval{A_\nu(t)A_\nu} = \frac{1}{q} \tr\left[\mathcal{M}_{\nu}^{2t}(a)a\right],
\end{align}
where $\nu=0\,(1)$ corresponds to $\mathcal{M}_+ (\mathcal{M}_-)$.
A detailed analysis is possible when choosing $a= a_{\lambda}$ as a Hermitian eigenoperator of $\mathcal{M}_{\nu}$ with eigenvalue $\lambda$. Writing $|\lambda| = e^{-\gamma}$ with $\gamma > 0$, the autocorrelation function decays as $\expval{A_\nu(t)A_\nu} = e^{-\gamma t}$ (redefining $2t$ as $t$ for consistency with \cite{fritzsch_eigenstate_2021}). The corresponding spectral function follows as
\begin{align}
|f(\omega;a_{\lambda})|^2 = \frac{\sinh(\gamma)}{\cosh(\gamma)-\textrm{sgn}(\lambda) \cos(\omega)}\,.
\end{align}
For a slowly decaying operator with $\gamma \ll 1$ the spectral function is strongly peaked around $\omega =0$ ($\lambda >0$) or $\omega=\pi$ ($\lambda < 0$). For a fast decaying operator with $\gamma \gg 1$ the spectral function is essentially flat and $|f_A(\omega)|^2 \approx 1$, similar to the result for ultralocal observables with instantaneously decaying correlations.

\begin{figure}[tb!]
\includegraphics[width=0.9\columnwidth]{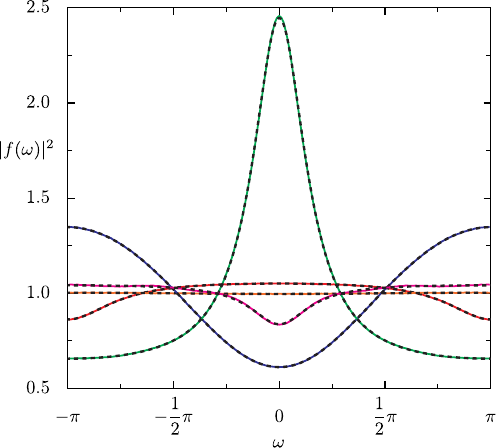}
\caption{Spectral function of an operator $A$ as an extensive sum of 1-local observables (cf.~Eq.~\eqref{eq:extensive_sum}) for five representative realizations of a dual-unitary circuit. The corresponding asymptotic expressions (cf.~Eq.~\eqref{eq:spectralfunction_asymptotic}) are depicted as dashed black lines. Adapted from~\cite{fritzsch_eigenstate_2021}.
\label{fig:spectralfunction}}
\end{figure}

For quantum channels with a single nonzero nontrivial eigenvalue an asymptotic form of the spectral function was proposed in \cite{fritzsch_eigenstate_2021} as
\begin{align}\label{eq:spectralfunction_asymptotic}
|f_A(\omega;a)|^2 = (1-\alpha)+ \alpha |f(\omega;a_{\lambda})|^2 
\end{align}
where $\alpha = \tr[\smash{a_{\lambda}^{\dagger}a}]$ is the overlap between $a$ and the eigenoperator $a_{\lambda}$ with nontrivial eigenvalue. Fig.~\ref{fig:spectralfunction} compares this asymptotic form with numerical results for a circuit of $2L=16$ lattice sites for five representative circuit realizations. These spectral functions show excellent agreement and cover the range of expected behaviors, from almost flat to strongly peaked.

The spectral function in dual-unitary circuits was also studied in \cite{pappalardi_general_2023} within the context of generalised ETH, as a numerical example showing how out-of-time-order correlation functions require correlations among matrix elements beyond ETH. \cite{suchsland_krylov_2023} used the spectral function of dual-unitary circuits to derive a representative example of the Krylov dynamics of generic unitary circuits.

\subsection{Nonergodic dual-unitary circuits}
\label{subsec:nonergodic}

Nonergodic systems are those where dual-unitarity imposes the strongest constraints. To see how this comes about we recall that nonergodicity on the level of 1-local observables means that there must exist another eigenoperator of $\mathcal{M}_{\nu}(\cdot)$, besides the identity, with unit-magnitude eigenvalue (see Sec.~\ref{sec:ergodicitycorrelations}).\footnote{In fact, $\phi\neq 0\pmod{2\pi}$ implies the slightly weaker property of non-mixing.} Namely
\begin{align}
\mathcal{M_{\nu}}(\tilde{a}) = e^{i\phi} \tilde{a}, 
\label{eq:solitonmap}
\end{align}
where $\tr[\tilde a]=0$ and $\phi\in \mathbb R$. Focusing on $\nu=+$ for concreteness, we have that \eqref{eq:solitonmap} is equivalent to 
\begin{align}
U^{\dagger}(\tilde{a} \otimes \mathbbm{1})U =  e^{i \phi} (\mathbbm{1} \otimes \tilde{a}),
\end{align}
because the conjugation with $U$ conserves the operator norm. Therefore, the full circuit dynamics acts on these operators as a simple translation to the right 
\begin{align}\label{eq:solitons_translation}
\mathbb{U}^{t}\tilde{a}_x \mathbb{U}^{-t} = e^{i2\phi t}\, \tilde{a}_{x+2t}.
\end{align}
Analogously, imposing $\mathcal{M_{-}}(\cdot)$ to have additional unit-magnitude eigenvalues produces operators that are translated to the left by the time evolution. These operators were first studied in~\cite{bertini_operator_ii_2020}, where they were termed \emph{solitons} (right or left moving) following the language of integrability. They have also been referred to as \emph{gliders} by~\cite{gombor_superintegrable_2022} following the language of Clifford circuits and quantum cellular automata~\cite{gutschow_time_2010,pivato_conservation_2002}. 
Explicit constructions of solitons in dual-unitary Clifford circuits were presented in \cite{sommers2023crystalline} and \cite{claeys_operator_2024}. 
\cite{bertini_operator_ii_2020} additionally showed that for $q=2$ any unitary circuit supporting such solitons is necessarily dual-unitary. 
Before continuing this discussion, we note that it is possible for a circuit to be ergodic on the level of 1-local observables but nonergodic for, e.g., multisite correlations, as explicitly discussed in~\cite{borsi_construction_2022}. Here we focus on circuits that are already nonergodic on the level of 1-local observables, consistent with Sec.~\ref{sec:ergodicitycorrelations}.

Eq.~\eqref{eq:solitons_translation} implies that  in the non-ergodic case $\phi=0$ one can construct local conserved charges of the form~\cite{bertini_operator_ii_2020}
\begin{align}
\sum_{y \in \mathbbm{Z}_L} \prod_{x_k} \tilde{a}_{x_k+y},
\end{align}
for any set of integers $\{x_k\}$. The number of such local conserved charges scales exponentially with the system size. Conversely, \cite{holden-dye_fundamental_2023} argued that any conserved quantity in dual-unitary circuits can be decomposed in linear combinations of finite-width solitons. 
The presence of a set of local conserved charges scaling super-extensively has led some authors to refer to nonergodic dual-unitary circuits as \emph{superintegrable}, see, e.g.,~\cite{gombor_superintegrable_2022, gombor_integrable_2024}. Importantly, however, this property does not imply Yang--Baxter integrability. In fact, nonergodic dual-unitary circuits are generically not Yang--Baxter-integrable~\cite{foligno2024nonequilibrium}. In essence, this is because the conservation laws are not sufficient to constrain the dynamics within the charge sectors, which are generically exponentially large. Yang-Baxter integrability is the generic situation only in the special cases where the charge sectors are all one-dimensional. More specifically, \cite{foligno2024nonequilibrium} showed that in nonergodic dual-unitary circuits with \emph{commuting} solitons the local gate $U$ can be decomposed as a direct sum over charge sectors. Explicitly, denoting by $n_\ltb/n_\rtb$ the number of linearly independent left/right moving solitons we have 
\be
U = \bigoplus_{\alpha=1}^{n_\rtb+1}  \bigoplus_{\beta=1}^{n_\ltb+1} U^{(\alpha,\beta)},
\label{eq:directsum}
\ee
where $U^{(\alpha,\beta)}$ are dual-unitary matrices\footnote{The blocks $U^{(\alpha,\beta)}$ implement unitary transformations with different domains and codomains. Since the latter are isomorphic, however, $U^{(\alpha,\beta)}$ can still be thought as matrices.} of dimension $q_\alpha^{{\rt}} q_\beta^{{\lt}}$ and ${q_\alpha^{(\ltb/\rtb)}> 0}$ are such that  
\be
\sum_{\alpha=1}^{n_\rtb+1} q_\alpha^{{\rt}} = \sum_{\beta=1}^{n_\ltb+1} q_\beta^{{\lt}}  = q\,. 
\ee
For generic choices of $U^{(\alpha,\beta)}$ the circuit is not Yang--Baxter integrable.

The simplest example of dual-unitary gate with solitons is the Trotterized XXZ circuit in Eq.~\eqref{eq:general_para_q2}. Indeed, it can be directly checked that
\begin{align}
V^{\dagger}(\sigma^z \otimes \mathbbm{1})V =  (\mathbbm{1} \otimes \sigma^z)\,.
\end{align}
In this case, however, $n_\rtb=n_\ltb=q_\alpha^{{\rt}}=q_\alpha^{{\lt}}=1$ and the blocks are all one-dimensional. This is consistent with the fact that the gate is indeed Yang--Baxter integrable. In fact, its connection with the XXZ Hamiltonian can be made explicit by noting that this gate can be expressed as
\begin{align}
    V \propto \exp\left[-i \frac{\pi}{4}\left(\sigma^x \otimes \sigma^x +\sigma^y \otimes \sigma^y + \Delta \sigma^z \otimes \sigma^z\right)\right],
\end{align}
up to an unimportant global phase, with anisotropy $\Delta = 1-4J/\pi$. This model was studied in different contexts, and will return multiple times throughout this review. For instance,~\cite{giudice2022temporal} characterised its temporal entanglement and~\cite{dowling2023scrambling} showed that its dynamics are scrambling but not chaotic on the level of the local-operator entanglement~\cite{bertini_operator_ii_2020}. Moreover,~\cite{rampp_hayden-preskill_2023} showed that in this system the Hayden-Preskill information recovery protocols exhibit dynamical signatures of integrability,~\cite{dowling_magic_2025} computed exactly its operator stabilizer R\'enyi entropies and~\cite{montana_lopez_exact_2024} its long-range stabilizer R\'enyi entropy,~\cite{alves_probes_2025} used it to establish certain dynamical predictions from free probability, and~\cite{vernier_integrable_2023} showed that the point $\Delta=1$ lies on a Trotter transition line. The eigenstates of the Trotterized XXZ circuit can generally be written as Bethe ansatz states~\cite{vanicat_integrable_2018,ljubotina2019ballistic}, but they take a particularly simple form if the model is additionally dual-unitary~\cite{alves_probes_2025}. The eigenstates can be constructed by observing that in this case the Floquet operator in Eq.~\eqref{eq:floquetoperator} can be factorised as a diagonal phase gate multiplied by a permutation circuit $\mathbb{S}$, where $\mathbb{S}$ corresponds to $\mathbb{U}$ with all gates given by SWAP gates. This permutation circuit additionally satisfies $\mathbb{S}^L = \mathbb{1}$, and the diagonalisation of the full Floquet operator reduces to the diagonalisation of $L \times L$ blocks for orbits of states obtained by the repeated action of $\mathbb{S}$ on an initial basis state. The full diagonalisation procedure is detailed in~\cite{alves_probes_2025}, which additionally identified a duality between the solitons and eigenstates. An even more special point corresponds to $\Delta=0$, where the system can be solved through a mapping to free fermions.

A more complicated example of non-ergodic dual-unitary circuit is obtained by considering $q=4$ and a local gate of the form~\cite{foligno2024nonequilibrium} 
\be
U = \begin{pmatrix}
U_A &&&\\
&&U_B&\\
&U_C&&\\
&&&U_D 
\end{pmatrix},
\label{eq:nonintegrableexample}
\ee
where $U_A, \ldots, U_D$ are $4 \times 4$ dual-unitary matrices. This gate has $n_{\ltb}=n_{\rtb}=1$ while both left and right moving solitons can be written as   
\be
\tilde a = {\rm diag}(1,1,-1,-1).   
\label{eq:asolitonnonintegrableexample}
\ee
This quantum circuit is not Yang--Baxter integrable for \emph{generic} choices of $U_A, \ldots, U_D$ (for instance if they are obtained by considering \eqref{eq:orbit} with $u_j$ drawn randomly from the Haar distribution). 

Other examples of Yang-Baxter integrable nonergodic dual-unitary gates have been constructed in 
\cite{gombor_superintegrable_2022} from solutions to the set theoretical (braid form of) Yang-Baxter equation, i.e.
\begin{align}
U_{12}U_{23}U_{12} = U_{23} U_{12} U_{23},
\end{align}
where $U_{ij}$ acts on sites $i$ and $j$. For such maps dual-unitarity is equivalent to the maps being non-degenerate. Such integrable dual-unitary maps that additionally satisfy $U_{12}^2=\mathbbm{1}$ were shown to induce a group action that is conjugate to the standard permutation group, i.e. dynamics generated by SWAP gates \cite{etingof_set-theoretical_1999}, which directly implies that the operator dynamics locally acts as a translation for the solitons. This connection was further extended in \cite{gombor_superintegrable_2022}, while \cite{gombor_integrable_2024} studied their deformations giving rise to Yang--Baxter integrable (but not superintegrable) spin chains.

To conclude this section, we note that it is also possible for dual-unitary circuits to exhibit weak ergodicity-breaking: \cite{logaric_quantum_2024} presented a systematic way of embedding quantum many-body scars in the spectrum of dual-unitary circuits.

\section{Quantum dynamics}
\label{sec:quantumdynamics}

In this section, we see how dual-unitarity can be used to characterise out-of-equilibrium dynamics. This includes \emph{state} or \emph{Schr\"odinger-picture} dynamics, i.e., quantum quench problems~\cite{calabrese2006time, calabrese2007quantum}, as well as \emph{operator} or \emph{Heisenberg-picture} dynamics, e.g. out-of-time-order correlation functions~\cite{larkin1969quasiclassical, lashkari2013towards, shenker2014multiple, shenker2014black, maldacena2016bound, roberts2018operator, swingle_unscrambling_2018}, and various kinds of operator entanglement~\cite{zanardi2001entanglement, prosen2007is, prosen2007operator, pizorn2009operator, hosur2016chaos}. The techniques used in these calculations tend to be largely graphical and differ from the usual approaches for systems in which exact results for the dynamics can be obtained through, e.g., mapping to free fermions~\cite{essler_quench_2016} or averaging over random variables~\cite{fisher_random_2023}. In most cases, however, the graphical derivations need to be supplemented with specific information about the choice of gate or ensemble of gates. This is intuitively given the fact that any calculation that can be done purely graphically should hold for any choice of dual-unitary gates, e.g.\ both for noninteracting SWAP gates and for generic ergodic gates, whereas the dynamical properties of these gates should clearly differ.  

\subsection{State dynamics}
\label{sec:statedynamics}

\subsubsection{Solvable states} A key concept in the study of state dynamics in dual-unitary circuits is that of \emph{solvable states}, whose dynamics can be characterised exactly in many of their aspects. Solvable states are a class of non-equilibrium initial states compatible with the dual unitarity condition. Intuitively, this means that they do not break the unitarity of the spatial evolution. More specifically, one can imagine to represent observables measured on a time-evolving state as the tensor network in Eq.~\eqref{eq:timeevolvingoperator}, where the top and bottom lines are contracted to some tensors representing the initial state. The solvable initial states are those characterised by tensors allowing to carry on the spatial-unitarity based simplifications discussed in Sec.~\ref{sec:correlationswithinLC}. At its heart this idea is similar to integrable (boundary) states in the context of integrability~\cite{ghoshal1994boundary, piroli2017what}. In this subsection we define the family of solvable states introduced by~\cite{piroli_exact_2020} and briefly describe its extension for nonergodic dual-unitary circuits put forward by~\cite{foligno2024nonequilibrium}. 

We begin by considering states in matrix product state (MPS) form, defined in terms of $q^2$ $\chi\times \chi$ matrices $\mathcal{M}^{i,j}$ as
\begin{align}
\!\!\ket{\Psi_0(\mathcal{M})}	\!=\!\!\!\!\!\! \sum_{i_1,\ldots, i_{2L}=1}^q \!\!\!\!\!\!\!\!\tr[{\mathcal{M}^{i_1,i_2}\!\!\!\!\ldots\mathcal{M}^{i_{2L-1},i_{2L}}}]\!\!\ket{i_1,\ldots, i_{2L}},
\label{eq:twositeMPS}
\end{align}
and denote its transfer matrix by  
\begin{align}
		\tau(\mathcal{M})=\sum_{i,j} \mathcal{M}^{i,j} \otimes \left(\mathcal{M}^{i,j}\right)^*\,. 
\label{eq:transfermatrix}
\end{align}
It also is useful to introduce the $q\chi \times q\chi$ matrix $\Gamma_\mathcal{M}$ whose matrix elements are specified by $\{\mathcal{M}^{i,j}\}$ as follows 
\begin{align}
\label{eq:matrixGamma}
\mel{i \alpha}{\Gamma_\mathcal{M}}{j \beta}= \sqrt{q} \mel{\alpha}{\mathcal{M}^{ij}}{\beta}. 
\end{align}
We consider MPS tensors $\{\mathcal{M}^{i,j}\}$ defined on pairs of adjacent sites rather than single sites as they better fit the brickwork geometry of Eq.~\eqref{eq:floquetoperator}. Namely, they directly encode the natural translational invariance of a homogeneous brickwork circuit, which is invariant under two-site shifts rather than one-site ones. We also recall that two MPS $\ket{\Psi_0(\mathcal{M})},\ket{\Psi_0(\mathcal{N})}$ are said to be \emph{equivalent} if, for every operator $\mathcal O$ with a finite support,
\begin{align}
\!\!\lim_{L\rightarrow\infty}	\!\!\mel{\Psi_0(\mathcal{N})}{\mathcal O}{\Psi_0(\mathcal{N})} 
\!-\!\! \mel{\Psi_0(\mathcal{M})}{\mathcal O}{\Psi_0(\mathcal{M})}=0. 
\end{align}

Having established the relevant framework we are now in a position to introduce solvable states. The MPS $\ket{\Psi_0(\mathcal{M})}$ is \emph{solvable} if the transfer matrix $\tau(\mathcal{M})$ has a unique maximal eigenvalue $1$ and there exists an operator $S$ acting on the auxiliary space such that
\be
\Gamma^\dagger_{\mathcal{M}} (\mathds{1}_q \otimes S) \Gamma_{\mathcal{M}}= \mathds{1}_q \otimes S. 
\label{eq:solvabilityformula} 
\ee
Starting from this definition~\cite{piroli_exact_2020} proved the following useful result.
\begin{theorem}[Equivalent MPS \cite{piroli_exact_2020}]
\label{th:solvabilityclassification}
For any given solvable MPS $\ket{\Psi_0(\mathcal{M})}$ one can always find an equivalent solvable MPS $\ket{\Psi_0(\mathcal{N})}$ such that 
\begin{align}
\Gamma^\dagger_{\mathcal{N}} \Gamma^{\phantom{\dagger}}_{\mathcal{N}}= \mathds{1}_\chi.
\label{eq:equivalentsolvability}
\end{align}
\end{theorem}

In essence, Theorem~\ref{th:solvabilityclassification} guarantees that one can always replace a solvable MPS with an  equivalent one where the matrix in Eq.~\eqref{eq:matrixGamma} is \emph{unitary}. Note that due to the unitarity of $\Gamma_{\mathcal{N}}$ one can immediately write the fixed points of $\tau(\mathcal{M}) $ as
\begin{align}
\bra{\mcirc_{\chi}}=\sum_{i=1}^{\chi} \bra{i} \otimes \bra{i}, \quad \ket{\mcirc_{\chi}}=\sum_{i=1}^{\chi} \ket{i} \otimes \ket{i}.
\label{eq:identitystate}
\end{align}	

Following Sec.~\ref{sec:diagrams}, we introduce the following graphical representation for the replicas of $\Gamma_{\mathcal{N}}$
\begin{align}
\left( \Gamma_{\mathcal{N}} \otimes \Gamma_{\mathcal{N}}^*\right)^{\otimes n} = \fineq[-0.8ex][1][1]{\MPSinitialstate[0][0][orange][topright][n]}, 
\label{eq:MPSstatediag}
\end{align}
where the thick line represents the replicated auxiliary space while the thin ones represent the replicated physical space. The condition in Eq.~\eqref{eq:equivalentsolvability} is then represented as 
\begin{align}
\!\!\!\!\!\!\!\!\!\fineq[-0.8ex][1][1]{
	\MPSinitialstate[0][0][orange][topright][n]
	\cstatep[.65][.65][\sigma] 
	\cstatep[.7][0][\sigma]
}&=
\fineq[-0.8ex][1][1]{
	\draw[thick] (.65,.65)--++(-1,0);
	\draw[very thick] (.7,0)--++(-1,0);
	\cstatep[.65][.65][\sigma] 
	\cstatep[.7][0][\sigma]
	 },\qquad\!\!
\fineq[-0.8ex][1][1]{
	\MPSinitialstate[0][0][orange][topright][n]	
	\cstatep[-.65][.65][\sigma]
	\cstatep[-.7][0][\sigma]
	 }=\fineq[-0.8ex][1][1]{
	\draw[ thick] (.65,.65)--++(+1,0);
	\draw[very thick] (.7,0)--++(+1,0);
	\cstatep[.65][.65][\sigma] 
	\cstatep[.7][0][\sigma]
	}, 
	\label{eq:solvabilitydiagram}
\end{align}
so that the tensor $\Gamma_{\mathcal{N}}$ fulfils the spatial unitary condition in Eq.~\eqref{eq:spaceunitarityfoldeddiagram}. As shown in Sec.~\ref{sec:subsystemrelaxation}, this property implies that all solvable states locally relax to the infinite temperature state regardless of the underlying dynamics.

The simplest set $\{\mathcal{N}^{i,j}\}$ fulfilling the solvability conditions are obtained for $\chi=1$ and 
\be
\mathcal{N}^{i,j} = \frac{1}{\sqrt{q}}\delta_{i,j}\,.
\label{eq:Bell}
\ee
This corresponds to a state composed by tensor products of Bell pairs among nearest neighbours, which we refer to as a `Bell-pair state'. In fact, any solvable state with $\chi=1$ can be written as a Bell-pair state on which an arbitrary product of one-qudit unitaries has been applied. 
This spatial unitarity of Bell-pair states also directly underpins their use in quantum teleportation~\cite{bennett_teleporting_1993}.
Note that, although our definition in Eq.~\eqref{eq:twositeMPS} implies that generic solvable states with $\chi>1$ are two-site shift invariant, this procedure allows one to define inhomogeneous solvable states with $\chi=1$. Inhomogeneous solvable states with $\chi =q^{2m}$ can be constructed by applying $m$ inhomogeneous time-evolution operators \eqref{eq:floquetoperator} on the Bell-pair state. For this special family of solvable states the spectrum of $\tau(\mathcal{M})$ is $\{0,1\}$.

\cite{foligno2024nonequilibrium} have shown that in the case of nonergodic dual-unitary circuits one can extend the family of solvable states by imposing a different solvability condition in each charge sector. This extended family has been dubbed \emph{charged solvable states} and generically shows relaxation to generalised Gibbs ensembles~\cite{dalessio_quantum_2016, vidmar2016generalised}. For example, for the the circuit generated by the nonergodic gate in Eq.~\eqref{eq:nonintegrableexample} one has a family of solvable states with $\chi=1$ and the following tensors~\cite{foligno2024nonequilibrium} 
\be
\mathcal{N}^{i,j} =\frac{1}{2} \left ( \sqrt{\alpha}\, \delta_{ij} + \sqrt{1-\alpha}\, \tilde a_{ij} \right)\,, \quad \alpha\in[0,1]\,,
\label{eq:chargedsolvablestates}
\ee
where $\tilde a$ is given in Eq.~\eqref{eq:asolitonnonintegrableexample}. One can explicitly verify that the state defined by these tensors does not fulfil the regular solvability condition in Eq.~\eqref{eq:solvabilityformula}.

\subsubsection{Subsystem relaxation}
\label{sec:subsystemrelaxation}

As demonstrated by~\cite{piroli_exact_2020}, whenever dual-unitary circuits are prepared in a solvable state the conditions in Eq.~\eqref{eq:solvabilitydiagram} allow to characterise many aspects of their non-equilibrium dynamics exactly. 

To illustrate how this happens let us consider the simplest setting of a subsystem $A$ consisting of a single contiguous block of $2 L_A$ qudits\footnote{For definiteness, we take $A$ to begin at a half-odd-integer site and comprise an even number of qudits. Other cases are treated in an analogous fashion.}. At time $t$ the density matrix reduced to $A$ can be represented diagrammatically as 
\begin{align}
\hspace{-.35cm} \rho_A(t) \!= \frac{1}{q^L}
\fineq[-0.8ex][.55][1]{
\foreach \j in {3,...,4}
{	
		\foreach \i in {1,...,5}
		{
			\roundgate[2*\i][2*\j+1.5][1][bottomright][orange][1]
			\roundgate[2*\i+1][2*\j+.5][1][bottomright][orange][1]
		}
}
\draw[very thick] (2,5.45)--++(9.5,0);
\foreach \i in {1,...,5}{
\MPSinitialstate[2*\i][5.5][orange][topright][1]
}
\foreach \i in {8,..., 11}{
\cstate[\i-0.5][10]			
}
\draw [decorate, thick, decoration = {brace}]   (1.25,10.125)--++(5.5,.0);
\node[scale=1.75] at (4,10.6) {$2 L_A$};
\draw [decorate, thick, decoration = {brace}]   (7,10.125)--++(4,.0);
\node[scale=1.75] at (9.25,10.6) {$2L-2L_A$};
}. 
\end{align}
Here we used the diagrammatic representation in Eq.~\eqref{eq:MPSstatediag} to write the initial MPS as the bottom row of this diagram (with appropriate normalisation) and the loose lines on the sides are connected to each other because of the periodic boundary conditions. Focussing on the limit where the rest of the system, $2L-2L_A$, is much larger than $t$ (we actually need only $2L \geq 2 L_A + 4 t$ because of the causal light cone) this diagram can be simplified using the unitarity conditions in Eq.~\eqref{eq:unitarityfoldeddiagram}. The result reads as follows 
\be
\label{eq:rhoredgen}
\rho_A(t)=\frac{1}{q^L}\hspace{-1.75cm}
\fineq[-0.8ex][.55][1]{
\foreach \j [evaluate=\j as \jplus using {\j+1}] in {1,...,2}
{	
\foreach \i in {-\j,...,\j}
{
\roundgate[2*\i][4-2*\j][1][bottomright][orange][1]
}
\foreach \i  [evaluate=\i as \ieval using {\i+.5}] in {-\jplus,...,\j}
{
\roundgate[2*\ieval][4-2*\j-1][1][bottomright][orange][1]
}
}
\draw[very thick] (-8.8,-2.05)--++(15,0);
\foreach \i in {-4,...,3}{
\MPSinitialstate[2*\i][-2][orange][topright][1]
}
\foreach \i in {-1,...,2}{
\cstate[\i-5.5][\i-.5]
\cstate[-\i+5.5][\i-.5]
}
\cstate[-8.5][-1.5]
\cstate[-7.5][-1.5]			
}
\hspace{-1cm},
\ee
where the MPS transfer matrix 
\be
\label{eq:MPSTMdiag}
\tau(\mathcal N) = \frac{1}{q} \fineq[-0.8ex][1][1]{\MPSinitialstate[0][0][orange][topright][1]
\cstate[-.5][0.5]
\cstate[.5][0.5]
},
\ee
is repeated $L-L_A-2t$ times along the horizontal line at the bottom. 

The simplification performed so far only relies on the unitarity of the gate and does not require dual-unitarity nor the solvability condition of the initial state: the latter conditions will be used from now on. 

First, we note that the conditions in Eq.~\eqref{eq:solvabilitydiagram} imply that the state in Eq.~\eqref{eq:identitystate} is an eigenstate of $\tau(\mathcal N)$ corresponding to eigenvalue $1$, which is unique and maximal by the definition of solvable states. This means that, in the thermodynamic limit, the reduced density matrix is represented by the following diagram  
 \be
 \label{eq:simplifiedrhoA}
\frac{1}{\chi q^{L_A+2t}}\hspace{-.5cm}
\fineq[-0.8ex][.55][1]{
\foreach \j [evaluate=\j as \jplus using {\j+1}] in {1,...,2}
{	
\foreach \i in {-\j,...,\j}
{
\roundgate[2*\i][4-2*\j][1][bottomright][orange][1]
}
\foreach \i  [evaluate=\i as \ieval using {\i+.5}] in {-\jplus,...,\j}
{
\roundgate[2*\ieval][4-2*\j-1][1][bottomright][orange][1]
}
}
\draw[very thick] (-6.5,-2.05)--++(12,0);
\foreach \i in {-3,...,3}{
\MPSinitialstate[2*\i][-2][orange][topright][1]
}
\foreach \i in {-1,...,2}{
\cstate[\i-5.5][\i-.5]
\cstate[-\i+5.5][\i-.5]
}
\cstate[-6.5][-2.05]
\cstate[6.5][-2.05]			
}
\hspace{-1cm},
\ee
 where the circles on the auxiliary space line denote the state in Eq.~\eqref{eq:identitystate}. We now observe that whenever $t\geq L_A/2$ the diagram can be fully contracted by repeated use of the solvability conditions in Eq.~\eqref{eq:solvabilitydiagram}, the spatial-unitarity conditions in Eqs.~\eqref{eq:spaceunitarityfoldeddiagram}, and, finally, the second of the unitarity conditions in Eq.~\eqref{eq:unitarityfoldeddiagram}. The result reads as  
\be
\hspace{-.25cm}\lim_{L\to\infty} \rho_A(t) =  \frac{1}{q^{2 L_A}} \fineq[-0.8ex][.55][1]{
\foreach \i in {0,...,5}{
\draw[thick] (\i,0)--++(0,.5);
}
\foreach \i in {0,...,5}{
\cstate[\i][0]
}			
}= \left(\frac{\mathds{1}}{q}\right)^{\otimes 2 L_A}\hspace{-.5cm}. 
\label{eq:stationarystateA}
\ee
This result has three remarkable features. First, it represents a proof that the reduced state of a subsystem approaches a stationary state --- the infinite temperature state --- at large times after a quantum quench. This statement is generically extremely hard to prove in interacting quantum many-body systems and, besides dual-unitary circuits evolving from solvable states, there are only a handful of known examples where an analogous rigorous proof can be achieved~\cite{klobas2021exact, klobas2021exact2, yu_hierarchical_2024, bertini_east_2024, liu_solvable_2023, wang2024exact, wang2024hopf, foligno2024nonequilibrium}. Second, it implies that relaxation occurs in a finite number of steps. This result is in contrast to what happens in generic systems, where there are exponential corrections and, strictly speaking, the stationary state is only reached in the infinite time limit. This fact can be explained by noting that in the setting under discussion the bath created by the rest of the system on the subsystem $A$ is \emph{purely Markovian} --- this property has led \cite{lerose2021influence} to refer to dual-unitary circuits as `perfect dephasers'. Finally, Eq.~\eqref{eq:stationarystateA} shows that in any dual-unitary circuit initialised in a solvable state, local subsystems relax to the infinite temperature state. This is true regardless of the nature of their dynamics, i.e., it also holds for non-ergodic dual-unitary circuits. The difference between the two is that in ergodic dual-unitary circuits Eq.~\eqref{eq:stationarystateA} is expected to be asymptotically true for all states~\cite{foligno_growth_2023}, while in non-ergodic cases when evolving from non-solvable initial states subsystems relax to exotic generalised Gibbs ensembles. This has been proven exactly by \cite{foligno2024nonequilibrium} in the case of charged solvable initial states (e.g.\ the state in Eq.~\eqref{eq:chargedsolvablestates}).

\subsubsection{Finite-time dynamics}
\label{subsec:entanglement_dyn}

The remarkable simplifications leading to Eq.~\eqref{eq:stationarystateA} do not exhaust what is possible in dual-unitary circuits evolving from solvable states. As shown by~\cite{piroli_exact_2020}, the diagram in Eq.~\eqref{eq:simplifiedrhoA} can also be simplified when $t\le L_A/2$, i.e., dual-unitarity and solvability can also be used to access finite-time dynamics. 

Specifically, by repeated application of the solvability conditions in Eq.~\eqref{eq:solvabilitydiagram} and the spatial-unitarity conditions in Eqs.~\eqref{eq:spaceunitarityfoldeddiagram} one finds 
 \be
 \label{eq:simplifiedrhoADU}
\hspace{-.45cm}\lim_{L\to\infty} \!\!\rho_A(t)\!=\!
\frac{1}{\chi q^{L_A+2t}}\hspace{-.45cm}
\fineq[-0.8ex][.55][1]{
\foreach \j [evaluate=\j as \jplus using {\j+1}] in {1,...,1}
{	
\foreach \i in {-\jplus,...,\jplus}
{
\roundgate[2*\i][2*\j+2][1][bottomright][orange][1]
}
\foreach \i  [evaluate=\i as \ieval using {\i+.5}] in {-\jplus,...,\j}
{
\roundgate[2*\ieval][2*\j+1][1][bottomright][orange][1]
}
}
\draw[very thick] (-2.5,1.95)--++(5,0);
\foreach \i in {-1,...,1}{
\MPSinitialstate[2*\i][2][orange][topright][1]
}
\foreach \i in {0,1}{
\cstate[\i+3.5][\i+2.5]
\cstate[-\i-3.5][\i+2.5]
}
\cstate[-2.5][1.95]
\cstate[2.5][1.95]			
}
\hspace{-.25cm}.
\ee
This is another significant result: it shows that in the thermodynamic limit the reduced density matrix at time $t$ is \emph{unitarily equivalent} to 
\be
\label{eq:sigmaADU}
\sigma_A(t) =  \left(\frac{\mathds{1}}{q}\right)^{\otimes 2 t} \otimes \lim_{L\to\infty} \rho_{A\setminus\{4t\}}(0) \otimes \left(\frac{\mathds{1}}{q}\right)^{\otimes 2 t}\hspace{-.5cm},
\ee
where $A\setminus\{4t\}$ is obtained from $A$ by removing the first and last $2t$ qudits and we noted 
\be
\lim_{L\to\infty} \!\!\rho_{[1,x]}(0) = \frac{1}{\chi q^{x}} \fineq[-1.7ex][.65][1]{
\draw[very thick] (-2.5,1.95)--++(5,0);
\foreach \i in {-1,...,1}{
\MPSinitialstate[2*\i][2][orange][topright][1]
}
\cstate[-2.5][1.95]
\cstate[2.5][1.95]	
\draw [decorate, thick, decoration = {brace}]   (-2.5,2.55)--++(5,0);
\node[scale=1.25] at (0,3) {$2x$};		
},
\ee
which again follows from the properties of the transfer matrix $\tau(\mathcal N)$. By unitarily equivalent we mean that, as one can verify by direct inspection, Eqs.~\eqref{eq:simplifiedrhoADU} and \eqref{eq:sigmaADU} can be obtained from one another by conjugation with the unitary matrix 
\be
W =
\fineq[-0.8ex][.55][1]{
\foreach \j [evaluate=\j as \jplus using {\j+1}] in {1,...,1}
{	
\foreach \i in {-\jplus,...,\jplus}
{
\roundgate[2*\i][2*\j+2][1][topright][red6][-1]
}
\foreach \i  [evaluate=\i as \ieval using {\i+.5}] in {-\jplus,...,\j}
{
\roundgate[2*\ieval][2*\j+1][1][topright][red6][-1]
}
}
\draw[ thick] (4.5,2.5)--++(0,1);
\draw[ thick] (-4.5,2.5)--++(0,1);
}\,. 
\label{eq:matrixW}
\ee
Since conjugation does not affect the spectrum, one can read off the spectrum of $\lim_{L\to\infty}\rho_A(t)$ directly from Eq.~\eqref{eq:sigmaADU}. This is remarkable as the latter quantity --- also referred to as the `entanglement spectrum'~\cite{li2008entanglement} --- completely characterises the dynamics of the entanglement between $A$ and its complement. Specifically, noting that $\lim_{L\to\infty}\rho_{x}(0)$ has at most $\chi=O(t^0)$ non-zero eigenvalues we have that 
\be
\label{eq:renyisolvablestate}
\!\!\!  \lim_{L\to\infty}\! S^{(\alpha)}(\rho_A(t))= \min(4t, 2 L_A) \log q + O(t^0),  \quad \forall \alpha,
\ee
where the symbol $S^{(\alpha)}(\rho)$ denotes the $\alpha$-R\'enyi entropy of $\rho$, i.e., 
\be
S^{(\alpha)}_A(\rho)=\frac{1}{1-\alpha}\log[\rho^\alpha]\,, \qquad \alpha\in \mathbb R\,, 
\label{eq:Srenyidef}
\ee 
which is a well-known bipartite entanglement monotone~\cite{vidal2000entanglement} and recovers the standard measure of bipartite entanglement --- the entanglement entropy --- in the limit $\alpha\to1$~\cite{amico2008entanglement}. In fact~\cite{piroli_exact_2020} showed that the solvability conditions imply
\be
\label{eq:sloperenyisolvablestate}
\lim_{L_A\to\infty}\lim_{L\to\infty}\!S^{(\alpha)}(\rho_A(t)) = 4t \log q + 2\log\chi, \quad \forall \alpha. 
\ee
A particularly simple result is obtained for the Bell pair state in Eq.~\eqref{eq:Bell}, where $\rho_{x}(0)$ is a projector and one has 
\be
\label{eq:Bellpairentanglement}
S^{(\alpha)}(\rho_A(t)) = \min(4t, 2 L_A) \log q, \quad L\geq L_A+2t\,. 
\ee
The growth of R\'enyi entropies described by this formula saturates the minimal cut bound formulated by~\cite{casini2016spread}, meaning that Eq.~\eqref{eq:Bellpairentanglement} describes the fastest growth allowed by the local structure of the time-evolution operator in Eq.~\eqref{eq:floquetoperator}. Dual-unitary circuits are said to have maximal \emph{entanglement velocity} (rate of entanglement growth). Remarkably, \cite{zhou_maximal_2022} showed that also the converse is true: if a brickwork circuit has maximal entanglement velocity and hence produces the maximal growth of entanglement entropy in a given time step, then it must be dual-unitary. Note that states might exist that do not satisfy the solvability condition but for which the entanglement dynamics remains tractable: \cite{claeys_operator_2024} presented a family of states and dual-unitary models in which the entanglement velocity could be exactly obtained and tuned from zero to its maximal value.

A remarkable feature of Eqs.~\eqref{eq:renyisolvablestate}, \eqref{eq:sloperenyisolvablestate}, and \eqref{eq:Bellpairentanglement} is that they are again completely independent of all the other properties of the local gate besides dual-unitarity. This means in particular that they take the same form for ergodic and non-ergodic dual-unitary circuits. This surprising feature is immediately broken whenever one goes beyond the spectral properties of $\rho_A(t)$ and the unitary matrix $W$ (cf.\ Eq.~\eqref{eq:matrixW}) starts to play a role. For example, one can consider the entanglement of the disjoint block obtained as the subset of $A$ that includes the first and last $2\ell$ qudits with $\ell<L_A/2$. In this case, \cite{foligno2024entanglement} have shown that the growth of entanglement from solvable states depends on the nature of the microscopic dynamics. Generic dual-unitary circuits follow the entanglement membrane picture put forward by \cite{jonay2018coarse, Zhou2020}, while non-ergodic dual-unitary circuits follow the quasiparticle picture of \cite{calabrese2005evolution}. Interestingly, this is the case despite the fact that charged dual unitary circuits are generically not Yang--Baxter-integrable. {Instead,~\cite{fraenkel_entanglement_2024} have considered the dynamics of entanglement in dual-unitary circuits in the presence of a non-dual-unitary impurities. They showed that also in this case the predictions of quasiparticle and membrane picture differ and agree with the exact calculations for free and maximally chaotic dual unitary circuits. Surprisingly, however, they found that dual-unitary circuits that are neither free nor maximally chaotic deviate from both predictions.} The entanglement growth from non-solvable states was instead considered by \cite{foligno_growth_2023}, who showed that in dual-unitary circuits with sufficiently large entangling power, the rate of entanglement growth of \emph{any} pair product initial state approaches the maximal one for large times.

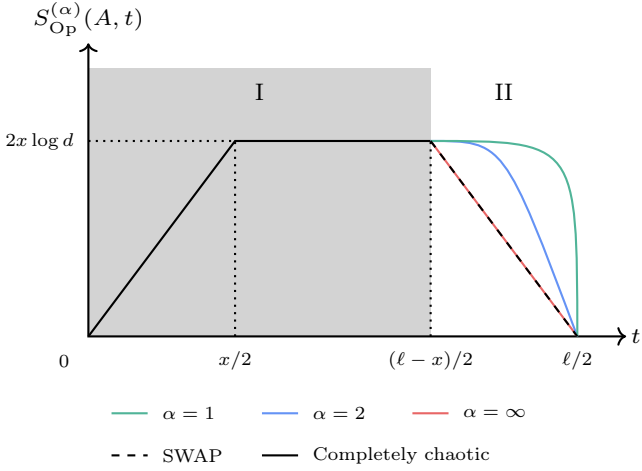
\begin{figure}[t]
    \centering
    \begin{tikzpicture}[baseline=(current bounding box.center), scale=0.65]

        \fill[grey2] (0,0) rectangle (7,5.5);

        \draw[thick, <->] (0,6) -- (0,0) -- (11,0);
        \draw[thick] (0,0) -- (3,4) -- (7,4);
        \draw[thick,red6] (7,4) -- (10,0);
        \draw[thick,dotted] (0,4)--(3,4);
        \draw[thick,dotted] (3,0)--(3,4);
        \draw[thick,dotted] (7,0)--(7,4);
        \draw[thick,blue6] (7,4) .. controls (8.5,4) .. (10,0);
        \draw[thick,green1] (7,4) .. controls (9.995,4) .. (10,0);
        \draw[thick,black, dashed] (7,4) -- (10,0);

        \node at (3.5,5) {\small I};
        \node at (8.5,5) {\small II};
        
        \node at (3,-0.5) {\scriptsize${x}/{2}$};
        \node at (7,-0.5) {\scriptsize${(\ell-x)}/{2}$};
        \node at (10,-0.5) {\scriptsize${\ell}/{2}$};
        \node at (-1.,4) {\scriptsize$2x \log d$};
        \node at (-.5,-.5) {\scriptsize$0$};
        \node[scale=1.2] at (0,6.5) {\scriptsize$S^{(\alpha)}_{\rm Op}(A, t)$};
        \node[scale=1.2] at (11.2,0) {\scriptsize$t$};
    \end{tikzpicture}

    \vspace{0.75em}
    \begin{tikzpicture}[scale=0.8]
        \draw[thick, green1] (0,0) -- (0.6,0); \node[right] at (0.7,0) {\scriptsize$\alpha=1$};
        \draw[thick, blue6] (2.5,0) -- (3.1,0); \node[right] at (3.2,0) {\scriptsize$\alpha=2$};
        \draw[thick, red6] (5,0) -- (5.6,0); \node[right] at (5.7,0) {\scriptsize$\alpha=\infty$};
        \draw[thick, black, dashed] (0,-0.7) -- (0.6,-0.7); \node[right] at (0.7,-0.7) {\scriptsize SWAP};
        \draw[thick, black] (2.5,-0.7) -- (3.1,-0.7); \node[right] at (3.2,-0.7) {\scriptsize Completely chaotic};
    \end{tikzpicture}

    \caption{Operator R\'enyi entropies $S^{(\alpha)}_{\rm Op}(A, t)$ of the reduced density matrix $\rho_A(t)$ after a quench from the Bell state in Eq.~\eqref{eq:Bell}. In Region I, ${t \leq (\ell-x)/2}$, the evolution is universal. In Region II, ${(\ell-x)/2\leq t \leq \ell/2}$, the behaviour depends on the dual-unitary gate and R\'enyi index $\alpha$. Adapted from~\cite{reid2021entanglement}.}
    \label{fig:OSEErho}
\end{figure}

The solvability conditions in Eq.~\eqref{eq:solvabilitydiagram} also allow for the characterisation of the entanglement of $\rho_A(t)$, rather than that of the (pure) state of the full system. This can be done both by viewing it as a genuine mixed state  --- computing, e.g., its logarithmic negativity following \cite{plenio2005logarithmic} --- or as a pure state in a doubled space --- computing its operator entanglement following~\cite{zanardi2001entanglement}. These quantities have been considered by \cite{bertini2022entanglement} and \cite{reid2021entanglement} respectively. Interestingly, they both show two distinct regimes as a function of time: a universal linear growth phase, which is independent of the gate-choice, and where both quantities behave exactly as the state R\'enyi entropies in Eq.~\eqref{eq:renyisolvablestate} and a decay regime which is strongly sensitive to the ergodicity properties of the local gate. Specifically, \cite{reid2021entanglement} have provided an exact characterisation of the latter two regimes for the operator entanglement of $\rho_A(t)$ in two prototypical classes of dual-unitary gates: the noninteracting family, represented by the SWAP gate, and the completely chaotic family (described more in detail in Sec.~\ref{subsec:operator_growth}), which represent the two extreme points of the ergodicity spectrum. The results are summarised in Fig.~\ref{fig:OSEErho}. Interestingly, the SWAP circuits give entanglement dynamics that coincides with that of rational conformal field theories~\cite{dubail2017entanglement}, while the completely chaotic family give results in quantitative agreement with holographic conformal field theories~\cite{wang2019barrier}, showing a longer barrier followed by a sharp drop for the von Neumann operator entanglement entropy. Moreover, for each fixed R\'enyi index, the barrier is longer for the completely chaotic case. This confirmed in a concrete class of lattice systems the proposal of~\cite{wang2019barrier}, i.e., that the length of the entanglement barrier provides a good indicator of quantum chaos. Interestingly, the completely chaotic family gives a nontrivial R\'enyi spectrum such that the barriers for noninteracting and completely chaotic circuits converge in the limit of an infinite number of replicas.

Finally, we should mention that, although the matrix $W$ in Eq.~\eqref{eq:matrixW} makes the dynamics of generic local observables difficult to access for $t<L_A/2$, one can still use the simplified form in Eq.~\eqref{eq:simplifiedrhoADU} to describe specific classes of observables. For example, \cite{piroli_exact_2020} have shown that a particularly simple result is obtained when studying the dynamics of equal-time two-point functions of 1-local operators. In this case, combining Eq.~\eqref{eq:simplifiedrhoADU} with the unitarity conditions in Eq.~\eqref{eq:unitarityfoldeddiagram} gives 
\begin{align}\label{eq:equaltimetwopoint}
&\hspace{-.25cm}C^{\alpha\beta}(r, t)=\expval{\mathbb{U}^{-t} a_0^{\alpha} a_{2r}^{\beta} \mathbb{U}^{t}}{\Psi_0(\mathcal{N})}=\notag\\
&\hspace{-.25cm}=\frac{1}{\chi q^{r+2t}}\hspace{-.25cm}
\fineq[-0.8ex][.65][1]{
\foreach \j in {0,...,1}
{
\roundgate[\j+3][\j+3][1][bottomright][orange][1]
\roundgate[-\j-3][\j+3][1][bottomright][orange][1]
}
\draw[very thick] (-2.5,1.95)--++(5,0);
\foreach \i in {-1,...,1}{
\MPSinitialstate[2*\i][2][orange][topright][1]
}
\foreach \i in {0,1}{
\cstate[\i+3.5][\i+2.5]
\cstate[\i+2.5][\i+3.5]
\cstate[-\i-3.5][\i+2.5]
\cstate[-\i-2.5][\i+3.5]
}
\cstate[-2.5][1.95]
\cstate[2.5][1.95]	
\cstate[1.5][2.5]
\cstate[0.5][2.5]
\cstate[-0.5][2.5]
\cstate[-1.5][2.5]		
\charge[4.5][4.5][black]
\charge[-4.5][4.5][black]
\node[scale=1.8] at (-4.5,5){$a^{\alpha}$};
\node[scale=1.8] at (4.5,5){$a^{\beta}$};
}
\hspace{-.25cm}.
\end{align}
We see that, as for dynamical correlations on the light cone (cf.\ Eq.~\eqref{eq:temp_corr_onlc_0}), this quantity is represented by a one-dimensional tensor network. Defining the unital quantum channels 
\begin{align}
&\mathcal{C}_+(a)= \frac{1}{\chi}\tr_2\!\!\left[\Gamma_\mathcal{M}^{\dagger}( \mathbbm{1} \otimes a)\,\Gamma_\mathcal{M}\right]\!, \\
&\mathcal{C}_-(a)= \frac{1}{q}\tr_1\!\!\left[\Gamma_\mathcal{M}^{\dagger}( a \otimes \mathbbm{1}_\chi)\,\Gamma_\mathcal{M}\right]\!, \\
&\mathcal{C}_0(a)= \frac{1}{\chi}\tr_2\!\!\left[\Gamma_\mathcal{M}^{\dagger}( a \otimes \mathbbm{1}_\chi)\,\Gamma_\mathcal{M}\right]\!, \\
&\mathcal{M}_0(a)= \frac{1}{q}\tr_2\!\!\left[\Gamma_\mathcal{M}^{\dagger}( \mathbbm{1} \otimes a)\,\Gamma_\mathcal{M}\right]\!, 
\end{align}
where the last one is the channel version of the MPS transfer matrix in Eq.~\eqref{eq:transfermatrix}, and recalling the definitions in Eqs.~\eqref{eq:mapMplus}, \eqref{eq:mapMminus}, we can then write 
\begin{align}
&C^{\alpha\beta}(r, t) \\
&\hspace{-0.25cm}=\!\!\begin{cases}
\tr[ \mathcal{M}_+^{2t}(\mathcal{C}_+(\mathcal{M}_0^{t-2r-1}(\mathcal{C}_-(\mathcal{M}_-^{2t}(a^{\alpha})))))a^{\beta} ] & t\!<\!r/2\\
\tr[\mathcal{M}_+^{2t}(\mathcal{C}_0(\mathcal{M}_-^{2t}(a^{\alpha})))a^{\beta}] & t\!=\!r/2\\
0 & t\!>\!r/2\\
\end{cases},
\notag
\end{align}
where the last equation follows immediately from Eq.~\eqref{eq:stationarystateA}.

\subsubsection{Measurement dynamics}

The interplay between unitary dynamics and measurements can give rise to dynamical phase transitions in the entanglement structure of their steady state, as recently reviewed for random quantum circuits by~\cite{potter2022entanglement, fisher_random_2023}. Such measurement-induced phase transitions (MIPTs) also occur in dual-unitary circuits and were numerically studied by~\cite{zabalo_operator_2022} for different classes of circuits with single-site measurements of Pauli operators. These transitions are governed by a nonunitary conformal field theory with central charge $c=0$, and different classes can be distinguished through their ``effective'' central charge $c_{\rm eff}$~\cite{ludwig_perturbative_1987}. It was argued that the resulting MIPT for dual-unitary Haar (Clifford) circuits lie in the same universality class as Haar (Clifford) circuits, for which $c_{\rm eff}^{\rm Haar} \approx 0.25(3)$ and $c_{\rm eff}^{\rm Cliff.} \approx 0.37(1)$ respectively. While dual-unitarity no longer allows for analytic results, constraints from space-time duality could be used to extract critical data including the effective central charge $c^{\rm DU}_{\rm eff} \approx 0.24(2)$ with higher precision.

\cite{manna_entangling_2024} considered the MIPT in circuits where gates with fixed entangling properties are combined with single-site Haar-random gates and measurements. The entanglement entropy, critical measurement rate, and exponent were calculated numerically for various classes of gates, including dual-unitary ones, and shown to be correlated with entangling power and gate typicality. For dual-unitary circuits it was argued that the steady-state entanglement in circuits with measurements exhibited a monotonic dependence on the entangling power. 

Dual-unitary circuits interspersed with projective measurements in the Bell-pair basis, preserving spatial unitarity, were studied by~ \cite{claeys_exact_2022}. While analytical results for generic dual-unitary circuits remain intractable, additional constraints leading to solvability can be imposed on the circuit, effectively preventing any operator scrambling of the projection operator following a measurement --- which is however a prerequisite for a phase transition between volume-law scaling and area-law entanglement scaling~\cite{choi_quantum_2020}. These restrictions allow the steady-state entanglement and the entanglement growth to be exactly characterised. The latter still exhibits a maximal entanglement growth but with a diffusive scaling of the steady-state subsystem entropy, indicating a critical purification phase. This behaviour can be understood by noting that after a sufficiently long time the dynamics reduce to a minimal model for the diffusive spreading of noninteracting Bell pairs previously studied by~\cite{ippoliti2021postselectionfree}.

\subsection{Operator dynamics}
\label{subsec:operator_growth}

The simplest probe of operator dynamics is given by the dynamical correlations considered in Sec.~\ref{sec:corr_functions}. The latter, however, only give very limited information. One can get a sense of this restriction by expanding the operator $\mathbb{U}^{-t} a_0^{\beta} \mathbb{U}^{t}$ in the basis of tensor products of $\{a^{\alpha}\}_{\alpha=0}^{q^2-1}$ (cf.\ Sec.~\ref{sec:corr_functions}) and noting that  the dynamical correlations only measure the component along $a^\alpha_x$, i.e. one out of $q^{4t}$ components. To attain a more comprehensive measure of operator spreading, and more generally of quantum information scrambling~\cite{hayden2007black, sekino2008fast, swingle_unscrambling_2018}, recent research has proposed a number of more refined quantities. Two popular choices are the \emph{local operator entanglement} (LOE)~\cite{prosen2007is, prosen2007operator, pizorn2009operator} and \emph{out-of-time-order correlation functions} (OTOCs)~\cite{larkin1969quasiclassical, lashkari2013towards, shenker2014multiple, shenker2014black, maldacena2016bound, roberts2018operator}. Both these quantities can be effectively characterised in dual unitary circuits, as we now review. We also show that the latter are closely related to another measure of quantum information scrambling, the tripartite information introduced by~\cite{hosur2016chaos}, which can also be characterised in dual-unitary circuits.

Before moving to the definition and characterisation of these quantities, it is instructive to first consider the effect of dual-unitarity on operator growth. Applying a two-site gate on a single-site operator $a$ and expanding in the operator basis we have 
\begin{align}
U(a\otimes \mathbbm{1}) U^{\dagger} = \sum_{\alpha,\beta=1}^{q^2} c_{\alpha \beta}\, a^{\alpha} \otimes a^{\beta},
\end{align}
where
\begin{align}\label{eq:expansion_coeffs_operatorgrowth}
c_{\alpha \beta} = \frac{1}{q^2} \tr \left[U(a\otimes \mathbbm{1}) U^{\dagger} \left(a^{\alpha} \otimes a^{\beta}\right)\right]\,.
\end{align}
We can distinguish different contributions to operator growth. All nonzero coefficients $c_{\alpha 0}$ represent probability amplitudes for the operator not to move, i.e., $a \otimes \mathbbm{1}$ gets mapped to $a^{\alpha} \otimes \mathbbm{1}$, whereas the coefficients $c_{0\alpha}$ and $c_{\alpha \beta}$ with $\alpha, \beta \neq 0$ result in a growing operator front. The former are directly related to the dynamical correlation functions, $c_{0 \alpha} = \tr[\mathcal{M}_+(a)a^{\alpha}]/q$.

Dual-unitarity enforces the operator to strictly grow at each time step: $c_{\alpha 0}=0$ whenever $\alpha \neq 0$. In this case we can graphically express Eq.~\eqref{eq:expansion_coeffs_operatorgrowth} as
\be
\frac{1}{q^2}
\fineq[-0.8ex][1][1]{
\roundgate[0][0][1][bottomright][orange][1]
\cstate[+.5][-.5]
\cstate[+.5][+.5]			
\charge[-.5][-.5][black]
\charge[-.5][.5][black]
\node[scale=1.1] at (-.75,-.65){$a$};
\node[scale=1.1] at (-.75,.65){$a^\alpha$};
} = \frac{1}{q^2}\fineq[-0.8ex][1][1]{
\draw[ thick] (-.5,.5)--++(+1,0);
\draw[ thick] (-.5,-.5)--++(+1,0);
\cstate[+.5][-.5]
\cstate[+.5][+.5]			
\charge[-.5][-.5][black]
\charge[-.5][.5][black]
\node[scale=1.1] at (-.75,-.65){$a$};
\node[scale=1.1] at (-.75,.65){$a^\alpha$};
} = 0\,, 
\ee
where in the first equality we have used dual-unitarity and in the second we have used that $a^{\alpha}$ is traceless. This result already indicates that in dual-unitary circuits the support of the initially 1-local operator will growth ballistically in time with maximal so-called \emph{butterfly velocity} $v_B=1$~\cite{claeys_maximum_2020}. This is in contrast to what happens for generic unitary circuits, where there is a nonzero probability that the operator front does not move and the operator growth can be described as a biased random walk with $v_B<1$ and a diffusively broadening front~\cite{nahum2017quantum,nahum_operator_2018,von_keyserlingk_operator_2018}. The above argument was also proposed in \cite{hosur2016chaos} to argue that circuits composed out of perfect tensors exhibit ballistic operator growth.

For a more quantitative characterisation we study two quantities: LOE and OTOCs. 
For the latter we consider OTOCs defined as
\begin{align}\label{eq:OTOC_def}
O_{\alpha \beta}(x,t) = \langle a_x^{\beta} \mathbb{U}^{-t} a_0^{\alpha} \mathbb{U}^{t}a_x^{\beta} \mathbb{U}^{-t} a_0^{\alpha} \mathbb{U}^{t}\rangle\,,
\end{align}
where again $\expval{\bullet} \equiv \tr[\bullet ]/\tr[\mathds{1}]$ and $x$ is an integer. The case of $x$ half-odd-integer in either Eq.~\eqref{eq:LOE_def} or in Eq.~\eqref{eq:OTOC_def} changes minor details of the discussion below but can be treated with an analogous reasoning. These OTOCs characterise scrambling in many-body systems.

The LOE presents a measure of dynamical complexity and quantum chaos and is defined as the entanglement of the local operator $a^{\alpha}_0$ evolved up to time $t$ and mapped to a pure state through a folding mapping. More formally, one associates a state $\ket{O}\in \mathcal H\otimes\mathcal H$ to an operator $O \in {\rm End}(\mathcal H)$ as follows 
\begin{align}
O \,\,\longmapsto\,\, \ket{O}&\equiv \sum_{n,m} \braket{n}{O|m} \ket{n}\otimes\ket{m}^*\,.
\label{eq:mapping}
\end{align}
For convenience we arrange the states $\ket{n}\otimes\ket{m}^*$ in $\mathcal H\otimes\mathcal H$ in such a way that the time evolution generated by $\mathbb U^\dag\otimes  \mathbb U^{\dag\ast}$ is local. Specifically,
\be
\ket{i_1 \ldots i_{2L}}\otimes\ket{j_1 \ldots j_{2L}}^*=\ket{i_1 \, j_1}\otimes\cdots\ket{i_{2L} \, j_{2L}}\,,
\ee
where ${\{\ket{i};\,i=1,2,\ldots,q\}}$ is the computational basis in Eq.~\eqref{eq:computationalbasis}. 

Since the operation in Eq.~\eqref{eq:mapping} maps operators into pure states, their entanglement is conveniently measured by the R\'enyi entanglement entropies
\be
S^{(n)}_A(a^{\alpha}, t)=S^{(n)}(\mathcal P_A(a^{\alpha},t)),
\label{eq:LOE_def}
\ee
where 
\begin{align}
\mathcal P_A(a^{\alpha},t) & = \frac{\tr_{\bar A}[{\ketbra{\mathbb{U}^{-t} a^{\alpha}_0 \mathbb{U}^{t}}}]}{\|\ket{\mathbb{U}^{-t} a^{\alpha}_0 \mathbb{U}^{t}}\|^2}\\ 
&= \frac{1}{q^{2L}} \tr_{\bar A}[(\mathbb{U}^{-t} \otimes \mathbb{U}^{t}) {\ketbra{ a^{\alpha}_0}} (\mathbb{U}^{t} \otimes \mathbb{U}^{-t})],\notag 
\end{align}
is the state $\ket{\mathbb{U}^{-t} a^{\alpha}_0 \mathbb{U}^{t}}/\|\ket{\mathbb{U}^{-t} a^{\alpha}_0 \mathbb{U}^{t}}\|$ reduced to the region $A$. Here we consider the case in which $A$ corresponds to ${[x,L/2]}$, with $x$ integer.

Performing all simplifications granted by unitarity one can easily see that --- as for dynamical correlations --- both $S^{(n)}_A(a^{\alpha}_0, t)$ and $O_{\alpha \beta}(x,t) $ are non-trivial only within a causal light cone. Specifically, \cite{bertini_operator_i_2020, bertini_operator_ii_2020} have shown that the LOE R\'enyi entropies in Eq.~\eqref{eq:LOE_def} can expressed as 
\be
S^{(n)}_A(a^{\alpha}_0, t)= \frac{1}{1-n}\log\tr[ (\mathcal C[a^{\alpha}]^\dag \mathcal C[a^{\alpha}])^n ],
\ee
where, recalling the diagrammatic representation introduced in Sec.~\ref{sec:diagrams}, we can represent the $q^{2(t+x)}\times q^{2(t-x)}$ (rectangular) matrix $\mathcal C[a]$ through the following diagram
\begin{align}
\mathcal C[a^{\alpha}]=\frac{1}{q^{t+1/2}} \fineq[-2.5ex][0.6][1]{
\begin{scope}[rotate around={-45:(0,0)}]
\foreach \i in {0,...,3}{
\foreach \j in {0,...,4}{
\roundgate[\i+\j][\i-\j][1][bottomright][orange][1]
}
}
\foreach \i in {0,...,3}{
\cstate[\i+4.5][\i-4.5]	
}
\foreach \i in {0,...,4}{
\cstate[\i-.5][-\i-.5]	
}
\charge[-0.5+4][-0.5-4][black]
\end{scope}
\draw [decorate, thick, decoration = {brace}]   (-0.35,1)--++(5,0);
\node[scale=1.5] at (2.15,1.5) {$t+x$};
\draw [decorate, thick, decoration = {brace}]   (5.15,.15)--++(0,-6);
\node[scale=1.5, rotate = 90] at (5.55,-2.75) {$t-x+1$};
\node[scale=1.5] at (-.6,-5.25){$a^\alpha$};
},
\label{eq:LOE_graphical}
\end{align}
where the open legs along the horizontal line are input indices and those along the vertical line are output ones.

Analogously, \cite{claeys_maximum_2020} have shown that the OTOC can be expressed as
\be
\hspace{-.25cm}O_{\alpha \beta}(x,t) = \tr[ \mathcal C[a^{\alpha}]^\dag \mathcal C[a^{\alpha}] (a^\beta\!\otimes_r\! a^{\beta*})\otimes \mathds{1}_{q^2}^{\otimes (t-x)}  ], 
\ee
where $\otimes_r$ is the tensor product among different replicas. The latter quantity is compactly represented diagrammatically by performing a further folding operation around the horizontal symmetry axis, i.e., 
\be
\label{eq:OTOC_graphical}
O_{\alpha \beta}(x,t) =\frac{1}{q^{2t+1}} \fineq[-2.4ex][.6][1]{
\begin{scope}[rotate around={-45:(0,0)}]
\foreach \i in {0,...,3}{
\foreach \j in {0,...,4}{
\roundgate[\i+\j][\i-\j][1][bottomright][orange][2]
}
}
\foreach \i in {0,...,3}{
\sqrstate[\i-.5][\i+.5]	
\cstate[\i+4.5][\i-4.5]	
}
\foreach \i in {0,...,4}{
\cstate[\i-.5][-\i-.5]	
\sqrstate[\i+3.5][-\i+3.5]
}
\charge[-0.5+4][-0.5-4][black]
\charge[3.5][3.5][black]
\end{scope}
\draw [decorate, thick, decoration = {brace}]   (-0.35,1)--++(5,0);
\node[scale=1.5] at (2.15,1.5) {$t+x$};
\draw [decorate, thick, decoration = {brace}]   (5.25,.15)--++(0,-6);
\node[scale=1.5, rotate = 90] at (5.65,-2.75) {$t-x+1$};
\node[scale=1.5] at (-.5,-5.25){$a^\alpha$};
\node[scale=1.5] at (5.15,.5){$a^\beta$};
},
\ee
and now the operator states are defined on four replicas as 
\begin{align}
\fineq{\draw[thick] (0,0)--++(-.5,0);\cstate[0][0][][black]} &= (a^\alpha\otimes \mathds{1})^{\otimes 2}\fineq{\draw[thick] (0,0)--++(-.5,0);\cstate[0][0]}\,,
\label{eq:circlestateblack}\\
\fineq{\draw[thick] (0,0)--++(-.5,0);\sqrstate[0][0][][black]} &= (a^\alpha\otimes \mathds{1})^{\otimes 2} \fineq{\draw[thick] (0,0)--++(-.5,0);\sqrstate[0][0]}\,.
\label{eq:sqrtstateblack}
\end{align}
A similar diagram can also be used to represent the product of $\mathcal C[a^{\alpha}]^\dag$ and $\mathcal C[a^{\alpha}]$, i.e., 
\be
\label{eq:CCdag_graphical}
\mathcal C[a^{\alpha}]^\dag \mathcal C[a^{\alpha}]  =\frac{1}{q^{2t+1}} \fineq[-2.4ex][.6][1]{
\begin{scope}[rotate around={-45:(0,0)}]
\foreach \i in {0,...,3}{
\foreach \j in {0,...,4}{
\roundgate[\i+\j][\i-\j][1][bottomright][orange][2]
}
}
\foreach \i in {0,...,3}{
\sqrstate[\i-.5][\i+.5]	
\cstate[\i+4.5][\i-4.5]	
}
\foreach \i in {0,...,4}{
\cstate[\i-.5][-\i-.5]	
}
\charge[-0.5+4][-0.5-4][black]
\end{scope}
\draw [decorate, thick, decoration = {brace}]   (-0.35,1)--++(5,0);
\node[scale=1.5] at (2.15,1.5) {$t+x$};
\draw [decorate, thick, decoration = {brace}]   (5.25,.15)--++(0,-6);
\node[scale=1.5, rotate = 90] at (5.65,-2.75) {$t-x+1$};
\node[scale=1.5] at (-.5,-5.25){$a^\alpha$};
}.
\ee
Due to the different contractions at the boundaries, the diagrams in Eqs.~\eqref{eq:OTOC_graphical} and~\eqref{eq:CCdag_graphical} cannot be evaluated only using the diagrammatic rules of dual-unitarity, indicating that both LOE and OTOCs exhibit nontrivial behaviour inside the light cone. This is in contrast, e.g., with the OTOC of a random operator considered by \cite{dowling2023scrambling} where dual-unitarity allows for a full contraction. 

To make progress \cite{bertini_operator_i_2020} introduced the light-cone transfer matrix
\begin{align}\label{eq:OTOC_transfermatrix}
T_x = \frac{1}{q}
\fineq[-2.4ex][.65][1]{
\begin{scope}[rotate around={45:(0,0)}]
\foreach \j in {0,...,4}{
\roundgate[\j][-\j][1][bottomright][orange][2]
}
\sqrstate[-.5][.5]	
\cstate[4.5][-4.5]	
\end{scope}
\draw [decorate, thick, decoration = {brace}]   (-0.15,.75)--++(5.9,0);
\node[scale=1.5] at (2.8,1.15) {$x$};
},
\end{align}
and noted that, while dual-unitarity does not allow Eq.~\eqref{eq:OTOC_graphical} and~\eqref{eq:CCdag_graphical} to be evaluated in general, it can be used to construct a set of leading eigenvectors for $T_x$, which govern the asymptotic behaviour of both LOE and OTOCs. 

Specifically, \cite{bertini_operator_i_2020} have shown that, although it is generically not normal, the matrix $T_x$ has its spectrum contained in or on the unit circle of the complex plane. Moreover, they have shown that Eqs.~\eqref{eq:unitaritynfoldeddiagram} and \eqref{eq:spaceunitaritynfoldeddiagram} give the following set of $x+1$ right eigenvectors of Eq.~\eqref{eq:OTOC_transfermatrix} with eigenvalue 1
\begin{align}
\ket{e_{x,k}} = \fineq[.8ex][.65][1]{
\foreach \i in {0,...,4}{
\draw[thick] (\i,0)--++(0,.75);
}
\foreach \i in {0,...,2}{
\sqrstate[\i][0]	
}
\foreach \i in {3,...,4}{
\cstate[\i][0]
}	
\draw [decorate, thick, decoration = {brace}]   (2.25,-.25)--++(-2.5,0);
\node[scale=1.5] at (1,-.75) {$k$};
\draw [decorate, thick, decoration = {brace}]   (2+2.25,-.25)--++(-1.5,0);
\node[scale=1.5] at (3.5,-.75) {$x-k$};
},\qquad k=0,\ldots,x,
\end{align}
as can be straightforwardly checked using unitarity in time (space) to contract from the right (left). Their Hermitian conjugates are left eigenvectors with eigenvalue 1 (interestingly this is true despite the matrix being not normal). These vectors are not orthonormal and their orthonormalisation returns the following set of eigenstates
\begin{align}
\label{eq:orthonormalrainbows}
\hspace{-.25cm}\ket{\tilde{e}_{x,0}} = \frac{1}{q^x}\ket{e_{x,0}},\,\,\, \ket{\tilde{e}_{x,k \neq 0}} = \frac{q \ket{e_{x,k}}-\ket{e_{x,k-1}}}{q^x \sqrt{q^2-1}}\,,
\end{align}
and similarly for $\bra{\tilde{e}_{x,k}}$, which satisfy $\braket{\tilde{e}_{x,k}}{\tilde{e}_{x,l}} = \delta_{kl}$. 

In dual-unitary circuits without additional symmetries these eigenvectors are expected to exhaust the leading eigenspace. The class of dual-unitary circuits for which this is the case was termed `completely chaotic' by \cite{bertini_operator_i_2020}. The completely chaotic property has been proven by \cite{bertini2020scrambling} in the case of \emph{random} dual-unitary circuits. Namely, for dual-unitary circuits where the one-qudit matrices in Eq.~\eqref{eq:orbit} are independently distributed random unitary matrices. Instead, in non-ergodic circuits (and even in the self-dual kicked Ising model discussed in Sec.~\ref{sec:KIM}) the eigenspace spanned by the vectors in Eq.~\eqref{eq:orthonormalrainbows} is known to be not exhaustive \cite{bertini_operator_i_2020}.
 
In the limit of large $x+t$ and fixed $t-x$, the transfer matrix can be replaced by a projector on its leading eigenspace and, assuming that the states in Eq.~\eqref{eq:orthonormalrainbows} exhaust the leading eigenspace one finds 
\begin{align}
\lim_{m \to \infty} (T_x)^m =  \sum_{k=0}^x \ketbra{\tilde{e}_{x,k}}\,.
\end{align}
Plugging this projector in Eq.~\eqref{eq:CCdag_graphical} and evaluating the overlap of the eigenvectors with the left boundary \cite{bertini_operator_i_2020} found  
\be
\hspace{-.25cm} \lim_{t+x \to \infty } \hspace{-.25cm}\mathcal C[a^{\alpha}]^\dag \mathcal C[a^{\alpha}] \!=\! \frac{q^{2(x-t)}}{q^2-1}\overbrace{\mathds{1}_{q^2}\!\otimes\! \cdots \!\otimes\! \mathds{1}_{q^2}}^{t-x}\!\otimes(\mathds{1}_{q^2}\!-\!P),
\ee 
where we introduced the one-dimensional projector
\be
P=\ketbra{\mcirc}/q.
\ee
This gives 
\be
\label{eq:LOErightedge}
\lim_{t+x \to \infty }  \!\!\!\!S^{(n)}_A(a^{\alpha}_0, t)= 2(t-x) \log q + \log(1-1/q^2)\,.
\ee
This expression shows that the entanglement through bipartitions close to the right edge of the causal light cone of $\mathbb{U}^{-t} a_0^{\alpha} \mathbb{U}^{t}$ is always maximal and the entanglement spectrum is flat (no dependence on $n$ is observed). Similarly, \cite{claeys_maximum_2020} have shown that
\begin{align}
\lim_{ t+x \to \infty } O_{\alpha \beta}(x,t) = 
\begin{cases}
\displaystyle -\frac{1}{q^2-1} \qquad &\text{if} \quad x=t,\\
0 \qquad &\text{if} \quad x < t.
\end{cases}
\end{align}
This expression means that the operator support completely fills the region close to the right edge of the causal light cone of  $\mathbb{U}^{-t} a_0^{\alpha} \mathbb{U}^{t}$. In particular, this implies that the right butterfly velocity is maximal ($v_B=1$).

In the opposite limit, i.e., large $t-x$ and fixed $x+t$, the problem can be treated via a transfer matrix very similar to the one in Eq.~\eqref{eq:OTOC_transfermatrix} --- the only difference is that the local gate $U$ is exchanged with its dual $\tilde U$, which does not change the dual-unitary properties. In this case \cite{bertini_operator_i_2020} found 
\be
 \lim_{(t-x) \to \infty } \mathcal C[a^{\alpha}] \mathcal C[a^{\alpha}]^\dag  = \sum_{k=0}^{x+t}  c_k(a^{\alpha}) O_k, 
\ee
where we introduced the operators 
\begin{align}
& O_{k\geq 1} =    \frac{q^{2-2k}}{q^2-1} \overbrace{P\otimes \cdots \otimes P}^{t+x-k} \otimes (\mathds{1}_{q^2}-P),\\
& O_0= \overbrace{P\otimes \cdots \otimes P}^{t+x}.
\end{align}
and the coefficients
\begin{align}
\hspace{-0.25cm}c_{k\geq 1}(a) =&  (\tr[\smash{(\mathcal{M}_{+}^{\dagger}(a))^{t+x-k} \mathcal{M}_{+}^{t+x-k}(a)}]\notag\\ 
&-\tr[\smash{(\mathcal{M}_{+}^{\dagger}(a))^{t+x+1-k} \mathcal{M}_{+}^{t+x+1-k}(a)}]),\\
\hspace{-0.25cm}c_0(a) =& \tr[\smash{(\mathcal{M}_{+}^{\dagger}(a))^{t+x} \mathcal{M}_{+}^{t+x}(a)}],
\end{align}
depend on the light cone channel for dynamical correlations (cf.\ Eq.~\eqref{eq:mapMplus}). This gives the following richer behaviour for the LOE entropies 
\begin{align}
&\hspace{-0.25cm}\lim_{t-x \to \infty }  \!\!\!\!S^{(n)}_A(a^{\alpha}_0, t) = \notag\\
&\hspace{-0.25cm} =\frac{1}{1-n} \log\left[\sum_{k=1}^{x+t}  c^n_k (a^{\alpha}) \left(\frac{q^{2-2k}}{q^2-1}\right)^{n-1}+c^n_0(a^{\alpha})\right]\!.
\end{align}
A similar rich behaviour is observed for OTOCs. For instance, considering for simplicity the case of $x$ half-odd integer, \cite{claeys_maximum_2020} found  
\begin{align}
&\lim_{t-x \to \infty } O_{\alpha \beta}(x,t) = c_0(a^\alpha) - \frac{1}{q^2-1} c_{1}(a^\alpha)\,. 
\end{align}
Assuming that $\mathcal{M}_+(\cdot)$ has a non-degenerate subleading eigenvalue $\lambda$ (the eigenvalue with maximal magnitude that is strictly smaller than one), for large $x+t$ the above expressions are approximated as 
\be
\hspace{-0.3cm}\lim_{t-x \to \infty }  \!\!\!\!S^{(n)}_A(a^{\alpha}_0, t) \!\simeq\! 
\begin{cases}
 2 (t+x) \log|\lambda|^{\frac{n}{1-n}}, & \!\!\!\!\!|\lambda| \geq q^{\frac{1-n}{n}}\!, \\ 
2 (t+x) \log q, & \!\!\!\!\!|\lambda| \leq q^{\frac{1-n}{n}}\!,
\end{cases}
\label{eq:LOEleftedge}
\ee
and 
\begin{align}
    \lim_{t-x \to \infty } O_{\alpha \beta}(x,t) \simeq C |\lambda|^{2(x+t)}\,. 
\end{align}
The first expression shows that the entanglement spectrum for bipartitions close to left edge of $\mathbb{U}^{-t} a_0^{\alpha} \mathbb{U}^{t}$ is controlled by $|\lambda|$. For small enough $|\lambda|$, i.e., for more ergodic circuits, the entanglement spectrum is again flat. Note that the von Neumann entropy is always maximal, which means that the entanglement across the bipartition is again the largest possible. Analogously, we see that although dual unitarity enforces $v_B=1$ also close to the left edge, the decay when moving inside the light cone is not sudden but exponential and controlled by $|\lambda|$. One can understand this behaviour by noting that $|\lambda|$ is setting the `leaking' of operators into the causal light cone, which in dual-unitary circuits corresponds to single-site operators being mapped to two-site operators. Note that there is nothing special about the left edge: considering a spreading operator prepared in a half-odd-integer site, e.g., $\mathbb{U}^{-t} a_{1/2}^{\alpha} \mathbb{U}^{t}$, the roles of left and right light cones are reversed.  

Going beyond the limits $t\pm x\to \infty$, and hence exploring the full light cone for both LOE and OTOC, is generically too hard even for dual-unitary circuits, and analytic progress was only made in the case of random dual-unitary circuits by \cite{bertini2020scrambling}. Interestingly, however, \cite{bertini_operator_i_2020} have shown that the leading order behaviour of LOE for arbitrary ``rays''  inside the causal light cone (i.e.\ for $|x|, t \gg 1$ and any $|x|/t\leq 1$) seems to be well described by summing the two asymptotic results in Eqs.~\eqref{eq:LOErightedge} and \eqref{eq:LOEleftedge}.   

A phenomenon implied by the grow of support of local operators, observed by both the OTOC and LOE, is that as time elapses quantum many-body dynamics cause the information that is initially stored locally (in space) to become increasingly more delocalised, i.e., `scrambled'.
This scrambling was studied by considering the decoding error of Hayden-Preskill information recovery protocols in~\cite{rampp_hayden-preskill_2023}, which observed perfect information transport on the light cone (absent in generic unitary circuits) and relaxation of the decoding error to a universal value inside the light cone. \cite{lopez-piqueres_operator_2021} studied the dynamics of the OTOC front using matrix product operators, exactly capturing the dual-unitary case, and identified signatures of many-body quantum chaos in this operator front.

In the presence of randomness the dynamics of the OTOC and LOE can be mapped to a classical Markov process, and~\cite{akhtar_dual-unitary_2024,song_monte_2025} studied operator dynamics in random dual-unitary circuits from this perspective. \cite{akhtar_dual-unitary_2024} presented a resulting picture for Pauli weight dynamics in terms of chiral quasi-particles propagating with maximal velocity, which can fuse and emit quasi-particles with a probability set by the entangling power (cf. Sec.~\ref{sec:entpow}), and argued that dual-unitary circuits in classical shadow tomography could be particularly advantageous for predicting large operators (as compared to shallow brickwork Clifford circuits). Monte Carlo simulations of this operator dynamics were performed in \cite{song_monte_2025}, confirming analytical predictions for the OTOC and observing a volume-law scaling of the LOE across different subregions, with a transition from maximal to sub-maximal entanglement growth depending on the gate's entangling power, consistent with predictions from~\cite{bertini_operator_i_2020,andreadakis_operator_2024}.

Scrambling can also be measured more directly by studying quantities related to the entanglement of the evolution operator itself. Specifically~\cite{hosur2016chaos} proposed to consider the following partition of $\mathbb{U}^t$
\be
\mathbb{U}^t=\fineq[-0.8ex][.65][1]{	
\foreach \j in {0,...,2}{
\foreach \i in {0,...,4}{
\roundgate[2*\i][2*\j][1][topright][red6][-1]
\roundgate[2*\i+1][1+2*\j][1][topright][red6][-1]}}	
\draw [decorate, thick, decoration = {brace}]   (0.5,5.65)--++(3,0);
\node[scale=1.5] at (2,6.15) {$C$};
\draw [decorate, thick, decoration = {brace}]   (4.5,5.65)--++(5,0);
\node[scale=1.5] at (7,6.15) {$D$};
\draw [decorate, thick, decoration = {brace}]   (4.5,-.65)--++(-5,0);
\node[scale=1.5] at (2,-1.15) {$A$};
\draw [decorate, thick, decoration = {brace}]   (8.5,-.65)--++(-3,0);
\node[scale=1.5] at (7,-1.15) {$B$};
},
\ee
and measure its `tripartite information' defined as 
\begin{align}
    I_3^{(n)}(A:C:D) &:= I^{(n)}(A:C)+I^{(n)}(A:D) \nonumber\\
    &-I^{(n)}(A:CD),
\end{align}
where 
\begin{align}
    I^{(n)}(X:Y) = S_X^{(n)}+S_Y^{(n)}-S_{XY}^{(n)},
\end{align}
is the R\'enyi-$n$ mutual information. From its definition we see that $-I_3^{(n)}(A:C:D)$ quantifies the amount of information on the input region $A$ that can be recovered by global measurements in $C\cup D$, but can not be obtained by probing $C$ and $D$ individually. Therefore a large and negative tripartite information signals the efficient scrambling of quantum information.

Tripartite information was studied in the context of quantum circuits by \cite{bertini2020scrambling}. Considering the thermodynamic limit and choosing the bipartitions as $A = (-\infty,0]$, $B = (0,+\infty)$, $C = (-\infty, x]$ and $D = (x,+\infty))$, \cite{bertini2020scrambling} showed that $\exp[\smash{I_3^{(n=2)}(A:C:D)}]$ can be evaluated by considering diagrams very similar to those in Eqs.~\eqref{eq:OTOC_graphical} and~\eqref{eq:CCdag_graphical}, but with different boundary conditions. In particular, evaluating these diagrams for completely chaotic dual-unitary circuits \cite{bertini2020scrambling} found that for fixed $x$ and large $t$ one has   
\begin{align}
\label{eq:I3du}
	I_3^{(2)}(x,t) \simeq -2t \ln q+\mathcal{O}(1),
\end{align}
while in random unitary circuits (non-dual-unitary) 
\begin{align}
\label{eq:I3ru}
	I_3^{(2)}(x,t) \simeq -2t \ln \left(\frac{q^2+1}{2q}\right)+\mathcal{O}(1). 
\end{align}
Since ${q^2+1}/({2q}) < q$ this shows that completely chaotic dual-unitary circuits scramble information faster than random ones. Note that, although the result in Eq.~\eqref{eq:I3ru} is exact, the one in Eq.~\eqref{eq:I3du} is obtained by summing together the results in the limits $t-x\gg t+x$ and $t+x\gg t-x$ (cf.\ the previous discussion for the LOE). \cite{bertini2020scrambling} showed that the result in Eq.~\eqref{eq:I3du} becomes exact when averaging over random dual-unitary circuits.

To conclude this section, we note that higher-order OTOCs were recently considered in \cite{chen_free_2025}. These can be represented in the form in Eq.~\eqref{eq:OTOC_graphical}, but with the folded gates now describing an arbitrary number of replicas. The leading eigenstates of the transfer matrix in Eq.~\eqref{eq:OTOC_transfermatrix} can similarly be constructed from the  permutation states in Eq.~\eqref{eq:permutation_states}, which are now, however, indexed by paths of nondecreasing noncrossing permutations. While the orthonormalisation of these eigenstates could no longer be performed exactly in general, in the limit $q \to \infty$ it was possible to establish that these higher-order OTOCs again spread with maximal butterfly velocity and vanish at late times. 

\subsection{Quantum complexity}
\label{subsec:Complexity}

In our discussion so far, we considered exact calculations of dynamical quantities such as equal-time correlation functions, entanglement, and scrambling. In all cases we found exact results, suggesting that dual-unitary circuits can be efficiently simulated classically -- and hence a quantum computation realised using dual-unitary gates would be classically simulatable. However, this connection breaks down for \emph{finite system sizes at long times}, as can be proven by interpreting dual-unitary dynamics as a quantum computation. As shown in \cite{suzuki_computational_2022}, dual-unitary brickwork circuits can perform universal quantum computation at late times, i.e. for circuit depths polynomial in system size. While, e.g., the expectation value of a local observables is classically simulatable at early times (linear in system size), at late times (polynomial in system size) this problem becomes a BQP-complete and can no longer be efficiently simulated classically.  

The proof is specific to qubits and hinges on showing that dual-unitary brickwork circuits form a \emph{universal gate set}. Such universal gate sets are an important concept in quantum computation, where a set of unitary gates is said to form a universal gate set if any unitary operation can be approximated to arbitrary accuracy by a quantum circuit constructed from such a set~\cite{nielsen2010quantum}. One choice of universal gate set is given by the two-site CZ gate and arbitrary one-site gates~\cite{barenco_elementary_1995}. Here, both the CZ gate and (products) of arbitrary one-site gates can be represented as a dual-unitary brickwork circuit, where we take periodic boundary conditions for concreteness, such that any unitary operator can be approximated to arbitrary precision as a dual-unitary brickwork circuit with periodic boundary conditions. The proof is straightforward: A CZ gate acting on two arbitrary sites can be represented as a brickwork circuit constructed out of dual-unitary SWAP gates and a single SWAP $\cdot$ CZ gate. For instance, a CZ gate acting on an arbitrary even and odd site can be constructed as
\begin{align}
	\textrm{CZ}_{x+y,x-y+1/2} = (\mathbb{S}_e\mathbb{S}_o)^{L-y} \textrm{CZ}_{x,x+1/2}(\mathbb{S}_e\mathbb{S}_o)^{y},
\end{align}
where we have defined
\begin{align}
\mathbb{S}_e=\bigotimes_{x=0}^{L-1} S_{x,x+1/2},\qquad \mathbb{S}_o=\bigotimes_{x=1}^{L} S_{x-1/2,x}\,.
\end{align}
This result can be directly verified graphically, since the swap circuit satisfies $(\mathbb{S}_e\mathbb{S}_o)^L = \mathbbm{1}$ such that the full circuit is diagonal and implements a CZ gate as
\begin{align}
  \hspace{-.5cm}\vcenter{\hbox{\includegraphics[width=0.5\linewidth]{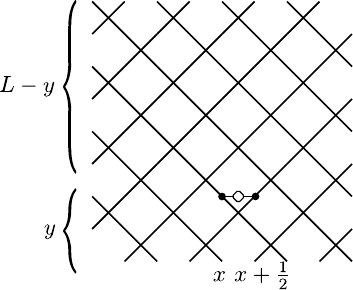}}} \,\,=\,\,\vcenter{\hbox{\includegraphics[width=0.32\linewidth]{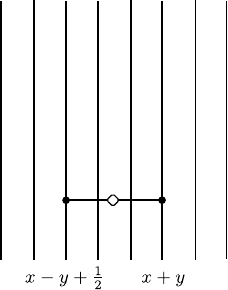}}} 
\end{align}
One can also extend this derivation to apply various CZ gates in parallel.
A similar construction can be found for (products of) single-site operators by substituting some of the SWAP gates with dual-unitary $(u_1 \otimes u_2) S$ gates. From this construction, it follows that any unitary circuit on $L$ gates can be approximated by a dual-unitary brickwork circuit with a depth that is polynomial in $L$. Because of this correspondence any decision problem involving dual-unitary gates, such as e.g. bounding the expectation value of an observable, is as hard as the corresponding problem for general unitary circuits and hence BQP-complete. Dual-unitary circuits hence present a class of models which can be efficiently classically simulated at short times but not at late times. This issue is more generally apparent in the fact that all exact results for dual-unitary circuits require an infinite system size, effectively restricting all dynamics to the short-time regime.

\section{Spectral statistics}
\label{sec:spectral}
Although in Sec.~\ref{sec:quantumdynamics} we linked quantum chaos to certain dynamical properties --- for instance the linear growth in time of LOE --- the traditional way to identify chaotic/non-integrable quantum systems is through their spectral statistics, i.e., the distribution of the eigenvalues of their Hamiltonian (for continuous times) or  time-evolution operator (for discrete times). Namely, a system is considered quantum chaotic whenever it has \emph{universal} spectral correlations matching those of the ensemble of random matrices with the same anti-unitary symmetry~\cite{haake2001quantum}. One of the key advances brought by dual-unitary circuits has been to put this approach on a firmer mathematical ground. 

This `spectral' approach to quantum chaos --- developed in the 1980s --- originates from the observation that in quantum systems there is no direct way to quantify the most intuitive property of chaos, i.e., the sensitivity of the dynamics to the initial conditions. 
Therefore, the idea was to identify quantum chaotic systems as those whose classical limit is classically chaotic The key realisation has been that the latter could be characterised through their spectral correlations. Specifically, two fundamental conjectures were put forward: (i) The Bohighas-Giannoni-Schmidt Conjecture (or Quantum Chaos Conjecture)~\cite{Casati1980,Bohigas1984}, asserting that systems with a chaotic and ergodic classical limit should have spectral statistics described by one of the maximal entropy (Gaussian or Circular) ensembles of Random Matrix Theory (RMT)~\cite{mehta2004random}. This implies strong statistical correlations among energy or quasi-energy levels, both at short distances --- spectral repulsion --- and at large distances --- spectral rigidity. (ii) The Berry-Tabor Conjecture~\cite{BerryTabor1977} asserting that the (quantum) spectrum of a classically integrable system is essentially a random Poisson process, i.e. there is no statistical correlation among levels at short enough distances. Despite being corroborated by extensive numerical and experimental data, none of these conjectures have so far been rigorously proven. A main triumph of semiclassical quantum chaos has been the (non-rigorous) derivation of the Quantum Chaos conjecture for uniformly hyperbolic systems achieved by~\cite{Muller2004}.

For quantum many-body systems that lack a meaningful classical or semiclassical description, such as spin-$1/2$ or fermionic systems, the Quantum Chaos conjecture can be rephrased by saying that, in the sectors where all the local conservation laws are fixed, the spectral correlations of non-integrable quantum many body systems are described by the appropriate ensemble of random matrices~\cite{Montambaux,Prosen1999,RigolSantos2010}. This conjecture is again supported by a large amount of empirical data, however, it is very far from being intuitive as it states that the spectra of \emph{structured} time-evolution operators with low Fock-space connectivity (i.e.\ sparse matrices) should behave indistinguishably from those of dense random matrices. Until recently, no satisfactory theoretical explanation for this behaviour was available. 

In line with the rest of the review, here we consider discrete-time (Floquet) systems for which the object of interest for the spectral statistics is 
\be
{\rm spect}(\mathbb{U}) = \{e^{i\varphi_j}\,|\,j=1,2\ldots \mathcal N\},
\ee
i.e., the spectrum of the Floquet operator $\mathbb{U}$ (cf.~Eqs.~\eqref{eq:floquetoperator} and \eqref{eq:floquetoperator_graphical}). We assume for simplicity that there are no conservation laws so we can consider spectral properties in the entire Hilbert space of dimension $\mathcal N = q^{2L}$. The simplest non-trivial measure of spectral correlations is the so-called spectral form factor (SFF), which is defined as the Fourier transform of the spectral 2-point density-density correlation function~\cite{mehta2004random}, and reads as
\be
K(t) = \mathbb E\!\left[\,|{\rm tr}\, \mathbb U^t|^2\,\right] = \mathbb E\left[\sum_{j,j'=1}^{\mathcal N} e^{i (\varphi_j-\varphi_{j'}) t}\right]\,.
\label{eq:SFF}
\ee
As the SFF is not a self-averaging quantity~\cite{prange1997the}, e.g.\ in RMT its variation is on the order of the average for any $\mathcal N$, it needs to be defined with an explicit averaging ($\mathbb E$) over an ensemble of similar systems, or even a moving average over time. 

Within semiclassical periodic orbit theory, and in the leading order in $t/{\cal N}$, Berry derived the so-called diagonal approximation of SFF \cite{BerryDiag}, which gives 
\be
K(t) = \beta t + {\mathcal O}(t^2/{\mathcal N}), 
\ee
for classically chaotic systems with ($\beta=2$) and without ($\beta=1$) time-reversal symmetry. This exactly matches the leading order result of RMT. The complete RMT expansion of the SFF to all orders has been derived from semiclassics two decades later by~\cite{Muller2004}. Importantly, the linear ramp of the SFF has to terminate by the so-called Heisenberg time $t_{\rm H}=\mathcal N$ after which the SFF saturates $K(t > t_{\rm H})\simeq \mathcal N$ as all energy levels are individually resolved for larger times. On the other hand, for  integrable systems the Berry-Tabor conjecture yields a drastically different behavior of the SFF, namely saturation at essentially all time scales $K(t) \simeq \mathcal N$, except possibly at very short times when the SFF is dictated by the shortest periodic orbits. Analogous results extend to many-body semiclassical systems such as highly excited bosonic chains \cite{Dubertrand2016,Richter2022}. 

\begin{figure}[tb!]
\includegraphics[width=0.9\columnwidth]{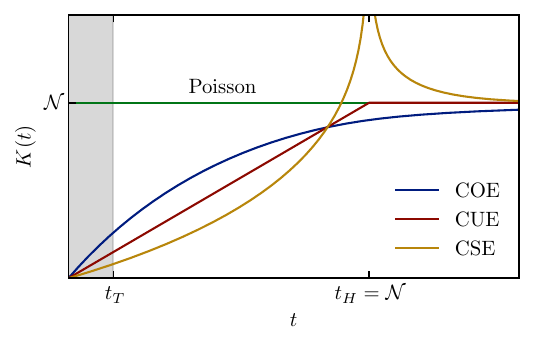}
\caption{Schematic behavior of the SFF for the three circular (Floquet) universality classes of RMT (COE/CUE/CSE), and the behavior for a typical quantum chaotic system (dashed) with Thouless and Heisenberg time-scales annotated.
\label{fig:SFF}}
\end{figure}

The first results for many-body systems away from the semiclassical regime only appeared very recently. The kicked Ising chain with long-range Ising interaction with power-law falloff has been studied in~\cite{Kos2018}, where the RMT SFF was derived to the first two orders in $t/t_{\rm H}$. This, however, required a random phase assumption in order to express the SFF in terms of a simple one-dimensional Ising-type transfer matrix. The method was later extended to more general kicked spin and kicked fermionic systems in~\cite{RoyProsen2020,Roy2022,KumarRoy2024}. On the other hand \cite{chan2018solution} studied a Floquet circuit with Haar-random gates and derived, within large-$q$ asymptotics, the leading order (linear ramp) behavior of the SFF. A more structured model with random-phase nearest neigbour interacting gates was studied in~\cite{ChanSpectral2018}. The behavior of the SFF at very short times is typically observed to be non-universal also in many-body systems, and agreement with RMT behavior is only expected after the so-called Thouless time scale, $K(t)\simeq K_{\rm RMT}(t)$ for $t>t_{\rm T}$. Determining the scaling of $t_{\rm T}(L)$ with the system size $L$ is an interesting problem in itself and the result depends on the locality of the interactions and the conservation-law content of a given system. For example, in all-to-all coupled (e.g. SYK) models~\cite{altland2018} as well as in long-range coupled kicked Ising models~\cite{Kos2018}, one finds $t_{\rm T}\propto \log L$, while in models with $U(1)$ symmetries one has $t_{\rm T}\propto L^2$ as a consequence of diffusion~\cite{Friedman2019}. In locally interacting spin models without extra symmetries the behavior of $t_{\rm T}(L)$ does not appear to be universal (see, however,~\cite{shivam2023many}). 

The first exact and rigorous result on the spectral statistics for a quantum many-body system away from the perturbative and/or semiclassical regime was found by~\cite{bertini_exact_2018}, who proved the emergence of the linear ramp of the SFF in the thermodynamic limit $L$ (though, for odd time steps $t$ only) of the self-dual kicked Ising model (cf.~Sec.~\ref{sec:KIM}). Later this result has been extended and generalized to a large class of dual-unitary Floquet circuits by~\cite{bertini2021random}. Below we briefly outline the main ideas of the latter work, while for details of the proofs we direct the reader to the original reference.

We consider local gates of the form  
\begin{align}
    U_{x+\frac{1}{2},\frac{1}{2}} = (u_{x} \otimes u_{x+\frac{1}{2}})\, U =&   \fineq[-0.6ex][.85][1]{
		\tsfmatV[0][-0.5][r][1][][][red6]
		\circgate[-.35][.85][blue6][left]
		\circgate[.35][.85][blue7][right]
		},\\
U_{x,1} = (w_{{\rm mod}(x-\frac{1}{2},L)} \otimes w_{x})\, W =& \fineq[-0.6ex][.85][1]{
		\tsfmatV[0][-0.5][r][1][][][red7]
		\circgate[-.35][.85][blue10][left]
		\circgate[.35][.85][blue9][right]
		},
		\label{eq:SFFgates}
\end{align}
where ${x\in\mathbb Z_L}$, $U,W$ is an arbitrary pair of dual-unitary gates parameterised as in Eqs.~\eqref{eq:orbit} and ~\eqref{eq:general_para_q2} ($\sigma^z/2$ represents the component of a quantum spin $(q-1)/2$ for $q>2$), and $u_x,w_x$ are arbitrary single-qudit gates, i.e. elements of $U(q)$. Note that for $q=2$ the parameterisation in Eqs.~\eqref{eq:orbit} and ~\eqref{eq:general_para_q2} covers \emph{all} dual-unitary gates. Moreover, we stress that all the results stated below could also be extended to position-dependent interactions, but we keep them spatially homogeneous for simplicity.

We now consider the dynamics generated by a Floquet brickwork circuit built using the gates in Eq.~\eqref{eq:SFFgates} with periodic boundary conditions, such that the single time-step Floquet operator is of the form
\be
\!\!\!\mathbb{U} =
\fineq[-0.6ex][.6][1]{
\draw[thick, dotted] (-9.5,-1.7) -- (0.4,-1.7);
\draw[thick, dotted] (-9.5,-1.3) -- (0.4,-1.3);
\foreach \i in {3,...,13}{
\draw[gray, dashed] (-12.5+\i,-2.1) -- (-12.5+\i,0.3);
}
\foreach \i in {3,...,5}{
\draw[gray, dashed] (-9.75,3-\i) -- (.75,3-\i);
}
\foreach \i in{0.5}
{
\draw[ thick] (0.5,1+2*\i-0.5-3.5) arc (-45:90:0.17);
\draw[ thick] (-10+0.5,1+2*\i-0.5-3.5) arc (270:90:0.15);
}
\foreach \i in{1.5}{
\draw[thick] (0.5,2*\i-0.5-3.5) arc (45:-90:0.17);
\draw[thick] (-10+0.5+0,2*\i-0.5-3.5) arc (90:270:0.15);
}
\foreach \i in {0,1}{
\Text[x=1.25,y=-2+2*\i]{\scalebox{1.5}{$\i$}}
}
\foreach \i in {1}{
\Text[x=1.25,y=-2+\i]{\scalebox{1.75}{$\frac{\i}{2}$}}
}
\foreach \i in {1,3,5}{
\Text[x=-7.5+\i+1-3,y=-2.6]{\scalebox{1.75}{$\frac{\i}{2}$}}
}
\foreach \i in {1,2,3}{
\Text[x=-7.5+2*\i-2,y=-2.6]{\scalebox{1.5}{${\i}$}}
}
\foreach \i in {0,...,4}{
\tsfmatV[-2*\i][-1.5][r][1][][][red7]
}
\foreach \i in {1,...,5}
{
\tsfmatV[-2*\i+1][-2.5][r][1][][][red6]
}
\circgate[-.5-9][-1][blue1][left]
\circgate[-.5-8][-1][blue2][right]
\circgate[-.5-7][-1][blue3][left]
\circgate[-.5-6][-1][blue4][right]
\circgate[-.5-5][-1][blue5][left]
\circgate[-.5-4][-1][blue6][right]
\circgate[-.5-3][-1][blue7][left]
\circgate[-.5-2][-1][blue8][right]
\circgate[-.5-1][-1][blue9][left]
\circgate[-.5][-1][blue10][right]
\circgate[-.5+1][0][blue11][right]
\circgate[-.5-8][0][blue12][left]
\circgate[-.5-7][0][blue13][right]
\circgate[-.5-6][0][blue14][left]
\circgate[-.5-5][0][blue15][right]
\circgate[-.5-4][0][blue16][left]
\circgate[-.5-3][0][blue2][right]
\circgate[-.5-2][0][blue4][left]
\circgate[-.5-1][0][blue5][right]
\circgate[-.5][0][blue1][left]
\Text[x=-2,y=-2.6]{$\cdots$}
}.
\ee
The SFF in Eq.~\eqref{eq:SFF} can then be written as a two-replica partition sum of a 2-dimensional complex weight vertex with periodic boundary conditions
\begin{align}
\!\!\!\!K(t) \!= 
& \mathbb E\!\!\left[
\fineq[-0.6ex][.55][1]{
\foreach \i in {1,...,5}{
\draw[thick, dotted] (2*\i+2-12.5+0.255,-1.75-0.1) -- (2*\i+2-12.5+0.255,4.25-0.1);
\draw[thick, dotted] (2*\i+2-11.5-0.255,-1.75-0.1) -- (2*\i+2-11.5-0.255,4.25-0.1);}
\foreach \i in {1,...,5}{
\draw[thick] (2*\i+2-11.5,4) arc (-45:175:0.15);
\draw[thick] (2*\i+2-11.5,-2) arc (315:180:0.15);
\draw[thick] (2*\i+2-0.5-12,-2) arc (-135:0:0.15);
}
\foreach \i in {2,...,6}{
\draw[thick] (2*\i+2-2.5-12,4) arc (225:0:0.15);
}
\foreach \j in {0,1,2}{
\foreach \i in{0.5}
{
\draw[ thick] (0.5,1+2*\i-0.5-3.5+2*\j) arc (-45:90:0.17);
\draw[ thick] (-10+0.5,1+2*\i-0.5-3.5+2*\j) arc (270:90:0.15);
}
\foreach \i in{1.5}{
\draw[thick] (0.5,2*\i-0.5-3.5+2*\j) arc (45:-90:0.17);
\draw[thick] (-10+0.5+0,2*\i-0.5-3.5+2*\j) arc (90:270:0.15);
}
\foreach \i in {0,...,4}{
\tsfmatV[-2*\i][-1.5+2*\j][r][1][][][red7]
}
\foreach \i in {1,...,5}
{
\tsfmatV[-2*\i+1][-2.5+2*\j][r][1][][][red6]
}
}
\foreach \j in {0,1,2}{
\circgate[-.5-9][-1+2*\j][blue1][left]
\circgate[-.5-8][-1+2*\j][blue2][right]
\circgate[-.5-7][-1+2*\j][blue3][left]
\circgate[-.5-6][-1+2*\j][blue4][right]
\circgate[-.5-5][-1+2*\j][blue5][left]
\circgate[-.5-4][-1+2*\j][blue6][right]
\circgate[-.5-3][-1+2*\j][blue7][left]
\circgate[-.5-2][-1+2*\j][blue8][right]
\circgate[-.5-1][-1+2*\j][blue9][left]
\circgate[-.5][-1+2*\j][blue10][right]
\circgate[-.5+1][2*\j][blue11][right]
\circgate[-.5-8][2*\j][blue12][left]
\circgate[-.5-7][2*\j][blue13][right]
\circgate[-.5-6][2*\j][blue14][left]
\circgate[-.5-5][2*\j][blue15][right]
\circgate[-.5-4][2*\j][blue16][left]
\circgate[-.5-3][2*\j][blue2][right]
\circgate[-.5-2][2*\j][blue4][left]
\circgate[-.5-1][2*\j][blue5][right]
\circgate[-.5][2*\j][blue1][left]
}
\def\shifty{-6.5}
\foreach \i in {1,...,5}{
\draw[thick, dotted] (2*\i+2-12.5+0.255,-2+\shifty) -- (2*\i+2-12.5+0.255,4+\shifty);
\draw[thick, dotted] (2*\i+2-11.5-0.255,-2+\shifty) -- (2*\i+2-11.5-0.255,4+\shifty);}
\foreach \i in {1,...,5}{
\draw[ thick] (2*\i+2-1.5-11,4+\shifty) arc (135:-0:0.15);
\draw[ thick] (2*\i+2-.5-11,4+\shifty) arc (-325:-180:0.15);
\draw[ thick] (2*\i+2-1.5-11,-2+\shifty) arc (-45:180:-0.15);
\draw[ thick] (2*\i+2-0.5-11,-2+\shifty) arc (45:-180:0.15);
}
\foreach \j in {-.75,-1.75,-2.75}{
\foreach \i in{0.5}
{
\draw[ thick] (0.5,1+2*\i-0.5-3.5+2*\j) arc (-45:90:0.17);
\draw[ thick] (-10+0.5,1+2*\i-0.5-3.5+2*\j) arc (270:90:0.15);
}
\foreach \i in{1.5}{
\draw[thick] (0.5,2*\i-0.5-3.5+2*\j) arc (45:-90:0.17);
\draw[thick] (-10+0.5+0,2*\i-0.5-3.5+2*\j) arc (90:270:0.15);
}
\foreach \i in {0,...,4}{
\tsfmatV[-2*\i][-3.5+2*\j][r][1][][][blue6][bottomright]
}
\foreach \i in {1,...,5}
{
\tsfmatV[-2*\i+1][-2.5+2*\j][r][1][][][blue7][bottomright]
}
}
\foreach \j in {-.75,-1.75,-2.75}{
\circgate[-.5-9][-2+2*\j][grey1][leftf]
\circgate[-.5-8][-2+2*\j][grey2][rightf]
\circgate[-.5-7][-2+2*\j][grey3][leftf]
\circgate[-.5-6][-2+2*\j][grey4][rightf]
\circgate[-.5-5][-2+2*\j][grey5][leftf]
\circgate[-.5-4][-2+2*\j][grey6][rightf]
\circgate[-.5-3][-2+2*\j][grey7][leftf]
\circgate[-.5-2][-2+2*\j][grey8][rightf]
\circgate[-.5-1][-2+2*\j][grey9][leftf]
\circgate[-.5][-2+2*\j][grey10][rightf]
\circgate[-.5+1][-3+2*\j][grey11][rightf]
\circgate[-.5-8][-3+2*\j][grey12][leftf]
\circgate[-.5-7][-3+2*\j][grey3][rightf]
\circgate[-.5-6][-3+2*\j][grey4][leftf]
\circgate[-.5-5][-3+2*\j][grey6][rightf]
\circgate[-.5-4][-3+2*\j][grey9][leftf]
\circgate[-.5-3][-3+2*\j][grey8][rightf]
\circgate[-.5-2][-3+2*\j][grey5][leftf]
\circgate[-.5-1][-3+2*\j][grey1][rightf]
\circgate[-.5][-3+2*\j][grey2][leftf]
}
}
\right]\notag\\
=& \mathbb E\!\!\left[\fineq[-0.6ex][.55][1]{
\foreach \i in {1,...,5}{
\draw[thick, dotted] (2*\i+2-12.5+0.255,-1.75-0.1) -- (2*\i+2-12.5+0.255,4.25-0.1);
\draw[thick, dotted] (2*\i+2-11.5-0.255,-1.75-0.1) -- (2*\i+2-11.5-0.255,4.25-0.1);}
\foreach \i in {1,...,5}{
\draw[thick] (2*\i+2-11.5,4) arc (-45:175:0.15);
\draw[thick] (2*\i+2-11.5,-2) arc (315:180:0.15);
\draw[thick] (2*\i+2-0.5-12,-2) arc (-135:0:0.15);
}
\foreach \i in {2,...,6}{
\draw[thick] (2*\i+2-2.5-12,4) arc (225:0:0.15);
}
\foreach \j in {0,1,2}{
\foreach \i in{0.5}
{
\draw[ thick] (0.5,1+2*\i-0.5-3.5+2*\j) arc (-45:90:0.17);
\draw[ thick] (-10+0.5,1+2*\i-0.5-3.5+2*\j) arc (270:90:0.15);
}
\foreach \i in{1.5}{
\draw[thick] (0.5,2*\i-0.5-3.5+2*\j) arc (45:-90:0.17);
\draw[thick] (-10+0.5+0,2*\i-0.5-3.5+2*\j) arc (90:270:0.15);
}
\foreach \i in {0,...,4}{
\roundgate[-2*\i][-.5+2*\j][1][bottomright][orange][1]
}
\foreach \i in {1,...,5}
{
\roundgate[-2*\i+1][-1.5+2*\j][1][bottomright][orange2][1]
}}
\foreach \j in {0,1,2}{
\circgate[-.5-9][-1+2*\j][yellow1][left]
\circgate[-.5-8][-1+2*\j][yellow2][right]
\circgate[-.5-7][-1+2*\j][yellow3][left]
\circgate[-.5-6][-1+2*\j][yellow4][right]
\circgate[-.5-5][-1+2*\j][yellow5][left]
\circgate[-.5-4][-1+2*\j][yellow6][right]
\circgate[-.5-3][-1+2*\j][yellow7][left]
\circgate[-.5-2][-1+2*\j][yellow8][right]
\circgate[-.5-1][-1+2*\j][yellow9][left]
\circgate[-.5][-1+2*\j][yellow10][right]
\circgate[-.5+1][2*\j][yellow6][right]
\circgate[-.5-8][2*\j][yellow9][left]
\circgate[-.5-7][2*\j][yellow10][right]
\circgate[-.5-6][2*\j][yellow4][left]
\circgate[-.5-5][2*\j][yellow3][right]
\circgate[-.5-4][2*\j][yellow2][left]
\circgate[-.5-3][2*\j][yellow1][right]
\circgate[-.5-2][2*\j][yellow7][left]
\circgate[-.5-1][2*\j][yellow10][right]
\circgate[-.5][2*\j][yellow1][left]
}}\right]\!\!\!, \label{eq:SFFdiag}
\end{align}
where in the second step we adopted the folded representation (cf.~Sec.~\ref{sec:diagrams}) and introduced
\be
\fineq{
\begin{scope}[rotate around={45:(0,0)}]
\draw[thick] (-0.25,-0.25) -- (0.25,0.25);
\circgate[0][0][grey3][right]
\draw[thick] (-0.25-.15,-0.25-.075) -- (0.25-.15,0.25-.075);
\circgate[-.15][-.075][blue3][right]
\end{scope}
}=\fineq{
\begin{scope}[rotate around={45:(0,0)}]
\draw[thick] (-0.25,-0.25) -- (0.25,0.25);
\circgate[0][0][yellow3][right]
\end{scope}}.
\ee
The expectation value is taken over disordered single-site gates (e.g.\ external magnetic fields) where $u_x,w_x$ are random and  independently distributed according to some smooth, but arbitrary distribution in $U(q)$. As we shall see later, this disorder can be made arbitrarily weak for the general results to apply, as long as we take a large enough system. Since the disorder is spatially uncorrelated the expectation value can now be factorised along the spatial direction 
\be
K(t) =
\fineq[1.6ex][.55][1]{
\foreach \i in{1.5,2.5,3.5}{
\draw[thick] (-10+0.5+0,2*\i-0.5-3.5) arc (45:270:0.15);
}
\foreach \i in{0.5,1.5,2.5}
{
\draw[thick] (-10+0.5,1+2*\i-0.5-3.5) arc (315:90:0.15);
}
}
\mathbb E\!\!\left[\fineq[-0.6ex][.55][1]{
\foreach \i in {5}{
\draw[thick, dotted] (2*\i+2-12.5+0.255,-1.75-0.1) -- (2*\i+2-12.5+0.255,4.25-0.1);
\draw[thick, dotted] (2*\i+2-11.5-0.255,-1.75-0.1) -- (2*\i+2-11.5-0.255,4.25-0.1);}
\foreach \i in {5}{
\draw[thick] (2*\i+2-11.5,4) arc (-45:175:0.15);
\draw[thick] (2*\i+2-11.5,-2) arc (315:180:0.15);
\draw[thick] (2*\i+2-0.5-12,-2) arc (-135:0:0.15);
}
\foreach \i in {6}{
\draw[thick] (2*\i+2-2.5-12,4) arc (225:0:0.15);
}
\foreach \j in {0,1,2}{
\foreach \i in {0}{
\roundgate[-2*\i][-.5+2*\j][1][bottomright][orange][1]
}
\foreach \i in {1}
{
\roundgate[-2*\i+1][-1.5+2*\j][1][bottomright][orange2][1]
}}
\foreach \j in {0,1,2}{
\circgate[-.5][-1+2*\j][yellow10][right]
\circgate[-.5+1][-1+2*\j][yellow6][left]
\circgate[-.5+1][2*\j][yellow10][right]
\circgate[-.5][2*\j][yellow1][left]
}}\right]
\!\!\cdot\!
\mathbb E\!\!\left[\fineq[-0.6ex][.55][1]{
\foreach \i in {5}{
\draw[thick, dotted] (2*\i+2-12.5+0.255,-1.75-0.1) -- (2*\i+2-12.5+0.255,4.25-0.1);
\draw[thick, dotted] (2*\i+2-11.5-0.255,-1.75-0.1) -- (2*\i+2-11.5-0.255,4.25-0.1);}
\foreach \i in {5}{
\draw[thick] (2*\i+2-11.5,4) arc (-45:175:0.15);
\draw[thick] (2*\i+2-11.5,-2) arc (315:180:0.15);
\draw[thick] (2*\i+2-0.5-12,-2) arc (-135:0:0.15);
}
\foreach \i in {6}{
\draw[thick] (2*\i+2-2.5-12,4) arc (225:0:0.15);
}
\foreach \j in {0,1,2}{
\foreach \i in {0}{
\roundgate[-2*\i][-.5+2*\j][1][bottomright][orange][1]
}
\foreach \i in {1}
{
\roundgate[-2*\i+1][-1.5+2*\j][1][bottomright][orange2][1]
}}
\foreach \j in {0,1,2}{
\circgate[-.5][-1+2*\j][yellow10][right]
\circgate[-.5+1][-1+2*\j][yellow6][left]
\circgate[-.5+1][2*\j][yellow10][right]
\circgate[-.5][2*\j][yellow1][left]
}}\right]
\!\!\cdots\!
\mathbb E\!\!\left[\fineq[-0.6ex][.55][1]{
\foreach \i in {5}{
\draw[thick, dotted] (2*\i+2-12.5+0.255,-1.75-0.1) -- (2*\i+2-12.5+0.255,4.25-0.1);
\draw[thick, dotted] (2*\i+2-11.5-0.255,-1.75-0.1) -- (2*\i+2-11.5-0.255,4.25-0.1);}
\foreach \i in {5}{
\draw[thick] (2*\i+2-11.5,4) arc (-45:175:0.15);
\draw[thick] (2*\i+2-11.5,-2) arc (315:180:0.15);
\draw[thick] (2*\i+2-0.5-12,-2) arc (-135:0:0.15);
}
\foreach \i in {6}{
\draw[thick] (2*\i+2-2.5-12,4) arc (225:0:0.15);
}
\foreach \j in {0,1,2}{
\foreach \i in {0}{
\roundgate[-2*\i][-.5+2*\j][1][bottomright][orange][1]
}
\foreach \i in {1}
{
\roundgate[-2*\i+1][-1.5+2*\j][1][bottomright][orange2][1]
}}
\foreach \j in {0,1,2}{
\circgate[-.5][-1+2*\j][yellow10][right]
\circgate[-.5+1][-1+2*\j][yellow6][left]
\circgate[-.5+1][2*\j][yellow10][right]
\circgate[-.5][2*\j][yellow1][left]
}}\right]
\fineq[1.6ex][.55][1]{
\foreach \i in{1,2,3}{
\draw[thick] (0.5,2*\i-0.5-3.5) arc (45:-90:0.15);
}
\foreach \i in{0,1,2}
{
\draw[thick] (0.5,1+2*\i-0.5-3.5) arc (-45:90:0.15);
}
}.
\label{eq:SFFTM}
\ee
This equation means that the SFF can be expressed in terms of traces of powers a transfer matrix $\mathbb T$ acting on a temporal lattice of $2t$ sites of local dimension $q^2$ (a single column on the r.h.s. of Eq.~\eqref{eq:SFFTM}), i.e.
\begin{equation}
K(t) = \tr[\mathbb T^L]\,.
\label{eq:SFFtrace}
\end{equation}
Note that $\mathbb T$ has periodic boundary conditions in the temporal direction due to the trace in the definition of the SFF.

Evaluating Eq.~\eqref{eq:SFFtrace} in the thermodynamic limit amounts to finding the leading eigenvalue of the operator $\mathbb T$ --- which turns out to be $1$ ---, counting its multiplicity, and, to prove convergence, also establish the positivity of its spectral gap for any $t$. The final result, i.e.\ the proof of the linear RMT ramp, is given by
\begin{theorem} [Theorem 1~\cite{bertini2021random}]
For i.i.d. local random 1-qubit gates $u_x,w_x$ with an arbitrary {\em smooth} (and nonsingular) distribution over $U(2)$, and for any dual-unitary 2-qubit gate $U$ other than the SWAP, we have
\be 
\lim_{L\to\infty} K(t) = {\rm dim}\, {\mathcal M}' = t\,.
\label{eq:mainresultSFF}
\ee
\end{theorem}
Here $\mathcal M'$ is the commutant of the following set of operators
\begin{align}
& \mathcal M = \left\{ 
M_{a,\iota},M_{ab,\iota}\,\,:\,\, 
a,b\in\{x,y,z\}, \iota\in\{0,1\}\right\} \\
& M_{a,\iota} =\sum_{\tau=0}^{t-1}\!\sigma^a_{\tau+\frac{\iota}{2}},\quad M_{ab,\iota}=\sum_{\tau=0}^{t-1}\!\sigma^a_{\tau+\frac{\iota}{2}}\sigma^b_{\tau+\frac{\iota+1}{2}}, \label{setM}
\end{align}
where
\be
\sigma^\alpha_\tau =\mathbb{1}_{2\tau} \otimes \sigma^\alpha \otimes \mathbb{1}_{2t - 2\tau - 1},\quad \tau\in \frac{1}{2}\mathbb Z_{2t}.
\ee
and, although~\cite{bertini2021random} considered generic $q\geq 2$, here we restricted to the case $q=2$ for the sake of simplicity. We keep this restriction until the end of the section. 

Note that $M_{a,\iota}$ represents a magnetization operator (in the $a$ direction) on the integer $(\iota=0)$ and half-odd-integer $(\iota=1)$ temporal sub-lattices. Analogously, $M_{ab,\iota}$ can be thought of as a `2-site magnetisation' operator or a nearest-neighbour Hamiltonian coupling the two sublattices. Both $M_{a,\iota}$ and  $M_{ab,\iota}$ are translationally invariant under an integer shift of $\tau$ as we have $\tau \equiv \tau+t$. Therefore, the {\em minimal} set of generators of the commutant is spanned by $t$ integer-site translation operators. The crux is to show that there are no others.

The proof amounts to establishing a collection of instructive lemmas, which we now describe. Let us begin by writing the SFF transfer matrix as a unitary operator acting on the temporal lattice multiplied with a non-expansive (disorder-averaging) map
\be
\begin{aligned}
&\mathbb T = (\tilde{\mathbb U} \otimes \tilde{\mathbb U}^*) \mathbb O^\dag_1 (\tilde{\mathbb W} \otimes \tilde{\mathbb W}^*) \mathbb O^{\phantom{\dag}}_0\,,  \\
&\tilde{\mathbb{U}} = \tilde{U}^{\otimes t}\,, \qquad
\tilde{\mathbb{W}} = \tilde{W}^{\otimes t}\,, \\
&\mathbb O_{\iota'} = \mathbb O_{0\iota'} \mathbb O_{1\iota'}=  \mathbb O_{1\iota'} \mathbb O_{0\iota'}\,,\\  
&\mathbb O_{\iota\iota'} \!\!=\! \!\!\int\!\!{\rm d}^{3}\boldsymbol{\theta}\,g_{\iota\iota'}(\boldsymbol{\theta})
\exp\left[i\boldsymbol\theta\!\cdot\! (\boldsymbol{M}_{\iota}\!\otimes\! \mathbb{1}_{2t}\!-\!\mathbb{1}_{2t}\!\otimes\! \boldsymbol{M}^*_{\iota})\right]\!.
\end{aligned}
\ee
Here $\{g_{\iota\iota'}(\boldsymbol{\theta})\}_{\iota,\iota'=0,1}$ are the disorder distributions (along the integer and half-odd-integer temporal sublattices at either integer or half-odd-integer positions) and we assume them to be smooth.  
We can then state a useful lemma that characterises the leading eigenvalues of $\mathbb T$.
\begin{lemma}[Leading Eigenvalues~\cite{bertini2021random}]
\label{lemma:leadingeig}
For dual-unitary circuits, $\mathbb T$ fulfills:
\begin{itemize}
\item[(i)] $|\lambda|\leq 1$ for all $\lambda\in {\rm spect}(\mathbb T)$.
\item[(ii)] If $\mathbb T\ket{A} =e^{i \phi} \ket{A}$, with $\phi\in\mathbb R$, then
\begin{eqnarray*}
&&(\tilde{\mathbb U} \otimes \tilde{\mathbb U}^*)\cdot (\tilde{\mathbb W} \otimes \tilde{\mathbb W}^*)\ket{A}=e^{i \phi}\!\ket{A}\,,\\
&&
( M_{a, \iota}\otimes \mathbb{1}_{2t}-\mathbb{1}_{2t}\otimes  M_{a, \iota}^*)(\tilde{\mathbb W} \otimes \tilde{\mathbb W}^*) \ket{A} = 0\,, \\
&&
( M_{a, \iota}\otimes \mathbb{1}_{2t}-\mathbb{1}_{2t}\otimes  M_{a, \iota}^*)\ket{A} = 0, \\
&&\iota\in\{0,1\},\;a\in\{x,y,z\}\,.
\end{eqnarray*}
\item[(iii)] For any unimodular eigenvalue $\lambda$, $|\lambda| =1$, its algebraic and geometric multiplicities coincide. 
\end{itemize}
\end{lemma}

Upon un-vectorizing the notation, it is easy to realise that (ii), for each eigenvector/eigenoperator $\ket{A}$ of unimodular eigenvalue, is equivalent to the following commutation identities
\begin{align}
&\tilde{\mathbb U} \tilde{\mathbb W} A \tilde{\mathbb W}^\dag \tilde{\mathbb U}^\dag = e^{i \phi} A, \nonumber\\
&[ M_{a, \iota},A ] = 0, \label{comm}\\
&[\tilde{\mathbb W}^\dag 
M_{a, \iota} \tilde{\mathbb W},A] = 0, \nonumber
\end{align}
where we introduced $\ket{A}$ as the un-vectorized form of $A$. For the explicit parametrization of dual-unitary gates in Eqs.~\eqref{eq:orbit} and ~\eqref{eq:general_para_q2}, the above condition can be rephrased as 
\begin{lemma}[Algebraic Reformulation~\cite{bertini2021random}]
\label{lemma:rephrasing}
For $J,J' \neq 0\, (\,{\rm mod}\,\pi)$ the conditions in Eq.~\eqref{comm} cannot be met, unless $\phi=0$  with resulting eigenvalue 1. In this case it is equivalent to:
\be
\begin{aligned}
&[A, M_{a,\iota}]=0, \quad [A, M_{ab,\iota}]=0\,, \\ 
& a,b\in\{x,y,z\}, \; \iota\in\{0, 1\}\,.
\end{aligned}
\ee
\end{lemma}
This means that the leading eigenvectors of $\mathbb T$ are in 1-to-1 correspondence with the elements of $\mathcal K'$: the commutant of the multiplicative algebra $\mathcal K \subset  {\rm End}((\mathbb C^2)^{\otimes 2t})$ spanned by a set of generators $\mathcal M$ in Eq.~\eqref{setM}. One can prove by explicit construction (see Lemma 3 of \cite{bertini2021random}) that the representations of $\mathcal K$ over the eigenspaces of the 2-site temporal shift operator (shifting for a full time-step) are all {\em irreducible} and {\em inequivalent}. This is accomplished by providing an explicit `path' between an arbitrary pair of states in each `temporal-momentum' sector of $(\mathbb C^2)^{\otimes 2t}$ via sequence of products of elements $\mathcal K$. The commutant $\mathcal K' = \mathcal M'$ is hence exactly spanned by the cyclic group $C_t$ of $2\tau$-site shifts, $\tau=0,1\ldots,t-1$, on the temporal lattice $(\mathbb C^2)^{\otimes 2t}$. In other words ${\rm dim}\, {\mathcal K}'={\rm dim}\, {\mathcal M}'=t$. 

From this point of view the linear ramp of the SFF is a simple consequence of the time-translation symmetry of Floquet systems, similarly to what periodic orbit theory predicts in the continuous time (Hamiltonian) case. The key to this statement is the ergodicity of the space evolution generated by the transfer matrix $\mathbb T$, meaning that there are no nontrivial eigenvectors of $\mathbb T$ beyond the trivial time translations. It is remarkable that this claim does not depend on the details of the disorder distributions $\{g_{\iota\iota'}(\boldsymbol{\theta})\}_{\iota,\iota'=0,1}$: the latter can be arbitrarily narrow (weak disorder) or wide (strong disorder) as long as they are smooth and the thermodynamic limit is taken first. We also stress that the time-translation symmetry of $M_{a,\iota}$ and $M_{ab,\iota}$ is what makes the proof of Lemma 3 of \cite{bertini2021random} particularly convoluted: because of this invariance, the multiplicative algebra $\mathcal K$ can be irreducible only modulo temporal translations, i.e., only in the quotient space ${\rm End}((\mathbb C^2)^{\otimes 2t})/C_t$. This means that the problem loses the tensor product structure and cannot be reduced to a single-particle one. Finally, we emphasise that the result in Eq.~\eqref{eq:mainresultSFF} implies that dual-unitary circuits are quantum chaotic at all time scales in the thermodynamic limit, i.e.\ their Thouless time is formally vanishing $t_{\rm T} = 0$. Thus they represent an example of what one may dub {\em critical quantum chaos}.

\cite{bertini2021random} also made a similar statement for dual-unitary circuits with time-reversal symmetry. In this case one can prove the analogue of Lemmas \ref{lemma:leadingeig} and \ref{lemma:rephrasing} to claim that the thermodynamic limit of the SFF is equal to the dimension of the commutant of the set $\mathcal M_{\rm T}$ generated by $\{M_{a,\iota}+R_{2t} M_{a,\iota} R_{2t}\}$ and $\{M_{ab,\iota}+R_{2t} M_{ab,\iota} R_{2t}\}$, where $R_{2t}$ implements a reflection of the temporal lattice around its centre. It is immediate to see that $\mathcal M_{\rm T}$ is invariant under reflections and translations of the temporal lattice, therefore ${\rm dim}\,\mathcal M'_{\rm T}\geq 2 t$. The proof that  ${\rm dim}\,\mathcal M'_{\rm T}= 2 t$, however, is still missing in generic dual-unitary circuits. Such a statement has only been proven (by~\cite{bertini_exact_2018}) for the self-dual kicked Ising model.  

A similar space-transfer-matrix-based approach can also be used to characterise the higher moments of the spectral form factor
\be
K_n(t) = \mathbb E\!\left[\,|{\rm tr}\, \mathbb U^t|^{2n}\,\right],
\label{eq:nSFF}
\ee 
for which the RMT prediction in the non-time-reversal-symmetric case reads 
\be
K_n(t) = n! t^n \left(1 + \mathcal O(t/\mathcal N)\right). 
\label{eq:nSFFRMT}
\ee
In this case \cite{bertini2021random} have shown that the leading eigenvectors of the relevant transfer matrix are in 1-to-1 correspondence with $(\mathcal K^{\otimes n})'$, where $\mathcal K$ is again the algebra generated by $\mathcal M$ in Eq.~\eqref{setM}.  This means that the thermodynamic limit of Eq.~\eqref{eq:nSFF} is again lower bounded by the one of Eq.~\eqref{eq:nSFFRMT} but proving the equality is again an open problem. Note that, because of the time-translation symmetry of $\mathcal K$, this problem goes beyond a direct application of the Schur--Weyl duality~\cite{fulton1991representation}. 

As a different extension of the SFF in Eq.~\eqref{eq:SFF}, \cite{fritzsch_eigenstate_2025} considered the partial spectral form factor (PSFF), defined as
\be
K_A(t) = \mathbb E\!\left[\mathrm{tr}_A \left(\mathbb U_A(t) \mathbb U_A(-t)\right)\right] \,,
\label{eq:PSFF}
\ee
where $\mathbb{U}_A(t) = \mathrm{tr}_{\bar{A}} [ \mathbb{U}^t]$. The PSFF encodes correlations between both eigenstates and eigenvalues~\cite{joshi_probing_2022,gong_classification_2020}. For dual-unitary circuits the PSFF at short times was shown to deviate from the RMT prediction (constant in dual-unitary circuits, linearly growing in RMT), whereas at later times the PSFF quickly approaches the random matrix result. Deviations from random matrix theory were also observed for the eigenstate entanglement in dual-unitary circuits  by~\cite{herrmann_deviations_2024}. 

We conclude this section by noting that the space-transfer matrix approach described here has been applied to characterise spectral correlations also beyond the dual-unitary case. Indeed, the connection in Eq.~\eqref{eq:SFFtrace} between SFF and the space transfer matrix $\mathbb T$ does not require dual-unitarity: the latter is only used to determine the leading eigenvalues of $\mathbb T$. Away from dual-unitarity, the spectrum of the space transfer matrix has been characterised by other (non-rigorous) means. Specifically, by linking the spectral properties of the space transfer matrix $\mathbb T$ to those of the system's orbits in the many-body Hilbert space, \cite{garratt_local_2021} conjectured that away from the dual-unitary point the leading eigenvalues of $\mathbb T$ should acquire an exponentially decaying corrections in $t$. When plugged back into Eq.~\eqref{eq:SFFtrace} these corrections introduce a finite Thouless time $t_{\rm T}(L) \propto \log L$. \cite{riddell2024structural} came to a similar conclusion developing a perturbative expansion of the leading eigenvalues of $\mathbb T$ around the dual unitary point. This expansion is convergent provided the validity of an exponential bound on certain multi-point correlation functions involving the unperturbed SFF transfer matrix. Finally, \cite{yoshimura_operator_2025} considered the (P)SFF and other probes of operator dynamics in Floquet circuits and showed, amongst others, how the nonuniversal behaviour of the SFF at times shorter than the Thouless time could be related to late-time tails of autocorrelation functions.

\section{Self-dual kicked Ising model}
\label{sec:KIM}

In this section we discuss the properties of the self-dual kicked Ising model (KIM), a particular dual-unitary circuit that occurs naturally in the study of periodically driven systems. Specifically, we focus on the (time-dependent) Hamiltonian dynamics where evolution under a classical Ising Hamiltonian is periodically alternated with a kick along the transverse direction~\cite{Prosen2000,prosen2002general, prosen2007chaos}
\begin{align}
H_{\textrm{KI}}(t;\textbf{h}) = H_{\textrm{I}}[\textbf{h}] + \sum_{n \in \mathbb{Z}} \delta(t-n) H_{\textrm{K}}\,,
\end{align}
where we defined
\begin{align}
    H_{\textrm{I}}[\textbf{h}] &= J \sum_{j=1}^{L} \sigma^z_j \sigma^z_{j+1} +\sum_{j=1}^L h_j \sigma^z_j, \\
    H_{\textrm{K}} &=  b \sum_{j=1}^L \sigma_j^y. \label{eq:HK}
\end{align}
Here $\sigma_j^{\alpha}$ with $\alpha \in \{x,y,z\}$ are the Pauli matrices, $L$ is the length of the chain and we adopted periodic boundary conditions\footnote{The kick is chosen to be along the $y$-direction, rather than the $x$-direction as in most of the literature, because this choice simplifies the graphical language used in the rest of the section.}. The parameters $J$ and $b$ represent respectively the strengths of the Ising interaction and of the transverse kick, while $\textbf{h} = \{h_1, h_2, \dots, h_L\}$ describes a (possibly inhomogeneous) longitudinal field. Considering stroboscopic dynamics after a discrete number $t$ of evolution cycles, the dynamics can be expressed in terms of a Floquet unitary $\mathbb U_{\textrm{KI}}[\textbf{h}]$ as
\begin{align}
\label{eq:KIM:U_Floq}
 \mathbb U(t) = (\mathbb U_{\textrm{KI}}[\textbf{h}])^t, \qquad \mathbb U_{\textrm{KI}}[\textbf{h}] = e^{-i H_{\textrm{K}}} e^{-i H_\textrm{I}[\textbf{h}]}\,.
\end{align}
All time scales have been absorbed in the definition of the Hamiltonians.

For some specific choices of the Ising interaction and transverse field strength, specifically 
\be
|J| = |b| = \pi/4,
\label{eq:selfdualcondition}
\ee
the KIM corresponds to a dual-unitary circuit. In fact, it corresponds to a special dual-unitary circuit whose spatial evolution operator takes exactly the form in Eq.~\eqref{eq:KIM:U_Floq}. For this reason, when the parameters fulfil the condition in  Eq.~\eqref{eq:selfdualcondition} the model is referred to as the \emph{self-dual} KIM. The duality of the KIM was originally pointed out in \cite{Gutkin}, following the observation of a similar duality in coupled cat maps~\cite{gutkin_classical_2016}, and later used to exactly characterise the spectral form factor~\cite{bertini_exact_2018} and entanglement dynamics~\cite{bertini2019entanglement} of the model. The latter results popularised the kicked Ising model as a `minimal model for many-body quantum chaos'. This model was recast as a dual-unitary circuit in~\cite{gopalakrishnan_unitary_2019}, which identified the dual-unitary condition (cf.~Eqs.~\eqref{eq:dualunitaritynonfolded}) and recast its entanglement dynamics in a quantum channel approach.

Besides its intrinsic interest as a solvable kicked spin-chain, the self-dual KIM is popular in the dual-unitary literature because it allows for various technical simplifications compared to general dual-unitary models. For instance, in this model solvable states are given by (possibly rotated) products of single-site states~\cite{bertini_exact_2018}, rather than two-site states. Moreover, the dynamics are translationally invariant over a single site rather than over two sites. Various results on dual-unitarity also require explicit parameterisations, and the self-dual KIM presents a particularly simple parameterisation that can be used to either fine-tune the model to return Clifford or integrable dynamics at isolated points or chaotic dynamics away from these points. 
For these reasons, most of the results in Secs.~\ref{sec:quantumdynamics} and \ref{sec:spectral}  (but also in the upcoming Sec.~\ref{sec:quantumcomputation}) were first obtained for the self-dual KIM.

The kicked Ising model is most naturally described in a graphical tensor network language that is not based on dual-unitary gates, but rather on Ising phases and quantum kicks that are interchangeable under space-time duality.
This graphical language is closely related to the ZX calculus~\cite{van_de_wetering_zx-calculus_2020}, which has as building blocks dephased delta tensors in the Z- and X-basis that can be mapped into one another through a Hadamard transformation. This representation has also led to a wide range of generalisations of these kicked Ising models, both within the class of dual-unitary models and beyond.

\subsection{Dual-unitarity}

The dual-unitarity of the self-dual KIM can be established by introducing a diagrammatic notation that makes explicit how the Ising dynamics and kicks are related through a space-time duality. Following the notation of~\cite{ho_exact_2022,stephen_universal_2024}, we introduce the following two building blocks
\be
\label{eq:KIM:diagrams}
\begin{aligned}
&\vcenter{\hbox{\includegraphics[height=0.16\linewidth]{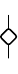}}}
 =\frac{1}{\sqrt{2}}\begin{pmatrix}
1 & 1 \\ 
1 & -1
\end{pmatrix} = H,\\ 
& \vcenter{\hbox{\includegraphics[height=0.16\linewidth]{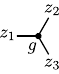}}} = \delta_{z_1,z_2,z_3} e^{-ig (1-2 z_1)}.
\end{aligned}
\ee 
The former represents the Hadamard gate $H$, whereas the latter is a delta tensor fixing all indices $\{z_1,z_2,z_3\}$ to be identical and applying a phase $e^{\pm ig}$ depending on the value of $z_i \in \{0,1\}$. We use this notation to denote dephased delta tensors with an arbitrary number of indices corresponding to the number of legs.

Focusing for concreteness on the self-dual point $J=b=\pi/4$, the two-site Ising interactions and the single-site kicks can be represented as
\begin{align}§
e^{-i \frac{\pi}{4} \sigma_1^z \sigma_2^z} = \sqrt{2}\,\,\,\vcenter{\hbox{\includegraphics[height=0.16\columnwidth]{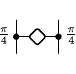}}}\,, \quad
 e^{-i \frac{\pi}{4} \sigma^y} = \vcenter{\hbox{\includegraphics[height=0.16\columnwidth]{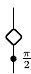}}}\,,
\end{align}
up to an irrelevant global phase. The inhomogeneous longitudinal field can be similarly represented as a dephased delta tensor
\begin{align}\label{eq:KIM:phasegate}
\vcenter{\hbox{\includegraphics[height=0.08\columnwidth]{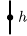}}} = 
\begin{pmatrix}
e^{-i h} & 0 \\
0 & e^{ih}
\end{pmatrix}  = e^{-i h \sigma^z} = P(h),\,
\end{align}
where we have defined the phase gate $P(h)$.

In this notation the full many-body evolution operator can be written as
\begin{align}\label{eq:KIM:unitary}
\!\!\!\!\!\!\! \mathbb U_{\textrm{KI}}[\textbf{h}]^t \!\!=\,\vcenter{\hbox{\includegraphics[width=0.75\columnwidth]{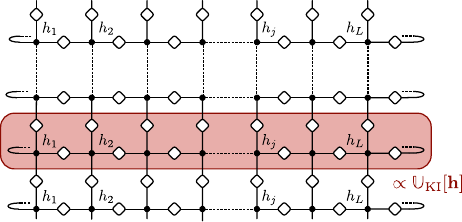}}}\!\!\!\!\!,
\end{align}
again up to an irrelevant global phase and a multiplication with Pauli matrices at the top boundary (which however do not change the bulk physics).

This construction is explicitly symmetric upon swapping the roles of space (horizontal direction) and time (vertical direction). More specifically, considering the case of homogeneous longitudinal fields, i.e.~${h_j = h}$ for all $j$, and denoting the time-evolution operator by $\mathbb U_\textrm{KI}[\textbf{h}] \equiv \mathbb U_\textrm{KI}[h]$, \cite{Gutkin} have noted 
\begin{align}
\mathrm{Tr}[(\mathbb U_\textrm{KI}[h])^t] = \mathrm{Tr}[(\tilde{\mathbb U}_\textrm{KI}[h])^L],
\end{align}
where the space evolution operator $\tilde{\mathbb U}_{\textrm{KI}}[h]$ has exactly the same form as $\mathbb U_{\textrm{KI}}[h]$ but acts on a lattice of $t$ (temporal) qubits.

\begin{figure}[tb!]
\includegraphics[width=0.9\columnwidth]{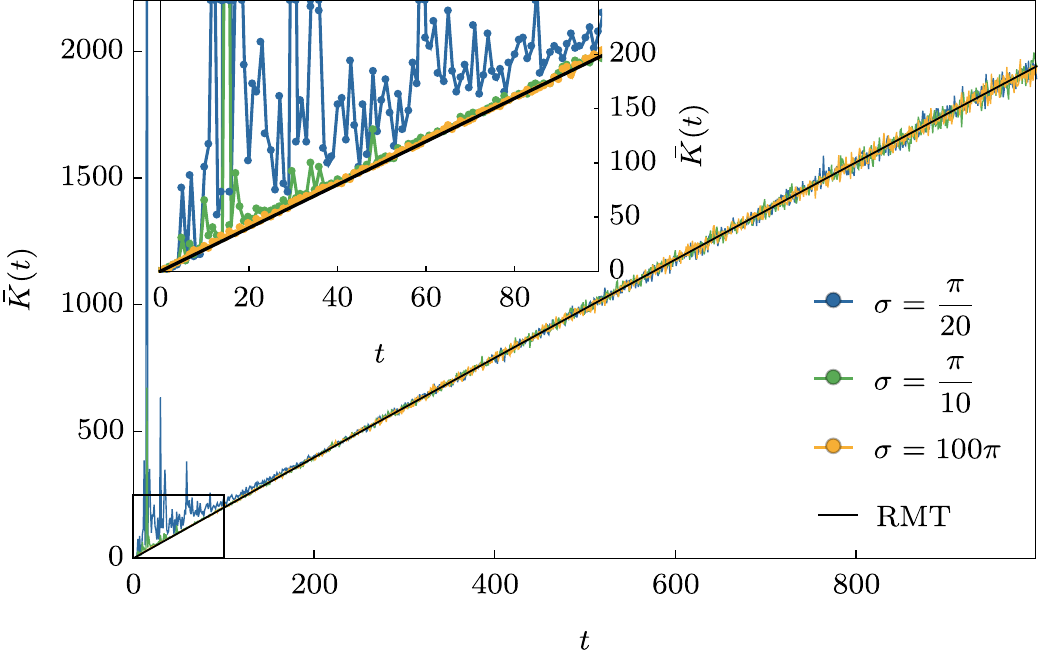}
\caption{Spectral form factor of the disordered kicked Ising model with $J=b=\pi/4$, $L=15$ and $h_j$ independently distributed Gaussian variables with mean value $\bar{h} = 0.6$ and variance $\sigma^2$. Inset: short-time window. The large-time fluctuations are due to the finite number $(N = 9490)$ of disorder realizations. Reproduced from~\cite{bertini_exact_2018}.
\label{fig:KIM:SFF}}
\end{figure}

As discussed in Sec.~\ref{sec:spectral}, one advantage of this duality is that averages over uncorrelated spatial disorder can be accommodated in a simple way, leading to a factorisable product of space-evolution operators. The basic observation is that for a general $\textbf{h}=\{h_1, h_2, \dots, h_L\}$, one can write
\begin{align}
&\mathrm{Tr}[(\mathbb U_\textrm{KI}[\textbf{h}])^t] = \mathrm{Tr}\left[\tilde{\mathbb U}_\textrm{KI}[h_1]\tilde{\mathbb U}_\textrm{KI}[h_2]\dots \tilde{\mathbb U}_\textrm{KI}[h_L]\right] \nonumber\\
&=\vcenter{\hbox{\includegraphics[width=0.75\columnwidth]{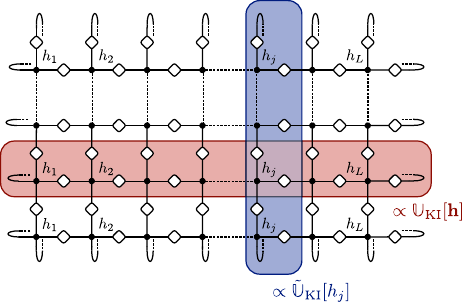}}}\!\!.
\end{align}
Using this decomposition, the spectral form factor of this model can be calculated in the same way as in Sec.~\ref{sec:spectral} and is illustrated in Fig.~\ref{fig:KIM:SFF}~\cite{bertini_exact_2018}. \cite{flack_statistics_2020} further studied the SFF in this model and found that, while the average SFF reproduces the prediction from the circular orthogonal ensemble (COE), the appropriate ensemble is not the COE but rather an ensemble of random matrices on a more restricted symmetric space (either $Sp(N)/U(N)$ or $O(2N)/O(N) \times O(N)$ depending on the parity of the number of sites), which reproduces the same averaged SFF as the COE but shows enhanced fluctuations. This distinction originates from an additional anti-unitary symmetry in the KIM originally observed by~\cite{braun_transition_2020}. 

The self-dual KIM can be recast in a brickwork-quantum-circuit form by introducing the local gate
\begin{align}\label{eq:KIM:gate}
 U_{\rm KI}[h] = \fineq[-0.8ex][1][1]{
		\tsfmatV[0][-0.5][r][1][][][red6]} = 2 \vcenter{\hbox{\includegraphics[width=0.13\columnwidth]{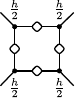}}} \,,
\end{align}
where we have fixed $h_j = h$ for concreteness. Written out in matrix elements, this gate reads
\begin{align}
\label{eq:KIM:matrixelements}
\!\! U_{\rm KI}[h]  = \frac{1}{2}\begin{pmatrix}
e^{-i2h} & e^{-ih} & e^{-i h} & -1 \\
e^{-ih} & -1 & +1 & e^{ih} \\
e^{-ih} & 1 & -1 & e^{ih} \\
-1 & e^{ih} & e^{ih} & e^{i2h}
\end{pmatrix}\!.
\end{align}
and one can explicitly verify that it fulfils the dual-unitarity conditions in Eqs.~\eqref{eq:dualunitaritynonfolded}. A dual-unitary brickwork circuit constructed from these gates corresponds to the unitary dynamics generated by the operator in Eq.~\eqref{eq:KIM:unitary} up to additional two-site gates at the space-time boundaries
\begin{align}\label{eq:KIM:brickwork}
\vcenter{\hbox{\includegraphics[width=0.6\columnwidth]{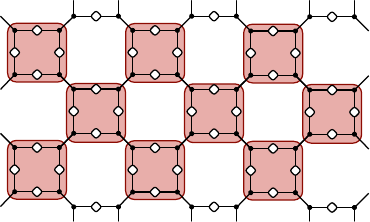}}},
\end{align}
where we have suppressed the explicit dependence on $h_j$.

\subsection{Solvable states}

The kicked Ising model allows for two classes of solvable states, originally dubbed `separating states' by \cite{bertini2019entanglement}, which differ from the solvable states discussed in Sec.~\ref{sec:quantumdynamics} because of the additional two-site gates at the top and bottom boundaries of Eq.~\eqref{eq:KIM:brickwork}. The first class of separating states is given by product states in the computational basis
\begin{align}\label{eq:KIM:solvablestates_1}
| z_1, z_2, \dots, z_L \rangle, \qquad z_j \in \{0,1\},
\end{align}
whereas the second class consists of equal-weight superpositions
\begin{align}\label{eq:KIM:solvablestates_2}
\bigotimes_{j=1}^L \left(\frac{1}{\sqrt{2}}|0\rangle + \frac{e^{i \varphi_j}}{\sqrt{2}} |1\rangle\right),  \qquad \varphi_j \in [0, 2\pi]\,.
\end{align}
This second class includes product states in the $X$- and $Y$-basis. For these states the subsystem entanglement dynamics follows the results of Sec.~\ref{sec:statedynamics}: Initial linear growth with maximal velocity is followed by saturation to the maximal (volume-law scaling) value corresponding to a maximally mixed state \cite{bertini2019entanglement}.

The separating states can be related to the solvable states of Sec.~\ref{sec:quantumdynamics} by noting that acting with the kicked Ising gate returns solvable states corresponding to maximally entangled Bell pairs between neighboring sites~\cite{piroli_exact_2020}. These states can be naturally incorporated in the graphical language, since
\begin{align}
|\varphi \rangle \equiv  \frac{1}{\sqrt{2}}|0\rangle + \frac{e^{i \varphi}}{\sqrt{2}} |1\rangle= \frac{e^{i \varphi/2}}{\sqrt{2}} \,\vcenter{\hbox{\includegraphics[height=0.12\columnwidth]{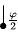}}}\,.
\end{align}
Acting with the two-site Ising gates appearing in the boundaries of Eq.~\eqref{eq:KIM:brickwork} on a pair of such states, we have that
\be
\begin{aligned}
 & \vcenter{\hbox{\includegraphics[height=0.12\columnwidth]{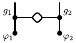}}}=\vcenter{\hbox{\includegraphics[height=0.12\columnwidth]{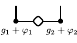}}} \\
 &= \sum_{z_1,z_2 \in \{0,1\}} \mathcal{N}^{z_1,z_2} |z_1\rangle \otimes |z_2 \rangle,
\end{aligned}
\ee
corresponding to a solvable state (cf.\ Eq.~\eqref{eq:equivalentsolvability}) with
\begin{align}
\mathcal{N} =  P\left(g_1+\varphi_1\right) H P\left(g_2+\varphi_2\right),
\end{align}
which is a unitary matrix decomposed as the product of two phase gates and a Hadamard gate (Eqs.~\eqref{eq:KIM:diagrams} and \eqref{eq:KIM:phasegate}). The state locally `injects' a phase in the circuit. As a specific example, for $\varphi=0$ we recover $| \varphi=0\rangle = \ket{+}$, which can hence be locally absorbed in the circuit. Namely, we have
\begin{align}
 \vcenter{\hbox{\includegraphics[height=0.12\columnwidth]{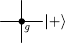}}}= \vcenter{\hbox{\includegraphics[height=0.12\columnwidth]{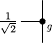}}}\,.
\end{align}

A similar argument applies when acting with the kicked Ising model on the product states in the computational basis, which return states of the form in Eq.~\eqref{eq:KIM:solvablestates_2} after a single discrete time step. Gates satisfying the property that they map a basis of product states to a basis of maximally entangled states are also called special perfect entanglers~\cite{rezakhani_characterization_2004}.

\subsection{Matchgate and Clifford points}
\label{subsec:KIM:specialpoints}

For special values of the longitudinal fields the kicked Ising model in Eq.~\eqref{eq:KIM:U_Floq} exhibits additional structure besides dual-unitarity that allows for an efficient classical simulation of its dynamics. 

First, in the absence of a longitudinal field, i.e.\ $h_j =0$ for all $j$, the system corresponds to the kicked transverse-field Ising model introduced by~\cite{Prosen2000}, i.e., the `kicked' Floquet analogue of the transverse-field Ising model. The latter is a textbook example of a many-body system that can be solved through a mapping to free fermions \cite{LSM,essler_quench_2016}, therefore generating noninteracting integrable dynamics. Analogously, the kicked transverse-field Ising model can also be mapped to free fermions~\cite{bertini_exact_2018} and its quantum circuit representation is a \emph{matchgate} circuit~\cite{terhal_classical_2002,jozsa_matchgates_2008}, i.e., it can be efficiently simulated on a classical computer via Gaussian methods. At the self-dual point the free fermions acquire a perfectly linear dispersion relation, since dual-unitarity restricts their dynamics to occur at the maximal velocity (cf.\ the non-dispersive solitons discussed in Sec.~\ref{subsec:nonergodic}) and the dual-unitary gate in Eq.~\eqref{eq:KIM:matrixelements} can be decomposed as
\begin{align}
U_{\rm KI}[h]  =
(H \otimes H) \begin{pmatrix}
1 & 0 & 0 & 0 \\
0 & 0 & 1 & 0 \\
0 & 1 & 0 & 0 \\
0 & 0 & 0 & -1
\end{pmatrix}
(H \otimes H).
\end{align}
The Hadamard matrices cancel in the full brickwork circuit, and the remaining two-site gate can be directly verified to satisfy the matchgate condition of~\cite{terhal_classical_2002,jozsa_matchgates_2008}.

Another special point in parameter space is attained for $h_j \in \pi \mathbbm{Z}/8$ for all  $j$, when the self-dual KIM reduces to a Clifford circuit, i.e., maps products of Pauli operators into products of Pauli operators without creating superpositions. This can be established by noting that, for $|J|=|b|=\pi/4$, Ising phases and kicks correspond to CZ gates and Hadamard gates, both Clifford gates, and for $h_j \in  \pi \mathbbm{Z}/8$ the (possibly inhomogeneous) action of the longitudinal fields can also be represented through Clifford gates --- their composition is therefore Clifford. Once again, this implies that the circuit admits efficient classical simulation~\cite{gottesman_heisenberg_1998}. 
 
\begin{figure}[tb!]
\includegraphics[width=\columnwidth]{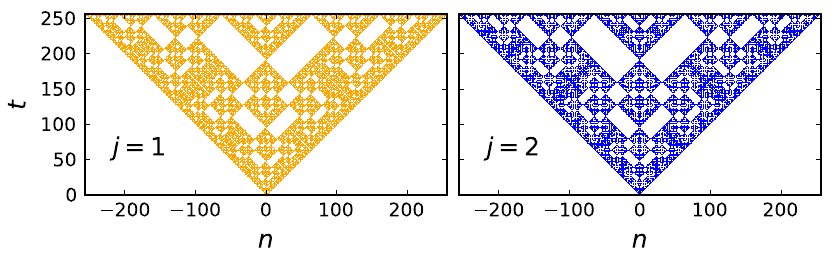}
\caption{Fractal operator spreading in the dynamics of an initial single-site $Z$ operator split into odd (left) and even (right) sites. The image on odd sites is only $Z$’s (orange), while the image on even sites is only $X$’s (blue). Reproduced from~\cite{sommers2023crystalline}.
\label{fig:KIM_fractal}}
\end{figure}

 \cite{sommers2023crystalline} studied the dynamics of more general classes of dual-unitary Clifford circuits using the formalism of Clifford quantum cellular automata (CQCA)~\cite{gutschow_time_2010,gutschow_fractal_2010} (see also \cite{farrelly_review_2020} for a recent review on the topic). In the context of the kicked Ising model \cite{sommers2023crystalline} discussed two inequivalent classes of translationally invariant Clifford dynamics: $h_j = 0$, corresponding to the integrable limit described above, and $h_j = \pi/4$. The former limit was termed a `poor scrambler' and the latter a `good scrambler', based on whether or not the Page curve reached its maximal slope under the corresponding circuit dynamics. Under the `good scrambler' Pauli strings exhibit a fractal structure in space-time, as illustrated in Fig.~\ref{fig:KIM_fractal}. The fractal dimension is given by $d_f = \log_2[(3+\sqrt{17})/2]=1.8325 \dots $, as determined in~\cite{gutschow_fractal_2010} for an equivalent CQCA. Instead, for the `poor scrambler' no such fractal structure appears and the full operator dynamics is rather characterised by gliders/solitons. These two models can also be distinguished by looking at the locality of their generating Hamiltonian. Indeed, the Floquet unitary evolution operator can always be formally written as the exponential of a (non-unique) time-independent Floquet Hamiltonian $H_F$ as $\mathbb{U}^t = \exp[- i H_F t]$~\cite{bukov_universal_2015}. The absence of a local Floquet Hamiltonian is necessary for maximally ergodic dynamics and thermalisation to the infinite-temperature state. Consistently with this, \cite{sommers2023crystalline} observed that results from \cite{zimboras_does_2022} imply that for the `good scrambler' all generating Floquet Hamiltonians are fully non-local, in the sense that interactions do not decay with the distance, whereas for the `poor scrambler' the Floquet unitary can be generated by a time-independent Hamiltonian with algebraic decay of interactions. Furthermore, \cite{yao_temporal_2024} studied the behaviour of these two models in the presence of spatial-unitarity-preserving Clifford measurements. Probing the temporal correlations by means of the so called `temporal entanglement' (see Sec.~\ref{sec:beyond_du} for more details), they showed that introducing Bell-pair measurements results in either a diffusive (poor scrambling) or ballistic (good scrambling) growth of temporal entanglement.

\subsection{Generalisations}
\label{subsec:KIM:generalizations}

Over the last few years the literature has proposed several generalisations of the self-dual kicked Ising model. In an effort of categorisation, the latter can be conveniently arranged in two main classes. The first class is based upon replacing the Hadamard matrix in Eq.~\eqref{eq:KIM:diagrams} --- which is the building block of both the Ising phases and kicks --- with a more general \emph{complex Hadamard matrix}. This results in classes of dual-unitary models that go beyond qubits, while still retaining the salient properties of the self-dual kicked Ising model. The second class, instead, is based upon the idea of modifying the square-lattice geometry of Eq.~\eqref{eq:KIM:unitary}. The resulting models in this case are no longer dual-unitary but continue to retain some form of solvability through different symmetries in space-time. These models can be incorporated in the framework of \emph{hierarchical dual-unitarity} (see Sec.~\ref{subsec:hierarchical}) but exhibit additional structure that is best understood in the graphical language of the kicked Ising model.

\subsubsection*{Complex Hadamard matrices}

The dual-unitarity of the kicked Ising model follows from the observation that the Ising phases are interchanged with the kicks under space-time duality. This idea can be extended through the notion of \emph{complex Hadamard matrices}, a class of matrices that generalise the Hadamard matrix of Eq.~\eqref{eq:KIM:diagrams}~\cite{Tadej_Karol_2006}.
A complex Hadamard matrix of order $q$ is a $q \times q$ matrix $H$ that satisfies the properties of (i) unimodularity, i.e. the modulus of each entry is one, $|H_{ab}| = 1, \forall a,b$ and (ii) orthogonality, i.e. the matrix is proportional to a unitary matrix, $H^{\dagger}H = H H^{\dagger} = q \mathbb{1}$.

Because of the unimodularity, a Hadamard matrix $H$ can be used to construct a two-site diagonal unitary gate with matrix elements
\begin{align}
\mathcal{I}_{ab,cd} = \delta_{ac} \delta_{bd} H_{ab},
\end{align}
acting a generalization of the Ising phase gate, and a single-site unitary gate 
\begin{align}
\mathcal{K}_{ab} = \frac{1}{\sqrt{q}} H_{ab}\,,
\end{align}
acting as the generalization of the kick gate. In analogy with Eq.~\eqref{eq:KIM:gate}, a dual-unitary gate can then be constructed as
\begin{align}\label{eq:KIM:gate_CHM}
U = \mathcal{I} (\mathcal{K} \otimes \mathcal{K})\mathcal{I} = q\, \vcenter{\hbox{\includegraphics[width=0.13\columnwidth]{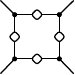}}}
\end{align}
where we have redefined the graphical calculus as
\begin{align}\label{eq:KIM:diagrams_CHM}
    \vcenter{\hbox{\includegraphics[height=0.16\linewidth]{fig_Had}}}
 = \frac{H}{\sqrt{q}}, \qquad 
\vcenter{\hbox{\includegraphics[height=0.16\linewidth]{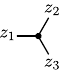}}} = \delta_{z_1 z_2 z_3}.
\end{align} 
The unitary evolution operator of Eq.~\eqref{eq:KIM:U_Floq} remains dual-unitary for any choice of complex Hadamard matrix. Furthermore, the dual-unitarity is preserved even if all complex Hadamard matrices are not identical. As long as each matrix represented by Eq.~\eqref{eq:KIM:diagrams_CHM} is a complex Hadamard matrix, the resulting circuit is dual-unitary. Choosing four Hadamard matrices  $H_1,H_2,H_3,H_4$ in Eq.~\eqref{eq:KIM:gate_CHM}, a dual-unitary gate can e.g.\ be constructed as
\begin{align}
U_{ab,cd} = (H_1)_{ab} (H_2)_{bd} (H_3)_{dc} (H_4)_{ca} / q\,.
\end{align}
Complex Hadamard matrices can additionally be `permuted' and `dephased', i.e.\ a complex Hadamard matrix $H_2$ can be obtained from a complex Hadamard matrix $H_1$ as
\begin{align}\label{eq:KIM:equivalent_CHM}
H_2 = D_1 P_1 H_1 P_2 D_2,
\end{align}
where $D_{1,2}$ are diagonal unitary matrices and $P_{1,2}$ are permutation matrices. The dephasing introduced by these diagonal unitary matrices is equivalent to the phases in the delta tensor \eqref{eq:KIM:diagrams} for $q=2$, but can be used more generally away from the case of a two-dimensional Hilbert space (for which all Hadamard matrices are of the form \eqref{eq:KIM:equivalent_CHM} with $H$ the Hadamard matrix of Eq.~\eqref{eq:KIM:diagrams}). Complex Hadamard matrices exist for all values of $q$, a particularly simple class existing in any dimension are the Fourier matrices with $F_{ab} = \exp[2 \pi i a b / q]$. Different choices of complex Hadamard matrices can give rise to different dynamics, e.g.\ Clifford vs.\ non-Clifford or non-ergodic vs.\ ergodic.

The idea to use complex Hadamard matrices to construct dual-unitary kicked spin chains was put forward by \cite{gutkin_exact_2020}, and constructions of dual-unitary gates in terms of complex Hadamard matrices were proposed in \cite{borsi_construction_2022} and \cite{claeys_emergent_2022}. A systematic study of operator dynamics under circuits constructed out of complex Hadamard matrices was undertaken by \cite{claeys_operator_2024}, making connections to Clifford cellular automata \cite{schlingemann2008structure} and to the classical spatiotemporal cat model of many body chaos in the limit $q \to \infty$~\cite{gutkin2021linear}. This work additionally identified the kicked Potts model of \cite{lotkov_floquet_2022} as a specific realization of a dual-unitary kicked spin chain using complex Hadamard matrices, and showed how the long-range entanglement protocol discussed therein could be reinterpreted in purely graphical terms. This model is additionally integrable, as conjectured in \cite{lotkov_floquet_2022} and later proven in \cite{miao_integrable_2023}, and \cite{claeys_operator_2024} established this integrability graphically for both this model and a large family of generalizations including the kicked Ising model at the free point.

Dual-unitarity and Hadamard matrices both fit within the framework of biunitarity, originating from quantum information theory and encompassing various objects having multiple notions of `unitarity'~\cite{reutter2016biunitary,claeys2023dual}. The biunitarity of complex Hadamard matrices was originally identified in \cite{jones1999planaralgebrasi}. From this perspective, the construction of dual-unitary gates from Hadamard matrices is underlaid by the more general result that biunitaries can be composed into different biunitaries, first used by \cite{claeys_emergent_2022} in the construction of dual-unitary gates and later fully developed in \cite{claeys2023dual}. These constructions were later extended in \cite{rampp_geometric_2024} to accommodate different space-time geometries. Biunitarity will be discussed in detail in Sec.~\ref{subsec:biunitarity}.

\subsubsection*{Different space-time geometries}

The dual-unitarity of the kicked Ising model follows directly from the symmetry of the unitary evolution operator as represented in Eq.~\eqref{eq:KIM:unitary}: The space-time duality follows from the symmetry of the square lattice in space-time.
A variant of the kicked Ising model called the ``alternating kicked Ising model'' (AKIM) was introduced by \cite{liu_solvable_2023}. In this model the dynamics alternates between two Ising models which act only on odd or even bonds with strength $J$, subject to dimerised longitudinal fields $g_0$ and $g_1$, with global transverse kicks of strength $b$. The corresponding Floquet unitary reads
\begin{align}
\mathbb U = e^{-i H_{\textrm{e}}} e^{-i H_\textrm{K}} e^{-i H_{\textrm{o}}}e^{-i H_{\textrm{K}}},
\end{align}
where the different Ising Hamiltonians are given by
\begin{align}
    &H_{\textrm{e/o}} = J \!\!\!\!\! \sum_{j \in \textrm{even/odd}}^{L} \!\!\!\!\!\!  \sigma^z_j \sigma^z_{j+1} +\sum_{j=1}^L \left(\frac{g_{j\, \textrm{mod}\, 2}}{2} + \frac{\pi}{4}\right) \sigma^z_j, 
\end{align}
and the kick Hamiltonian is again defined as in Eq.~\eqref{eq:HK}. For $J=b= \pi/4$ the unitary operator for $t$ steps can be graphically represented analogously to the one of the self-dual KIM, where the Hadamard matrices are now arranged on a honeycomb lattice rather than a square lattice
\begin{align}\label{eq:KIM:honeycomb}
\!\!U(t) = \!\!\!\vcenter{\hbox{\includegraphics[width=0.75\columnwidth]{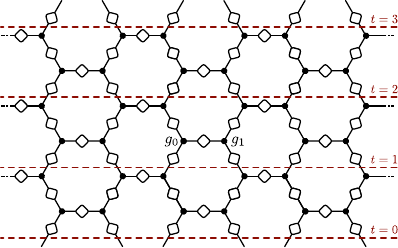}}},
\end{align}
As shown by \cite{liu_solvable_2023}, in this model the dynamics of entanglement can be solved for all times, using a similar graphical calculus as for the KIM. The entanglement dynamics exhibit an initial linear growth at half the maximal entanglement velocity, $v_E = 1/2$, followed by saturation to a final entropy that can either be the maximal value or an extensive but sub-maximal entropy, depending on the choice of initial state. This model can be reinterpreted as a brickwork circuit with two-site gate
\begin{align}
\label{eq:KIM:gate_honeycomb}
 U &= \sqrt{2} \, \vcenter{\hbox{\includegraphics[width=0.16\columnwidth]{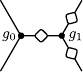}}} \notag\\
 &= e^{-i g_0}  |0\rangle \langle 0 | \otimes U_0 + e^{i g_0} |1 \rangle \langle 1 | \otimes U_1,
\end{align}
which acts as a controlled gate with
\begin{align}
&U_0 = \begin{pmatrix}
  \cos g_1 & -i \sin g_1  \\
  -i \sin g_1 &  \cos g_1 
  \end{pmatrix}\!,\notag\\
&U_1 = \begin{pmatrix}
-i  \sin g_1 &\cos g_1 \\
\cos g_1  & -i \sin g_1
\end{pmatrix}\!.
\end{align}
For ${g_0 = g_1=0}$ the unitary gate in Eq.~\eqref{eq:KIM:gate_honeycomb} reduces to the CNOT gate and the corresponding circuit is also known as the Floquet Quantum East model at the deterministic point~\cite{bertini_east_2024, klobas_exact_2024}. The latter is a special point in the parameter space of the Floquet version~\cite{bertini_localised_2023} of the Quantum East model~\cite{pancotti_quantum_2020} and has recently gained attention as a minimal model for kinetically constrained dynamics.

The above model and all following models in this subsection exhibit the same solvability of dual-unitary circuits for any choice of complex Hadamard matrices as building blocks, such that in the following all phases in the delta tensors are made implicit and the white squares in these diagrams can be interpreted as arbitrary complex Hadamard matrices. 
An alternative circuit with the same three-fold rotation symmetry was proposed by \cite{rampp_entanglement_2023}, which can be graphically represented by placing Hadamard matrices on the links of a triangular lattice
\begin{align}\label{eq:KIM:triangular}
\!\!\!U(t) \!=\!\!\vcenter{\hbox{\includegraphics[width=0.75\columnwidth]{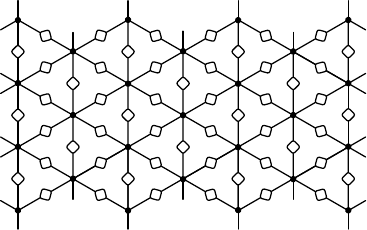}}}.
\end{align}
This evolution operator is equivalent to a brickwork circuit with local two-site gates of the form
\begin{align}\label{eq:KIM:gate_triangular}
U = \vcenter{\hbox{\includegraphics[width=0.13\columnwidth]{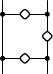}}} \, = \, \vcenter{\hbox{\includegraphics[width=0.13\columnwidth]{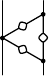}}}\,,
\end{align}
In the same work, \cite{rampp_entanglement_2023} additionally proposed two-site gates constructed out of two complex Hadamard matrices as
\begin{align}\label{eq:KIM:gate_sheared}
U = \vcenter{\hbox{\includegraphics[width=0.13\columnwidth]{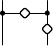}}}\,.
\end{align}
The resulting brickwork circuit is of the form
\begin{align}\label{eq:KIM:sheared_1}
\!\!\!\!U(t) = \!\!\vcenter{\hbox{\includegraphics[width=0.75\columnwidth]{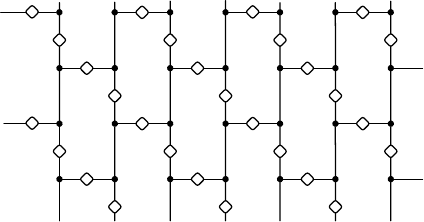}}}\,,
\end{align}
and can again be realised as a kicked Ising model by choosing appropriate alternating Ising interactions and kicks.
By merging delta tensors, this circuit can be brought in a form that is locally equivalent to the dual-unitary square lattice
\begin{align}\label{eq:KIM:sheared_2}
U(t) = \!\!\!\!\vcenter{\hbox{\includegraphics[width=0.75\columnwidth]{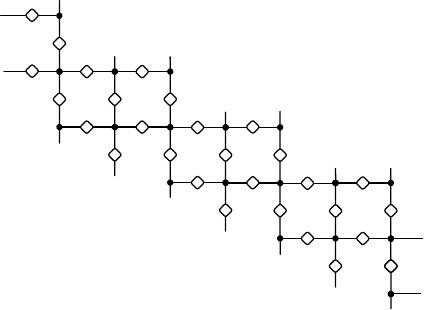}}}\!\!\!.
\end{align}
A Clifford representative of this class of gates was previously presented by~\cite{sommers_zero-temperature_2024}, where the equivalence with a `sheared' dual-unitary circuit was also explicitly observed in the operator dynamics. 

These latter two classes of circuits where first studied through the lens of entanglement membrane theory, as developed in~\cite{Zhou2020}, since the so-called \emph{entanglement line tension} can be exactly obtained in these models. The entanglement line tension $\mathcal{E}(v)$ is the fundamental quantity characterising the entanglement membrane and can be defined as the operator entanglement of the time evolution operator across a bipartition in space-time with boundary given by the line $x = vt$. From this entanglement line tension the butterfly and entanglement velocity can be directly obtained~\cite{Zhou2020}.
It was argued in~\cite{sommers_zero-temperature_2024} that the existence of multiple (generalised) unitary directions forces the entanglement membrane tension to be a piecewise linear function of the orientation of the membrane. This result is consistent with the results from~\cite{rampp_entanglement_2023,foligno_quantum_2024}, which studied the entanglement line tension in generalised dual-unitary circuits.

\begin{figure}[tb!]
\includegraphics[width=1.\columnwidth]{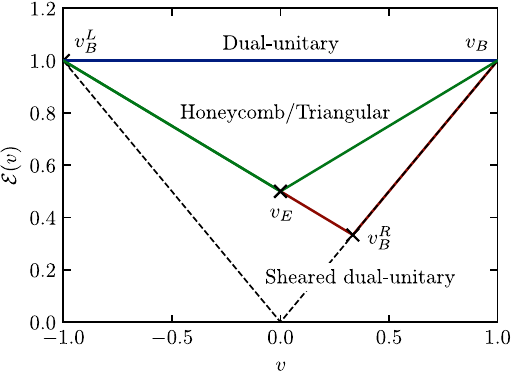}
\caption{Entanglement line tension $\mathcal{E}(v)$ for different space-time geometries as a function of the slope $v$ in space-time. Entanglement velocity $v_E$ and left/right butterfly velocities $v_B^{L/R}$ are indicated. Reproduced from~\cite{rampp_entanglement_2023}.
\label{fig:ELT}}
\end{figure}

In the same way that the kicked Ising model presents a special class of dual-unitary circuits, the above models present special classes of the generalised dual-unitary circuits discussed in Sec.~\ref{subsec:hierarchical}.

\section{Dual-unitarity in quantum computation}
\label{sec:quantumcomputation}

In this section we discuss dual-unitarity in the context of quantum computation. A dual-unitary circuit can be directly interpreted as a gate-based quantum computation, where the number of discrete time steps corresponds to the circuit depth and the gates are restricted to dual-unitary gates arranged in a brickwork structure. With this interpretation one can establish a direct connection between the complexity of certain quantum computations and the ergodicity of dual-unitary dynamics. As discussed in Sec.~\ref{subsec:Complexity}, this connection can be used to show that, while certain dynamical quantities can be evaluated exactly in dual-unitary circuits, other calculations remain provably hard -- highlighting the nontriviality of dual-unitary dynamics. Alternatively, spatial unitarity allows dual-unitary dynamics to be interpreted as measurement-based quantum computation along the spatial direction. This interpretation has the advantage that sampling over measurement outcomes can be related to sampling over quantum computations, leading to exact results on e.g.\ deep thermalisation and Hilbert space delocalisation. Most results in this section were initially obtained for the self-dual kicked Ising model and later extended to generic dual-unitary circuits. For consistency with the original literature, we follow this approach in the rest of this section. 

\subsection{Measurement-based quantum computing} 
\label{subsec:mbqc}

\begin{figure*}[tb!]
\includegraphics[width=1.5\columnwidth]{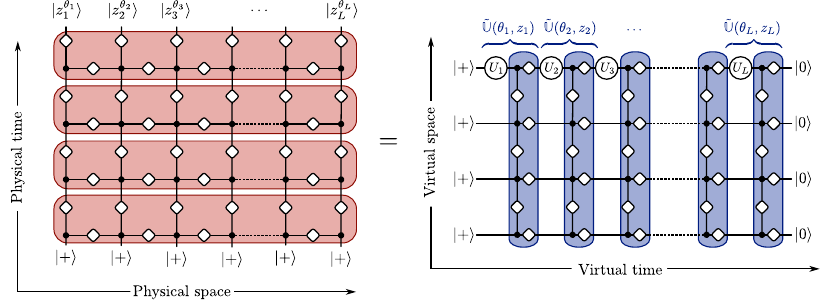}
\caption{Setup for measurement-based quantum computation under self-dual kicked Ising dynamics. Left: An initial product state $\otimes_{j=1}^L |+\rangle_j$ is evolved in time and locally measured in a local basis $|z_j^{\theta_j}\rangle$. Right: This protocol can be interpreted as a quantum computation where an initial product state $\otimes_{j=1}^t |0\rangle_j$ is evolved under a series of unitary transformations $U_j$ set by the corresponding measurement outcomes and measured in a final state $\otimes_{j=1}^L |+\rangle_j$. Based on a similar figure in~\cite{stephen_universal_2024}.
\label{fig:dumbqc}}
\end{figure*}

Rather than considering dual-unitary circuits as \emph{gate-based} quantum computations in time, it is possible to interpret them as \emph{measurement-based} quantum computations in space~\cite{stephen_universal_2024}. Measurement-based quantum computing (MBQC) is a paradigm of quantum computing in which an initial `resource' state is prepared using a finite-depth circuit, after which the state is sequentially measured and, through this measurement, quantum information is propagated along the spatial direction realising a particular quantum computation~\cite{raussendorf_measurement-based_2003,raussendorf_quantum_2012} (note again the similarity to quantum teleportation~\cite{bennett_teleporting_1993}). 

As an example of this, \cite{stephen_universal_2024} considered a setup where each qubit is initially prepared in a $|+\rangle$ state, evolved under dual-unitary kicked Ising dynamics at the Clifford point $h = \pi/4$, and measured from left to right in a rotated basis $\{|0^{\theta}\rangle, |1^{\theta}\rangle\}$ where $|z^{\theta}\rangle = e^{-i \theta X} |z\rangle$. The corresponding circuit is illustrated in the l.h.s.\ of the equation in Fig.~\ref{fig:dumbqc}. The output of the computation is given by the probabilities of obtaining different measurement outcomes, i.e.\ $|\langle z_1^{\theta_1 } \dots z_n^{\theta_n}|\psi_t\rangle|^2$. Due to the solvability of the initial state and the measurement basis, this probability can be recast as 
\begin{align}
\langle z_1^{\theta_1 } \dots z_n^{\theta_n}|\psi_t\rangle = \langle R | \tilde{\mathbb{U}}(\theta_n ,z_n) \dots  \tilde{\mathbb{U}}(\theta_1, z_1)| L \rangle
\end{align}
where the boundary vectors are defined as $|L \rangle = \otimes_{i=1}^t |+\rangle_i$ and $|R\rangle = \otimes_{i=t}^t |0\rangle_i$ and the unitary operator is illustrated in Fig.~\ref{fig:dumbqc}, which provides a graphical representation of the identity above. This mapping uses that the measured states act as solvable states for the kicked Ising dynamics and can be mapped to single-site unitary gates in the spatial direction (see discussion in Sec.~\ref{sec:KIM}). The measurement basis can be written as
\begin{align}
\langle z_j^{\theta_j}| &= \langle 0 | (\sigma^x)^{z_j} e^{-i \theta_j \sigma^x} \nonumber\\
&= \langle + | (\sigma^z)^{z_j} e^{-i \theta_j \sigma^z} H = \langle + | U_j  H,
\end{align}
where the Hadamard matrix $H$ transforms between the $Z$- and $X$-basis and satisfies $\langle 0| H = \langle + |$. In Fig.~\ref{fig:dumbqc} the Hadamard matrix cancels with the Hadamard matrix in the top row, and the remaining term only injects a phase, which translates into the (diagonal) unitary operation
\begin{align}
U_j = (\sigma^z)^{z_j} e^{-i \theta_j \sigma^z}\,,
\end{align}
 in the spatial direction. Graphically, this manipulation corresponds to
\begin{align}
    \vcenter{\hbox{\includegraphics[width=0.7\columnwidth]{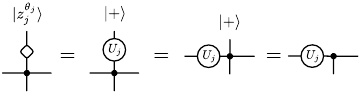}}}\,.
\end{align}
In this way the measurement protocol can be interpreted as a unitary transformation in the temporal Hilbert space of $t$ qubits, with the selection of gates based on the measurement outcomes. In order for this protocol to be controllable and universal, it should however fulfil two conditions: (i) it should be able to realise any specific unitary by choosing appropriate measurement outcomes and (ii) it should perform a full projective measurement rather than projecting on a fixed boundary vector. These conditions can be fulfilled simultaneously because it is possible to select the unitaries by `correcting' unwanted measurement outcomes. This is achieved by adjusting future measurement bases depending on past measurement outcomes, which simultaneously maps the right boundary condition into a random product state. Note that, however, the latter protocol is specific to Clifford dynamics (for details, see \cite{stephen_universal_2024}).

\subsection{Deep thermalisation. Emergent state designs.} 
\label{sec:deepthermal}

The interpretation of dual-unitary dynamics as a measurement-based quantum computation also led to an exact characterisation of `deep thermalisation'~\cite{ho_exact_2022}. The latter concept extends the notion of thermalisation to multilinear functions of the state of a subsystem and it is closely connected to the idea of designs from quantum computation~\cite{ambainis_quantum_2007,gross_evenly_2007}.

More specifically, consider an $L$-qubit lattice partitioned in a system $A$ and a bath $B$, consisting of $L_A$ and $L_B$ sites respectively (we fix $L_B \gg L_A$ in order to interpret $B$ as a bath). For any given quantum state $\ket{\psi}$ of the whole system one can define an ensemble of states in $A$ --- the \emph{projected ensemble} ---  following~\cite{cotler_emergent_2023} and \cite{choi_preparing_2023}. The idea is to consider projective measurements of $B$: performing one such measurement, in e.g.\ the computational basis, we obtain the outcome $z_B = (z_{1},z_2,\dots,z_{N_B}) \in \{0,1\}^{L_B}$ with probability $p(z_B)$ and an associated pure state $\ket{\psi(z_B)}$ on $A$ given by
\begin{align}
&p(z_B) = \mathrm{Tr}\left[ \left(\mathbbm{1}_A \otimes |{z_B}\rangle\langle z_B | \right) \ket{\psi}\bra{\psi} \right],\\
&\ket{\psi_A(z_B)} = \left(\mathbbm{1}_A \otimes \bra{z_B}\right)\ket{\psi}/\sqrt{p(z_B)}\,.
\end{align}
The projected ensemble is then defined as the set of all such states with the corresponding probabilities, i.e., 
\be
\mathcal{E}_A=\{\ket{\psi_A(z_B)}, p(z_B)\}.
\ee
Deep thermalization then posits that, in systems without any conservation laws and at late times, this ensemble of states is statistically indistinguishable from the Haar-random ensemble of states, 
\be
\mathcal{E}_{\textrm{Haar}} = \{\ket{\phi},\phi \sim \textrm{Haar}\}.
\label{eq:Haarensemble}
\ee
Physically, this means that a measurement of the bath returns a purely random state in the system, in line with a notion of entropy maximisation. 

The reduced density matrix for $A$ can be obtained as the first moment of $\mathcal{E}_A$, i.e.
\begin{align}
\rho_A = \mathrm{Tr}_B\big(|\psi \rangle\langle \psi|\big) = \sum_{z_B} p(z_B) |\psi_A(z_B) \rangle \langle \psi_A(z_B)|,
\end{align}
and higher moments are defined as
\begin{align}
\rho_{A}^{(k)} = \sum_{z_B}p(z_B) \left(|\psi_A(z_B) \rangle \langle \psi_A(z_B)|\right)^{\otimes k}\,.
\end{align}
Deep thermalization is quantified by comparing these moments to those of the ensemble of uniformly (Haar-random) distributed states on $A$, which admit a representation in terms of permutation operators~\cite{roberts_chaos_2017} as
\begin{align}\label{eq:moments_Haar}
\rho_{\textrm{Haar}}^{(k)} &= \int_{\phi_A \sim \textrm{Haar}} d\phi_A \left(\ket{\phi_A}\bra{\phi_A}\right)^{\otimes k} \nonumber\\
&= \frac{\sum_{\sigma \in S_k}P(\sigma)}{q_A(q_A+1)\dots(q_A+k-1)}\,.
\end{align}
The states $\ket{\phi_A}$ are Haar random states in a $q_A=2^{L_A}$ dimensional Hilbert space, and $P(\sigma)$ is an operator acting on $k$ copies of this Hilbert space, permuting them according to $\sigma \in S_k$, i.e.
\begin{align}\label{eq:def_P_pi}
P(\sigma)\ket{i_1, i_2,\dots,i_k} = \ket{i_{\sigma(1)},i_{\sigma(2)}, \dots, i_{\sigma(k)}}.
\end{align}
The permutation states of Eq.~\eqref{eq:permutation_states} correspond to the vectorization of these operators.
Note that standard thermalisation only probes the reduced density matrix --- the first moment of this ensemble --- for which $\rho_{\textrm{Haar}}^{(1)} = \mathbbm{1}/q_A$, while quantities involving multiple copies of the system will depend on the higher moments. 

\begin{figure*}[tb!]
\includegraphics[width=1.78\columnwidth]{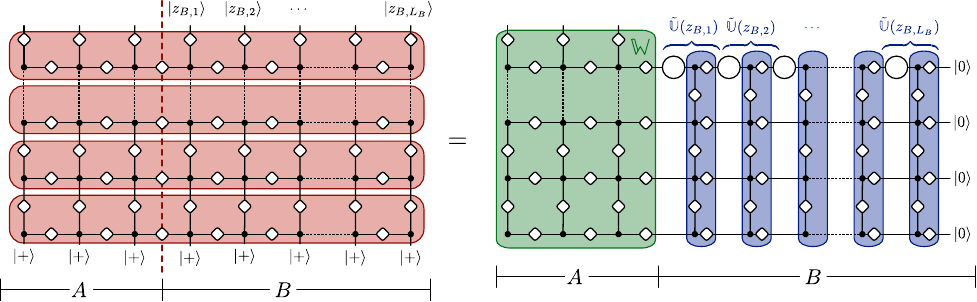}
\caption{Setup for deep thermalisation under self-dual kicked Ising dynamics. 
Left: An initial product state $\otimes_{j=1}^L |+\rangle_j$ is evolved in time, after which part of the system is measured in the computational basis, returning a bath state $|z_{B,1}, z_{B,2} ,\dots ,z_{B,L_B}\rangle$. Right: This protocol can be interpreted as a quantum computation where an initial product state $\otimes_{j=1}^t |0\rangle_j$ is evolved under a series of unitary transformations $\tilde{U}(z_{B,j})$ set by the corresponding measurement outcomes and subsequently mapped by an isometry $W$ from the temporal Hilbert space to the physical Hilbert space. Based on a similar figure in~\cite{ho_exact_2022}.
\label{fig:deepthermalization}}
\end{figure*}

A solvable model for deep thermalisation was proposed by \cite{ho_exact_2022} and is illustrated in Fig.~\ref{fig:deepthermalization}. An initial product state $\ket{\psi(t=0)} = \otimes_{i=1}^L \ket{+}_i$ is evolved under self-dual kicked Ising dynamics, and after $t$ discrete time steps the bath is measured in the computational basis. While at short times the projected ensemble of the time-evolved state $\ket{\psi(t)}$ is far from Haar-random, after a fixed number of time steps scaling with subsystem size, i.e. $t = L_A$, the projected ensemble exactly returns a Haar-random distribution in the limit of an infinite bath size $L_B \to \infty$. The projected ensemble is then said to form a quantum state design~\cite{ambainis_quantum_2007}. 

The proof of this statement is based upon interpreting the dual-unitary dynamics as a measurement-based quantum computation (see also Sec.~\ref{subsec:mbqc}) and translating the sampling over measurement outcomes to sampling different unitaries along the spatial direction. For a given measurement outcome $z_B = (z_{B,1}, z_{B,2}, \dots, z_{B,L_B})$, the corresponding state on the subsystem can be represented as
\begin{align}
\!\!\!\ket{\psi_A(z_B)} \propto \mathbb W \left[\tilde{\mathbb U}(z_{B,1}) \tilde{\mathbb U}(z_{B,2}) \dots \tilde{\mathbb U}(z_{B,L_B})\ket{R}\right],
\end{align}
where $\tilde{\mathbb U}(z_{B,j})$ is the unitary along the spatial direction obtained from measurement outcome $z_{B,j}$ on site $j$, $\ket{R} = \otimes_{i=1}^t \ket{0}_i$ is the spatial boundary vector on the right, and $\mathbb W$ is a linear transformation between the temporal Hilbert space and the physical Hilbert space of subsystem $A$. This decomposition is illustrated in Fig.~\ref{fig:deepthermalization}. The proportionality constant is the normalisation factor corresponding to the probability of obtaining this particular measurement outcome. 

\begin{figure}[tb!]
\includegraphics[width=\columnwidth]{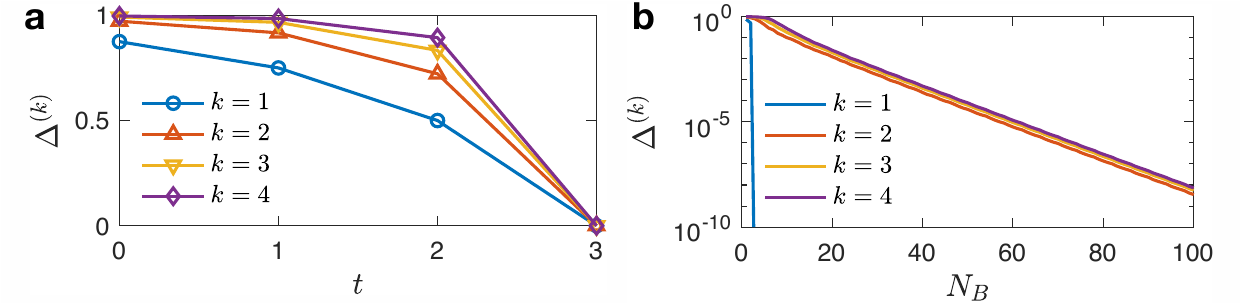}
\caption{Trace distance $\Delta^{k} = ||\rho_A^{(k)} - \rho_{\textrm{Haar}}^{(k)} ||_1$ of the $k$th moment of projected ensemble to a Haar random ensemble versus (a) time and (b) projected subsystem size $L_B$, for $h = \pi/9$ and $L_A=3$. For (a), $N_B=100$. For (b) $t = L_A = 3$. Reproduced from~\cite{ho_exact_2022}.
\label{fig:res_deepthermalization}}
\end{figure}

The important observation is the following:
\begin{theorem}[Theorem 2, \cite{ho_exact_2022}]
\label{th:universality}
The set of spatial unitary operators $\{\tilde{\mathbb U}(0),\tilde{\mathbb U}(1)\}$ constitutes a universal gate set away from the Clifford points, i.e. for $h \notin \mathbbm{Z} \pi/8$.
\end{theorem}
Furthermore, an infinite product of unitary gates sampled from a universal gate set acting on an initial wave function is indistinguishable from a Haar-random state. Specifically, we can again define a spatial transfer matrix acting on $k$ copies as
\begin{align}
\mathbbm{T}_k = \frac{1}{2} \sum_{z_B \in \{0,1\}} \left(\tilde{\mathbb U}(z_B)\otimes \tilde{\mathbb U}(z_B)^*\right)^{\otimes k},
\end{align}
and this transfer matrix satisfies
\begin{align}
\lim_{L_B \to \infty} \left(\mathbbm{T}_k\right)^{L_B}  = \int_{U \sim \textrm{Haar}} dU\, (U \otimes U^*)^{\otimes k},
\end{align}
where the integral is taken over the Haar measure for unitary matrices acting on a $2^t$-dimensional Hilbert space. From this result it follows that
\begin{align}
\lim_{L_B \to \infty} \rho_{A}^{(k)} = \int_{\phi \sim \textrm{Haar}} d\phi_t \left(\mathbb W|\phi_t\rangle \langle\phi_t| \mathbb W^{\dagger}\right)^{\otimes k}\,,
\end{align}
where $\phi_t$ is a Haar-random state in a $2^t$-dimensional Hilbert space. 
Furthermore, $\mathbb W$ is an isometry due to the dual-unitarity, mapping states in a $2^t$-dimensional temporal Hilbert space to a $2^{|A|}$-dimensional physical Hilbert space. For $t > L_A$ this isometry satisfies $\mathbb W \mathbb W^{\dagger} = \mathbbm{1}_A$, such that the Haar-random states in a $2^t$-dimensional Hilbert space get mapped to Haar-random state in a $2^{L_A}$-dimensional Hilbert space
\begin{align}
\label{eq:momentsequivalence}
\lim_{L_B \to \infty} \rho_{A}^{(k)} = \rho_{\textrm{Haar}}^{(k)} \quad \textrm{for} \quad t \geq L_A.
\end{align}
Therefore for $L_B\to \infty$ and $t \geq L_A$ the projected ensemble forms a state design. At earlier times the ensemble rather corresponds to a so-called Scrooge ensemble~\cite{jozsa_lower_1994,choi_preparing_2023}, also known as the Gaussian adjusted projected ensemble~\cite{goldstein_distribution_2006,goldstein_universal_2016}. The Scrooge ensemble is a deformation of the Haar ensemble returning a fixed first moment, and can be realised as $\{\mathbb W \ket{\phi}, \phi \sim \textrm{Haar}\}$, where the reduced density matrix is decomposed as $\rho = \mathbb W \mathbb W^{\dagger}$. The reduced density matrix directly follows from the first moment of the Haar ensemble, since $\rho_{\textrm{Haar}}^{(1)} = \mathbbm{1}_A/q_A$ (see Eq.~\eqref{eq:moments_Haar}). Note that the fact that $\mathbb W$ is an isometry directly implies the entanglement dynamics discussed in Sec.~\ref{subsec:entanglement_dyn}. The distance to the Haar-random ensemble for different times and bath sizes is shown in Fig.~\ref{fig:res_deepthermalization}.

This argument was extended to more general dual-unitary circuits in \cite{claeys_emergent_2022}, which presented an alternative derivation using a replica trick and also introduced the notion of `solvable measurement scheme', i.e.\ measurement bases for which every measurement outcome preserves spatial unitarity such as the Bell pair basis --- these bases are also known as unitary error bases in the quantum information literature~\cite{knill_non-binary_1996}. In this more general setting the rigorous proof of Eq.~\eqref{eq:momentsequivalence} for almost all dual-unitary gates was found by \cite{ippoliti_dynamical_2023} (theorem in Sec.~3C). \cite{shrotriya_nonlocality_2023} studied the effect of (spatial) boundary conditions on deep thermalisation in the self-dual KIM, showing that the rate at which a state design is formed is twice as fast for periodic boundary conditions compared to open boundary conditions, highlighting the nonlocality of deep thermalisation.

Away from the dual-unitary point, different time scales are necessary for the formation of higher-$k$ designs, i.e.\ it takes longer for higher moments of the projected ensemble to agree with the moments of the Haar-random ensemble. \cite{ippoliti_dynamical_2023} argued that the absence of dynamical purification in the space-time dual dynamics yields the collapse of all design times and the emergence of quantum state designs after a finite number of time steps. A distinct solvable random matrix model of deep thermalisation was introduced by \cite{ippoliti_solvable_2022}, where it was argued that the appearance of distinct design times in more general models is due to imperfect thermalisation, i.e. $\rho_A \approx \mathbbm{1}_A/q_A$ up to some small correction. Conversely, the collapse of all design times in dual-unitary models was argued to follow from the perfect thermalisation exhibited by these models, where $\rho_A = \mathbbm{1}_A/q_A$ after a finite number of time steps. Note, however, that for measurements that do not preserve spatial unitarity the projected ensemble will no longer correspond to the Haar-random ensemble, but the reduced density matrix will still exhibit exact thermalisation, since it is independent of the choice of measurement scheme. \cite{ippoliti_solvable_2022} also identified the Scrooge ensemble as the relevant ensemble at late times. The Scrooge ensemble was later argued to be the relevant ensemble for systems with conservation laws \cite{mark_maximum_2024, chang2025deep}.

\subsection{Anticoncentration. Porter-Thomas distribution.}

Following previous results on the interpretation of dual-unitary dynamics as a universal quantum computation in the spatial direction, \cite{claeys_fock-space_2024} studied the dynamics of the self-dual kicked Ising model focusing on Hilbert space delocalisation, also known as anticoncentration in the context of quantum information. Rather than considering local measures of quantum ergodicity, it is possible to quantify the delocalisation of a quantum state in Hilbert space through the inverse participation ratios (IPRs)~\cite{wegner_IPR_1980,evers_IPR_2000},
\begin{align}
    I_\alpha(t) = \sum_{z_j \in \{0,1\}} |\langle z_1, z_2, \dots z_L|\psi(t) \rangle|^{2\alpha}\,,
\end{align}
and corresponding participation entropies $S_{P,\alpha}(t) = \ln[I_{\alpha}(t)]/(1-\alpha)$~\cite{baumgratz_quantifying_2014}.
For a fully localised state $I_\alpha = 1$ and $S_{P,\alpha}=0, \forall \alpha$, whereas for a fully delocalised state that spans the entire space homogeneously, $|\langle z|\psi(t)\rangle| = 2^{-L/2}$ and $I_\alpha = 2^{(1-\alpha)L}$ resulting in an extensive value of $S_{P,\alpha} = L \ln(2)$. 

Under ergodic many-body dynamics an initially localised state is expected to delocalise in the Hilbert space, becoming increasingly indistinguishable from a Haar-random state. 
For an initial product state $|\psi(t=0)\rangle = | 00 \dots 0\rangle$ and dual-unitary kicked Ising dynamics, in the thermodynamic limit ($L\to\infty$) the time-dependent IPRs can be recast as the expectation value of a random unitary along the spatial direction (using the approach of~\cite{ho_exact_2022})
\begin{align}\label{eq:KIM:IPR_to_ULR}
     I_\alpha=  2^{L(1-\alpha)} \times \mathbbm{E}_{U \sim \textrm{Haar}}\left[|\langle L|U|R\rangle|^{2 \alpha}\right],
\end{align}
where $U$ is a Haar-random unitary acting on a $2^{t-1}$ dimensional Hilbert space\footnote{The appearance of $t-1$ rather than $t$ compared to the previous sections is due to the initial state being a product state in the $Z$-basis rather than in the $X$-basis.}, $|L \rangle = \otimes_{i=1}^{t-1} |+\rangle_i$ and $|R\rangle = \otimes_{i=t}^{t-1} |0\rangle_i$. These expectation values can be directly evaluated to return
\begin{align}\label{eq:KIM:IPR_DU}
    I_{\alpha}(t+1) =2^{L(1-\alpha)}\frac{\alpha! \, 2^{\alpha t}}{2^t (2^t+1)\dots (2^t+\alpha-1)}\,,
\end{align}
with participation entropies
\begin{align}\label{eq:KIM:S_q_DU}
&S_{P,\alpha}(t+1) = L \ln(2) \nonumber\\
    &\qquad + \frac{1}{1-\alpha} \ln\left[\frac{\alpha!\, 2^{\alpha t}}{2^t (2^t+1)\dots (2^t+\alpha-1)}\right]\,.
\end{align}
The long-time limit directly follows as $S_{P,\alpha}(t \to \infty) = L \ln(2) + \ln(\alpha!)/(1-\alpha)$, corresponding to the IPRs for a Haar-random state. This convergence happens on a time scale $t \propto \ln(\alpha!)$, independent of system size, as opposed to the random circuit case where this time scales as $\log(L)$~\cite{turkeshi_hilbert_2024}.

\begin{figure}[t]
  \includegraphics[width=\columnwidth]{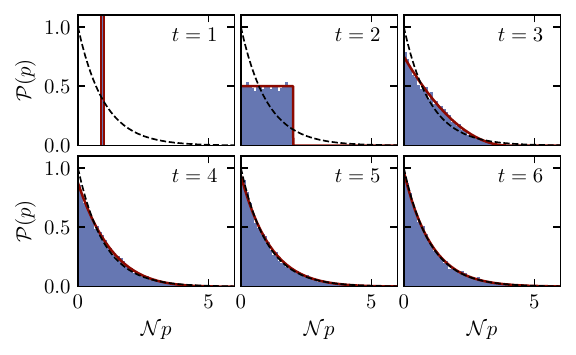}
  \caption{Distribution of bit-string probabilities/overlaps $p = |\langle z|\psi(t) \rangle|^2$ at different times.
  Full red lines indicate analytical results and blue histograms indicate numerical results for $L=14$ and $h =\pi/3$. The distribution rapidly approaches the Porter-Thomas distribution (dashed line). Adapted from ~\cite{claeys_fock-space_2024}.
  \label{fig:KIM:PorterThomas}}
\end{figure}
In the context of quantum computation, this delocalisation underlays quantum random sampling, which is arguably the most promising route to test quantum advantage~\cite{boixo_characterizing_2018,arute2019quantum,hangleiter_computational_2023}. Quantum random sampling characterises the distribution of measurement outcome probabilities $\{p_z = |\langle z|\psi \rangle|^2\}$, also known as bit-string probabilities for $z = z_1, z_2, \dots z_L$ with $z_j \in \{0,1\}$. For Haar-random states this distribution returns the Porter-Thomas (or exponential) distribution~\cite{porter_fluctuations_1956,haake2001quantum,mark_maximum_2024}. Eq.~\eqref{eq:KIM:IPR_DU} can be used to directly reconstruct this distribution at all times for self-dual kicked Ising dynamics, returning
\begin{align}\label{eq:P_p}
    \mathcal{P}(p;t+1) = \mathcal{N} (1-2^{-t}) \theta(2^t-\mathcal{N}p)\left(1-\frac{\mathcal{N} p}{2^t}\right)^{2^t-2}\!\!\!\!\!\!\!\!,
\end{align}
where $\mathcal{N}=2^L$ and $\theta(x)$ is the Heaviside function. This distribution is illustrated in Fig.~\ref{fig:KIM:PorterThomas} and rapidly approaches the Porter-Thomas distribution $\mathcal{P}(p) \propto e^{-\mathcal{N}p}$~\cite{claeys_fock-space_2024}.

\subsection{State and unitary designs in finite systems}

The results discussed in the previous subsections are concerned with the production of random states (state designs) in thermodynamically large systems. \cite{suzuki2024global} and \cite{riddell2025quantum}, however, have pointed out that dual-unitary circuits can efficiently produce quantum state designs also in finite systems.

To this end one considers a quantum circuit with finite width of $2L$, where some of the gates defining the evolution are random. Then, one introduces the following ensemble
\be
    \mathcal{E}_S = \left\{ \prod_{j=1}^t\mu({\boldsymbol{\alpha}_j}),\quad \mathbb U(\boldsymbol \alpha_t) \cdots \mathbb U(\boldsymbol \alpha_1) \ket{\psi_0} \right\},
\ee
where $\ket{\psi_0}$ is the initial state, $\boldsymbol \alpha = (\alpha_1, \ldots, \alpha_{2L})$ is a vector of random matrices, $\mathbb U(\boldsymbol \alpha)$ is the time evolution operator of a brickwork circuit (cf. Eq.~\eqref{eq:floquetoperator}) constructed with the matrices $\boldsymbol \alpha$, and $\mu({\boldsymbol{\alpha}})$ is the probability measure of $\boldsymbol{\alpha}$. Analogously, one can consider the ensemble of unitary matrices produced by the circuit
\be
\mathcal{E}_U = \left\{ \prod_{j=1}^t\mu({\boldsymbol{\alpha}_j}),\quad \mathbb U(\boldsymbol \alpha_t) \cdots \mathbb U(\boldsymbol \alpha_1) \right\}.
\ee

This setting has been studied extensively in the recent literature, see, e.g.,~\cite{hunterjones2019unitary, haferkamp2022random, hearth2024unitary, chen2024incompressibility, liu2024unitary, yada2025non}, however, the prevalent choice has been to consider the case of `random circuits', where all two-site gates are independently drawn from the Haar measure. Instead, \cite{suzuki2024global} and \cite{riddell2025quantum} proposed two `less random' settings: (i) The two-qudit gates are fixed but there are random one-qudit  gates everywhere (structured random circuit)~\cite{suzuki2024global}; (ii) a fixed brickwork quantum circuit with a single one-qudit random matrix at a boundary (minimally random circuit)~\cite{riddell2025quantum}.

Adapting the theorem of \cite{ippoliti_dynamical_2023} (cf.~Sec.~\ref{sec:deepthermal}), \cite{riddell2025quantum} have shown that both the settings (i) and (ii) produce ensembles $\mathcal{E}_S$ that approach the Haar distribution (cf.~Eq.~\eqref{eq:Haarensemble}) in the limit of large circuit depth for almost all choices of local gates. A similar general statement has not been proven for $\mathcal{E}_U$, however, its moments do appear to agree with those of the Haar distribution (for matrices), suggesting that both (i) and (ii) generate unitary $k$-designs for large depths. Specifically, \cite{suzuki2024global} proved that this is so for the second moment ($k=2$) focussing on case (i) and a special family of gates fulfilling a `solvability condition' that allows to map the problem to a free fermion calculation. Instead, \cite{riddell2025quantum} presented numerical evidence suggesting that the first three moments of $\mathcal{E}_U$ ($k=1,2,3$) approach those of the Haar distribution for almost all two-site gates in both cases (i) and (ii). By adopting the entanglement membrane picture (cf.~\cite{Zhou2020}), it was argued that this conclusion holds for all $k$. In all these cases, the moments of the ensembles $\mathcal{E}_S$ and $\mathcal{E}_U$ approach those of the Haar ensemble in a timescale of order $L$.

Interestingly, for a fixed value of the entangling power (cf.~Sec.~\ref{sec:entpow}) the relaxation of the second moment ($k=2$) to the Haar ensemble appears to be fastest when the fixed two-site gates are dual-unitary. This seems to be the case both for the solvable family of \cite{suzuki2024global} --- which includes a dual-unitary gate with entangling power $e_p(U) =1-1/q^2$ for each local dimension $q>2$ --- in general~\cite{riddell2025quantum}. A proof of this statement, however, is currently still lacking. The maximal relaxation rate is reached for dual unitary gates with entangling power~\cite{Jonay2024, riddell2025quantum}
\be
e_p(U) \geq 1-\frac{1}{q}\,.
\ee
This is consistent with the observation of \cite{brahmachari_dual_2024} that for a fixed entangling power the distance of (local) gates from product gates is maximised by dual-unitary matrices (proven for $q=2$ and conjectured for $q >2$). 
\cite{riddell2025quantum} also argued that the rate of production of higher moments $k>2$ does not only depend on the entangling power, however, dual-unitary circuits with maximal entangling power --- perfect tensors --- give the optimal $k$-design preparation for any $k$.

Finally, it is worth mentioning that recently~\cite{schuster2024random} and \cite{laracuente2024approximate} proposed a different setting --- dubbed patchwork circuits --- where Haar moment preparation is achieved on a $\log(L)$ time scale rather than $L$. These circuits can be thought of as brickwork circuits where the local gates are themselves brickwork circuits of width $\log (L)$, or as standard brickwork circuits where some of the local gates are set to be identities. Building such circuits out of dual-unitary gates might further speed up Haar moment preparation.

\section{Experimental realisations}

Dual-unitary gates have been experimentally realised in several quantum computing setups, since their efficiency at generating entanglement and randomness makes them a valuable resource for quantum computation and simulation. Google's quantum supremacy experiment~\cite{arute2019quantum}, e.g.,  prepared quantum states using a circuit consisting of iSWAP gates followed by a controlled phase gate and dressed with random one-site unitary gates --- dual-unitary gates of the form in Eq.~\eqref{eq:paraSD} --- to create quantum states considered hard to sample from classically. Rather than listing all the works where dual-unitary gates appear incidentally, here we focus on experimental realisations that directly probe dual-unitarity and its dynamical consequences.

\begin{figure}[tb!]
\includegraphics[width=\columnwidth]{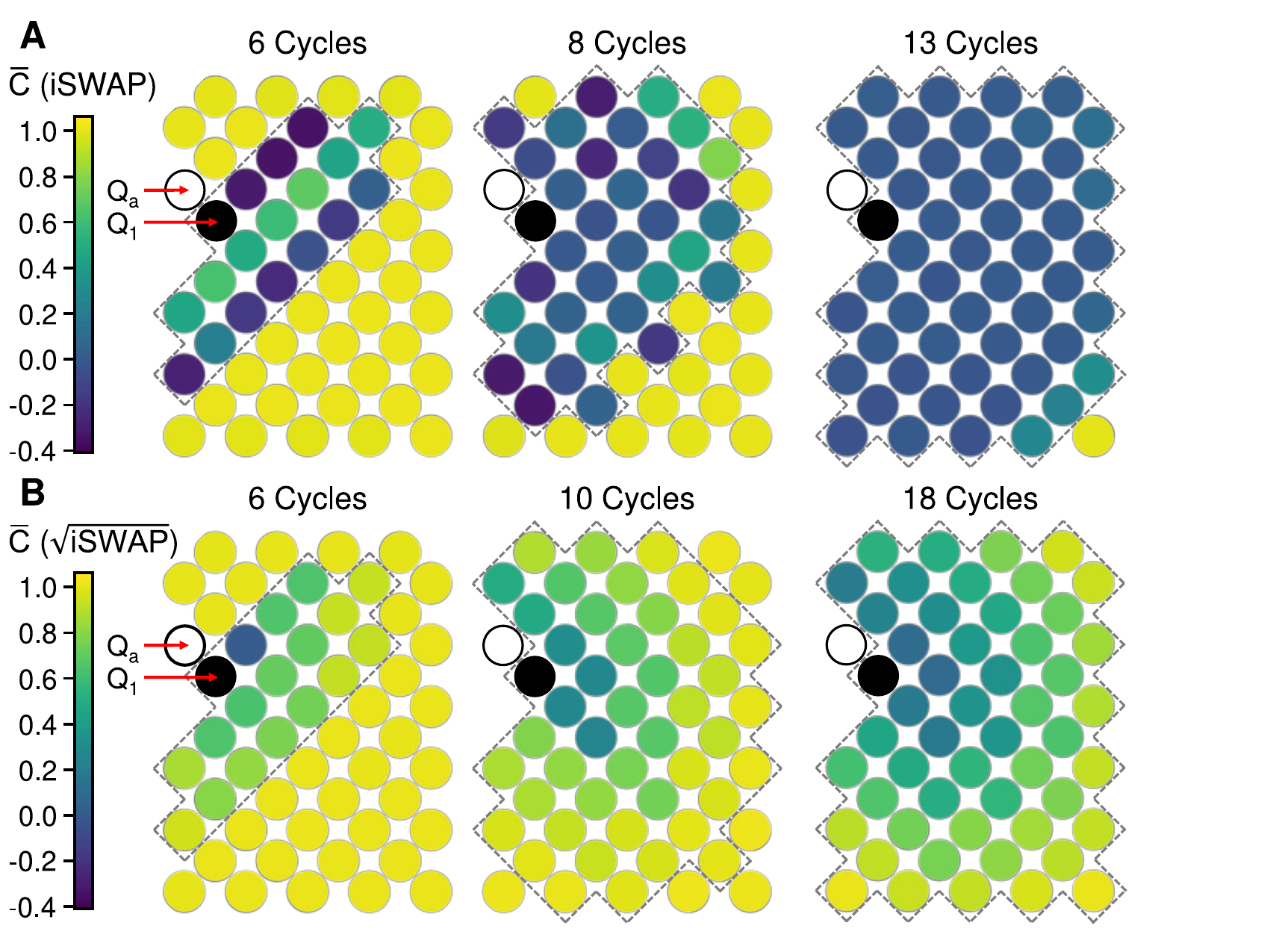}
\caption{Spatial profiles of the averaged OTOC, $\overline{C}$, in two spatial dimensions measured on the 53-qubit Sycamore processor. The rows correspond to different choices of entangling gates and the columns correspond to a different number of cycles (discrete time steps), with every circle indicating a single qubit in the Sycamore processor. The colours of the filled circles represent the averaged OTOC between the corresponding site and an initial operator on the site indicated by a black circle, with the white circle indicating the ancilla qubit.  The dashed lines delineate the light cone of the initial operator. 
(A) The two-qubit gates are iSWAP, resulting in dual-unitary dynamics and a maximal spreading of the OTOC.  
(B) The two-qubit gates are $\sqrt{\textrm{iSWAP}}$, resulting in a slower spreading of the OTOC. 
Adapted from~\cite{mi_information_2021}.
\label{fig:experiment_mi}}
\end{figure}

Out-of-time-order correlators (Sec.~\ref{subsec:operator_growth}) were experimentally investigated in \cite{mi_information_2021} on Google's Sycamore processor for a system of 53 qubits.
The OTOC was considered for different circuit realizations comprising random single-qubit gate and fixed two-qubit gates, which were chosen to be either the iSWAP gates, giving rise to dual-unitary dynamics for each circuit realisation, or the $\sqrt{\textrm{iSWAP}}$ gates, resulting in non-dual-unitary dynamics.
The experimental observation of the OTOC is performed using an interferometric protocol originally proposed in \cite{swingle_measuring_2016}: An initial wave function is evolved forward in time, perturbed with a local operator, evolved back in time, after which the OTOC between this local operator and an operator on an initial site can be measured by performing a projective measurement on an ancilla qubit initially entangled with this site. The geometry of the processor and the dynamics of the OTOCs averaged over different circuit realisations are illustrated in Fig.~\ref{fig:experiment_mi}. While the more complicated lay-out of this processor does not correspond to the one-dimensional lattice considered so far, the same qualitative behaviour is observed: Dual-unitary gates result in maximal operator growth, with the OTOC taking a nontrivial value everywhere inside the geometric light cone, to be contrasted with the slower spreading and broadening front of OTOCs in generic circuits. 

\begin{figure}[tb!]
\includegraphics[width=0.9\columnwidth]{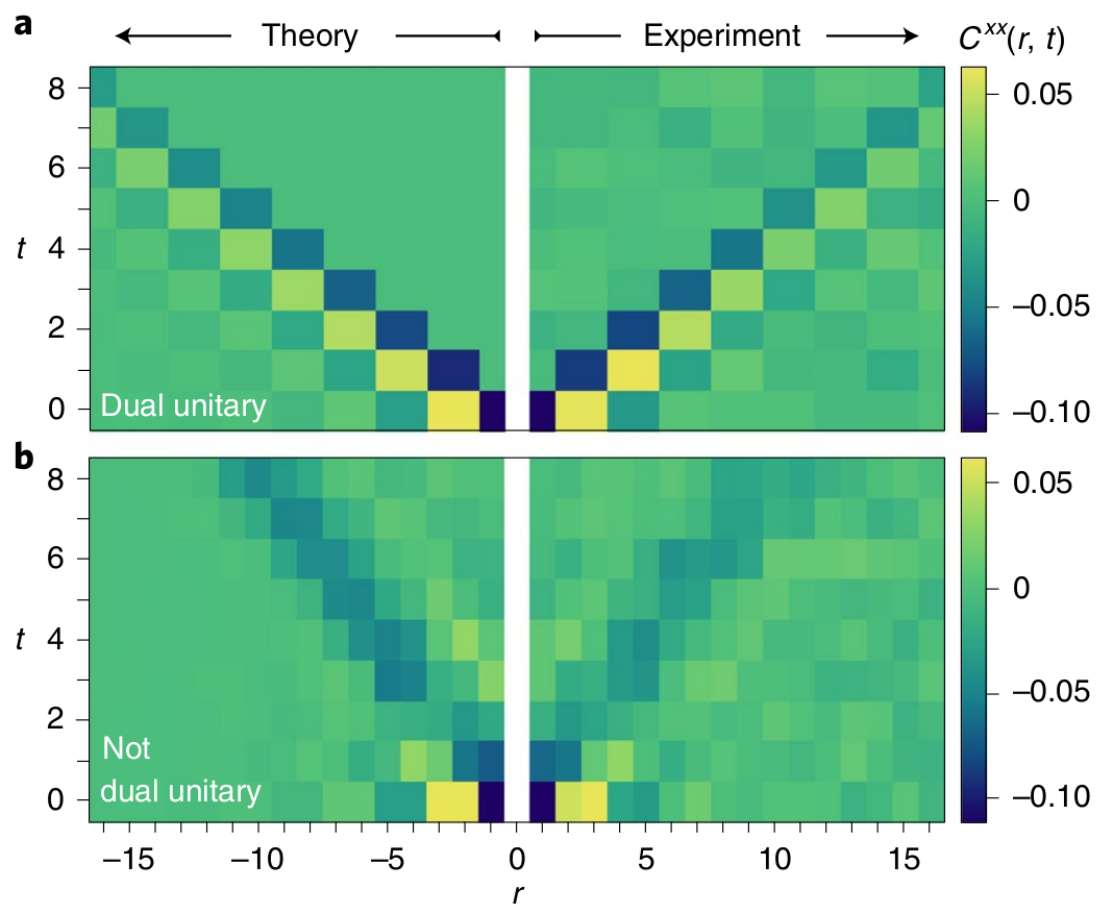}
\caption{a) Equal-time spin–spin correlators for the quench dynamics of a solvable MPS, showing theoretical results (left side, $r < 0$) and experimental results (right, $r > 0$). Correlations spread with the maximum velocity along the light cones of the circuit and decay exponentially along the light cones. The data in this plot are aggregated into bins $r \in \{2j, 2j + 1\}$ for $j > 0$ to smooth and reduce statistical fluctuations. b) The same correlator for a circuit which is not dual unitary.
Reproduced from~\cite{chertkov_holographic_2022}.
\label{fig:experiment_chertkov}}
\end{figure}

\begin{figure*}[htb!]
\includegraphics[width=0.9\textwidth]{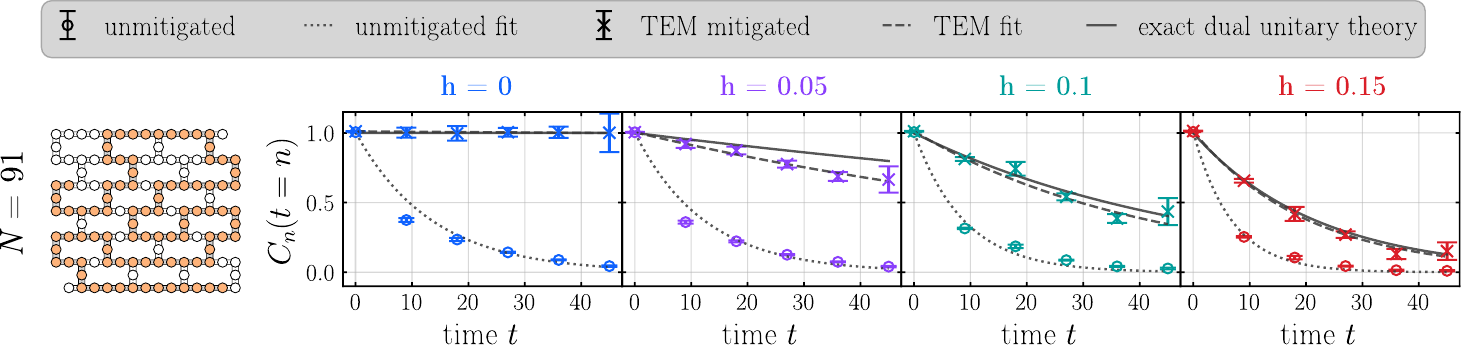}
\caption{Left: Layout of a one-dimensional chain of 91 qubits embedded within the \emph{ibm strasbourg} quantum processor. Right: Experimental correlation function on the light cone $C_n(t = n)$ for dual-unitary kicked Ising dynamics in this chain at increasing values of the longitudinal field strength $h$. For $h = 0$ the dynamics are Clifford and used to calibrate the noise model, such that the mitigated signal matches the theoretical curves nearly exactly. Away from this point the theoretical prediction of $C_n(t=n)=[\cos(2h)]^t$ is recovered after noise mitigation.
Adapted from~\cite{fischer_dynamical_2024}.
\label{fig:experiment_fischer}}
\end{figure*}

Quantum state dynamics were experimentally observed in a trapped-ion quantum computer using 11 qubits in \cite{chertkov_holographic_2022}. 
This work measured the equal-time two-point correlation functions in a quantum state evolved using dual-unitary dynamics (Sec.~\ref{sec:statedynamics}) corresponding to the self-dual kicked Ising model (Sec.~\ref{sec:KIM}). 
These correlation functions were experimentally obtained using an algorithm for simulating the quantum dynamics of initial matrix product states in the thermodynamic limit through a holographic approach involving qubit reuse and partial measurements \cite{foss-feig_holographic_2021}. Note that this algorithm did not use the dual-unitarity of the gate, but rather proposed the use of dual-unitary circuits for benchmarking quantum algorithms. 
The experimental results are illustrated in Fig.~\ref{fig:experiment_chertkov} and show excellent agreement with exact theoretical results in the thermodynamic limit, highlighting the spreading of correlations along the light cone and the exponential decay to the ergodic value. A larger scale version of this experiment was later realised by~\cite{moses2023a} using the more recent Quantinuum's H2 quantum computer, which at the time of the experiment involved 32 qubits and was later upgraded to 56 qubits~\cite{decross2025computational}.

Infinite-temperature autocorrelation functions for dual-unitary kicked Ising dynamics were experimentally simulated on a system of 91 superconducting transmon qubits on the \emph{ibm strasbourg} quantum processor \cite{fischer_dynamical_2024}. In this experiment an initial solvable Bell-pair state was prepared, with a single $\sigma^x$ at an initial site, this wave function was evolved using dual-unitary kicked Ising dynamics, and a projective measurement in the $\sigma^x$ basis was performed at a final site. Through space-time duality, the corresponding expectation value returns the dynamical correlations between the two Pauli matrices at the initial and final site, such that these results can be interpreted as either probing the infinite-temperature autocorrelation functions or dynamical correlation functions following a quench.
The correlation functions were probed for one-dimensional chains up to 91 qubits, and the vanishing of correlation functions outside the edge of the causal light cone was directly observed. The theoretically predicted exponential decay of the light-cone correlation functions was matched with experimental results through the use of noise characterisation and the tensor-network error mitigation methods of~\cite{filippov_scalable_2023,filippov_scalability_2024}, showing excellent agreement as illustrated in Fig.~\ref{fig:experiment_fischer}.

\section{Perturbed dual-unitarity}

As detailed in previous sections, generic dual unitary circuits represent an exactly solvable paradigm of quantum chaotic and ergodic dynamics. In a loose sense, they may be considered quantum many-body analogues of the uniformly hyperbolic dynamical systems with Markov partitions --- Axiom A systems --- that are tractable models of chaos in dynamical systems~\cite{cvitanovic_chaos_2016}.
It is known that such hyperbolic classical deterministic dynamics is \emph{structurally stable}~\cite{robbin_structural_1971,robinson_structural_1974}, i.e., a small perturbation of these systems has similar (or topologically identical) dynamics. One could therefore hope that something similar is true for dual-unitary circuits. In other words, one may conjecture that dual-unitary systems with exponentially decaying local spatiotemporal correlators represent an ideal setup where many-body perturbation theory is extremely well behaved, possibly showing finite radii of convergence. This is of course in sharp contrast with the typical behaviour of perturbation series around non-interacting points, where the radius of convergence vanishes. Developing a systematic strategy to answer the above question, however, proved to be very challenging. To date, precise answers have been obtained only for a few quantities, which we discuss in the rest of this section. 

We begin by following~\cite{kos_correlations_2021} and consider what is possibly the simplest non-trivial observable to analyse: two-point functions of local operators. Under a suitable stability condition on the parameters of the gate, one can show that in perturbed dual-unitary circuits the dynamical correlations of Sec.~\ref{sec:corr_functions} can be expressed in terms of a path-integral (or path-sum) formula, where the role of the free propagator is played by the exponentially decaying two-point function of the unperturbed dual-unitary circuit.

More specifically, we focus on the following two-point space-time correlation function (written in the folded/operator-space representation, cf.~Sec.~\ref{sec:diagrams})
\be
\label{eq:p2bc}
\!\!\!\!\!\!\!\!\langle b_x | a_0(t)\rangle \!=\! \fineq[1.5ex][.55][1]{
\begin{scope}[rotate around={-45:(0,0)}]
\foreach \i in {0,...,6}{
\foreach \j in {0,...,4}{
\roundgate[\i+\j][\i-\j][1][bottomright][orange][1]
}
}
\foreach \i in {0,...,2}{
\foreach \j in {0,...,2}{
\roundgate[3*\i+2*\j][3*\i-2*\j][1][bottomright][green1][1]
}
}
\foreach \i in {0,...,6}{
\cstate[\i-.5][\i+.5]	
\cstate[\i+4.5][\i-4.5]	
}
\foreach \i in {0,...,4}{
\cstate[\i-.5][-\i-.5]	
\cstate[\i+6.5][-\i+6.5]
}
\charge[-0.5+4][-0.5-4][black]
\charge[6.5][6.5][black]
\end{scope}
\draw [decorate, thick, decoration = {brace}]   (8.75,-6.65)--++(-4.75,0);
\node[scale=1.5] at (6.5,-7.15) {$\nu_+$};
\draw [decorate, thick, decoration = {brace}]   (9.5,.15-2.75)--++(0,-3.25);
\node[scale=1.5, rotate = 0] at (10.15,-4.25) {$\nu_-$};
\node[scale=1.5] at (-.75,-5.25){$a$};
\node[scale=1.5] at (9.15,.5){$b$};
}\!\!\!\!\!\!\!\!, 
\ee
where $a,b$ are two local operators and $x,t$ are the spatial and temporal distance. Note that the diagram has been simplified maximally using the unitarity conditions in Eq.~\eqref{eq:unitarityfoldeddiagram} and tilted by $45^\circ$ clockwise for easier visualisation. Orange boxes denote a generic folded dual unitary gate $U_{\rm du}$, i.e.\ $U_{\rm du}\otimes U_{\rm du}^*$, and green boxes denote a perturbed gate (or a generic, non-dual unitary gate) which can be written as
\be
U_\eta = U_{\rm du}e^{i \varepsilon P},
\ee
where $P$ is some hermitian two-site operator. For simplicity we assume that all green gates and all orange gates are the same, and that the arrangement of perturbed (green) gates is a regular sub-lattice with spacing $\nu_{\pm}$ in the horizontal/vertical direction. Both of these assumptions, however, can be lifted, and the only important parameter is the density of the perturbed gates $\delta=1/(\nu_+ \nu_-)$. One can then study two perturbative regimes, with either the density $\delta$ or the perturbation strength $\varepsilon$ being a small parameter. It turns out that the small $\delta$ analysis is technically much simpler, although numerical simulations suggested~\cite{kos_correlations_2021} that even $\delta=1$ and small $\varepsilon$ yield a similar stability estimation. In the following we only assume that $\nu_\pm$ are both large. Under certain conditions specified below, it was shown that the perturbed dynamical correlation function in Eq.~\eqref{eq:p2bc} can be written as a sum of all path contributions of the type
\be
\fineq[1.5ex][.55][1]{
\begin{scope}[rotate around={-45:(0,0)}]
\foreach \j in {2,...,4}{
\roundgate[\j][-\j][1][bottomright][orange][1]
}
\foreach \i in {0,...,3}{
\roundgate[\i+2][\i-2][1][bottomright][orange][1]
}
\foreach \j in {0,...,2}{
\roundgate[3+\j][3-\j][1][bottomright][orange][1]
}
\foreach \i in {3,...,6}{
\roundgate[\i][\i][1][bottomright][orange][1]
}
\roundgate[4][-4][1][bottomright][green1][1]
\roundgate[2][-2][1][bottomright][green1][1]
\roundgate[3][3][1][bottomright][green1][1]
\roundgate[6][6][1][bottomright][green1][1]
\roundgate[3+2][3-2][1][bottomright][green1][1]
\foreach \i in {4,...,6}{
\cstate[\i-.5][\i+.5]	
\cstate[\i+.5][\i-.5]	
}
\foreach \i in {0,...,1}{
\cstate[\i+2.5][\i-.5]	
\cstate[\i+3.5][\i-1.5]	
}
\cstate[3-.5][3+.5]
\foreach \i in {3,...,4}{
\cstate[\i-.5][-\i-.5]	
\cstate[\i+.5][-\i+.5]
}
\cstate[1.5][-1.5]
\cstate[4.5][-4.5]
\cstate[1.5][-2.5]
\cstate[4+1.5][3-2.5]
\cstate[4+1.5][4-2.5]
\cstate[3+1.5][5-2.5]
\cstate[2+1.5][5-3.5]
\cstate[1+1.5][6-3.5]
\charge[-0.5+4][-0.5-4][black]
\charge[6.5][6.5][black]
\end{scope}
\node[scale=1.5] at (-.75,-5.25){$a$};
\node[scale=1.5] at (9.15,.5){$b$};
}\!\!\!\!\!\!. 
\vspace{1cm}
\label{eq:onepath}
\ee
Here the strings of horizontal/vertical orange gates correspond to the iterated quantum channels $\mathcal M_\pm $ characterising the light-cone correlations in dual-unitary circuits (Eqs.~\eqref{eq:mapMplus} and \eqref{eq:mapMminus}). Parameterising all possible paths by the lengths of horizontal/vertical segments $\{l^\pm_j\}$ one finds the so-called \emph{skeleton approximation} or path-sum formula for the dynamical correlation function in Eq.~\eqref{eq:p2bc}, i.e.
\begin{widetext}
\be
\langle b_x | a_0(t)\rangle \simeq
\begin{cases}
\displaystyle \sum_{n\ge1}\sum_{\{l^\pm_j\}}
\langle b|
(\mathcal M_+)^{l^+_{n+1}} \mathcal E_1
\cdots
(\mathcal M_+)^{l^+_2} \mathcal E_1
(\mathcal M_-)^{l^-_1} \mathcal E_2
(\mathcal M_+)^{l^+_1}|a\rangle,
& x\in\mathbb Z\\
\displaystyle \sum_{n\ge1}\sum_{\{l^\pm_j\}}
\langle b|
(\mathcal M_-)^{l^-_{n}} \mathcal E_2
\cdots
(\mathcal M_+)^{l^+_2} \mathcal E_1
(\mathcal M_-)^{l^-_1} \mathcal E_2
(\mathcal M_+)^{l^+_1}|a\rangle,
& x\in\mathbb Z+\frac{1}{2}\\
\end{cases}\,.
\label{eq:skeleton}
\ee
\end{widetext}
In this expansion $\mathcal E_{1,2}$ are the `right-turn' and `left-turn' channels, formed from the perturbed (green) gates as
\be
\mathcal E_{1}= \fineq[-.75ex][.85][1]{
\roundgate[0][0][1][bottomright][green1][1]
\cstate[-.5][-.5]
\cstate[-.5][.5]
},
\qquad\quad
\mathcal E_{2}= \fineq[-.75ex][.85][1]{
\roundgate[0][0][1][bottomright][green1][1]
\cstate[.5][-.5]
\cstate[.5][.5]
},
\ee
and representing defects at which the dual-unitary (orange) propagators can scatter, while $n$ denotes the total number of scattering events. From the parameterisation of the path we have the following constraints 
\begin{align}
&\sum_{j=1}^{n+\nu} l_j^+ = x+t-n,\\
&\sum_{j=1}^n l_j^- = x-t-n-\nu, \quad x\in\mathbb Z +\frac{\nu}{2}, \quad\nu\in\{0,1\}.
\end{align}
We stress that there is no general principle guaranteeing that an object as the one in Eq.~\eqref{eq:p2bc} should be well approximated by the skeleton formula in Eq.~\eqref{eq:skeleton}. In fact, whenever this happens one has a drastic reduction in computational complexity since the diagram in Eq.~\eqref{eq:p2bc} is \emph{exponentially hard} to compute, while Eq.~\eqref{eq:skeleton} can be efficiently evaluated for any finite order $n$ as it is a sum of terms (paths) obtainable via one-dimensional transfer matrices. As discussed by~\cite{kos_correlations_2021}, the key ingredient to attain the approximation in Eq.~\eqref{eq:skeleton} is to make a precise spectral characterisation of the $4^x\times 4^x$ transfer matrices in Eq.~\eqref{eq:AC}. These transfer matrices can be used to contract regions between the defects in the low density regime $\delta\ll 1$ (or, equivalently, for weak perturbations $\varepsilon \ll 1$). While their leading eigenvalue is always $1$ with the trivial eigenvector, \cite{kos_correlations_2021} observed that --- remarkably --- for some regions in the parameter space of the dual-unitary gates their spectral gap is independent of $x$ for all $x$. When this happens, the `loop corrections' to Eq.~\eqref{eq:skeleton} can be neglected and the latter provides an asymptotically accurate expression. In turn, this result implies that, in the stable regions, dynamical correlations are continuous functions of the perturbation parameters ($\delta$, $\varepsilon$) uniformly in $x$ and $t$. While this statement is based on (extensive) numerical simulations for dual-unitary circuits, it is possible to provide a rigorous threshold between stable and unstable regions in the special case of so-called \emph{reduced} dual-unitary circuits \cite{kos_correlations_2021}, which can also be understood as dual-bistochastic Markov circuits where each gate is obtained by $U(1)$ averaging dual-unitary circuits over i.i.d.\ random and noisy longitudinal fields (say in the $z-$direction). These results can also be reinterpreted following \cite{nahum_real-time_2022}, who argued that these dynamics correspond to a `bound phase' in which the support of operator trajectories determining the two-point dynamical correlation functions remains bounded. In a separate but related direction, \cite{kos_thermalization_2021} have shown that one can write a path summation formula to describe the thermalisation of a subsystem evolved with perturbed dual-unitary gates and immersed in a sea of dual-unitary gates. 

An analogous derivation can be performed for more sophisticated observables. To show the example of a multi-replica quantity, here we briefly describe the analysis of OTOCs of nearly dual-unitary circuits, following~\cite{rampp_dual_2023}. The aim of this work was to recover the ballistic spreading and diffusive broadening of the operator front in generic circuits (cf.~\cite{von_keyserlingk_operator_2018,nahum_operator_2018}) through a perturbative expansion around the dual-unitary point.

We remind the reader of the 4-replica diagrammatic representation of the OTOC in Eq.~\eqref{eq:OTOC_graphical}, which holds for general unitary circuits. In the dual-unitary limit the OTOC was calculated from the $x_-+1$ ($x_-=x-t$) leading eigenvectors with eigenvalue $1$ of the light-cone transfer matrix \eqref{eq:OTOC_transfermatrix}, where these eigenvectors are exhausted by the states of Eq.~\eqref{eq:orthonormalrainbows} in the case of a completely chaotic unperturbed gate $U_{\rm du}$, which is the case considered here. For small enough density $\delta$, or perturbation strength $\varepsilon$, one can argue that the transfer matrix $T_{x_-}$ for the perturbed dual-unitary circuits can be approximated by its $(x_-+1) \times (x_-+1)$ projection to the subspace spanned by the unperturbed right and left eigenvectors of Eq.~\eqref{eq:orthonormalrainbows}. Denoting these eigenvectors as $|k\rangle = |e_{x_i,k}\rangle $ with $k=0,\ldots x_-$, the matrix elements $T_{kl}=\langle k|T_{x_-}|l \rangle$ of the projected transfer matrix can be characterised. Unitarity implies a tridiagonal form for this transfer matrix, with $T_{00}=1$ and 
\begin{align}
&T_{0k}=\frac{z_k}{q^{k-2}}, \qquad T_{k\geq 1 k\geq 1}=1-z_1, & &k\ge 1, \\
&T_{lk}=(q^2 z_{k-l}-z_{k-l+1}), \qquad T_{kl}=0, & &k>l, 
\end{align}
where
\be
z_1=\frac{B_1-1}{q^2-1}, \qquad z_k=\frac{B_k-B_{k-1}}{q^2-1},
\ee
are expressed in terms of quantities
\be
B_k = \frac{1}{q^{k+1}} \fineq[1.5ex][.7][1]{
\begin{scope}[rotate around={-45:(0,0)}]
\foreach \i in {0,...,4}{
\foreach \j in {0,...,0}{
\roundgate[\i+\j][\i-\j][1][bottomright][orange][2]
}
}
\foreach \i in {0}{
\sqrstate[\i-.5][-\i-.5]	
\cstate[\i+4.5][-\i+4.5]
}
\foreach \i in {0,...,4}{
\cstate[\i-.5][\i+.5]	
\sqrstate[\i+.5][\i-.5]	
}
\end{scope}
\draw [decorate, thick, decoration = {brace}]   (5.75,-1)--++(-6,0);
\node[scale=1.5] at (2.75,-1.5) {$k$};
}. 
\ee
This expression generalises the linear entropy of Eq.~\eqref{eq:linearentropy_U} to diagonal compositions of the two-site gate and trivialises for dual-unitary gates: $z_k = 0$ for all $k$ if and only if the gates are dual-unitary. Hence, for the OTOC these quantify the proximity to dual-unitarity and represent a `small parameter' for perturbed dual-unitary circuits.

The triangular form of $T_{kl}$ immediately implies a path-sum formula akin to Eq.~\eqref{eq:skeleton} for the OTOC where the off-diagonal elements in the matrix power $T^{x_+}$ expanded by multiple insertions of $\sum_k |k \rangle \langle k|=1$, all being of sub-leading order $O(\varepsilon^2)$, determine the OTOC within the light-cone and indicate a sub-maximal butterfly velocity $v_B<1$. Quantitative analytic results can be obtained in the large $q$ regime, where the summation is dominated by paths where the steps inside the light cone are at most of size $1$. Indeed, in this case one can simplify $T_{kl}$ by only retaining the near-diagonal matrix element $z_1$ since others are suppressed by higher powers of $1/q$. This leads to the expected asymptotic form describing ballistic propagation and diffusive broadening
\be
O_{\alpha\beta}(x,t) \simeq 
\frac{1}{2}\left(1+\erf\left(\frac{x-v_{B,1} t}{\sqrt{2D_1t}}\right)\right),
\ee
with butterfly velocity and diffusion constant
\be 
v_{B,1}=\frac{1-z_1}{1+z_1},\quad
D_1 = v_{B,1}(1-v_{B,1}^2).
\ee
This approximation works remarkably well even for qubit circuits $q=2$, where it can be also analytically improved by including 2-step paths. In the dual-unitary limit $z_1 \to 0$ and hence $v_B \to 1$ and $D_1 \to 0$, recovering the phenomenology of OTOCs in dual-unitary circuits.

A similar structure for transfer matrices in OTOCs was proposed in \cite{huang_out--time-order_2023} for generic circuits, where this transfer matrix was termed a `light-like generator'. Universal properties of this matrix were conjectured, one on the algebraic degeneracy of its eigenvalues and another on the geometric degeneracies of subleading eigenvalues, consistent with the perturbed dual-unitary result. It was additionally proposed that the OTOC can be approximated from the leading singular state of this transfer matrix, and a product-state variational ansatz for this state was postulated and computed variationally.

\section{Extensions and generalisations of dual-unitarity}
\label{sec:generalisations}

A variety of extensions and generalisations of dual-unitarity have appeared in the literature.
Moving away from unitary dynamics, these extensions have unveiled both open and classical solvable dynamics by identifying appropriate conditions that can be represented in the graphical form of the dual-unitary conditions (cf.~Eq.~\eqref{eq:unitarityfoldeddiagram} and~\eqref{eq:spaceunitarityfoldeddiagram}).
For unitary dynamics, conversely, generalisations of dual-unitarity can be broadly classified in two (interconnected) categories: (i) extensions to different geometries in space-time, generalising the duality between space and time in 1+1 dimensions to different space-time symmetries, or (ii) identifying conditions satisfied by local gates that relax the dual-unitary conditions while retaining some degree of solvability. The former fall under the umbrella term `multi-unitary dynamics', or `multi-directional unitary operators' as suggested by \cite{mestyan_multi-directional_2024}, whereas the latter were dubbed `hierarchical dual-unitarity' by \cite{yu_hierarchical_2024}. An alternative approach, again closely connected to the aforementioned directions, consists of extending the graphical calculus of dual-unitarity to biunitarity, which allows for the graphical representation of various solvable unitary circuits in a unified manner. We review these different approaches in the following subsections.

\subsection{Open and classical dynamics}

Dual-unitarity has been extended away from unitary dynamics to both open systems, by~\cite{kos_circuits_2023}, and classical dynamics, by \cite{christopoulos_dual_2024} and \cite{lakshminarayan_solvable_2024}. In all these cases the extensions consist of pinpointing a local evolution operator which can be represented in the same way as the folded gate of Eq.~\eqref{eq:foldedgatepicturen1} and identifying conditions that extend the graphical identities encoding unitarity in time (cf.~Eqs.~\eqref{eq:unitarityfoldeddiagram} and~\eqref{eq:spaceunitarityfoldeddiagram}).

Specifically, \cite{kos_circuits_2023} proposed a family of `circuits of quantum channels' were the folded gate in Eq.~\eqref{eq:foldedgatepicturen1} represents a local quantum channel $\mathcal C$ and the analogue of Eqs.~\eqref{eq:unitarityfoldeddiagram} corresponds to $\mathcal C$ being (i) unital and (ii) trace preserving. While for unitarity the two conditions of Eq.~\eqref{eq:unitarityfoldeddiagram} are equivalent (unitarity for, e.g., left multiplication implies unitarity for right multiplication) for quantum channels these conditions are independent. Similarly, the unitarity and trace preserving conditions can be independently imposed along the spatial direction, returning the analogue of Eqs.~\eqref{eq:spaceunitarityfoldeddiagram}. Different families of models that satisfy different combinations of unitality constraints and corresponding solvability can hence be defined. When all conditions are satisfied the resulting dynamics display the phenomenology of dual-unitary circuits, and solvable initial states can again be defined as matrix-product density operators~\cite{kos_circuits_2023}.

Space-time duality was utilised in discrete-time classical dynamics for a chain of coupled Arnold cat maps by~\cite{fouxon_local_2022}. The quantisation of this model returns a dual-unitary circuit closely related to the kicked Ising model of Sec.~\ref{sec:KIM}, and techniques similar to those discussed there were used to characterise the dynamical correlations --- which are again restricted to light rays --- in the classical case. The specific choice of model here led to additional simplifications in the calculation of correlations for operators with larger support.
\cite{christopoulos_dual_2024} and \cite{lakshminarayan_solvable_2024} subsequently extended the general concept of dual-unitarity to the classical realm by identifying appropriate local conditions on the dynamical evolution. Specifically, \cite{christopoulos_dual_2024} considered `classical circuits' consisting of local symplectic maps on a classical phase space and introduced the notion of \emph{dual symplectic classical circuits}. Symplecticity implies the invariance of the uniform (flat) phase-space distribution under symplectic transformations, which is here the equivalent of unitality. For dual symplecticity the constituting maps remain symplectic when exchanging the role of space and time, such that Eqs.~\eqref{eq:unitarityfoldeddiagram} and \eqref{eq:spaceunitarityfoldeddiagram} remain satisfied when the folded gate corresponds to a symplectic transformation and the folded identities correspond to uniform phase-space distributions. These conditions alone, however, do not suffice to produce solvable dynamics. The change to the dual picture introduces a change of integration variables with a non-trivial Jacobian: the latter has to be equal to one in order for dual symplecticity to return the dual-unitary phenomenology. For instance, dual-symplectic maps appeared in \cite{krajnik_kardarparisizhang_2020} in the study of integrable rotationally symmetric classical dynamics. In that case, however, the unit Jacobian is not satisfied, and the correlations did not display dual-unitary phenomenology (they are nonvanishing inside the causal light cone). A dual-symplectic classical Ising SWAP model for which this condition is satisfied, and the correlations can be exactly determined, was introduced in \cite{christopoulos_dual_2024}. A closely related extension of dual-unitarity to the classical domain was proposed in \cite{lakshminarayan_solvable_2024} through the Koopman approach to classical mechanics, in which functions on a classical phase space are evolved in time using a unitary transformation. Circuits of dual-unitary Koopman operators, or \emph{dual-Koopman circuits} in short, were introduced, where unitality in the both the time and space direction were similarly shown to result in the characteristic light-cone dynamics of correlation functions. Note that in this case there is no additional condition on the Jacobian as the latter is already implied by the dual-Koopman property.

\subsection{Different space-time geometries}

In Sec.~\ref{subsec:KIM:generalizations} we discussed various extensions of dual-unitarity to different geometries in space-time in the context of the kicked Ising model. Here we will focus on generalisations that do not use the particular structure of that model.

\emph{Triunitary circuits} were introduced by \cite{jonay_triunitary_2021} as a class of quantum circuits that exhibit three `arrows of time' (unitary directions), as opposed to the two unitary directions of dual-unitarity. A triunitary circuit is constructed out of three-site unitary gates, typically represented as hexagons in order to highlight their space-time symmetries, and can be represented graphically as
\begin{align}\label{eq:triunitary:circuits}
    \!\!\!\vcenter{\hbox{\includegraphics[width=0.4\columnwidth]{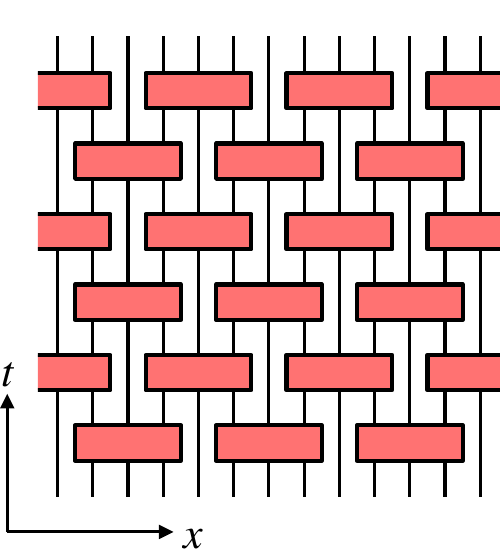}}}\,\, = \!\!\! \vcenter{\hbox{\includegraphics[width=0.43\columnwidth]{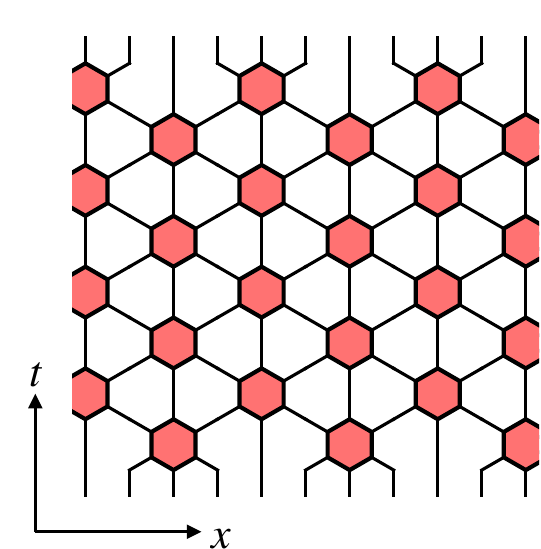}}}
\end{align}
These three-site gates are fixed to be unitary for contractions of any three neighbouring sites, resulting in three independent unitarity conditions 
\begin{align}\label{eq:triunitary:condition}
&\vcenter{\hbox{\includegraphics[width=0.15\columnwidth]{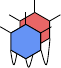}}} = 
\vcenter{\hbox{\includegraphics[width=0.14\columnwidth]{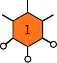}}} = 
\vcenter{\hbox{\includegraphics[width=0.14\columnwidth]{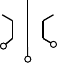}}}\\
&\vcenter{\hbox{\includegraphics[width=0.15\columnwidth]{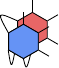}}} = 
\vcenter{\hbox{\includegraphics[width=0.14\columnwidth]{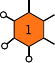}}} = 
\vcenter{\hbox{\includegraphics[width=0.14\columnwidth]{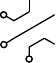}}}\\
&\vcenter{\hbox{\includegraphics[width=0.15\columnwidth]{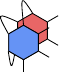}}} = 
\vcenter{\hbox{\includegraphics[width=0.14\columnwidth]{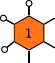}}} = 
\vcenter{\hbox{\includegraphics[width=0.14\columnwidth]{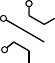}}}
\end{align}
Using triunitarity, one can find exact results on dynamical correlations and entanglement growth proceeding along the lines of Secs.~\ref{sec:corr_functions} and \ref{sec:quantumdynamics}. There are, however, interesting physical differences. First, two-point correlation functions among ultralocal operators nonvanishing on three rays in space-time, the light-cone edges $x = \pm t$ \emph{and} the static worldline $x=0$, reflecting the three-fold rotation symmetry of the triunitary gate. Second, even when considering compatible initial states, their entanglement dynamics of a finite block (cf.~Sec.~\ref{sec:quantumdynamics}) can only be solved exactly in the early time regime, i.e., for $t\leq L_A/2$ in the language of Sec.~\ref{sec:quantumdynamics} --- after this time the results begin to be gate dependent. This shows that the `degree of solvability' of triunitary circuits is lower than that of dual-unitary ones. Finally, the entanglement growth produced by these circuits is not the maximal potentially allowed by their (three site gate) geometry.

An alternative generalisation of dual-unitarity to three-site gates is given by \emph{dual-unitarity round-a-face}~\cite{prosen2021many}. The three-site unitary gates are chosen to be 1-site gates with two control qudits, i.e., the two outer wires act as control parameters for the central 1-site unitary. The gates can be represented as either a three-site gate with matrix elements $U_{abc,def} = \delta_{ad}\delta_{cf} \left(U^{ac}\right)_{be}$, where $U^{ac}$ denotes a family of single-site unitary gates, or as  face plaquettes
\begin{align}\label{eq:duirf:gates}
\vcenter{\hbox{\includegraphics[width=0.14\columnwidth]{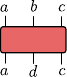}}}=
\vcenter{\hbox{\includegraphics[width=0.14\columnwidth]{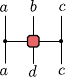}}}=
\vcenter{\hbox{\includegraphics[width=0.14\columnwidth]{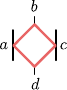}}}
\end{align}
The resulting circuit can be represented as
\begin{align}\label{eq:duirf:circuit}
\vcenter{\hbox{\includegraphics[width=0.45\columnwidth]{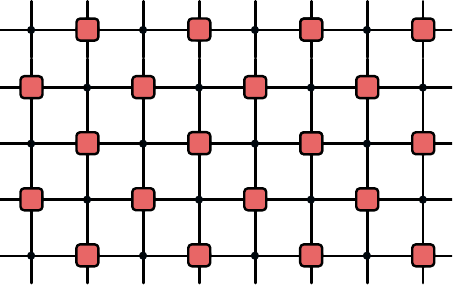}}} = 
\vcenter{\hbox{\includegraphics[width=0.32\columnwidth]{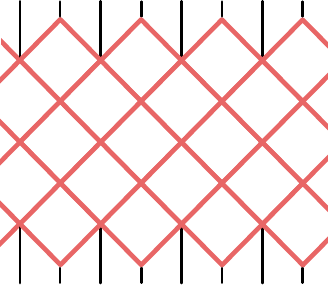}}}\,.
\end{align}
The latter representation corresponds to so-called `interactions round-a-face', see, e.g.,~\cite{baxter1982exactly}. Dual-unitarity round-a-face corresponds to the condition that the local gate remains unitary when the control indices are exchanged with the indices of the 1-site unitary gates, i.e. the matrices $(\tilde{U})^{bd}$ with matrix elements $(\tilde{U}^{bd})_{ac} = (U^{ac})_{bd}$ are also unitary. The resulting circuits exhibit the phenomenology of dual-unitary circuits, where all dynamical correlation functions of 2-local\footnote{These operators are 2-local rather than 1-local due to the specific interaction structure of these circuits, with control wires on alternating sites.} operator vanish except on the edge of the causal light cone. Solvable states for this class were introduced in \cite{claeys2023dual} (see also Sec.~\ref{subsec:biunitarity}), where it was shown that these states lead to maximal entanglement growth and exact thermalisation after a finite number of time steps.

As originally anticipated by \cite{bertini_exact_2019}, dual-unitarity can also be directly extended to higher dimensions. \cite{suzuki_computational_2022} presented the first result on dual-unitarity in (2+1) dimensions, where dynamics on a square lattice of qudits is generated by `glueing' (1+1)-dimensional dual-unitary circuits along one direction. For initial solvable states, consisting, e.g., of rows of Bell pairs, the dynamics of expectation values of local observables remain tractable and the system exhibits thermalisation to the infinite-temperature thermal state. The dynamics in these models is, however, highly anisotropic since the space-time duality only applies along a single direction. As a different example, \cite{jonay_triunitary_2021} argued that triunitary gates could be used to realise multi-unitary dynamics in (2+1) dimensions by placing triunitary gates at the vertices of a cubic lattice, where the three directions now correspond to two spatial directions and a single time direction. The resulting two-point correlation functions were similarly shown to be nonvanishing only on one-dimensional rays in space-time, where they could be calculated using a similar quantum channel approach as in dual-unitary dynamics.

A general formulation of dual-unitarity in (2+1) dimensions was considered by \cite{milbradt_ternary_2023} and termed \emph{ternary unitarity}. The dynamics is generated by 4-qudit gates
\begin{align}\label{eq:ternary:gates}
U = \vcenter{\hbox{\includegraphics[width=0.16\columnwidth]{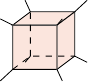}}}, \quad 
U^{\dagger} = \vcenter{\hbox{\includegraphics[width=0.16\columnwidth]{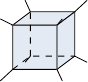}}} \qquad
\vcenter{\hbox{\includegraphics[width=0.08\columnwidth]{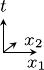}}}
\end{align}
which exhibit unitarity along the vertical direction, 
\begin{align}
\vcenter{\hbox{\includegraphics[width=0.14\columnwidth]{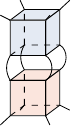}}} = 
\vcenter{\hbox{\includegraphics[width=0.14\columnwidth]{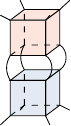}}} = 
\vcenter{\hbox{\includegraphics[width=0.14\columnwidth]{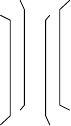}}}\,\,, 
\end{align}
as well as the two horizontal directions. Along the $x_1$ direction this condition reads
\begin{align}\label{eq:ternary:condition}
\vcenter{\hbox{\includegraphics[width=0.185\columnwidth]{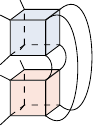}}} = 
\vcenter{\hbox{\includegraphics[width=0.116\columnwidth]{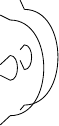}}}\,\,, \quad 
\vcenter{\hbox{\includegraphics[width=0.185\columnwidth]{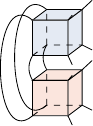}}} = 
\vcenter{\hbox{\includegraphics[width=0.116\columnwidth]{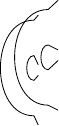}}}\,\,,
\end{align}
and similar for the $x_2$ direction, resulting in three independent conditions.
The dynamics of correlation functions is again restricted to one-dimensional light rays. \cite{milbradt_ternary_2023} additionally introduced solvable projected entangled pair states (PEPSs) as a two-dimensional extension of solvable MPSs (cf.~Sec.~\ref{sec:statedynamics}) and showed how to efficiently evaluate equal-time correlation functions. These solvable states were further studied by \cite{haag_correlations_2023}, which considered realisations of such solvable MPS and PEPS and exactly characterised their two-point equal-time correlation functions (cf.~Eq.~\eqref{eq:equaltimetwopoint}) when averaged over initial solvable states constructed from Haar-random tensors. These correlation functions vanish on average, while higher moments can be obtained through a transfer matrix calculation.

Moving from Euclidean to non-Euclidean geometries, \cite{breach_solvable_2025} considered unitary circuit dynamics when the spatial geometry was given by a tree graph and proposed an extension of dual-unitarity termed \emph{tree-unitarity}. For a tree with coordination number $z$, the relevant gates are $z$-site gates, which in addition to unitarity need to satisfy the $z$ different conditions (here illustrated for $z=3$) 
\begin{align}
\vcenter{\hbox{\includegraphics[height=0.16\columnwidth]{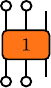}}}\,=\,
\vcenter{\hbox{\includegraphics[height=0.16\columnwidth]{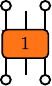}}}\,=\,
\vcenter{\hbox{\includegraphics[height=0.16\columnwidth]{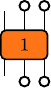}}}\,=\,q^{z-2}\,
\vcenter{\hbox{\includegraphics[height=0.127\columnwidth]{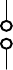}}}\,.
\end{align}
The resulting dynamics of correlation functions and entanglement remains tractable, but due to the expander properties of the tree the OTOC no longer needs to spread with maximum butterfly velocity. The tree geometry results in a proliferation of light cone directions, where imposing the additional constraint of a maximum butterfly velocity in one of these directions results in vanishing correlation functions along all other directions.

All multi-unitary gates can be related to multi-party quantum states which have maximal entanglement for all bipartitions that follow from the reflection symmetries of the given geometry. This connection was explored in \cite{mestyan_multi-directional_2024}, which presented explicit constructions of such gates for hexagonal, cubic, and octahedral geometries of dual-unitary gates, with the additional constraint that the gates are self-dual, i.e., invariant with respect to their geometrical symmetry group.

Next to the regular geometries considered so far, dual-unitarity was extended to random geometries in \cite{kasim_dual_2023} by considering a setup where dual-unitary gates are placed at the intersections of random arrangements of straight lines in two dimensions --- a `mikado' geometry. For random gates arranged in two mikado random geometries, one with a preferred space-time direction and the other with a random isotropic space-time direction, the average correlator was shown to vanish and the variance was explicitly characterized.

\subsection{Hierarchical dual-unitarity}
\label{subsec:hierarchical}

An alternative extension of dual-unitarity was proposed by \cite{yu_hierarchical_2024} with the aim to unify the solvability of dual-unitary circuits with other classes of gates solvable by different methods, such as the identity, CNOT, and CZ gates. These gates were shown to be part of a class of \emph{generalized} or \emph{hierarchical dual-unitary} circuits. Hierarchical dual-unitarity is expressed in terms of conditions involving multiple gates, where dual-unitarity is a condition on a single gate, corresponding to the first level of the hierarchy.
For the second level in the hierarchy, the two-gate conditions read
\begin{align}\label{eq:DU2_condition}
\vcenter{\hbox{\includegraphics[width=.9\columnwidth]{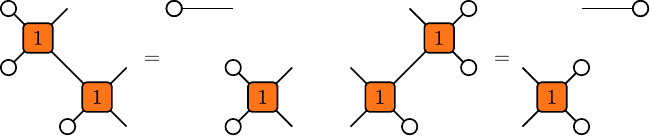}}}\!\!\!,
\end{align}
and it is also known as the `DU2 condition'. Note that, unlike the case of regular dual-unitarity, the left and right conditions are generally independent. It can be directly checked that gates that satisfy the dual-unitary condition~\eqref{eq:spaceunitarityfoldeddiagram} also satisfy Eq.~\eqref{eq:DU2_condition}, but not all gates that satisfy this condition are dual-unitary, i.e., this condition is \emph{less restrictive} than the dual-unitary condition. A complete parameterisation of non-dual-unitary DU2 gates for qubits was found by \cite{yu_hierarchical_2024} to be
\begin{align}
U = e^{i \frac{\pi}{4}\sigma^z \otimes \sigma^z} (v_1 \otimes v_2),
\end{align}
where $v_{1,2} \in U(2)$ are rotations that map $\sigma^z$ to the $x-y$ plane.

As shown by~\cite{foligno_quantum_2024}, the DU2 condition implies that the space-time swapped gate $\tilde U$ (cf.~Eq.~\eqref{eq:Utilde}) is an isometry, which should be contrasted with the unitary $\tilde U$ of dual-unitary circuits. Nevertheless, ~\cite{yu_hierarchical_2024} showed that the dynamical correlations in brickwork circuits of DU2 gates remain exactly solvable, and the correlation functions are only nonvanishing on the edges of the causal light cone $x = \pm t$, as in dual-unitary circuits, and at the static wordline $x=0$, where they can be efficiently calculated using a quantum channel approach. Moreover,~\cite{foligno_quantum_2024} showed that the entanglement dynamics of DU2 gates is exactly solvable at early times ($t\leq L_A/2$) for a family of compatible initial states. Interestingly, for non dual-unitary DU2 gates the entanglement spectrum continues to be flat (cf.~Sec.~\ref{subsec:entanglement_dyn}) but the entanglement velocity is sub-maximal, i.e.\ 
\be
\label{eq:entvelDU2}
v_E = \frac{\log r}{2 \log d}, 
\ee 
where ${r<d^2}$ is the rank of $\tilde U \tilde U^\dag$. We emphasise that the aforementioned properties of DU2 gates (three light-ray correlations; exactly solvable entanglement dynamics at early times; submaximal entanglement growth) are similar to those of triunitary gates, however, triunitarity and hierarchical dual-unitarity are generically independent conditions. Interestingly there exist models that can be formulated both in terms of triunitaries and DU2s (see, e.g.,~\cite{rampp_geometric_2024} and Sec.~\ref{subsec:biunitarity}). 

Various realizations of DU2 gates have appeared in the literature, and different results on exact solvability can be reinterpreted in terms of hierarchical dual-unitarity. The unitary gates of Eqs.~\eqref{eq:KIM:gate_honeycomb} and \eqref{eq:KIM:gate_triangular}, extending the self-dual kicked Ising model to honeycomb and triangular space-time respectively, both satisfy the DU2 condition. Note however that, in its simplest form, the condition in Eq.~\eqref{eq:DU2_condition} requires all gates in the brickwork circuit to be identical, which is not the case for the models discussed in the examples given in Sec.~\ref{sec:KIM}.

Moving up the hierarchy, however, the situation becomes more complicated as there is increasingly more arbitrariness in how the conditions are imposed and solvability is generically lost. For instance, \cite{yu_hierarchical_2024} proposed the following three gate (DU3) conditions   
\begin{align}\label{eq:DU3_condition}
\vcenter{\hbox{\includegraphics[width=0.65\columnwidth]{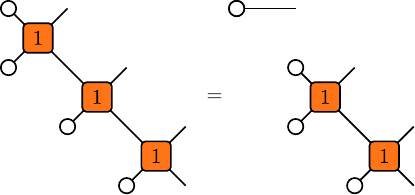}}},
\end{align}
which, however, are not sufficient to produce solvable dynamics. It is also possible for gates to satisfy mixed conditions with one level of the hierarchy satisfied from the left and a different one from the right. For instance, the unitary gate in Eq.~\eqref{eq:KIM:gate_sheared} fulfils the DU2 condition from the left and the DU3 condition from the right leading to the kicked Ising model on the sheared square lattice. 

The entanglement line tension of hierarchical circuits, which can be calculated using an approach analogous to that of Sec.~\ref{subsec:operator_growth}, was considered by \cite{foligno_quantum_2024}, \cite{rampp_entanglement_2023}, and \cite{sommers_zero-temperature_2024}. The entanglement line tension was shown to be piecewise linear, as illustrated in Fig.~\ref{fig:ELT}, with maximal butterfly velocity and submaximal but quantised entanglement velocity, in agreement with Eq.~\eqref{eq:entvelDU2}.

\subsection{Biunitarity}
\label{subsec:biunitarity}
Dual-unitary gates present a specific realisation of {\it biunitary connections}, or, in short, biunitaries. These objects encompass a variety of algebraic structures that satisfy multiple notions of unitarity, typically referred to as `vertical' and `horizontal' unitarity~\cite{reutter2016biunitary}. Examples include dual-unitary gates and dual-unitary interactions round-a-face, but also complex Hadamard matrices, quantum Latin squares, and unitary error bases. Different biunitaries can be expressed in a unified way through the so-called shaded calculus~\cite{jones1999planaralgebrasi,reutter2016biunitary}, generalising the tensor network representation of unitary gates, and satisfy graphical identities that in turn generalise the dual-unitary condition in Eqs.~\eqref{eq:DU}. For this reason, any graphical proof found for dual-unitary circuits can be directly extended to more general biunitary circuits. Biunitarity hence presents a unified picture for dual-unitarity and dual-unitary interactions round-a-face, as well as a systematic way of constructing solvable quantum circuits~\cite{claeys2023dual}. 

Biunitarity originates from the study of subfactors~\cite{Ocneanu1989} and planar algebras~\cite{krishnan1996OnBP, jones1999planaralgebrasi, jones1997introduction}, and has found different applications in, e.g., quantum error correction and teleportation~\cite{knill_non-binary_1996,werner_all_2001}. Biunitaries can be systematically combined to return different biunitaries~\cite{reutter2016biunitary}, a property that was used by~\cite{claeys_emergent_2022} for the construction of dual-unitarity gates. While original discussions of biunitarity did not consider dual-unitarity and dual-unitary interactions round-a-face (as `quantum crosses' in this context), these can be naturally included in the framework of biunitarity~\cite{claeys2023dual}. A pedagogical introduction to biunitary constructions in quantum information can be found in \cite{reutter2016biunitary}, and here we focus on their application in a quantum many-body setting, following~\cite{claeys2023dual} and \cite{rampp_geometric_2024}.

\begin{figure*}[tb!]
\includegraphics[width=2.\columnwidth]{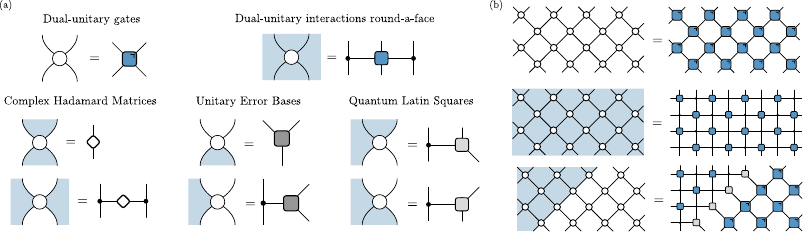}
\caption{ (a) Different shadings correspond to different algebraic objects, all of which allow a quantum gate representation that is unitary irrespective of the orientation. (b) Different shadings of the brickwork (square) lattice return different quantum circuits that are unitary in both the horizontal and vertical direction.
\label{fig:biunitaries}}
\end{figure*}

The shaded calculus extends the diagrammatic Penrose notation for tensor networks. Every tensor remains represented by a geometrical shape, but indices can now be represented by both the \emph{legs} connected to the tensor and the \emph{shaded regions} bordering it. Closed shaded regions represent an index that is implicitly summed over, similar to connecting legs. In this context, following~\cite{reutter2016biunitary}, one can write a biunitary tensor $U$ as 
\begin{align}
U =\,\vcenter{\hbox{\includegraphics[width=0.2\columnwidth]{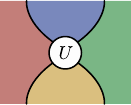}}}\,,
\end{align}
where the different colours indicate that every region can be either shaded or not. A conjugate tensor $U^{\dagger}$ is defined in such a way to satisfy a notion of (vertical) unitarity
\begin{align}\label{eq:biunitarity_VU}
\!\!\vcenter{\hbox{\includegraphics[width=0.4\columnwidth]{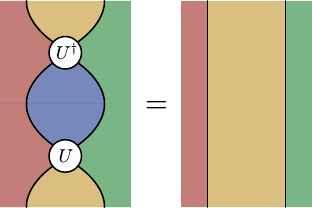}}}\quad
\vcenter{\hbox{\includegraphics[width=0.4\columnwidth]{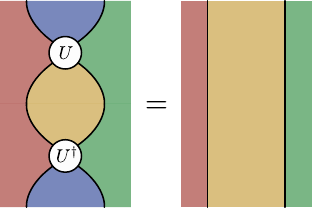}}}\,,
\end{align}
as well as a notion of horizontal unitarity
\be
\label{eq:biunitarity_HU}
\begin{aligned}
&\vcenter{\hbox{\includegraphics[width=0.47\columnwidth]{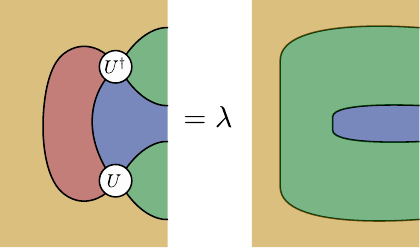}}}\,,\\
&\vcenter{\hbox{\includegraphics[width=0.47\columnwidth]{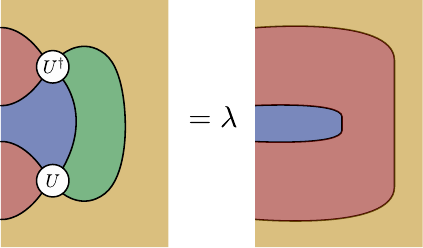}}}\,.
\end{aligned}
\ee
The constant $\lambda$ can be uniquely determined from the biunitary and generally corresponds to a power of the local Hilbert space dimension $q$. In these diagrams every region of $U$ ($U^{\dagger}$) can be either shaded or not, with the restriction that the shading of $U$ uniquely fixes the shading of $U^{\dagger}$. For different choices of shading these graphical identities correspond to different algebraic identities, which can deviate from unitarity. 

All the possible choices of biunitaries are illustrated in Fig.~\ref{fig:biunitaries}. In the absence of any shading these diagrams reduce to standard tensor network diagrams, biunitaries return dual-unitary two-site gates, and horizontal and vertical unitarity reduce to the standard unitarity in the horizontal and vertical direction. A fully shaded biunitarity returns the dual-unitary interactions round-a-face introduced by~\cite{prosen2021many}. Biunitaries with two opposing shaded regions correspond to complex Hadamard matrices~\cite{jones1999planaralgebrasi}, where horizontal unitarity corresponds to the unitarity of the matrix and vertical unitarity fixes all matrix elements to have the same modulus. Biunitarity hence underpins the use of complex Hadamard matrices as either a single-site unitary gate or a two-site controlled phase gate in kicked Ising dynamics (Sec.~\ref{subsec:KIM:generalizations}). Biunitaries with two neighbouring shaded regions correspond to `quantum Latin squares', i.e., a grid of states for which each row and each column forms an orthonormal basis~\cite{jones1999planaralgebrasi,musto_quantum_2016}. The two relations from vertical (horizontal) unitarity return the completeness and orthonormality of the states within each row (column). For a single shaded region, biunitarity returns a unitary error basis, i.e. a complete orthogonal family of unitary matrices, such as the Pauli matrices~\cite{knill_non-binary_1996,vicary_higher_2012,musto_quantum_2016}. These biunitaries can all be represented as unitary gates, where the horizontally orientated shaded regions correspond to control parameters.

For biunitarity connections arranged as a brickwork circuit (see Fig.~\ref{fig:biunitaries}), biunitarity guarantees that the corresponding dynamics will exhibit the same phenomenology as dual-unitary brickwork circuits, e.g., generating maximal entanglement growth and exhibiting dynamical correlations that are only nonvanishing on the causal light cone~\cite{claeys2023dual}. In Fig.~\ref{fig:biunitaries}, we illustrate how different shadings of the biunitary circuit result in different quantum circuits. Additionally, a periodic shading returns the self-dual kicked Ising model of Sec.~\ref{sec:KIM} in terms of complex Hadamard matrices
\begin{align}
\!\!\!\vcenter{\hbox{\includegraphics[width=0.4\columnwidth]{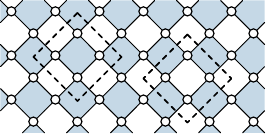}}}
=
\vcenter{\hbox{\includegraphics[width=0.4\columnwidth]{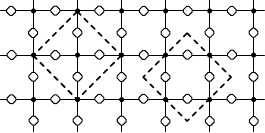}}}\,.
\end{align}
Two unit cells can be identified, where the rightmost unit cell corresponds to the KIM gate of Eq.~\eqref{eq:KIM:gate} and the leftmost unit cell corresponds to an alternative decomposition of the KIM dynamics in terms of dual-unitary interaction round-a-face gates~\cite{claeys_operator_2024}. Note that these shadings need not be homogeneous or periodic, leading to a rich family of biunitary circuits. 

Biunitaries can alternatively be arranged on the Kagome lattice to return a circuit which has a three-fold rotation symmetry in spacetime. It was already observed by \cite{jonay_triunitary_2021} that arranging dual-unitary gates on the vertices of the Kagome lattice returns a triunitary circuit, which was subsequently extended by \cite{rampp_geometric_2024} to general biunitary circuits. The resulting models remain exactly solvable and exhibit the same physics as triunitary and DU2 circuits, with correlation functions only nonvanishing on the light rays $x = \pm t$ and the static world line $x=0$. Additionally, for specific initial states this model exhibits entanglement growth with entanglement velocity $v_E = 1/2$. These similarities are not a coincidence as one can identify two unit cells on the Kagome lattice
\begin{align}
\vcenter{\hbox{\includegraphics[width=0.65\columnwidth]{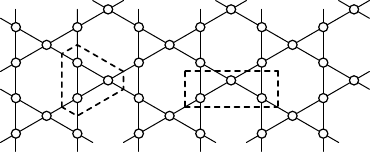}}}\,,
\end{align}
where the gate identified by the triangular unit cell satisfies the triunitary condition in Eq.~\eqref{eq:triunitary:condition} and the one identified by the rectangular choice satisfies the hierarchical DU2 condition in Eq.~\eqref{eq:DU2_condition}, both possibly shaded~ \cite{rampp_geometric_2024}. Different shadings of this model can now return triunitary or hierarchical DU2 gates, as well as different circuits that exhibit the same phenomenology. Triunitary circuits are realised in the absence of shading, returning the composite triunitary gates from \cite{jonay_triunitary_2021}, or alternatively as
\begin{align}
\vcenter{\hbox{\includegraphics[width=0.65\columnwidth]{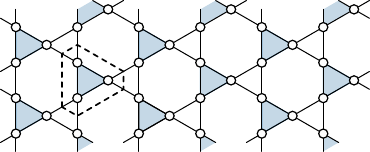}}}\,,
\end{align}
resulting in a triunitary circuit for which the gates are constructed out of three unitary error bases. Hierarchical DU2 circuits can be realised by, e.g., shading all triangles
\begin{align}
\vcenter{\hbox{\includegraphics[width=0.65\columnwidth]{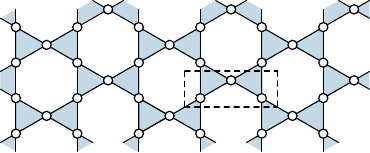}}}\,,
\end{align}
which returns the honeycomb lattice of complex Hadamard matrices considered by ~\cite{liu_solvable_2023} (cf.~Eq.~\eqref{eq:KIM:honeycomb}). Shading all hexagons conversely returns the triangular lattice considered by~\cite{rampp_entanglement_2023} (cf.~Eq.~\eqref{eq:KIM:triangular}). 

In these models the notions of triunitarity and hierarchical DU2 unitarity are equivalent, and a rich family of models can be systematically realised by choosing different shading patterns. Due to the biunitary conditions in Eqs.~\eqref{eq:biunitarity_VU} and \eqref{eq:biunitarity_HU}, all graphical proofs directly carry over. \cite{rampp_geometric_2024} additionally presented new families of models with solvable correlations and entanglement dynamics, solvable initial states, and introduced multilayer circuits with monoclinic symmetry and higher level hierarchical dual-unitary solvability with tunable entanglement velocity through the use of multilayer biunitary constructions.

\section{Space-time duality beyond dual-unitarity}
\label{sec:beyond_du}

As discussed in Sec.~\ref{sec:setting}, the basic idea behind the introduction of dual-unitary circuits is to seek systems that are `symmetric' under the exchange of space and time, i.e., systems maintaining the same basic feature (unitary evolution) when space and time switch roles. A natural question is then whether swapping space and time can give information about the system beyond this `symmetric' case. In fact, utilising the \emph{duality} between space and time to extract relevant information is a general idea that has found many fruitful applications in physics. Approaches based on this fundamental principle have been utilised, for instance, in the context of conformal~\cite{cardy1996scaling} and integrable~\cite{ghoshal1994boundary} field theory, string theory~\cite{arutyunov2007on}, tensor networks~\cite{banuls2009matrix}, and integrable spin chains~\cite{kluemper2004integrability, piroli2017what} --- all of these predate the advent of dual-unitary circuits. As we now discuss, however, dual-unitary circuits have brought new insights and energy to space-time duality approaches, and triggered many significant developments.

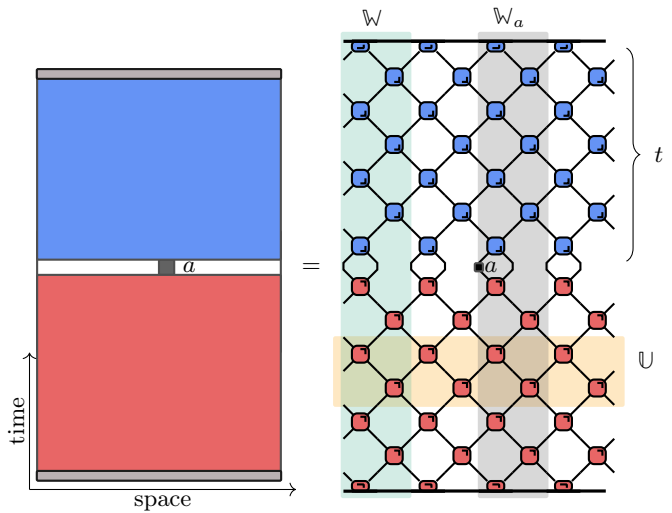
\begin{figure}
\begin{tikzpicture}[baseline={([yshift=-0.6ex]current bounding box.center)},scale=1.35]
\node at ({10.5*\dx},{10.25*\dx}){$\mathbb U$}; 
\node at ({1.5*\dx},{21.5*\dx}){$\mathbb W$}; 
\node at ({6*\dx},{21.5*\dx}){$\mathbb W_a$}; 
\draw [thick,colLines,fill=white,rounded corners=0.5] (-9.5*\dx,10.5*\dx) rectangle (-1.5*\dx,14.5*\dx);
\draw [thick,colLines,fill=grey1,rounded corners=0.5] (-5.5*\dx,13.05*\dx) rectangle (-5*\dx,13.55*\dx);
\draw [thick,yellow4,fill=yellow4,rounded corners=0.5, opacity=0.25] (0.25*\dx,8.75*\dx) rectangle (9.75*\dx,11*\dx);
(-4.5*\dx,12.75*\dx);
\draw [thick,green1,fill=green1,rounded corners=0.5, opacity=0.25] (.5*\dx,5.75*\dx) rectangle (2.75*\dx,21*\dx);
\draw [thick,grey1,fill=grey1,rounded corners=0.5, opacity=0.25] (5*\dx,5.75*\dx) rectangle (7.25*\dx,21*\dx);   
\TEsheetrotated{-13.55}{1-10.5}
\TECsheet{-6.3}{1-10.5} 
\node at ({-.5*\dx},{13.3*\dx}){$=$};
\node at ({-4.5*\dx},{13.3*\dx}){$a$};
\draw[->] ({(0.25-10)*\dx},{(6)*\dx}) -- ({(0.25-10)*\dx},{(10.5)*\dx}) node [midway,xshift=-5pt,rotate=90] {time};
\draw[->] ({(0.25-10)*\dx},{(6)*\dx}) -- ({(9-10)*\dx},{(6)*\dx}) node [midway,yshift=-5pt] {space};
\draw [decorate,decoration={brace,amplitude=5pt,mirror,raise=4ex}] (8.5*\dx,13.5*\dx) -- (8.5*\dx,20.5*\dx) node[midway,xshift=+3em]{$t$};
\foreach \j in {6.3,...,8.3}{
\foreach \i in {0.5,...,3.5}{
\roundgate[2*\i/3][2*\j/3][1/3][bottomright][blue6][-1]
\roundgate[2*\i/3+1/3][2*\j/3+1/3][1/3][bottomright][blue6][-1]}}
\foreach \i in {0.5,...,7.5}{
\draw[thick] (\i/3,3.96)--(\i/3,4.04);}		
\foreach \j in {3.7,...,5.7}
{\foreach \i in {0.5,...,3.5}
{\roundgate[2*\i/3][2*\j/3][1/3][topright][red6][-1]
\roundgate[2*\i/3+1/3][2*\j/3-1/3][1/3][topright][red6][-1]}
}	
\sSquare{-13.3}{5}{black}
\node at ({5.4*\dx},{13.3*\dx}){$a$};
\foreach \i in {1,...,4}{
\MPSinitialstate[2*\i/3-1/3][1.8][red6][topright][-1][1/3]
}
\foreach \i in {1,...,4}{
\begin{scope}[rotate around={180:(2*\i/3-1/3,6.2)}]
\MPSinitialstate[2*\i/3-1/3][6.2][blue6][topleft][-1][1/3]
\end{scope}
}
\draw [very thick,black] (0.25,6.219) -- (2.75,6.219);
\draw [very thick,black] (0.25,1.785) -- (2.75,1.785);
\end{tikzpicture}
\caption{One point function of the local operator $a$ on the time-evolving state $\ket{\Psi_t}$ and its diagrammatic representation. The shaded boxes highlight the time evolution operator $\mathbb U$ (orange), the space transfer matrix $\mathbb W$ (green), and the space transfer matrix $\mathbb W_a$ decorated with the local operator (grey). In these conventions $\mathbb W$ acts from left to right on a lattice of $2t$ sites.
}	
\label{fig:duality}
\end{figure}

In the context of non-equilibrium quantum many-body dynamics, the central asset of space-time duality is that it allows one to turn questions on \emph{non-equilibrium} properties of infinite systems into questions on \emph{equilibrium} properties of (different) finite systems. This fundamental observation was first formulated by~\cite{banuls2009matrix} --- see, however, also~\cite{muellerhermes2012tensor, hastings2015connecting, friasperez2022light} --- through the analysis of one-point functions of local operators after quantum quenches. More specifically, denoting by $a$ a one-site operator in a system evolving \emph{unitarily} from the non-equilibrium state $\ket{\Psi_0}$,~\cite{banuls2009matrix} considered
\be
\expval{a}{\Psi_t}=\expval{\mathbb U^{-t} a \mathbb U^t}{\Psi_0}. 
\label{eq:onepointfunction}
\ee
Expressing this quantity in a one-dimensional quantum circuit, i.e., writing $\mathbb U$ as in Eq.~\eqref{eq:floquetoperator} and $\ket{\Psi_0}$ as in Eq.~\eqref{eq:twositeMPS}, we can represent it diagrammatically --- an example is reported in Fig.~\ref{fig:duality}. Now comes the key insight of ~\cite{banuls2009matrix}: The one-point function in Eq.~\eqref{eq:onepointfunction} can be equivalently represented by contracting the circuit in Fig.~\ref{fig:duality} horizontally rather than vertically. Namely, one can express Eq.~\eqref{eq:onepointfunction} using a suitably defined \emph{space transfer matrix} $\mathbb W$ (green box in Fig.~\ref{fig:duality}), which acts on a doubled temporal lattice of $2t$ sites. Specifically, assuming for simplicity translational invariance of the problem we have 
\be
\expval{a}{\Psi_t}= \tr[\mathbb W^{L-1} \mathbb W_a],
\ee
where $\mathbb W_a$ is the space transfer matrix decorated with the operator $a$ (grey box in Fig.~\ref{fig:duality}). Therefore, in the thermodynamic limit $L\to\infty$ the object of interest is entirely determined by the leading eigenvalue of $\mathbb W$ --- which is one for normalised initial states~\cite{banuls2009matrix} --- and the corresponding (left and right) eigenvectors $\ket{R}, \bra{L}$~\footnote{\cite{klobas2021exact2} have shown that for unitary temporal dynamics $\ket{R}$, $\bra{L}$ are unique under mild conditions on the initial state.}. Putting everything together one has 
\be
\lim_{L\to\infty} \expval{a}{\Psi_t} =  \braket{L}{\mathbb W_a|R}.
\label{eq:evala} 
\ee
We emphasise that, generically, the local gate $\tilde U$ implementing the space evolution is not unitary. Moreover, as pointed out by~\cite{lerose2021influence}, the leading eigenvectors, or \emph{fixed points}, of the spatial evolution describe the action of an infinite environment on the subsystem where the observables act, in a way that is reminiscent of the Feynman-Vernon approach~\cite{feynman1963the}. To emphasise this physical interpretation, \cite{lerose2021influence} proposed to dub them \emph{influence matrices}. This idea was then further developed by \cite{lerose2021scaling, sonner2021influence, sonner2022characterising, lerose2023overcoming}. 
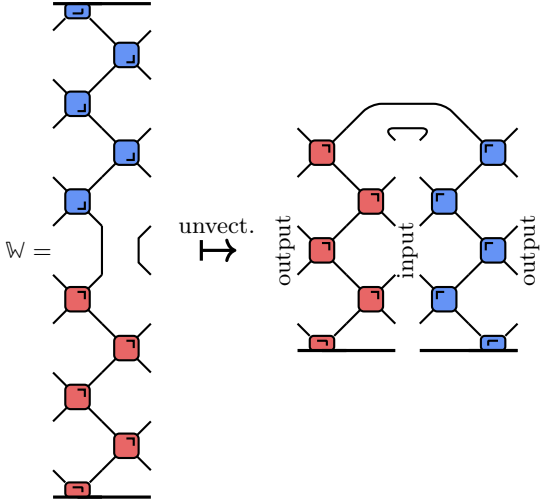
\begin{figure}
\begin{tikzpicture}[baseline={([yshift=-0.6ex]current bounding box.center)},scale=0.65]
\draw[very thick] (-1.5,-0.05) -- (0,-0.05);
\draw[very thick] (-1.5,10.05) -- (0,10.05);
\tsfmatV[-1.5][5][r][4][][][blue6][bottomright]
\tsfmatV[-1.5][0][l][4][][][red6]
\draw [thick] (-1,4.5) -- (-1,5.5);
\draw [thick] (0, 4.5) -- (-.25,4.75) -- (-.25,5.25) -- (0, 5.5);
\MPSinitialstate[-1.5][0][red6][topright][-1][1]
\begin{scope}[rotate around={180:(-1.5,10)}]
\MPSinitialstate[-1.5][10][blue6][topleft][-1][1]
\end{scope}
\draw[|->,very thick] (1,5) -- (1.75,5);
\node[rotate = 0] at (1.35,5.5){unvect.};
\tsfmatV[3.5][3][l][4][][][red6]
\tsfmatV[6][3][r][4][][][blue6][topleft]
\MPSinitialstate[3.5][3][red6][topright][-1][1]
\MPSinitialstate[7][3][blue6][topleft][-1][1]
\draw[very thick] (3.5,3-0.05) -- (5,3-0.05);
\draw[very thick] (5.5,3-0.05) -- (7.5,3-0.05);
\draw [thick, rounded corners] (1.5+3.5, 4.5+2.75) -- (1.25+3.5,4.75+2.75) -- (1.25+4.5,4.75+2.75) -- (1.5+4, 4.5+2.75);
\draw [thick, rounded corners] (4, 7.5) -- (4.5, 8) -- (6, 8) -- (6.5, 7.5);
\node[rotate = 90] at (5.25,5){input};
\node[rotate = 90] at (7.75,5){output};
\node[rotate = 90] at (2.75,5){output};
\node at (-2.5,5){$\mathbb W=$}; 
\end{tikzpicture}
\caption{Unvectorisation of the space transfer matrix. The matrix $\mathbb W$ acting on the tensor product space $(\mathbb C^{q})^{\otimes 2t} \otimes (\mathbb C^{q})^{\otimes 2t}$ can be equivalently represented as a super operator acting on ${\rm End}((\mathbb C^{q})^{\otimes 2t})$, i.e., on $q^{2t}\times q^{2t}$ matrices.}	
\label{fig:unvectorisation}
\end{figure}

In fact, \cite{klobas2021entanglement, bertini2022growth, bertini2022entanglement, bertini2023nonequilibrium, bertini2024dynamics} have shown that, besides describing the full evolution of local observables in the thermodynamic limit, the fixed points of the space evolution also control the growth of entanglement --- both pure and mixed state --- and conserved-charge fluctuations at intermediate times. More specifically, considering the `unvectorisation' operation that maps $\mathbb W$ into a quantum channel (super-operator) $\mathbb W[\cdot]$ acting on a chain of $t$ qubits, see Fig.~\ref{fig:unvectorisation}, one maps $\ket{R}, \bra{L}$ to Hermitian, non-negative operators $R$ and $L$ (the fixed points of the channel). Following~\cite{klobas2021entanglement, bertini2022growth} one can then show that the entanglement entropies between a block $A$ (of $2L_A$ qudits) and its complement $\bar A$ fulfill
\be
\lim_{L\to\infty}S^{(\alpha)}(\rho_A^\alpha(t)) = 2 S^{(\alpha)}(\tilde\rho_{{\rm stat}}),\quad  L_A\geq 2t, 
\label{eq:dualityRE}
\ee
where $S^{(\alpha)}(\rho)$ is the R\'enyi entropy defined in Eq.~\eqref{eq:Srenyidef} and we introduced the pseudo density matrix $\tilde\rho_{{\rm stat}}=R L$~\footnote{This matrix is generically not Hermitian but has a positive spectrum as, e.g., a thermal state of a non-Hermitian but PT-symmetric Hamiltonian.}. The factor 2 on the r.h.s.\ of Eq.~\eqref{eq:dualityRE} comes from the fact that one gets a contribution $S^{(\alpha)}(\tilde\rho_{{\rm stat}})$ for each edge of $A$. Similarly~\cite{bertini2022entanglement} have shown that measuring the entanglement between two blocks $A$ and $B$ in the mixed state $\rho_{AB}(t)$ through the \emph{logarithmic negativity} $\mathcal E(t)$ (see, e.g.,~\cite{vidal2002computable}) one finds 
\be
\mathcal E(t) = 2 S^{(1/2)}(\tilde\rho_{{\rm stat}}),\qquad  L_A, L_B \geq 2t\,. 
\ee
Moreover,~\cite{bertini2023nonequilibrium, bertini2024dynamics} showed that, in systems with continuous internal symmetries, $\tilde\rho_{{\rm stat}}$ can also be used to characterise the full counting statistics of the conserved charge $Q$ after a quench, i.e., 
\be
\frac{\tr[\rho_A(t)e^{- \beta Q}]}{\tr[\rho_A(0) e^{- \beta Q}]} = \tr[\tilde\rho_{{\rm stat}} e^{- \beta \tilde Q}]\tr[\tilde\rho_{{\rm stat}} e^{ \beta \tilde Q}],
\label{eq:dualitysymm}
\ee
where $\tilde Q$ is the charge of the space evolution, which, as shown by~\cite{bertini2024dynamics}, exists whenever the time evolution has a conserved charge. \cite{bertini2023nonequilibrium, bertini2024dynamics} have also shown that Eq.~\eqref{eq:dualitysymm} can be extended to describe \emph{symmetry resolved entanglement} --- see, e.g.,~\cite{goldstein2018symmetry, xavier2018equipartition, laflorencie2014spin, bonsignori2019symmetry, murciano2020entanglement} for a definition of the latter quantity. Finally, note that the r.h.s.\ of Eq.~\eqref{eq:evala} can be written as $\tr[a \tilde\rho_{{\rm stat}}]$ in terms of the pseudo density matrix. 

These equations allow one to interpret several non-equilibrium properties of the original system in terms of equilibrium properties for the `dual system' living on the temporal lattice. This significantly aids the qualitative understanding of non-equilibrium phenomena. For example, Eq.~\eqref{eq:dualityRE} gives a general explanation to the ubiquitous linear growth of entanglement observed in locally interacting quantum many-body systems, see, e.g.~\cite{calabrese2005evolution, alba2017entanglement, laeuchli2008spreading, kim2013ballistic}. From the point of view of space-time duality this phenomenon occurs because $R$ and $L$ are stationary states of the space evolution and, generically, stationary states have extensive thermodynamic entropy in 1D. 

Space-time duality, however, can provide more than just qualitative pictures: Whenever one is able to access the fixed points $R$ and $L$, Eqs.~(\ref{eq:evala}--\ref{eq:dualitysymm}) give a \emph{fully  quantitative} description of the system's dynamics in the regime $t\leq L_A/2$, i.e., when it is far from any equilibrium state. As expected, this is possible for dual-unitary circuits evolving from solvable states. Indeed, \cite{piroli_exact_2020} have shown that in that case
\be
L \propto R \propto \mathds{1}.
\label{eq:unvectorisedDUfixedpoints}
\ee
This result, which has lead to refer to dual-unitary circuits as `perfect dephasers' in the context of influence matrices, allows one to recover all the exact results discussed in Sec.~\ref{sec:statedynamics}. For the hierarchical DU2 circuits of Sec.~\ref{subsec:hierarchical} solvable states can similarly be identified for which these fixed points, while different from Eq.~\eqref{eq:unvectorisedDUfixedpoints}, still factorise~\cite{foligno_quantum_2024}. This factorisation also holds for infinite-temperature expectation values in (hierarchical) dual unitary circuits, although possibly only along specific cuts in space-time, which presents an alternative interpretation of their solvability~\cite{foligno_quantum_2024,rampp_entanglement_2023}.

Recent research has however identified families of systems with exactly solvable fixed points that, differently from those discussed above, have non-zero entanglement along the temporal lattice. Examples include interacting integrable circuits such as the (quantum) cellular automaton Rule 54 evolving from a family of compatible states~\cite{klobas2021exact, klobas2021exact2} and integrable dual-unitary circuits evolving from non-solvable states~\cite{giudice2022temporal}, but also non-integrable systems fulfilling special algebraic conditions~\cite{wang2024exact}. Finally, as shown by~\cite{thoennis2023nonequilibrium, thoenniss2023efficient, ng2023real}, fixed points can be conveniently evaluated in impurity problems.

All the exactly solved examples mentioned above involve fixed points with an area-law entangled matrix-product state structure, and a key question is whether one should expect this property to continue to (approximately) hold also when the conditions for exact solvability are violated. If so, this would make of space-time duality a very powerful numerical technique for simulating quantum many-body dynamics. A naive expectation is that the fixed points should always be simple for circuits with ergodic dynamics. Indeed, following~\cite{lerose2021influence} by adopting a Feynman-Vernon perspective, one interprets the fixed points of the space evolution as an encoding of the action that the environment exerts upon the subsystem of interest over time. It is then reasonable to expect it not to involve strong time correlations for sufficiently ergodic dynamics. The situation, however, turns out to be more complicated: \cite{foligno2023temporal} have shown that even in the case of ergodic dual-unitary circuits --- which for compatible initial states have fixed points with zero entanglement in the temporal direction (cf.~Eq.~\eqref{eq:unvectorisedDUfixedpoints}) --- a perturbation of the initial state generates a fixed point with volume law entanglement, signalling strong temporal correlations. On the other hand, some numerical results presented by \cite{giudice2022temporal} suggest that the fixed points might generically be simple (e.g.\ have logarithmically scaling entanglement) in the case of integrable dynamics. This line of research, however, is very much still ongoing at the time of writing.

We started this section by considering the application of the space-time duality approach for quantum quenches in unitary circuits to recast dynamical properties as static properties, however, the opposite perspective can also be adopted. For instance, \cite{ippoliti2021postselectionfree,ippoliti_fractal_2022,lu_spacetime_2021} used space-time duality to find information about systems with non-unitary dynamics but whose dual system is a unitary circuit. In non-unitary dynamics the interplay of unitary evolution and projective measurements can support new out-of-equilibrium entanglement phases and phase transitions --- this has been a very active research topic during the last few years and we refer the interested reader to the recent reviews by~\cite{potter2022entanglement} and \cite{fisher_random_2023}. The experimental observation of such phases is, however, plagued by the `postselection problem'. Namely, to estimate expectation values on states corresponding to a given record of measurement outcomes one has to realise a specific sequence of outcomes multiple times. The probability of this event occurring is exponentially suppressed in both width and depth of the circuit. \cite{ippoliti2021postselectionfree} proposed the use of space-time duality to avoid postselection, by observing that the space-time dual of a generic unitary circuit can be interpreted as unitary dynamics interspersed with specific types of forced projective measurements. Indeed, for a generic unitary gate $U$ its spacetime dual $\tilde{U}$ (as defined in Eq.~\eqref{eq:Utilde}) can be written as $\tilde{U} = 2 VH$, where $V$ is unitary and $H$ is a positive semidefinite matrix of unit norm which can be seen as a generalised measurement with forced measurement outcome. The purification dynamics of a mixed state under such nonunitary evolution can then be mapped to a particular correlation function in its space-time dual unitary circuit.

\begin{figure}[tb!]
\includegraphics[width=0.85\columnwidth]{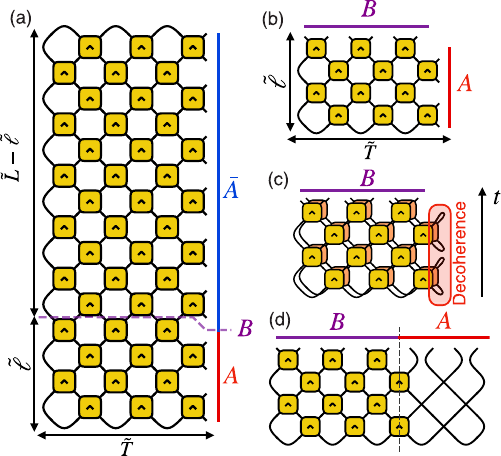}
\caption{Setup for mapping the growth of entanglement in a unitary circuit (vertical direction) to steady-state entanglement scaling in a space-time dual nonunitary circuit (horizontal direction). The depth of the unitary and non-unitary circuits are $\tilde{L}$ and $\tilde{T}$.
	(a) The input state is a product of Bell pairs (open boundary, left) and an entanglement cut on the output state (right boundary) between a subsystem $A$ and its complement $\bar{A}$ is considered. 
	(b) In the limit $\tilde{L}\to\infty$, the part of the circuit above cut $B$ can be elided when calculating the entropy.
	(c) Reduced density matrix on subsystem $B$. The tracing out of $A$ can be interpreted as the action of decoherence via a fully depolarizing channel. 
	(d) Equivalently, one can ``teleport'' subsystem $A$ from a timelike surface to a spacelike one by using ancillas initialized in Bell pair states and SWAP gates, and compute the entropy of the resulting pure state about the cut between $A$ and $B$ (dashed line). Reproduced from~\cite{ippoliti_fractal_2022}.
\label{fig:setup_ippoliti}}
\end{figure}

Building on this idea, \cite{ippoliti_fractal_2022} showed how steady-state phases of nonunitary dynamics with a rich variety of entanglement scaling with subsystem size could be realised as the space-time duals of unitary circuits with different entanglement growth in time. The basic setup is illustrated in Fig.~\ref{fig:setup_ippoliti} and was shown to result in logarithmic, extensive, or fractal entanglement scaling depending on the choice of unitary circuit. Space-time duality was here additionally used to analytically derive a universal subleading logarithmic correction to the entanglement of nonthermal volume-law steady states that was previously conjectured~\cite{li_statistical_2021,fan_self-organized_2021}. This duality mapping and the resulting different entanglement phases were subsequently experimentally realised in Google Quantum AI's quantum processors for system sizes up to 70 qubits~\cite{hoke_measurement-induced_2023}. 
A similar approach was used in \cite{lu_spacetime_2021} to show that the space-time dual of unitary circuits that display localisation-delocalisation transitions (including free fermion, Clifford, and many-body models) present nonunitary circuits which again exhibit a rich variety of entanglement phases and transitions. It was also recently observed by ~\cite{bastidas_complexification_2024} that the space-time dual of the Floquet operator for the kicked Ising model (cf.~Eq.~\eqref{eq:KIM:U_Floq}) away from the self-dual point can be interpreted as a quantum signal processing sequence with complex parameters, motivating a study of complexified quantum signal processing.

Besides these examples, space-time duality ideas have been used by~\cite{zhou_space-time_2022}, to connect the sensitivity of the dynamics to initial perturbations with non-Hermitian boundary effects in the study of OTOCs, by \cite{hamazaki_exceptional_2021}, to investigate dynamical phase transitions in the Kicked Ising model, and, as we discuss in Sec.~\ref{sec:spectral}, by \cite{bertini_exact_2018, flack_statistics_2020, braun_transition_2020, bertini2021random, garratt_local_2021, riddell2024structural, yoshimura_operator_2025, fritzsch_eigenstate_2025} to analyse the spectral statistics.

Finally, we stress that spatial-direction contractions analogous to those at the core of the space-time duality approach are standard in the theory of tensor networks, see, e.g.,~\cite{cirac2021matrix}. For instance, the availability of known left and right fixed points underlies the canonical form of generic matrix product states, where the constituting tensors can be recast as isometries in the spatial direction to obtain fixed points of the form in Eq.~\eqref{eq:unvectorisedDUfixedpoints}, see, e.g., \cite{perez-garcia_matrix_2007}. In this context, a concept that is particularly relevant to our discussion, and which was inspired by dual-unitarity, is that of dual-isometric projected entangled pair states (PEPS)~\cite{yu_dual-isometric_2024}. Isometric PEPS were originally proposed in~\cite{zaletel_isometric_2020} as two-dimensional tensor networks where the tensors satisfy an isometric condition that allows for an efficient contraction. Imposing an additional dual-isometry condition permits the calculation of observables and correlations inaccessible in standard isometric PEPS, while these states remain general enough to, e.g., represent a transition from topological to trivial order~\cite{yu_dual-isometric_2024}. Spatial unitarity was also recently exploited to show how a wide variety of tensor network states could be systematically prepared by combining Bell-pair measurements with local feedback~\cite{stephen_preparing_2024,sahay_finite-depth_2024,smith_constant-depth_2024,sahay_classifying_2025}. The spatial unitarity of Bell pairs here allows for `fusing' (entangling) spatially separated tensors, and the local feedback corrects for errors induced in the spatial direction by the probabilistic outcome of the measurements. Similarly,~\cite{mcginley_measurement-induced_2024} used transverse-direction contractions to analyse the entanglement structure generated by two-dimensional circuits of constant-depth followed by projective measurements.

\section{Summary and outlook}

In this review we have discussed several exact results concerning the dynamics of interacting many-body systems out of equilibrium, thus providing an introduction to the nascent field of `exactly solvable quantum many-body dynamics'. This is a very recent and rapidly developing field of research, which was barely existent only a decade ago and is now prospering and expanding to cover a wealth of different aspects of quantum statistical mechanics, quantum computation, and high-energy physics. Key progress was achieved in the last eight years and can be associated with two main methodological breakthroughs. First, \cite{nahum2017quantum, nahum_operator_2018,von_keyserlingk_operator_2018} have put forward quantum many-body systems in discrete time --- quantum circuits --- as a new paradigm for quantum dynamics providing enhanced analytical control. Second, \cite{bertini_exact_2018, bertini_exact_2019} have shown that one can exactly access non-trivial many-body properties, even in the presence of non-integrable interactions, by exploiting a duality between space and time. The combination of these two elements has led to the introduction --- by \cite{bertini_exact_2019,gopalakrishnan_unitary_2019} --- of \emph{dual-unitary circuits}: a class of interacting systems whose dynamics can be accessed by exchanging the roles of space and time, since such a space-time swap does not change their dynamical features. 
In contrast to random quantum circuits, dual-unitary circuits do not require randomness and ensemble-averaging for exact results and can hence realise Floquet dynamics. These systems are generically non-integrable (chaotic) but admit integrable points. 

We began by reviewing the most direct application of dual-unitary circuits, i.e., the characterisation of dynamical correlations. 
Two-point correlations between ultralocal (one-local) operators in the thermodynamic limit are exactly accessible in these circuits 
and the effect of dual-unitarity is particularly striking: dynamical correlations vanish everywhere except on the edge of the causal light cone. Different families of dual-unitary circuits can be systematically realised to display the full ergodic hierarchy~\cite{bertini_exact_2019,claeys_ergodic_2021,aravinda_dual-unitary_2021}. 
Two-point correlations between extended observables are more complicated to characterise, as they require determining the gap of a many-body transfer matrix, but have been proven to be ergodic and mixing in a class of dual-unitary circuits that is representative of the generic situation~\cite{foligno2024entanglement}. More general multipoint correlations have so far escaped any analytical description and their determination remains an open task. The same applies for all dynamical correlations in a finite volume, since there is no known way to introduce boundaries in dual-unitary circuits without destroying their solvability (at late times). 

We subsequently discussed the quantum many-body dynamics generated by dual-unitary circuits. These systems admit a class of initial states --- the solvable states proposed by~\cite{piroli_exact_2020} --- whose many-body dynamics can be characterised exactly in several aspects, while their full description remains a BQP-complete problem~\cite{suzuki_computational_2022}. Some of the observed properties, e.g.\ the maximally fast growth of entanglement when prepared in a solvable state, are characterising the full class of dual-unitary circuits, while others, like the behaviour of two-point functions, depend on the specific circuit under consideration. Operator dynamics (Heisenberg evolution) similarly offers many exactly solvable features in dual-unitary circuits, e.g.\ the butterfly velocity characterising the growth of out-of-time-order correlators~\cite{claeys_maximum_2020} or the growth of entanglement of local operators~\cite{bertini_operator_i_2020, bertini_operator_ii_2020} can be determined exactly. The situation becomes considerably more complicated when one studies hybrid quantum dynamics alternating unitary evolution and measurements. In this case exact results have so far been achieved only by imposing further strong restrictions on the circuit dynamics~\cite{claeys_exact_2022}. Analytical insight in measurement-induced phase transitions in dual-unitary circuits remains lacking.

Another important feature of dual-unitary circuits resides in their accessible (pseudo-)spectral correlations, allowing standard probes of quantum chaos to be exactly characterised. Specifically,~\cite{bertini2021random} have shown that for a representative class of dual-unitary circuits the spectral form factor can be computed exactly in the thermodynamic limit, and agrees exactly with result found in random (many-body) unitary matrices. This provides a full characterisation of two-point spectral correlations at all energy scales. \cite{fritzsch_eigenstate_2025} have then extended this treatment to the case of \emph{partial} spectral form factor, which also encodes certain statistical properties of the eigenstates. Here we note that all results on the dynamics avoided the use of eigenstates, which are generically out of reach in ergodic dual-unitary circuits.
Numerical investigations show that higher-point spectral correlations also display a distinctive random-matrix-like structure~\cite{flack_statistics_2020}, but their exact calculation remains an open problem. 

Due to their discrete-time structure, dual-unitary circuits can directly be interpreted as gate-based quantum computations. In fact, the duality of these systems allows one to use them to perform \emph{measurement-based} quantum computations by evolving unitarily in space~\cite{stephen_universal_2024}. Similarly, one can use space propagation to characterise the `projected ensemble' obtained via projective measurements on subsystems~\cite{ho_exact_2022, claeys_emergent_2022, ippoliti_solvable_2022} and Hilbert space delocalisation through the inverse participation ratio~\cite{claeys_fock-space_2024}. Both these quantities attain universal `random-matrix forms' for large systems. Furthermore, random-matrix behaviour generally appears at the fastest possible rate, consistent with the vanishing Thouless time in this models.
The fast dynamics of dual-unitary circuits can be exploited to produce many-body random states in finite circuits by adding small amounts of local randomness~\cite{suzuki2024global, riddell2025quantum}. An exact characterisation of the time-scale/space-scale for these universal forms to occur has not yet been achieved. 

We have also discussed the effect of duality-breaking perturbations to dual-unitary circuits. Since these systems are generically ergodic, and this property is maintained by generic perturbations, it is natural to expect many of their dynamical features to be robust. This suggests that one should be able to quantitatively describe the dynamical properties of generic quantum many-body systems by performing a perturbative expansion from the closest dual-unitary point. Given the expected weak dependence on the perturbation, this expansion should be extremely well-behaved (potentially convergent for some quantities), ideally providing a way to access generic non-equilibrium quantum many-body dynamics analytically~\cite{kos_correlations_2021}. This is of course in sharp contrast with the typical behaviour of perturbation series around non-interacting points, where the radius of convergence vanishes. Despite these promising premises, a general perturbative treatment of dual-unitarity breaking perturbations has not been achieved and the available results only concern specific quantities~\cite{kos_correlations_2021, kos_thermalization_2021, rampp_dual_2023, riddell2024structural}.

A related question concerns possible extensions and generalisations of dual unitarity, i.e., whether the special symmetry characterising dual-unitary circuits can be generalised, or weakened, retaining some degree of exact solvability. We discussed how the notion of dual-unitarity can be extended to open systems~\cite{kos_circuits_2023} and to the classical realm~\cite{fouxon_local_2022,christopoulos_dual_2024, lakshminarayan_solvable_2024}, can be implemented in different space-time geometries~\cite{bertini_exact_2019, jonay_triunitary_2021, suzuki_computational_2022, milbradt_ternary_2023}, and generalised to the concept of \emph{biunitarity}, which allows to treat a wealth of seemingly different systems with the same basic techniques~\cite{claeys2023dual}. We also discussed the so called `hierarchical generalisations' of dual-unitary, that provide a systematic way to weaken the dual-unitarity property~\cite{yu_hierarchical_2024}. The exact solvability, however, is lost at the second level of the hierarchy, and an open question is whether a more effective way to relax dual-unitarity and possibly obtain more general space-time symmetries can be found. 

Finally, we discussed how the idea of describing quantum many-body dynamics by exchanging the roles of space and time can be fruitful beyond the example of dual-unitary circuits. This approach allows both to find exactly solvable systems that do not pertain to the dual-unitary setting~\cite{klobas2021exact, bertini2022growth, wang2024exact} and to find general descriptions of many-body dynamics with~\cite{ippoliti2021postselectionfree} and without measurements~\cite{bertini2022growth}. 

Besides the questions discussed above, a more general open question concerns the scaling limit of dual-unitary circuits when both space and time become continuous --- which can be taken without breaking their space-time duality. The symmetry between space and time displayed by dual-unitary circuits suggests that this limiting theory should be conformally invariant. This connection is also indicated by the similarity between several dynamical behaviours in ergodic (non-ergodic) dual-unitary circuits and holographic (rational) conformal field theories~\cite{reid2021entanglement, foligno2024entanglement, carignano_loschmidt_2025}. Emergent dual-unitarity has also been predicted and confirmed in quantum quenches to a conformally invariant point~\cite{carignano_loschmidt_2025} and dual-unitary matrices can be used to implement discrete models of holography~\cite{evenbly_hyperinvariant_2017,masanes_discrete_2023, bistron_bulk-boundary_2025} and discrete conformal transformations~\cite{masanes_discrete_2023}. 
A complete understanding of this connection remains an open problem.

\section*{Acknowledgements}

The authors deeply acknowledge collaborations and fruitful discussions on the topics of the review with many colleagues, where special mention and thanks go to Pavel Kos, Austen Lamacraft, and Lorenzo Piroli.
We more generally thank and acknowledge collaborations and discussions with Gabriel Alves, Pasquale Calabrese, Soonwon Choi, Andrea De Luca,
Giuseppe De Tomasi, Alessandro Foligno, Felix Fritzsch, Juan P.\ Garrahan, Boris Gutkin, Wen Wei Ho, Curt von Keyserlingk, Katja Klobas, Roderich Moessner, Adam Nahum, Sid Parameswaran, Bal\'azs Pozsgay, Michael Rampp, Suhail Rather, Jonathon Riddell, Colin Rylands, Philippe Suchsland, Jamie Vicary, Jiangtian Yao, Tianci Zhou, and Marko \v Znidari\v c.

B.\ B.\ acknowledges financial support from the Royal Society through the University Research Fellowship No.\ 201101. P.\ W.\ C.\ acknowledges support from the Max Planck Society. T.\ P.\ acknowledges support by European Research Council (ERC) through Advanced grant QUEST (Grant Agreement No. 101096208), and Slovenian Research and Innovation agency (ARIS) through the Program P1-0402 and Grants N1-0219, N1-0368.

\bibliography{Library_DU}
\end{document}